\documentclass[12pt
]{article}

\usepackage{graphicx}
\usepackage{amsfonts}
\usepackage{amsmath}
\usepackage[usenames]{color}
\usepackage{amssymb}
\usepackage[mathscr]{eucal} 

\usepackage[all]{xy}

\usepackage{color}
\input{colordvi.tex}
 \def\kaku#1{{#1}}
 \def\kesu#1{}

\def\N{\mathbb{N}}
\def\Z{\mathbb{Z}}
\def\Q{\mathbb{Q}}
\def\R{\mathbb{R}}
\def\C{\mathbb{C}}
\def\P{\mathbb{P}}

\begin{document}
\baselineskip 0.6cm
\newcommand{\gsim}{ \mathop{}_{\textstyle \sim}^{\textstyle >} }
\newcommand{\lsim}{ \mathop{}_{\textstyle \sim}^{\textstyle <} }
\newcommand{\vev}[1]{ \left\langle {#1} \right\rangle }
\newcommand{\bra}[1]{ \langle {#1} | }
\newcommand{\ket}[1]{ | {#1} \rangle }
\newcommand{\Dsl}{\mbox{\ooalign{\hfil/\hfil\crcr$D$}}}
\newcommand{\nequiv}{\mbox{\ooalign{\hfil/\hfil\crcr$\equiv$}}}
\newcommand{\nsupset}{\mbox{\ooalign{\hfil/\hfil\crcr$\supset$}}}
\newcommand{\nni}{\mbox{\ooalign{\hfil/\hfil\crcr$\ni$}}}
\newcommand{\EV}{ {\rm eV} }
\newcommand{\KEV}{ {\rm keV} }
\newcommand{\MEV}{ {\rm MeV} }
\newcommand{\GEV}{ {\rm GeV} }
\newcommand{\TEV}{ {\rm TeV} }

\def\diag{\mathop{\rm diag}\nolimits}
\def\tr{\mathop{\rm tr}}

\def\Spin{\mathop{\rm Spin}}
\def\SO{\mathop{\rm SO}}
\def\O{\mathop{\rm O}}
\def\SU{\mathop{\rm SU}}
\def\U{\mathop{\rm U}}
\def\Sp{\mathop{\rm Sp}}
\def\SL{\mathop{\rm SL}}

\def\change#1#2{{\color{blue}#1}{\color{red} [#2]}\color{black}\hbox{}}
%\def\change#1#2{#2}

%%%%%%%%%%
%%%%%%%%%%      title page
%%%%%%%%%%

\begin{titlepage}
  
\begin{flushright}
%  LTH 829 \\
  UT-09-23 \\
  IPMU09-0125
\end{flushright}
  
\vskip 1cm
\begin{center}
  
{\large \bf  Flavor Structure in F-theory Compactifications}
  
\vskip 1.2cm
  
Hirotaka Hayashi$^1$, Teruhiko Kawano$^1$, Yoichi Tsuchiya$^1$ and 
Taizan Watari$^2$

\vskip 0.4cm
    
%  $^1$Division of Theoretical Physics, Department of Mathematical
%  Sciences, The University of Liverpool, Liverpool, L69 3BX, England,
%   U.K. \\[2mm]

{\it
  $^1$Department of Physics, University of Tokyo, Tokyo 113-0033, Japan  
  \\[2mm]
  
 $^2$Institute for the Physics and Mathematics of the Universe, University of Tokyo, Kashiwa-no-ha 5-1-5, 277-8568, Japan
  }
\vskip 1.5cm
  
\abstract{F-theory is one of frameworks in string theory where
 supersymmetric grand unification is accommodated, and all the Yukawa 
couplings and Majorana masses of right-handed neutrinos are generated. 
Yukawa couplings of charged fermions are generated at codimension-3 
singularities, and a contribution from a given singularity point 
is known to be approximately rank 1. Thus, the approximate rank of 
Yukawa matrices in low-energy effective theory of generic F-theory 
compactifications are minimum of either the number of generations  
$N_{\rm gen} = 3$ or the number of singularity points of certain types. 
If there is a geometry with only one $E_6$ type point and 
one $D_6$ type point over the entire 7-brane for SU(5) gauge 
fields, F-theory compactified on such a geometry would 
reproduce approximately rank-1 Yukawa matrices in the 
real world. We found, however, that there is no such geometry.
Thus, it is a problem how to generate hierarchical Yukawa eigenvalues 
in F-theory compactifications. A solution in the literature so far 
is to take an appropriate factorization limit. In this article, 
we propose an alternative solution to the hierarchical structure 
problem (which requires to tune some parameters) by studying how zero mode
wavefunctions depend on complex structure moduli. In this solution, 
the $N_{\rm gen} \times N_{\rm gen}$ CKM matrix is predicted to have 
only $N_{\rm gen}$ entries of order unity without an extra tuning 
of parameters, and the lepton flavor anarchy is predicted for the 
lepton mixing matrix. The hierarchy among the Yukawa eigenvalues of 
the down-type and charged lepton sector is predicted to be smaller than 
that of the up-type sector, and the Majorana masses of left-handed 
neutrinos generated through the see-saw mechanism have small hierarchy. 
All of these predictions agree with what we observe in the real world. 
We also obtained a precise description of zero mode wavefunctions near 
the $E_6$ type singularity points, where the up-type Yukawa couplings 
are generated. 
} 
  
\end{center}
\end{titlepage}

%%%%%%%%%%
%%%%%%%%%%      main part 
%%%%%%%%%%

\tableofcontents

%%%%%%%%%%%%%%%%%%%%%%%%%%%%%%%%%%%%%%%%%%%%%%%%%%%%%%%
\section{Introduction}
%%%%%%%%%%%%%%%%%%%%%%%%%%%%%%%%%%%%%%%%%%%%%%%%%%%%%%

Masses and mixing angles of fermions constitute large fraction of 
parameters of the Standard Model that includes neutrino masses. 
Such flavor parameters are free coefficients of various operators 
in a low-energy effective field theory on 3+1 dimensions. Although 
one could play a game of deriving the observed flavor structure 
from a flavor model with a symmetry and its small breaking, quantum 
field theory is such a flexible framework that we can construct many 
models for the flavor structure of the Standard Model. 

Superstring theory achieves unification of all the degrees of 
freedom including vector bosons and fermions, and has much stronger 
theoretical constraints than effective field theories. It is thus 
interesting what kind of insight string theory compactification 
could provide to understanding of flavor structure. 

In unified theories with $G'' = \SU(5)_{\rm GUT}$ gauge group, up-type 
Yukawa couplings come from the interaction of the form 
\begin{equation}
 \Delta {\cal L} \sim {\bf 10}^{ab} {\bf 10}^{cd} H({\bf 5})^e 
     \epsilon_{abcde},
\end{equation}
where $a,b,c,d,e$ are $\SU(5)_{\rm GUT}$ indices running from 1 to 5. 
An observation of \cite{TW-1} was that an underlying symmetry $G$ 
containing $G''$ not only determines 
\begin{itemize}
\item variety of representations of low-energy particles charged under $G''$ 
\end{itemize}
through the irreducible decomposition of $\mathfrak{g}$-${\bf adj.}$ 
under $G''$, but also 
\begin{itemize}
 \item interactions among them 
\end{itemize}
through the Lie algebra of $\mathfrak{g}$; 
\begin{itemize}
\item multiplicities ($\approx$ number of generations) of particles 
in a given representation,
\end{itemize}
however, are not determined from the underlying symmetry $G$ and 
its breaking pattern to $G''$. The multiplicities, often regarded 
as one of the most important clues in search for microscopic 
descriptions of elementary particles, are determined by topology 
in geometric compactifications of string theories, not purely 
from algebra (symmetry breaking). 
Thus, by ignoring the information of multiplicities and by focusing 
both on the type of representations and their interactions, one can  
determine the underlying symmetry $G$ and its breaking pattern to $G''$. 
This is a natural generalization of the determination 
of chiral $\SU(N_f)\times \SU(N_f)$ symmetry behind the physics of pions. 
In the case of $G'' = \SU(5)_{\rm GUT}$ unified theories, at least 
$E_6$ is necessary as an underlying symmetry. If one tries to 
generate all other particles in supersymmetric Standard Models and 
all the Yukawa couplings from a single underlying symmetry,\footnote{
In generic F-theory compactifications, though, this assumption does 
not have to be imposed. See \cite{Hayashi-2, Tsuchiya}, 
or discussion around eq. (\ref{eq:E8-limit}) in this article.}
$E_7$ is the minimal choice \cite{TW-1}. 
See also \cite{Bourjaily-1, Harvard-E8} for recent
articles. 

Although underlying symmetries can be inferred from low-energy data, 
other inputs, either experimental or theoretical, are necessary to 
find out how the underlying symmetries are implemented (like QCD!); 
without a firm theoretical implementation, underlying symmetry alone 
can not do very much.
In super string theory, we can spot three frameworks where $E_6$ 
and other exceptional type symmetries can be implemented. They are 
Heterotic $E_8 \times E_8$ string theory, M-theory compactified 
on $G_2$ holonomy manifolds and F-theory. Moduli space of these 
frameworks partially overlap with one another, but not entirely. 
Even in overlapping region of moduli space, analysis of low-energy 
physics may be easier in one framework than in the other. 
In F-theory compactifications (to be more precise, in region of 
moduli space where there is no Heterotic or M-theory dual, or 
where F-theory provides an easier way of analysis) charged matter 
fields are known to have wavefunctions localized in internal space, 
and flavor structure can be studied in a relatively intuitive way. 
This is why we study flavor structure in F-theory compactifications. 

Study of flavor structure begins with understanding how 
to read out the low-energy degrees of freedom and their 
properties (wavefunctions in the internal space) from 
geometric data for F-theory compactifications. There has 
been a considerable progress in this direction in the 
last two years \cite{DW-1, BHV-1, Hayashi-1, Hayashi-2, 
DW-3, Tsuchiya}. An F-theory compactification with ${\cal N} = 1$ 
supersymmetry in 3+1 dimensions is described by a Calabi--Yau
4-fold $X$ that is an elliptic fibration over a complex 3-fold 
$B_3$. It has been known since 90's that gauge fields are 
localized on complex surfaces in $B_3$, and charged matter 
fields are localized on complex curves in the surfaces, 
but now we further know that all the massless modes of 
charged matter fields\footnote{By ``all'' the charged matter fields, we 
only mean those in ${\bf 10}$, $\overline{\bf 10}$, $\bar{\bf 5}$ and 
${\bf 5}$ representation of $G''=\SU(5)_{\rm GUT}$,  
${\bf 16}$, $\overline{\bf 16}$ and ${\bf vect.}$ representations of 
$G'' = \SO(10)$, and ${\bf 27}$ and $\overline{\bf 27}$ representations 
of $G'' = E_6$. Those fields (and ${\bf adj.}$ of $G''$) are the minimal 
matter contents. When $G'' = \SU(5)_{\rm GUT}$, matter fields in 
${\rm Sym}^2 {\bf 5}$ appear when the $\SU(5)_{\rm GUT}$ irreducible piece 
of discriminant locus develops an extra singularity (see
e.g. \cite{Sadov}). }
 correspond to {\it smooth} wavefunctions 
on the complex curves; although there are multiple points of 
enhanced singularity types along the complex curves, none of the charged 
matter fields are specifically associated with the singularity 
points \cite{Hayashi-1}.\footnote{
This is a pedagogical expression of a statement that 
all the charged matter fields are global holomorphic sections of some 
line bundles, not of a sheaf with torsion components associated with 
singularity points. It was far from obvious whether this statement was 
true; see e.g. \cite{DI} or \cite{DW-1}.}\raisebox{5pt}{,}\footnote{This result has an immediate application to
phenomenology. The absence of torsion components was a crucial input 
in making predictions on the branching fraction of dimension-6 proton 
decay (\cite{Wijnholt-review} and version 3 of \cite{TW-2}).}  
It is also known how to determine the smooth wavefunctions 
on the complex curves \cite{Hayashi-1}; we will review the technique 
with some explicit examples in section \ref{sec:hol-wavefcn}, and
elaborate more on the technique in this article. 

Charged matter fields are regarded as M2-branes wrapped on vanishingly 
small 2-cycles along the complex curves (called matter curves), 
and Yukawa couplings are re-wrapping process of M2-branes that preserve 
the sum of topological 2-cycles. Examining the algebra of topological 
2-cycles, it turns out that the up-type Yukawa couplings and 
down-type Yukawa couplings are generated \cite{TW-1}, presumably\footnote{
Discussion in section \ref{ssec:F-Yukawa} will make it clear why we keep 
the word ``presumably'' here.}
at around singularity points in $B_3$ where the singularity 
is enhanced to $E_6$ and $D_6$, respectively \cite{BHV-1, DW-1, Hayashi-1}. 
Since F-theory does not have microscopic quantum formulation,  
it is hard to imagine how Yukawa couplings associated with 
singularity points can be analyzed quantitatively without microscopic 
theory describing geometry around the singularity points. 
References \cite{KV, DW-1, BHV-1}, however, developed an effective 
description of gauge-theory sector using field theory on 7+1 dimensions, 
and proposed to use it to study Yukawa couplings. Although microscopic 
formulation of F-theory is not known yet, the idea is that such 
unknown effects can be incorporated as higher-dimensional operators 
in the field theory Lagrangian, with unknown coefficients; as long as 
we are interested in far infra-red physics such as Yukawa couplings 
below the Kaluza--Klein scale, leading order terms will be the most 
important.  Despite unknown coefficients of the higher-order operators, 
such effects must be power suppressed, and influence on the low-energy 
observables (such as masses and mixing angles) should be estimated (by order 
of magnitude).  
In this program, geometry of $X$ needs to be translated properly 
into the background field configuration of the effective field theory 
on 7+1 dimensions. 
Physics and mathematics involved in this translation was (almost) 
clarified in \cite{Hayashi-2, DW-3}  
and references therein\footnote{
The process of translation is not trivial. For example, the linear 
configuration of $\varphi$ with $\Z_2$ quotient, adopted e.g. 
in \cite{HV-Nov08-rev} turns out not to be correct (see appendix 
of \cite{Tsuchiya} for more information). } We will add 
a little more material to this translation process 
in the appendix \ref{sec:Hitchin} in this article. 

Now we know how to calculate smooth wavefunctions of massless 
modes, how to calculate Yukawa couplings from individual regions 
around singularity points by using field theory models for the local
geometry, and how to combine these two techniques \cite{Hayashi-2}.  
It turns out that contribution to the up-type Yukawa matrix 
from a region around an $E_6$-type singularity point is approximately 
rank-1 \cite{Hayashi-2}, and that to the down-type Yukawa 
matrix from a region around a $D_6$-type singularity point is 
also approximately rank-1 (e.g., \cite{BHV-1}). 

Such studies, however, do not make a clear distinction between 
$H_d$ (down-type Higgs multiplet) and $L$ (left-handed lepton doublets). 
Any sensible theories of flavor structure in supersymmetric 
compactifications should have a framework where 
dimension-4 proton decay operators are brought safely under control.
Since the vanishing dimension-4 proton decay operators 
is about vacuum value of moduli, and since right-handed neutrinos 
are identified with fluctuations of moduli fields from vacuum, 
the dimension-4 proton decay problem and how to generate 
Majorana masses to right-handed neutrinos are deeply related 
issues. Three frameworks were proposed in \cite{Tsuchiya} 
(two of which were essentially carried over and generalized from \cite{TW-1}):
\begin{itemize}
\item matter parity: perhaps the most natural scenario to many people. 
   Geometry of $X$ and 4-form flux background $G^{(4)}$ should be arranged 
   so that a $\Z_2$ symmetry remains unbroken. Various matter fields  
   at low-energy should come out with the right assignment of the parity
   ($\Z_2$ transformation law). Majorana masses of right-handed
      neutrinos are generated automatically in flux compactifications. 
   Flux compactification of F-theory predicts the scale of the Majorana 
      mass somewhat below the GUT scale, 
   which is phenomenologically favorable. (section 4.1 of \cite{Tsuchiya})
\item factorized spectral surface with an unbroken U(1) symmetry:
    Complex structure of $X$ needs to be tuned so that the spectral 
    surface factorizes \cite{TW-1}. Ultimately the factorization has 
    to be specified in a global geometry, because the factorization only 
    at quadratic order may not be sufficient to get rid of the proton decay 
    operators (c.f. section 4.3.3 of \cite{Tsuchiya}). 
   Reference \cite{Caltech-0906} (implicitly) proposed to tune 
      complex structure parameters further so that an $E_8$ Higgs 
      bundle is defined globally on a complex surface $S$ for the 
      unified theories, so that the global factorization of spectral 
      surface can be discussed within the canonical bundle $K_S$.
     In the presence of an unbroken U(1) symmetry in the factorization limit, 
    the dimension-4 proton decay operators are forbidden, but at the same time, 
    the Majorana mass of right-handed neutrinos are also prohibited 
    at perturbative level \cite{Caltech-0906}, when the $\SU(5)_{\rm GUT}$ 
    symmetry is broken either by a Wilson line \cite{BHV-2} or by a flux 
      in the hypercharge \cite{BHV-2, TW-2, DW-2}.\footnote{
      Whether a vector boson associated with this 
   U(1) symmetry remains massless or not depends on global aspects of 
      compactification geometry (including fluxes). If it remains 
      massless down to low-energy, there are other phenomenological problems. 
   i) the U(1) vector boson has to be Higgsed at some scale above the
      experimental bound, ii) one needs an idea what sets the scale of
      the ``Higgs mechanism'' of the U(1) vector field, iii) 1-loop
      mixing between the extra U(1) and $\U(1)_Y$ of the Standard Model 
     change the running of gauge coupling constants and ruin the
      prediction of gauge coupling unification, if the extra U(1) vector
      field remains massless down to low-energy, and finally, iv) it is 
    Higgsed at high-energy, then the U(1) symmetry is not effectively
     solving the dimension-4 proton decay problem.} 
\item spontaneous $R$-parity violation: factorized spectral surface with 
  non-zero Fayet--Iliopoulos parameter. A safe way to break this 
  unwanted U(1) symmetry while keeping the proton decay operators 
  from being generated is to trigger spontaneous 
      breaking of the U(1) symmetry (and Higgsing of the vector field) 
      by non-zero Fayet--Iliopoulos parameter \cite{TW-1, KNW}. 
      Majorana mass of right-handed neutrinos are also generated. 
 (section 4.4 of \cite{Tsuchiya}) 
\end{itemize}

Now all the necessary theoretical tools and frameworks 
are available; time is ripe, and we are ready to study 
flavor structure in F-theory compactifications.\footnote{
This article does not study the flavor structure in the 
third framework, however, because we do not have 
enough theoretical tools yet.} 
Our goal is to clarify which parameters can be tuned\footnote{
We tune them by hand, for now; hoping someday that the tuning 
is justified by flux compactification or phenomenological/cosmological 
considerations.} to reproduce known flavor structure, and 
which aspects of flavor structure are derived theoretically. 

This article is organized as follows. Yukawa matrices of low-energy 
effective theory are given by summing up contributions from all 
the codimension-3 singularity points. An idea of \cite{HV-Nov08-rev} 
is to assume that there is only one codimension-3 singularity point 
of a given type to reproduce hierarchical structure among Yukawa 
eigenvalues. But this assumption is not satisfied generically, and 
furthermore, we found in section \ref{sec:inv} that this cannot be satisfied. 

Thus, we have to take account of all the contributions from 
different codimension-3 singularities, and we need to be able 
to evaluate which contribution is more important relatively to 
others. For this purpose, we take a moment in section
\ref{sec:hol-wavefcn} to present a technique of how to calculate 
wavefunctions of zero modes along the matter curves. 

The zero mode ``wavefunctions'' are not ``functions'', but are 
sections of some bundles. Thus, an extra care has to be taken 
in describing the ``wavefunctions''.
% not just in a local trivialization patch. 
In section \ref{sec:DandF}, a clear distinction is introduced 
among descriptions of the sections in holomorphic frame, unitary frame and 
diagonalization frame. In order to determine physical Yukawa couplings
in the low-energy effective theory, both the kinetic terms and 
tri-linear (Yukawa) couplings need to be expressed in terms of  
wavefunctions in a certain frame; clear distinction among different 
frames is crucial in writing down the expression for the 
kinetic terms. This conceptual clarification achieved in section 
\ref{sec:DandF} enables us to take on a problem of providing a better 
field-theory description of ``branch locus'' that appears in local 
geometry with enhanced $E_7$, $E_6$ and $A_6$ singularity. 
Progress beyond \cite{Hayashi-1, Hayashi-2} is presented 
in the appendix \ref{sec:Hitchin}.

We will study in section \ref{sec:unit-wavefcn} how 
the ``wavefunctions'' in unitary frame change when some 
complex structure parameters are changed. It is only 
a case study of simple examples, is not meant to be 
a thorough or extensive one. But, at least, the study shows 
that zero mode wavefunctions get localized within the matter 
curves for certain choices of complex structure parameters. 

Yukawa matrices of low-energy effective theory of F-theory 
compactifications are generically predicted not to have 
hierarchical structure. That is a problem. 
We show in section \ref{sec:flavor} that this problem can 
be solved by the localized wavefunctions found in 
section \ref{sec:unit-wavefcn}. Small mixing structure 
of the CKM matrix partially follows as a consequence of 
this localized wavefunctions, without an extra tuning 
of moduli parameters. 

Section \ref{sec:discuss} is devoted to summary and discussion. 
Two solutions to the hierarchical structure problem of Yukawa 
eigenvalues are given in this article: one is not written elsewhere 
in this article and the other is the one in section \ref{sec:flavor}. 
Their summary and comparison are given in this section. We also 
briefly comment on another solution to the hierarchical structure 
problem that is already mentioned in \cite{Caltech-0904, Caltech-0906}.

The appendix \ref{sec:Het-F} is a side remark. The moduli map 
between the Heterotic $E_8 \times E_8$ string theory and F-theory 
is often written down in a form that includes only the visible 
(or hidden) sector bundle moduli, but not the other. We wrote down 
a map that includes both, and show that the ``stringy corrections'' 
to the locus of $A_6$-type codimension-3 singularity points vanish 
for Heterotic--F dual models at any points in the moduli space.

{\bf Reading Guide}
Sections \ref{sec:flavor}, \ref{sec:discuss} and this Introduction 
are written in as plain language as possible. We believe that 
they are accessible to non-experts including phenomenologists. 
Section \ref{ssec:flavor-review}, on the other hand, is intended to 
be a review of basic things in flavor structure for string theorists. 

Section \ref{sec:DandF} provides conceptual clarification, while 
sections \ref{sec:hol-wavefcn} and \ref{sec:unit-wavefcn} solve 
technical problems, and overall prepare for section \ref{sec:flavor}.
Section \ref{sec:flavor} itself, however, can be understood in 
an intuitive way, and it is an option to skip these technical 
sections \ref{sec:hol-wavefcn}--\ref{sec:unit-wavefcn}. 

Basic concepts and ideas covered in section \ref{sec:hol-wavefcn} have 
already appeared in \cite{Hayashi-1, BHV-2, DW-3, Tsuchiya,
Caltech-0906, Blumenhagen-0908}. 
Section \ref{sec:hol-wavefcn} elaborates a little more on them 
by using explicit examples, and tries to make them more
accessible. 
Geometry associated with matter parity is discussed 
in section \ref{sssec:matter-parity} in much more detail than 
in \cite{Tsuchiya}.
Thus, we hope that some of the contents in 
section \ref{sec:hol-wavefcn} will be useful from the perspective 
of ``geometric engineering'' of the Standard Model,
although the engineering of the Standard Model is not the primary 
subject of this article.

The appendix \ref{sec:Hitchin} is a technical note for 
section \ref{sec:DandF}, but it will be interesting on its own right, 
especially for those who are interested in Hitchin equation.

{\bf Note}: Shortly before this article was completed, two articles 
\cite{Harvard-CS, Conlon} appeared. There may be some overlap in 
the subjects.

%%%%%%%%%%%%%%%%%%%%%%%%%%%%%%%%%%%%%%%%%%%%%%%%%%%%%%%
\section{Topological Invariants of Matter Curves}
\label{sec:inv}
%%%%%%%%%%%%%%%%%%%%%%%%%%%%%%%%%%%%%%%%%%%%%%%%%%%%%%%

In an F-theory compactification with the grand unification group 
$\SU(5)_{\rm GUT}$, the up-type Yukawa matrix gets 
contributions from all the $E_6$-type singularity points with 
approximately rank-1 \cite{Hayashi-2} from  each of the singularities.
On the other hand, all the $D_6$-type singularity points contribute to 
the down-type Yukawa matrix with approximately rank-1 \cite{BHV-2} from 
each of them. It suggests that in the F-theory compactification 
with the number of generations $N_{\rm gen}$, 
the up-type and down-type Yukawa matrices 
are of approximately rank  ${\rm min}(N_{\rm gen}, \# E_6)$ and 
${\rm min}(N_{\rm gen}, \# D_6)$, respectively, 
where $\# E_6$ and $\# D_6$ are the number of $E_6$-type and 
$D_6$-type singularity points, respectively, in the GUT divisor $S$. 

The number of points of each type singularity is a topological invariant,  
and cannot be tuned by hand. 
The number of $E_6$-type points is not generically 
one \cite{Hayashi-1}, but the up-type Yukawa matrix 
in the real world is known to be approximately rank-one, 
in that the top-quark Yukawa eigenvalue is of order one, 
and all others are much smaller. 
A proposal of \cite{HV-Nov08-rev} is that there must be 
an F-theory compactification where $\# E_6$ and 
$\# D_6$ are both one in $S$. In this section, however, we will raise 
a question whether such a geometry with such a topology ever exists.

We begin, however, with a brief review on matter curves in an F-theory 
compactification with the grand unification groups $SU(5)_{\rm GUT}$ and 
$SO(10)$, while setting up notations. 
Those familiar with the contents in \cite{6authors, Sadov, GM, TW-1, 
Hayashi-1, DW-3} can proceed to p.\pageref{pg:skip-convention}. 

An F-theory compactification to 3+1 dimensions is described by 
specifying a Calabi--Yau 4-fold $X$ that is an elliptic fibration 
on a complex 3-fold $B_3$:
\begin{equation}
 \pi_X: X \rightarrow B_3.
\end{equation}
Let the elliptic fibration be given by a Weierstrass model
\begin{equation}
 y^2 = x^3 + f x + g,
\end{equation}
where $f$ and $g$ are holomorphic sections of ${\cal O}(-4 K_{B_3})$ 
and ${\cal O}(-6 K_{B_3})$, respectively, when an unbroken ${\cal N} = 1$ 
supersymmetry is required below the Kaluza--Klein scale. 
To obtain an $\SU(5)_{\rm GUT}$ and $\SO(10)$ unification 
in the F-theory compactification, the set of the zero loci of the discriminant 
\begin{equation}
 \Delta = 4 f^3 + 27 g^2 \in \Gamma(B_3; {\cal O}(-12 K_{B_3}))
\label{eq:discr-in--12K}
\end{equation}
needs to contain an irreducible component $S$ with 
multiplicity 5 and 7, respectively, and the singularity of $X$ 
in the transverse direction of $S$ is $A_4$ and $D_5$, respectively,  
on a generic point of $S$. We call this $S$ in $B_3$ as the 
GUT divisor.

Charged matter chiral multiplets are localized on matter 
curves. For $G'' = \SU(5)_{\rm GUT}$ unified theory models, 
a 6D hypermultiplet in the representation ${\bf 10}+\overline{\bf 10}$ 
is localized on a curve $\bar{c}_{({\bf 10})}$ in $S$ and a 
6D hypermultiplet in the representation $\bar{\bf 5}+{\bf 5}$ on a curve 
$\bar{c}_{(\bar{\bf 5})}$ in $S$. 
For $G'' = \SO(10)$ unified theory models, a 6D hypermultiplet in 
the {spin and its conjugate representation is localized on a curve 
$\bar{c}_{({\bf spin})}$, 
and one in the vector representation 
on a curve $\bar{c}_{({\bf vect})}$ in $S$. 
It might appear at first sight that these curves may be 
in any topological classes of $S$, by arranging their ``7-brane''
configurations in $B_3$. In fact, there are not much freedom.
One can show in a generic F-theory compactification 
% (of course including those without a Heterotic dual) 
that 
once a normal bundle $N_{S|B_3}$ on $S$ is assumed, 
then no other freedom is left; this follows by requiring both 
box anomaly cancellation in any compact 2-cycles $\times \R^{3,1}$ 
and the topological condition of ``all the 7-branes'' 
(\ref{eq:discr-in--12K}) \cite{Sadov, GM, TW-1, DW-3}.\footnote{
In Calabi--Yau orientifold compactifications of Type IIB string theory 
with only D-branes and orientifold planes, the box anomaly cancellation
and the topological condition (\ref{eq:discr-in--12K}) both correspond
to a single condition; the Ramond--Ramond charge cancellation 
(Bianchi identity) associated with the Ramond--Ramond 0-form field. 
In F-theory in general, however, these two conditions are {\it not} 
the same. }
Conventionally a divisor $\eta$ on $S$ is defined 
\cite{Rajesh}\footnote{To our knowledge, 
this was the first reference to try to define $\eta$ only from 
geometric data around $S$, so that the definition does not depend 
on the global topological aspects of $B_3$. } from the normal bundle by 
\begin{equation}
 c_1(N_{S|B_3}) = 6K_S + \eta.
\end{equation}
Using this divisor $\eta$, topological classes of the matter curves 
are given by 
\begin{eqnarray}
 G'' = \SU(5)_{\rm GUT} & & \bar{c}_{({\bf 10})} \in |5K_S + \eta |,
  \qquad  \bar{c}_{(\bar{\bf 5})} \in |10 K_S + 3\eta |, 
  \label{eq:SU(5)-curve}\\
 G'' = \SO(10), & & \bar{c}_{({\bf spin})} \in |4K_S + \eta |, \qquad 
 \bar{c}_{({\bf vect})} \in |3K_S + \eta|.
  \label{eq:SO(10)-curve}
\end{eqnarray}

An easier way to see this is to take a local patch 
of $S$, so that there is a normal coordinate $z$ of $S$ in $B_3$. 
In order to obtain a split $A_4$ singularity along $S$, the defining 
equation of $X$ should locally look like \cite{6authors}
\begin{eqnarray}
 y^2 & = & x^3 + \left( a_5 + z a'_5 + \cdots \right) x y 
           + \left( a_4 z + a'_4 z^2 + \cdots \right) x^2 
           + \left( a_3 z^2 + a'_3 z^3 + \cdots \right) y
           \label{eq:def-eq} \\ 
     & & \qquad \quad 
           + \left( a_2 z^3 + a'_2 z^4 + \cdots \right) x 
           + \left( a_0 z^5 + a'_0 z^6 + \cdots \right).
    \nonumber 
\end{eqnarray}
In order to obtain a $D_5$ singularity, $a_5$ should also vanish. 
Because $x$ and $y$ are sections of ${\cal O}(-2 K_{B_3})$ and 
${\cal O}(-3 K_{B_3})$, respectively, to be glued together over $B_3$, 
they behave near $S$ as sections of 
\begin{equation}
 {\cal O}_S(-2 K_{B_3}|_S) = 
 {\cal O}_S(-2 K_S) \otimes N_{S|B_3}^{\otimes 2} = 
 {\cal O}_S(10 K_S + 2\eta), \quad {\rm and} \quad 
 {\cal O}_S(-3 K_{B_3}|_S) = {\cal O}_S(15K_S + 3\eta), 
\end{equation}
respectively. Thus, all the terms in (\ref{eq:def-eq}) should also be 
sections of ${\cal O}_S(6(5K_S + \eta))$. 
Since the normal coordinate $z$ is a section of $N_{S|B_3}$, one 
can see that 
\begin{equation}
 a_r \in \Gamma (S; {\cal O}_S(r K_S + \eta)), \qquad 
 a'_r \in \Gamma (S; {\cal O}_S ((r-6) K_S)), \qquad \cdots,
\end{equation}
for $r = 5,4,3,2,0$. 
Once the normal bundle $N_{S|B_3}$ (and hence the divisor $\eta$) 
is specified, all the line bundles for $a_r$'s, $a'_r$'s, and so on, 
are determined. 

% It is straightforward to cast the equation (\ref{eq:def-eq}) 
% into the Weierstrass form, and the result is found in \cite{Hayashi-1}.
The discriminant becomes\footnote{technical note: 
Let us explain how $\tilde{R}^{(5)}_{\rm mdfd}$ is related to 
$R^{(5)}$ in \cite{Hayashi-1}, $R^{(5)}_{\rm mdfd}$ in version 4 of 
\cite{Hayashi-1} and $\tilde{R}^{(5)}$ in \cite{Tsuchiya}.
First, 
\begin{equation}
 R^{(5)} = \tilde{R}^{(5)}|_{a'_{3,4,5} = 0}, \qquad 
 R^{(5)}_{\rm mdfd} = \tilde{R}^{(5)}_{\rm mdfd}|_{a'_{3,4,5} = 0}.
\end{equation}
$a'_{3,4,5}$ vanish in an F-theory compactification that has 
a Heterotic dual, and hence $\tilde{R}^{(5)}$ and 
$\tilde{R}^{(5)}_{\rm mdfd}$ become $R^{(5)}$ and $R^{(5)}_{\rm mdfd}$, 
respectively. See also the appendix \ref{sec:Het-F} for more details.
$R^{(5)}_{\rm mdfd}$ and $\tilde{R}^{(5)}_{\rm mdfd}$ are modified 
from $R^{(5)}$ and $\tilde{R}^{(5)}$, respectively, by 
\begin{equation}
R^{(5)}_{\rm mdfd} = a_5 R^{(5)} - (4a_4/a_5) P^{(5)}, \qquad 
\tilde{R}^{(5)}_{\rm mdfd} = a_5 \tilde{R}^{(5)} - (4a_4/a_5) P^{(5)}. 
\end{equation}
Since it is the location of zero points of $R^{(5)}$ or
$\tilde{R}^{(5)}$ on the curve $P^{(5)} = 0$ that is directly relevant 
to low-energy physics, the modification by $P^{(5)}$ above does not 
make a practical difference. As the ``branch locus'' of the spectral
{\it surface} $C_{(\bar{\bf 5})}$, however, $R^{(5)}_{\rm mdfd}$ and 
$\tilde{R}^{(5)}_{\rm mdfd}$ are the correct expression. When a limit 
(\ref{eq:E8-limit}) is taken, a result in \cite{DW-3} is obtained 
from $\tilde{R}^{(5)}_{\rm mdfd}$.} 
\begin{equation}
 \Delta  \propto  z^5 \left( \frac{1}{16}a_5^4 P^{(5)} 
   + \frac{z}{16} a_5^2 ((8 a_4 + 6 a_5 a'_5) P^{(5)} - a_5 \tilde{R}^{(5)}_{\rm mdfd})
   + {\cal O}(z^2) \right). 
\end{equation}
\begin{eqnarray}
 P^{(5)} & = & a_0 a_5^2 - a_2 a_5 a_3 + a_4 a_3^2, \\
 \tilde{R}^{(5)}_{\rm mdfd} & = & a_5 (a_2^2 - 4 a_4 a_0) 
 + \left(a_3^3 + a'_2 a_3 a_5^2 - a'_0 a_5^3\right) \nonumber \\
 & &  + (a_2 a_5 - 2 a_4 a_3) (a_5 a'_3 - a_3 a'_5) 
      - a_5 a_3^2 a'_4. \label{eq:R5}
\end{eqnarray}
The matter curves $\bar{c}_{({\bf 10})}$ and $\bar{c}_{(\bar{\bf 5})}$ 
correspond to $a_5 = 0$ and $P^{(5)} = 0$, respectively, and they are in the 
topological classes specified in (\ref{eq:SU(5)-curve}), because 
$a_5$ and $P^{(5)}$ are sections of the line bundles specified 
by the divisors in (\ref{eq:SU(5)-curve}). In $\SO(10)$ 
unified theory models, the discriminant indicates 
two matter curves, $a_3 = 0$ and $a_4 = 0$, on the divisor $S$. 
Similarly, one can see that the two curves are in the topological classes 
in (\ref{eq:SO(10)-curve}), because $a_3$ and $a_4$ are sections 
of the line bundles ${\cal O}(3 K_S + \eta)$ and ${\cal O}(4K_S + \eta)$, 
respectively. 

An important point is that all the topological classes of matter 
\label{pg:skip-convention}
curves have already been determined, when the complex surface $S$ 
of generic singularity type $G''$ and the first Chern class of the 
normal bundle $N_{S|B_3}$ are specified. 
Alternatively, one can specify a topological class of 
$\bar{c}_{({\bf 10})}$ for $G'' = A_4$ (or $\bar{c}_{({\bf spin})}$ 
for $G'' = D_5$), instead of the first Chern class of $N_{S|B_3}$.
In Table~\ref{tab:top-data}, we showed several examples of topological 
choice of the GUT divisor $S$ of $A_4$ singularity and the matter 
curve $\bar{c}_{({\bf 10})}$. All the topological invariants 
in the 3rd--7th column are determined, once the topological classes 
in the first and second columns are specified. 
%all the topological invariants from the 3rd to 7th column are derived. 
%once we assume topology of $S$ (the first column) and a topological 
%class of $\bar{c}_{({\bf 10})}$ (the second column). 
%%%%%%%%%%%%%%%%%%%%%%%%%%%%%%%%%%%%%%%%%
\begin{table}[tb]
\begin{center}
\caption{\label{tab:top-data}
Topological data of matter curves and codimension-3 singularity
 points in several F-theory compactifications with an unbroken 
$\SU(5)$ gauge theory on a complex surface $S$. 
In the example X, we present only the large $d$ behavior 
of the topological data.
A generic complex structure is assumed for all the examples below. 
None of those examples satisfy $\# E_6 = \# D_6 = 1$.} 
 \begin{tabular}{c||c|c||c|c|c|c|c|c}
  & $S$ & $\bar{c}_{({\bf 10})} \sim $ & 
    $g(\bar{c}_{({\bf 10})})$ & \# $E_6$ & \# $D_6$ & 
    $g(\tilde{\bar{c}}_{(\bar{\bf 5})})$ & \# $A_6$ & Ref. \\
\hline
I & $F_1$ & $f$ & 0 & 2 & 4 & 104 & 298 & \cite{Hayashi-1} \\
II & $F_1$ & $b$ & 0 & 0 & 1 & 89 & 262 & \cite{Hayashi-1} \\
III & $F_1$ & $f + b$ & 0 & 4 & 7 & 119 & 334 & \cite{Hayashi-1} \\
IV & $F_0$ & $f+b$ & 0 & 6 & 10 & 134 & 370 & \cite{DW-3} \\
V & $dP_2$ & $2H-E_1 - E_2$ & 0 & 6 & 10 & 124 & 340 & \cite{Caltech-0904} \\
VI & $\P^2$ & $H$ & 0 & 4 & 7 & 129 & 364 & \\
VII & $\P^2$ & $2H$ & 0 & 10 & 16 & 174 & 472 & \\
VIII & $\P^2$ & $3H$ & 1 & 18 & 27 & 226 & 594 & \\
IX & $\P^2$ & $4H$ & 3 & 28 & 40 & 285 & 730 & \\
\hline
X & $\P^2$ & $d H$ & $\sim d^2/2$ & $\sim d^2$ & $\sim d^2$ & 
  $\sim 7d^2/2$ & $\sim 7d^2$ 
 \end{tabular}
\end{center}
\end{table}
%%%%%%%%%%%%%%%%%%%%%%%%%%%%%%%%%%%%%%%%%%%%%%%%%%%%%%%%%%%%%%
The 4th and 5th columns give the numbers of 
codimension-3 singularity points on $S$ where the $A_4$ 
singularity is enhanced to $E_6$ and $D_6$, respectively. 
An $E_6$-type singularity point is given by a common zero of the pair 
($a_5$, $a_4$), while a $D_6$-type singularity point is given by a common 
zero of the pair ($a_5$, $a_3$), and their numbers are thus given by 
\begin{equation}
 \# E_6 = (5K_S + \eta) \cdot (4K_S + \eta), \qquad 
 \# D_6 = (5K_S + \eta) \cdot (3K_S + \eta).
\end{equation}
The genus of the matter curve $\bar{c}_{({\bf 10})}$ for 
the representation ${\bf 10}+\overline{\bf 10}$ and that of 
the covering matter curve $\tilde{\bar{c}}_{(\bar{\bf 5})}$ for the 
representation $\bar{\bf 5}+{\bf 5}$ are given by 
\begin{eqnarray}
  2g(\bar{c}_{({\bf 10})}) - 2 &  = & 
    (5K_S + \eta) \cdot (6K_S + \eta), \\
  2g(\tilde{\bar{c}}_{(\bar{\bf 5})}) - 2 & = & 
    (10K_S + 3\eta) \cdot (11K_S + 3 \eta) - 2 \# D_6, \\
 & = & 
  (10K_S + 3\eta) \cdot (11K_S + 3 \eta) - 2 (5K_S + \eta) \cdot (3K_S + \eta).
\label{eq:genus-5curve-BUP}
\end{eqnarray}
Since an $A_6$-type singularity point is a common zero of 
$\tilde{R}^{(5)}_{\rm mdfd}$ and $P^{(5)}$ and is not 
a $D_6$-type singularity \cite{Hayashi-1} (and its v4), 
\begin{eqnarray}
 \# A_6 & = & (4K_S + 2 \eta) \cdot (10 K_S + 3 \eta) + \# E_6, \\
        & = & (9K_S + 3 \eta) \cdot (10 K_S + 3 \eta) - 2 \# D_6.
\end{eqnarray}
%
%%%%%%%%%%%%%%%%%%%%%%%%%%%%%%%%%%%%%%%
\begin{figure}[tb]
 \begin{center}
   \includegraphics[width=.4\linewidth]{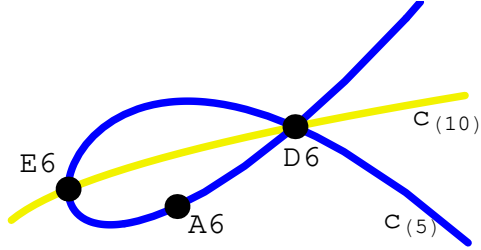}
\caption{\label{fig:schem-curve} (color online) 
A schematic figure shows various kinds of singularity enhancement 
on the GUT divisor $S$ of an $A_4$ singularity. Singularity is enhanced 
to $D_5$ on the matter curve $\bar{c}_{({\bf 10})}$ (yellow/light gray), 
and to $A_5$ on the matter curve $\bar{c}_{(\bar{\bf 5})}$ (blue/dark).
Singularity is enhanced to $E_6$, $D_6$ and $A_6$ at points on these 
matter curves. More realistic figures are found 
in Figure~\ref{fig:realistic-curve}.  }
 \end{center}
\end{figure}
%%%%%%%%%%%%%%%%%%%%%%%%%%%%%%%%%%%%%%%%
All these numbers are topological, and cannot be changed by tuning 
moduli parameters. 
In all the examples in Table \ref{tab:top-data}, the number of $E_6$-type 
and $D_6$-type singularities are not 1 in any one of the examples. 

In fact, from the expression of 
$\# E_6$ and $g(\bar{c}_{({\bf 10})})$, one can see that 
\begin{equation}
\# E_6 = (2g(\bar{c}_{({\bf 10})}) - 2) - 2 K_S \cdot (5K_S + \eta).
\end{equation}
The number of $E_6$-type singularities is always even; it cannot be 1.
The number of codimension-3 singularity points is 
a crucial element of flavor structure of Yukawa matrices of 
low-energy effective theory.
An idea of Ref. \cite{HV-Nov08-rev} was to assume a geometry with 
$\# E_6 = 1$ for a compactification, in order to realize an 
approximately rank-1 up-type Yukawa matrix of the real world. 
Now we know, however, that this idea does not work.
 
One might think of taking a limit of the complex structure by hand, 
so that the matter curve $\bar{c}_{({\bf 10})}$ factorizes into
irreducible pieces in the GUT divisor $S$. 
\begin{equation}
\bar{c}_{({\bf 10})} = \sum_i \bar{c}_{({\bf 10})i}. 
\end{equation}
Only one of the irreducible pieces, say, $\bar{c}_{({\bf 10})i_0}$, 
may be regarded as the support of $(u^c, q, e^c) = {\bf 10}$ fermions 
of the Standard Model. Similarly, the matter curve 
$\bar{c}_{(\bar{\bf 5})}$ may also factorizes into irreducible pieces, 
$\bar{c}_{(\bar{\bf 5})} = \sum_j \bar{c}_{(\bar{\bf 5})j}$, and only
one of them, say, $\bar{c}_{(\bar{\bf 5})j_0}$ is regarded as the
support of $(d^c, l)= \bar{\bf 5}$ fermions of the Standard Model. 
If the factorization limit is chosen properly, then there may be just 
one $E_6$-type and one $D_6$-type intersection points that involve the 
relevant pieces $\bar{c}_{({\bf 10})i_0}$ or/and 
$\bar{c}_{(\bar{\bf 5})j_0}$ \cite{Caltech-0904}.  
The idea of \cite{HV-Nov08-rev} may still be valid in this 
context \cite{Caltech-0904}.\footnote{Some articles 
(e.g., \cite{Harvard-nu}) derive predictions on flavor structure 
in the lepton sector under an assumption that there is only one point 
in the GUT divisor $S$ where the matter $\bar{\bf 5}=(d^c,l)$ curve 
and the Higgs curve intersect. The number of this type of intersection 
points, however, is also a topological invariant, and is expressed 
in terms of intersection numbers of divisors on $S$. In the 
appendix \ref{sec:4+1}, we determined the numbers of all sorts 
of codimension-3 singularities in terms of intersection numbers 
in the case a 5-fold spectral cover (that we state shortly in the main 
text) factorizes into a 4-fold cover and a 1-fold cover. } 

Such a factorization limit of the curves may exist, but a factorization 
of a matter curve does not always imply that independent massless fields 
can be classified into irreducible pieces of the reducible matter curve 
based on their support \cite{Tsuchiya}.
In order to make sure that the Standard Model fields 
e.g. in the representation ${\bf 10}$ localizes on only one irreducible piece 
of the factorized matter curve $\bar{c}_{({\bf 10})}$, 
not only the curve, but also its spectral surface needs to be factorized. 
Furthermore, in order to achieve a well-defined factorization 
(see section 4.3.3 of \cite{Tsuchiya}), 
one needs to take the limit $\epsilon \rightarrow 0$ in 
\begin{equation}
 a_2 = \epsilon^2 a_{2,0}, \quad a_3 = \epsilon^3 a_{3,0}, \quad 
 a_4 = \epsilon^4 a_{4,0}, \quad a_5 = \epsilon^5 a_{5,0}, 
\label{eq:E8-limit}
\end{equation}
to make sure that an $E_8$ Higgs bundle is well-defined globally on 
$S$, as implicitly done in \cite{Caltech-0906}. The higher order terms 
in the $z$-series expansion in (\ref{eq:def-eq}) surely become 
irrelevant in this limit,\footnote{Introduce a new set of coordinates 
($x', y', z'$) satisfying $x = \epsilon^{10} x'$, $y = \epsilon^{15} y'$
and $z = \epsilon^6 z'$. This is to look into a region near 
$(x, y, z) = (0,0,0)$ closely. In the new coordinates, 
the equation (\ref{eq:def-eq}) becomes 
\begin{equation}
(y')^2 \simeq (x')^3 + a_{5,0} x' y' + a_{4,0} z' (x')^2 
 + a_{3,0} (z')^2 y' + a_{2,0} (z')^3 x' + a_{0,0} (z')^5 + {\cal O}(\epsilon),
\end{equation}
which is an equation of a deformed $E_8$ singularity with negligible terms 
of order $\epsilon$. The two-cycles with the $E_8$
intersection form stay within the range of $x' \approx {\cal O}(1)$, 
$y' \approx {\cal O}(1)$ and $z' \approx {\cal O}(1)$, where higher 
order terms are negligible.} and we are sure that there is a set 
of 2-cycles with the intersection form of $E_8$ fibered globally over $S$.
In the presence of an unbroken U(1) symmetry in such 
a factorization limit, the Majorana mass of right-handed 
neutrinos is forbidden by this U(1) symmetry, if the up-type Higgs and 
down-type Higgs multiplets are vector-like in $E_8$, which is the case 
if the $\SU(5)_{\rm GUT}$ symmetry is broken either by a hypercharge 
Wilson line associated with $\pi_1(S) \neq \{ 1 \}$ \cite{BHV-2}, 
or by a hypercharge line bundle \cite{BHV-2, TW-2, DW-2}, as pointed 
out in \cite{Caltech-0906}. One also has to make sure that 
all kinds of physics associated with $\SU(5)_{\rm GUT}$-charged non-Standard 
Model matter fields on $\bar{c}_{({\bf 10}) i \neq i_0}$ and 
$\bar{c}_{(\bar{\bf 5}) j\neq j_0}$ do not conflict against 
low-energy phenomenology. 

If we want to maintain the successful prediction of 
Majorana masses of right-handed neutrinos in the matter 
parity scenario \cite{Tsuchiya}, on the other hand, we cannot rely 
on a globally defined $E_8$ Higgs bundle or vacuum with an unbroken 
U(1) symmetry. In this case, the up-type Yukawa matrix 
of the effective theory may have contributions from more than 
one $E_6$-type singularity points. Clearly, the idea of \cite{HV-Nov08-rev} 
does not work in this case. In order to evaluate the flavor structure 
of Yukawa matrices that involve contributions from multiple points 
in the GUT divisor $S$, we need to know how the zero mode 
wavefunctions behave along the matter curve. This is what we 
study in the next three sections. 

Before moving on to the next section, let us pause for a moment to 
pose the following question, 
which we think is interesting at least in the context of model building. 
In all the examples in Table~\ref{tab:top-data}, the matter curve 
for the representation $\bar{\bf 5}$--${\bf 5}$ 
has a very large genus.\footnote{It should be fair to mention that this large 
genus of the matter curve for the representation $\bar{\bf 5}$ was already 
known in \cite{Penn5}; the genus of a matter curve is determined only 
by the choice of $\eta$, in F-theory as well as in elliptically fibered 
compactifications of Heterotic strings, and hence the result in \cite{Penn5} 
should be regarded as the same phenomenon as in Table~\ref{tab:top-data}. }
The question is if that is a generic feature of F-theory compactifications.  
If it is generic, we should think of phenomenological consequences 
of this feature. 

To address this question, note that the right-hand side 
of (\ref{eq:genus-5curve-BUP}) can be reorganized as %follows: 
\begin{equation}
 2g(\tilde{\bar{c}}_{(\bar{\bf 5})})-2 = 
   7 (5K_S + \eta) \cdot (3K_S + \eta) - 9 K_S \cdot (5K_S + \eta) 
    + 20 K_S \cdot K_S. 
\end{equation}
Since the first term is proportional to 
$\# D_6 = (5K_S + \eta) \cdot (3K_S + \eta) \geq 0$, 
it is always zero or positive. The second term 
is positive when the anti-canonical divisor $(-K_S)$
is ample,\footnote{This is the case when $S= dP_k$ with 
$k \leq 8$. $F_2$ is marginal, in that the second term 
can also be zero.} because $(5K_S + \eta)$ containing $\bar{c}_{({\bf 10})}$ 
is effective. 
Finally, in the last term, $K_S^2$ is 8 for all the Hirzebruch 
surfaces $S = F_n$, and is $(9-k)$ for the del Pezzo surface $S = dP_k$. 
Thus, unless $k > 8$ for del Pezzo surfaces, the last term is 
always positive with the large coefficient 20. 
All the three terms are positive, with their relatively large coefficients 
7, 9 and 20.  
It is now easy to see why the (covering) matter curve of 
the representation $\bar{\bf 5}$--${\bf 5}$ tend to have very 
large genus in the examples of Table~\ref{tab:top-data}; 
the last term alone contributes by $160$ for $S = F_n$, and 
$20(9-k) = 180$ for $S = dP_{k=0} = \P^2$. 

We can also learn another lesson. The term $20 K_S^2$ 
becomes small, for example, for $S = dP_k$ with larger $k$. 
See Table~\ref{tab:top-data-B}. 
%%%%%%%%%%%%%%%%%%%%%%%%%%%%%%%%%%%%%%%%%
\begin{table}[tb]
\begin{center}
\caption{\label{tab:top-data-B} Some other examples of the topological 
 invariants for $S = dP_8$ or \kaku{a rational elliptic surface $dP_9$}. 
 Thus, $K_S^2$ is either 1 or 0 in all the examples in this table. 
Consequently, 
$g(\tilde{\bar{c}}_{(\bar{\bf 5})}) \approx {\cal O}(10)$ 
in this table, which is in contrast with 
$g(\tilde{\bar{c}}_{(\bar{\bf 5})}) \approx {\cal O}(100)$ in the
 examples in Table~\ref{tab:top-data}. }
 \begin{tabular}{c||c|c|c|c|c|c|c}
  & $S$ & $\bar{c}_{({\bf 10})} \sim $ & 
    $g(\bar{c}_{({\bf 10})})$ & \# $E_6$ & \# $D_6$ & 
    $g(\tilde{\bar{c}}_{(\bar{\bf 5})})$ & \# $A_6$ \\
\hline
XI & $dP_8$ & $H-E_1-E_2$ & 0 & 0 & 1 & 19 & 52 \\
XII & $dP_9$ & $H-E_1-E_2$ & 0 & 0 & 1 & 9 & 22 \\
XIII & $dP_8$ & $2H-\sum_{i=1}^5 E_i$ & 0 & 0 & 1 & 19 & 52 \\
XIV & $dP_9$ & $H-E_1$ & 0 & 2 & 4 & 24 & 58 \\
XV & $dP_9$ & $2H - \sum_{i=1}^4 E_i$ & 0 & 2 & 4 & 24 & 58 \\
XVI & $dP_9$ & $3H - \sum_{i=1}^7 E_i$ & 1 & 4 & 6 & 31 & 72 \\
XVII & $dP_9$ & $3H - \sum_{i=1}^8 E_i$ & ``1'' & 2 & 3 & 16 & 36  
 \end{tabular}
\end{center}
\end{table}
%%%%%%%%%%%%%%%%%%%%%%%%%%%%%%%%%%%%%%%%%%%%%%%%%%%%%%%%%%%%%%
By using $S= dP_8$ or $dP_9$, the genus of $\tilde{\bar{c}}_{(\bar{\bf 5})}$ 
can be reduced from ${\cal O}(100)$ to at least ${\cal O}(10)$. 
The large genus of the matter curve 
in the representation $\bar{\bf 5}$ of 
${\cal O}(100)$, therefore, is {\it not} a generic prediction 
of F-theory compactifications, but it is an artifact of 
$K_S^2 \sim {\cal O}(8)$ in the Hirzebruch series $S = F_n$ and 
the del Pezzo series $S = dP_k$ with small $k$. 

On a complex curve with a large genus $g$, a line bundle ${\cal O}(D)$
with a negative degree ${\rm deg} D < 0$ does not have non-zero $h^0$. 
Those with ${\rm deg} D > 2g-2$ do not have non-zero $h^1$, either. 
For bundles in the range $0 \leq {\rm deg} D \leq 2g - 2$, 
however, both $h^0$ and $h^1$ can be non-zero. Such massless fields 
in a pair of vector like representations may be identified with 
the two Higgs doublets of the supersymmetric Standard 
Models \cite{BHV-2} or messenger fields in gauge mediation of 
supersymmetry breaking.
For larger $g$, the window for $h^0 \neq h^1 \neq 0$ becomes larger. 
Too many extra pairs of chiral multiplets in the 
representation {\bf 5}$+\bar{\bf 5}$ \kaku{(See \cite{Penn5})} 
will change the running of the gauge coupling constants 
too much and are not acceptable to retain a perturbative gauge coupling 
unification. 
Thus, a large genus does not seem to be a favorable situation. 
The observation so far, therefore, might be taken as an indication 
that $S=dP_k$ with a large $k$ is a favorable choice phenomenologically. 

The authors are clearly aware, however, that things are more complicated 
than that, and we have just scratched the surface of possible choices of 
topology.
A choice of topological class of $\bar{c}_{({\bf 10})}$ and that of 
$S$ set a constraint on the smallest possible net chirality, and 
on the choices of fluxes to achieve it. The genus 
$g(\tilde{\bar{c}}_{(\bar{\bf 5})})$ is already determined from the 
topological data of $S$ and $\bar{c}_{({\bf 10})}$. The number of 
unnecessary ${\bf 5}+\bar{\bf 5}$ pairs is determined not by 
the genus alone but also by the choice of fluxes that has already been 
constrained tightly to reproduce the net chirality $N_{\rm gen}$.
This problem is a bit complicated, and we will not point to 
a particular direction to search for a geometry giving rise to the Standard Model. 
In this article, therefore, we just note that the genus of the curve 
depends strongly on $K_S^2$, and that the genus becomes smaller for smaller $K_S^2$.

%%%%%%%%%%%%%%%%%%%%%%%%%%%%%%%%%%%%%%%%%%%%%%%%%%
\begin{figure}[tb]
 \begin{center}
  \begin{tabular}{ccc}
   (a) & (b) & (c) \\
   \includegraphics[width=.3\linewidth]{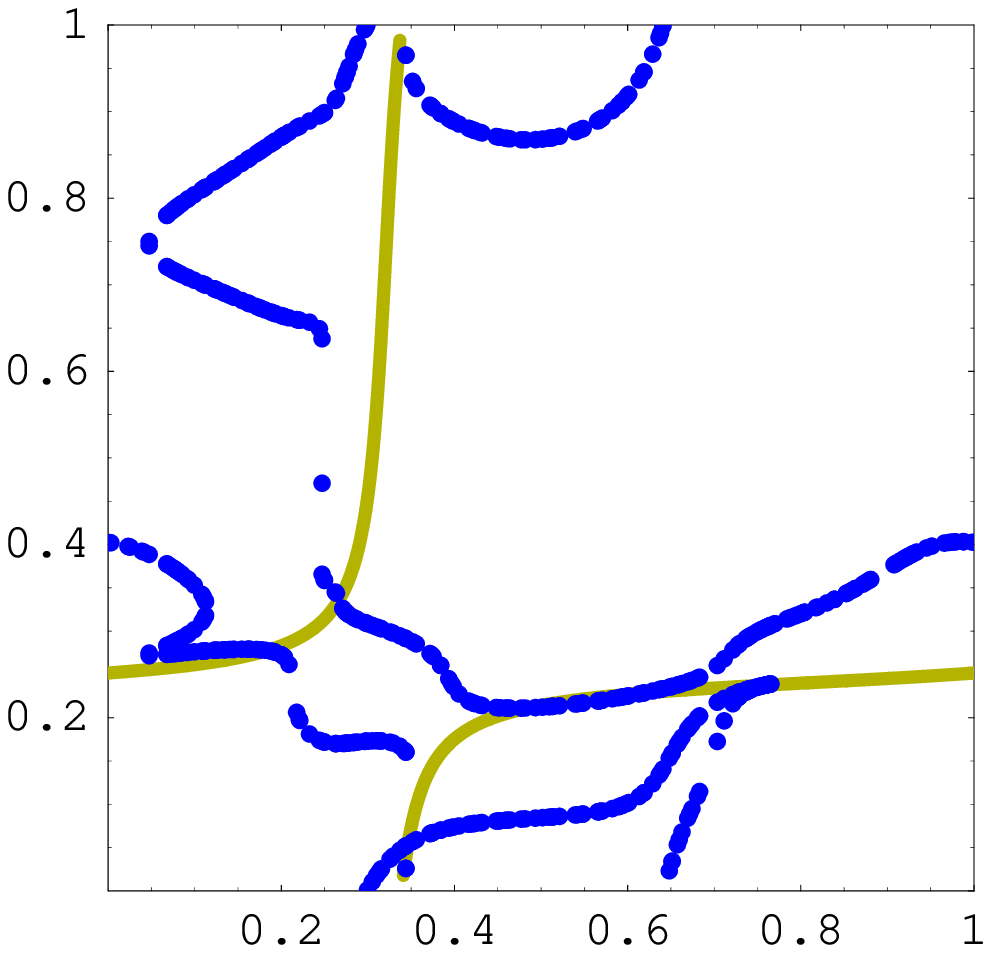} & 
   \includegraphics[width=.3\linewidth]{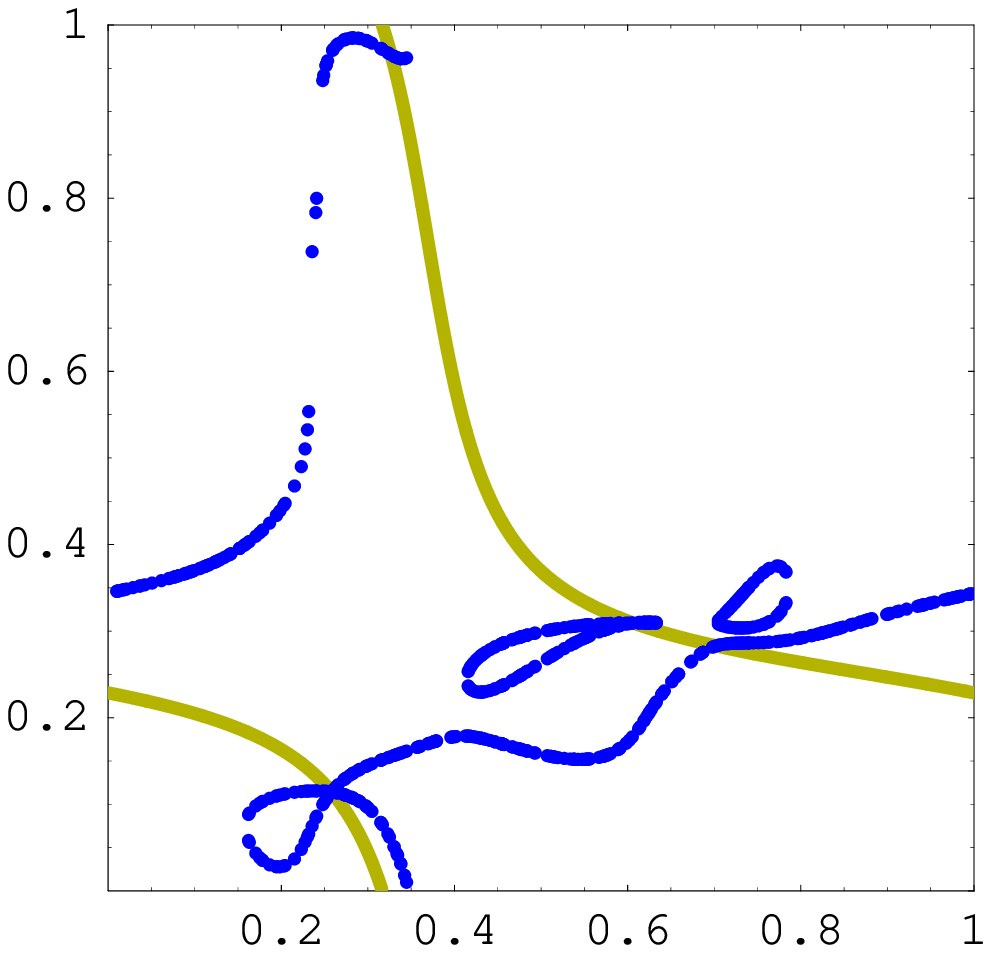} & 
   \includegraphics[width=.3\linewidth]{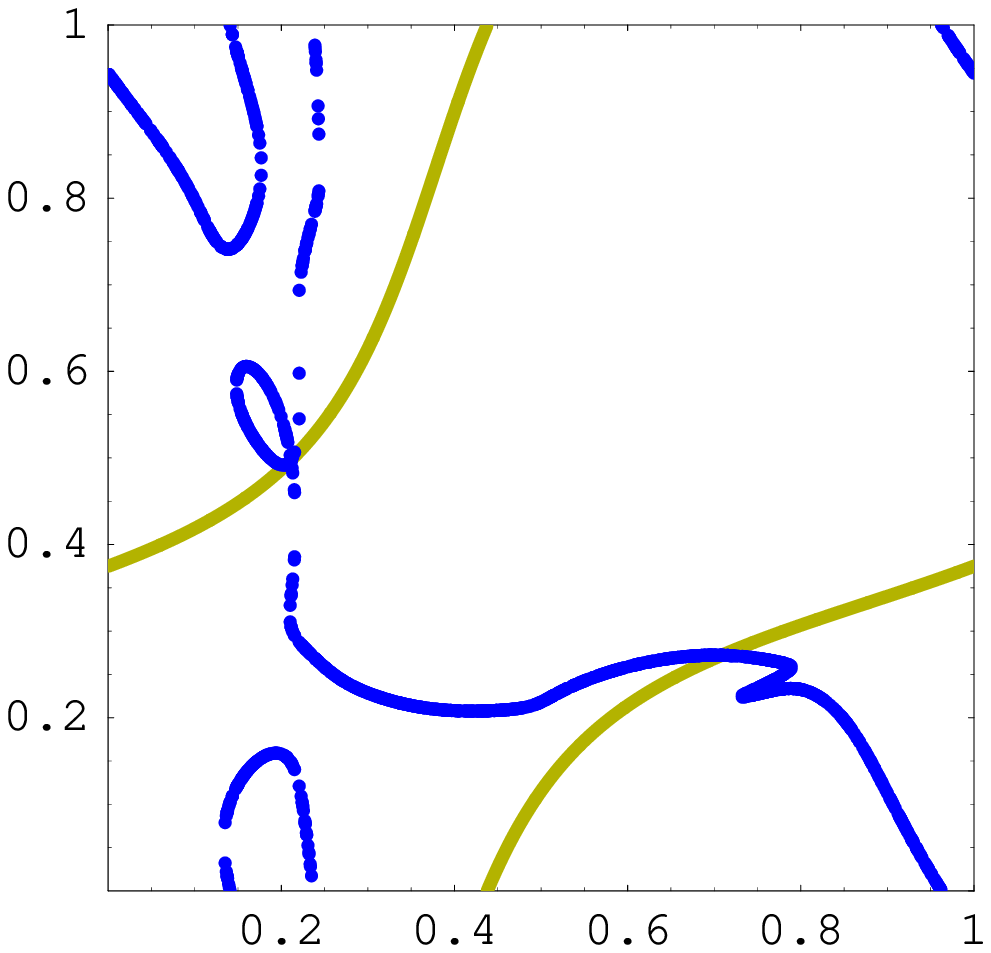}  
  \end{tabular}
  \caption{\label{fig:realistic-curve} (color online)
``Real''istic pictures of the matter curves in $S$ are obtained 
by numerically solving the equation $P^{(5)} = 0$. 
We used the example IV, and drew the curves $\bar{c}_{({\bf 10})}$ 
(yellow) and $\bar{c}_{(\bar{\bf 5})}$ (blue) for three different choices 
(a), (b) and (c), of the complex structure. In this example, 
$S = \P^1 \times \P^1$, and $a_5$ is a homogeneous function 
of bi-degree (1,1) , $a_4$ a homogeneous function of bi-degree (3,3), and 
$a_{r}$ ($r = 3,2,0$) homogeneous functions of bi-degree 
$(11-2r, 11-2r)$. 
In order to visualize the geometry of the complex curves in the complex 
surface $S$, we cut out the real locus of $S = \P^1 \times \P^1$, 
which is $S^1 \times S^1 = T^2$. $T^2$ was cut open and described 
as $[0, 1] \times [0,1]$ in the figure. We restricted all the 
coefficients to be real valued, so that the matter curves 
appear in the real locus as real 1-dimensional curves.
All the coefficients are chosen randomly from $[0, 1] \subset \R$ 
separately for (a), (b) and (c). }
 \end{center}
\end{figure}
%%%%%%%%%%%%%%%%%%%%%%%%%%%%%%%%%%%%%%%%%%%%%%%%%%%

%%%%%%%%%%%%%%%%%%%%%%%%%%%%%%%%%%%%%%%%%%%%%%%%%%%%%%%
\section{Holomorphic Wavefunctions on Matter Curves}
\label{sec:hol-wavefcn}
%%%%%%%%%%%%%%%%%%%%%%%%%%%%%%%%%%%%%%%%%%%%%%%%%%%%%%%

We have seen that the number of $E_6$-type codimension-3 
singularity points is more than one, and the up-type 
Yuakwa matrix of low-energy effective theory receives 
contributions from more than one local patches on $S$.
It is therefore necessary to calculate the zero-mode 
wavefunctions over the compact matter curve 
$\bar{c}_{({\bf 10})}$, so that one can argue relative 
importance of contributions to Yukawa couplings 
from multiple codimension-3 singularity points 
on $\bar{c}_{({\bf 10})}$.

A zero mode chiral multiplet is identified with 
a holomorphic section of a certain line bundle 
on a matter curve, and a divisor specifying 
the line bundle is determined from the loci 
of codimension-3 singularity points \cite{Curio, DI, Hayashi-1}. 
It must be therefore straightforward to calculate the 
number of independent zero modes, and to determine their wavefunctions 
as holomorphic sections on the covering matter curve. 
We think, however, that it is worthwhile to illustrate it with 
concrete examples, and that is what we will do in this section.

%%%%%%%%%%%%%%%%%%%%%%%%%%%%%%%%%%%%%%%%%%%%%%%%%%%%
\subsection{Global Holomorphic Sections}
\label{ssec:review-hol-wave}
%%%%%%%%%%%%%%%%%%%%%%%%%%%%%%%%%%%%%%%%%%%%%%%%%%%%

A zero mode chiral multiplet of a charged matter field 
correspond to a global holomorphic section of a line bundle 
on a matter curve. Here, we give a short brief review of 
relevant mathematical materials, partially as a guide to 
readers unfamiliar with them, and partially 
for the purpose of setting up notations. 

In order to describe global holomorphic sections of a 
line bundle ${\cal O}(D)$ on a manifold/variety $X$, one chooses 
an open covering $\{ U_a \}$ of $X$; $\cup_{a\in A} U_a = X$. 
A rational function $\varphi_a$ on each $U_a$ is chosen to describe the divisor $D$, 
so that 
\begin{equation}
D|_{U_a} = {\rm div} \; \varphi_a. 
\label{eq:Cartier}
\end{equation}
It follows that neither $\varphi_a/\varphi_b$ nor $\varphi_b/\varphi_a$ 
should have a pole or zero in $U_a \cap U_b$, so that (\ref{eq:Cartier}) 
in $U_a\cap U_b$ is consistent with the one in $U_b$.  
$\{(U_a, \varphi_a)\}_{a\in{A}}$ is called 
the Cartier divisor description of the divisor $D$. The open covering 
$\{ U_a \}_{a \in A}$ plays the role of trivialization patches 
of the line bundle ${\cal O}(D)$; although the cohomology group 
\begin{equation}
H^0(X; {\cal O}(D)) = 
  \{ \tilde{f} \in \C (X) | D + {\rm div} \tilde{f} \geq 0 \}
\label{eq:def-lin-sys}
\end{equation}
is primarily described by a set of rational functions on $X$, 
one can assign a holomorphic function 
$\tilde{f}_a \equiv \varphi_a\cdot\tilde{f}|_{U_a}$ on $U_a$ 
for an element $\tilde{f}$ of $H^0(X; {\cal O}(D))$. The holomorphic 
functions $\tilde{f}_a$ and $\tilde{f}_b$ in the overlapping patch 
$U_a\cap{U}_b$ are related by 
$\tilde{f}_a = (\varphi_a/\varphi_b) \tilde{f}_b$; thus, 
$g_{ab} = (\varphi_a/\varphi_b)$ plays the role of the transition function 
from $U_b$ to $U_a$ of the line bundle ${\cal O}(D)$. 
The holomorphic function $\tilde{f}_a$ on $U_a$ is the coefficient 
function in the component description in the individual trivialization 
patches (there is only one component because we are talking about 
a rank-1/line bundle). 
Thus, the zero mode wavefunction can be described by specifying 
an open covering $\{ U_a \}_{a\in A}$, transition functions $g_{ab}$
on $U_a \cap U_b $, and a holomorphic function $\tilde{f}_a$ on each $U_a$ 
glued together consistently by the transition functions. 

The description in terms of $U_a$ and $\tilde{f}_a$ does not change 
when the divisor $D$ is replaced by $D' = D + {\rm div} \varphi$  
with a rational function $\varphi$ on $X$, one that is linearly 
equivalent to the original divisor $D$. One can use 
$\varphi'_a \equiv \varphi_a \varphi$ in a Cartier divisor description 
of $D'$, but $g'_{ab} = \varphi'_a/\varphi'_b= \varphi_a/\varphi_b=g_{ab}$. 
Further, since $H^0(X; {\cal O}(D')) = \{ \tilde{f}' = \tilde{f} \varphi^{-1} \}$, 
$\tilde{f}'_a = \varphi'_a \tilde{f}' = \varphi_a \tilde{f}=\tilde{f}_a$.
 
The wavefunctions $\{(U_a, \tilde{f}_a)\}_{a\in{A}}$, however, do depend on 
the choice of divisor class (divisor modulo linear equivalence), not just 
on the first Chern class of the line bundle ${\cal O}(D)$. 
(Here, we now consider a case where $X$ of $H^0(X; {\cal O}(D))$ is a curve.) 

Let us see this in one of the simplest examples. 
We take an elliptic curve $X = E$ given by\footnote{The coordinates $x$ and $y$ 
have nothing to do with $x$ and $y$ in (\ref{eq:def-eq}).} 
\begin{equation}
 y^2 = x^3 + f x + g, 
\end{equation}
and consider a line bundle ${\cal O}(D) = {\cal O}(p)$ specified by 
a point $p \in E$. The divisors $p$ and $p'$ (both $p, p' \in E$) are 
not linear equivalent if $p \neq p'$. Let us see explicitly that 
a global holomorphic section of ${\cal O}(p)$ depends on the choice 
of the divisor class $(p \in E)$. 

For a given $p \in E$, the elliptic curve $E$ can be covered by 
three patches
\begin{equation}
 U_1 = E \backslash \{ p, \boxminus p, e \}, \qquad 
 U_2 = E \backslash \{ p, q, \boxminus (p \boxplus q) \} 
    \quad {\rm and} \quad
 U_3 = E \backslash \{ (p/2), \boxminus (p/2), e \}, 
\end{equation}
where $e \in E$ is the zero element in the Abelian group 
structure on $E$. $\boxplus$ is the summation of the group law 
on $E$, and $(\boxminus q)$ for $q \in E$ denotes the inverse 
element of $q \in E$ of the group law. An arbitrary point $q \in E$ 
can be used in defining $U_2$ above.
$(p/2) \in E$ in the definition of $U_3$ is a point satisfying 
$(p/2) \boxplus (p/2) = p$.

Let us construct a rational function $\varphi_{a}$ on each 
$U_a$ $(a=1, 2, 3)$ to give a Cartier divisor description of the 
divisor $D=p$. 
On the patches $U_{1, 2}$, where the point $p$ is removed, we can choose 
$\varphi_{1, 2}$ as constants, {\it i.e.}, $\varphi_{1}=1$, $\varphi_{2}=1$. 
On the other hand, $\varphi_{3}$ should have a zero of order one 
at $p$ and have no other zeros or poles anywhere on the patch $U_{3}$. 
In order to have a zero at $p$, we can choose $\varphi_{3} \propto x-x(p)$. 
$x-x(p)$ has a zero at $p$ and at $\boxminus p$, and a second order pole 
at $e$. 
But a second order pole $e$ is irrelevant since the patch $U_{3}$ does 
not include the origin $e$. On the other hand, we have to cancel the 
zero at $\boxminus p$. To this end, we divide $x-x(p)$ by 
$y+cx-(y(\boxminus p)+cx(\boxminus p))=y+cx-(-y(p)+cx(p))$ 
where $c=({y(p)+y(p/2)})/({x(p)-x(p/2)})$. 
$y+cx-(-y(p)+cx(p))$ has a zero of order one at $\boxminus p$, a zero 
of order two at $p/2$, and a pole or order three at $e$. 
The zero of order two at $p/2$ can be understood from the fact that 
the group-law sum of the zero points of an elliptic function is the same 
as that of the poles. The zero of order two at $p/2$ and the pole of
order three at $e$ are also irrelevant on the patch $U_{3}$, 
and the zero at $\boxminus p$ of the denominator $y+cx-(-y(p)+cx(p))$ cancels 
the zero at $\boxminus p$ of the numerator $x-x(p)$. 
Thus, a rational function $({x-x(p)})/({y+cx-(-y(p)+cx(p))})$ 
has a zero of order one at $p$, and no other zeros or poles in $U_{3}$. 
This function can be used for $\varphi_3$.

As a whole,  the rational functions $\varphi_a$ ($a = 1,2,3$) for 
the Cartier divisor description $(U_a, \varphi_a)$ of $D = p$ can be 
chosen as  
\begin{equation}
 \varphi_1 = 1, \qquad \varphi_2 = 1 \quad {\rm and} \quad 
 \varphi_3 = \frac{x - x(p)}{y + c x - (-y(p) + c x(p))}, 
\end{equation}
with $c=({y(p)+y(p/2)})/({x(p)-x(p/2)})$. 

Since $H^0(E; {\cal O}(p))$ in the sense of (\ref{eq:def-lin-sys}) 
consists of only one rational function (mod $\C^\times$), 
$\tilde{f} = 1$, the generator of the vector space of zero mode(s) 
correspond to a holomorphic wavefunction 
$(U_a, \tilde{f}_a) = (U_a, \varphi_a)$. 
Clearly the wavefunction $\tilde{f}_a$ depends on the choice 
of $p \in E$. 
% \kaku{the holomorphic section $\tilde{f}_a$ on $U_{a}$ of the line bundle 
% ${\cal O}(D)$ is given by $\tilde{f}_a=\varphi_a\cdot1$, and it does 
% depend on  the choice of $p \in E$.}

%%%%%%%%%%%%%%%%%%%%%%%%%%%%%%%%%%%%%%%%%%%%%%
\subsection{Example VII}
\label{ssec:Ex-7}
%%%%%%%%%%%%%%%%%%%%%%%%%%%%%%%%%%%%%%%%%%%%%

Let us choose the example VII in Table \ref{tab:top-data}, first. 
This is one of the easiest examples, and will be 
suitable for illustrative purpose. 

In the example VII, the GUT divisor is $S=\P^2$, 
and the matter curve $\bar{c}_{({\bf 10})}$ is in the topological class 
$|2H|$, where $H$ is a hyperplane of $\P^2$. The explicit 
choice of $a_5 \in \Gamma (S; {\cal O}(2H))$---a homogeneous 
function of degree two of the homogeneous coordinates---determines the curve 
$\bar{c}_{({\bf 10})}$ in $\P^2$. 
When $a_{4,3,2,0}$ are 
also chosen from their appropriate line bundles, 
the curve $\bar{c}_{(\bar{\bf 5})}$ is also determined 
by $P^{(5)} = 0$ as a subvariety of $S = \P^2$. 
Figure~\ref{fig:realistic-curve} illustrates the configurations 
of the matter curves in Example IV for different complex structures.

%%%%%%%%%%%%%%%%%%%%%%%%%%%%%%%%%%%%%%%%%%%%%
\subsubsection{Fluxes for Chirality: Set-up for Calculation}
%%%%%%%%%%%%%%%%%%%%%%%%%%%%%%%%%%%%%%%%%%%%

We cannot talk about the chiral matter contents in 
the low-energy effective theory below the 
Kaluza--Klein scale without introducing 4-form 
fluxes in the Calabi--Yau 4-fold $X$. The net 
chirality of a pair of Hermitian conjugate representations of 
unbroken symmetry $G''$ is determined by integrating 
the 4-form flux over the vanishing 2-cycle parametrized 
by the covering matter curve \cite{Hayashi-1}. 

An available 4-form flux depends on the choice of Calabi--Yau 4-fold. 
It cannot be determined only from the geometry of the GUT divisor $S$ and 
its infinitesimal neighborhood in the base 3-fold $B_3$. 
For a given 4-fold $X$, one needs to identify the available 
flux in $H^{2,2}(X; \Q)$, and further finds out how individual 
generators of this cohomology group contribute to the 
4-cycle of vanishing 2-cycles over (covering) matter curves 
of various representations in $S$. That is what one is 
supposed to do to search for the geometry describing the real world, and 
that is an area where more technical development is 
yet to be necessary.   

We know, however, that the 4-form flux background on $X$ is once 
encoded as line bundles on spectral surfaces of various representations.
Line bundles on matter curves are obtained from restriction of the line 
bundles on the spectral surfaces.
The wavefunctions of zero modes are determined by using the line bundles
on the matter curves. Thus, as an intermediate step, 
it is possible to assume a certain form of a line bundle on a spectral 
surface, and study the consequences; such approach does not 
guarantee that a set of a 4-fold $X$ and a 4-form $G^{(4)}$ 
exists for \kaku{the assumed line bundles on the spectral surfaces. }
In this intermediate step approach, the existence proof---also known 
as ``swampland program''---can be put aside as an open problem, and 
the remaining half of the problem can be addressed separately. 
We do not necessarily mean that this is the best strategy, 
but at least for the purpose of illustrating how to 
calculate the wavefunctions of the zero modes, we consider 
that it is wise to start from the intermediate step, 
instead of doing everything in a top-down approach from the first principle. 
We will thus consider consequences of a hypothetical 4-form flux $G^{(4)}$ 
on a hypothetical variety $X$ by assuming a line bundle on each of 
all the spectral surfaces. 

The GUT divisor $S$ is covered by open patches 
$U_\alpha$, where field theory local models are 
defined. When a local geometry of $X$ in a neighborhood of 
$U_\alpha \subset S$ is approximately described by an ALE fibration 
of type $G$ where $G$ is one of the $A-D-E$ series, then 
the physics associated with this ALE fibration is 
described by a field theory local model on $U_\alpha$ 
with the gauge group $G$. The choice of the gauge 
group may, in principle, be different for different 
patches $U_\alpha$ in $S$ \cite{Hayashi-2, Tsuchiya}. 
Here, however, for illustrative purposes, we choose 
a specific choice (limit) of the complex structure, so that 
the GUT divisor $S$ is covered by a single field theory 
local model, whose gauge group is $E_8$. That is the case, 
for example, when the limit (\ref{eq:E8-limit}) is chosen 
in an F-theory compactification (this includes an F-theory 
compactification with a Heterotic dual, and the complex 
structure moduli are in the stable degeneration limit.)
In that case, a spectral cover $C_{({\bf 10})}$ for the 
representation ${\bf 10}$ of the unbroken symmetry 
$\SU(5)_{\rm GUT}$ is a divisor of $\mathbb{K}_S$, and a 
5-fold cover over $S$. We assume a line bundle 
${\cal N}_{({\bf 10})}$ on $C_{({\bf 10})}$ in a form 
\begin{equation}
 {\cal N}_{({\bf 10})} = {\cal O}\left( \frac{1}{2} r_{({\bf 10})} 
  + \gamma \right), 
\end{equation}
where $r_{({\bf 10})}$ is the ramification divisor associated with the projection 
$\pi_{C_{({\bf 10})}}: C_{({\bf 10})} \rightarrow S$, and the divisor $\gamma$ on 
$C_{({\bf 10})}$ is given by 
\begin{equation}
 \gamma = \gamma_{FMW} = \lambda \left( 
   5 \bar{c}_{({\bf 10})} - \pi_{C_{({\bf 10})}}^{-1}(\bar{c}_{({\bf 10})}) \right). 
\label{eq:gamma-FMW}
\end{equation}
%
%here, $\gamma$ is a divisor on $C_{({\bf 10})}$, 
%$\pi_{C_{({\bf 10})}}: C_{({\bf 10})} \rightarrow S$ is a projection, 
%and $r_{({\bf 10})}$ is the ramification divisor associated with this projection. 
$\bar{c}_{({\bf 10})}$ is regarded as a 
divisor within $C_{({\bf 10})}$ in the first term of (\ref{eq:gamma-FMW}), 
while in the second term, it is regarded as a divisor on $S$, 
and is pulled back to $C_{({\bf 10})}$ by the projection $\pi_{C_{({\bf 10})}}$.
\kaku{In terms of a fiber coordinate $\xi$ of $\mathbb{K}_S$, 
the former $\bar{c}_{({\bf 10})}$ is given by $\xi=0$, and the latter by $a_5=0$.} 
This divisor is chosen so that $\pi_{C_{({\bf 10})} *} \gamma = 0$ in $S$.
% $\lambda$ has to be half-integer in order 
% for ${\cal N}_{({\bf 10})}$ to be a well-defined line bundle 
% on the spectral surface. 
This line bundle on the spectral surface was developed originally in 
a description of Heterotic string compactifications \cite{FMW}, but 
now we know that the spectral surface and a line bundle on it are 
readily used for F-theory compactifications as well \cite{Hayashi-2, DW-3}. 

We are fully aware that this is by no means a generic 
choice of the complex structure of $X$ or of flux on it. 
Higgs bundles via the extension construction, or 
Higgs sheaves in $\mathbb{K}_{U_\alpha}$ that are not 
represented as a pushforward of a line bundle on the spectral surface 
in $\mathbb{K}_{U_\alpha}$ \cite{Tsuchiya}\footnote{
An F-theory compactification using such a background may be 
used for a framework of phenomenologically viable R-parity 
violation.} cannot be covered in this way of incorporating a 
flux background. But for now, we will content ourselves with 
working out zero-mode wavefunctions for a limited class of choices 
of background. 

%%%%%%%%%%%%%%%%%%%%%%%%%%%%%%%%%%%%%%%%%%%%%%%%%%%%%
\subsubsection{Zero Mode Wavefunctions in the Representation 
${\bf 10}$--$\overline{\bf 10}$}
%%%%%%%%%%%%%%%%%%%%%%%%%%%%%%%%%%%%%%%%%%%%%%%%%%%%%

Now all the relevant data are set, and we are ready to calculate 
zero mode wavefunctions. We will first identify independent 
zero modes in ${\bf 10}$ (and possibly in $\overline{\bf 10}$) 
of $\SU(5)_{\rm GUT}$, and then determine their wavefunctions. 
The vector space of zero modes in the representation ${\bf 10}$ 
is given \cite{Curio, DI, Hayashi-1} by 
\begin{equation}
 H^0 \left( \bar{c}_{({\bf 10})}; {\cal F}_{({\bf 10})} \right)
=  H^0 \left( \bar{c}_{({\bf 10})} ; 
  {\cal O}\left(K_S|_{\bar{c}_{({\bf 10})}} + \frac{1}{2} p_{E6} 
    + \gamma|_{\bar{c}_{({\bf 10})}}  \right)\right).
\label{eq:H0-10-repr}
\end{equation}
The divisor $K_S$ on $S$ is restricted on $\bar{c}_{({\bf 10})}$, 
and so is the divisor $\gamma$ on $C_{({\bf 10})}$. $p_{E6}$ stands 
for the collection of all the codimension-3 singularity points of the 
$E_6$-type singularity enhancement (also called type (a) points 
in \cite{Hayashi-1}). All the information of the divisor class 
\kaku{of the line bundle ${\cal F}_{({\bf 10})}$} 
is retained\kaku{ in (\ref{eq:H0-10-repr})}, not just the degree of 
this line bundle; 
this expression can be used in calculating the wavefunctions. 

In order to study this line-bundle valued cohomology group, 
we need to know the divisor on the curve very well. There is no 
problem with the restriction of $K_S$ or $p_{E6}$, but 
we have a bit more work with the restriction of $\gamma$.

The divisor $\gamma$ in $C_{({\bf 10})}$ has two irreducible 
components. One is $\bar{c}_{({\bf 10})} \subset C_{({\bf 10})}$ 
itself, and the other is a collection of points of $C_{({\bf 10})}$ 
that is projected to $\bar{c}_{({\bf 10})}$, but are not on 
the zero section of $\mathbb{K}_S$. The first component has 
a coefficient $\lambda (5-1) = 4 \lambda$, and the second one 
$-\lambda$. 
In order to provide a more intuitive understanding of this divisor, 
let us focus on a region of $C_{({\bf 10})}$ around a point of 
$E_6$-type codimension-3 singularity. 
Let the local defining equation of the spectral surface $C_{({\bf 10})}$ 
be 
\begin{equation}
 \xi^2 + \tilde{a}_4 \xi + \tilde{a}_5 \simeq 0, 
\label{eq:spec-surf-E6}
\end{equation}
where $\tilde{a}_{4,5} = a_{4,5}/a_3$, and $\xi$ is a fiber coordinate 
of $\mathbb{K}_S$. $(\tilde{a}_4, \tilde{a}_5)$ can be chosen 
as a set of local coordinates of $S$, while it is more appropriate 
to take $(\xi, \tilde{a}_4)$ as the local coordinates on $C_{({\bf 10})}$ 
(\cite{Hayashi-1, Hayashi-2}). 
The divisor $\gamma$ in $C_{({\bf 10})}$ is described in this 
local region in this set of local coordinates as 
\begin{eqnarray}
\gamma_{FMW} & = & \lambda 
   \left( 5 \; {\rm div} (\xi) - \pi_{C_{(10)}}^{-1}({\rm div} \;
    \tilde{a}_5)\right)
   \nonumber \\ 
 & = & \lambda \left( 5 \; {\rm div}(\xi) - {\rm div}(\xi (\xi + \tilde{a}_4)) 
  \right) =
  \lambda \left( 4 {\rm div}(\xi) - {\rm div}(\xi + \tilde{a}_4) \right). 
\end{eqnarray}
The first component corresponds to ${\rm div} \; \xi$, and the second 
one to ${\rm div} \; (\xi + \tilde{a}_4)$. 
See Figure~\ref{fig:divisorsATtypeA}.
%%%%%%%%%%%%%%%%%%%%%%%%%%%%%%%%%%%%%%%%%%%%%%%%%%%%
\begin{figure}[tb]
 \begin{center}
\begin{tabular}{ccc}
   \includegraphics[width=.35\linewidth]{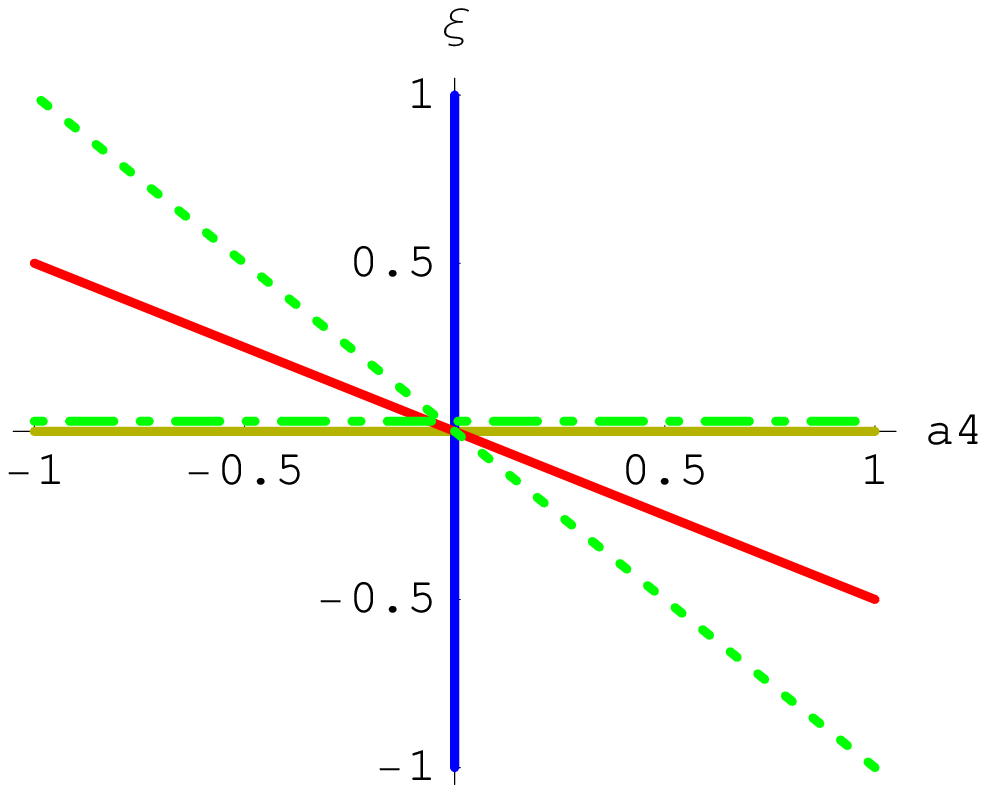}    
& & 
   \includegraphics[width=.35\linewidth]{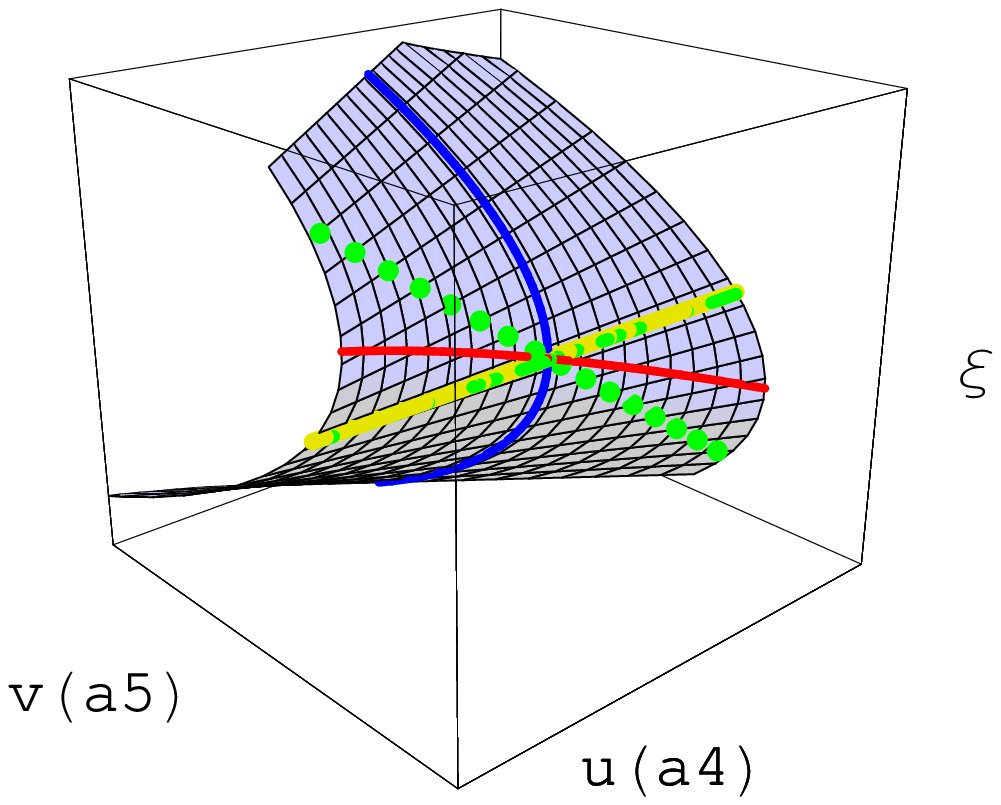}     \\
(a) & & (b)
\end{tabular}
\caption{\label{fig:divisorsATtypeA} (color online) 
Divisors on the spectral surface $C_{({\bf 10})}$ near 
a type (a) point (or an $E_6$-type point). $(\xi, \tilde{a}_4)$ are 
chosen as a set of local coordinates on $C_{({\bf 10})}$. 
The matter curve $\bar{c}_{({\bf 10})}$ is the $\xi = 0$ line 
(yellow). The $(2\xi + \tilde{a}_4) = 0$ line (red) is the 
ramification divisor  
$r_{({\bf 10})}$ of $\pi_{C_{({\bf 10})}}: C_{({\bf 10})} 
\rightarrow S$, and the curve $D$ on $C_{({\bf 10})}$ is locally given by 
$\tilde{a}_4 = 0$ (blue) near the $E_6$-type point,  
which is projected to $\bar{c}_{(\bar{\bf 5})}$.
The divisor $\gamma$ in (\ref{eq:gamma-FMW}) (green) consists of two 
components, one along $\bar{c}_{({\bf 10})}$ (dash-dot) and the 
other at $(\xi + \tilde{a}_4) = 0$ (dotted).
See the appendix B.2 of \cite{Hayashi-1}, if necessary. 
The right figure (b) shows the spectral surface in the total space 
of $K_S$, and the spectral surface is unfolded and presented in 
the left figure (a) with the local coordinates $(\tilde{a}_4,\xi)$.
In the right figure, it may be easy to see that the spectral surface 
is ramified indeed at the ramification divisor $r$, the green-dotted 
component of $\gamma$ is projected to the matter curve $\bar{c}_{({\bf
  10})}$ at $v = \tilde{a}_5 = 0$, and $D$ (blue) to the matter curve 
$\bar{c}_{(\bar{\bf 5})}$ at $u = \tilde{a}_4 = 0$.}
 \end{center}
\end{figure}
%%%%%%%%%%%%%%%%%%%%%%%%%%%%%%%%%%%%%%%%%%%%%%%%%%%%
The first component div($\xi$) is right on the matter curve 
$\bar{c}_{({\bf 10})}$, 
while the second one % component div($\xi+\tilde{a}_4$) 
intersects transversally with the matter 
curve $\overline{c}_{(10)}$ at the $E_6$-type singularity point 
$(\xi,\tilde{a}_4) = (0,0)$,  and hence at 
$(\tilde{a}_4, \tilde{a}_5) = (0,0)$ on $S$. 

When a divisor $D$ of $C_{({\bf 10})}$ is restricted on 
$\bar{c}_{({\bf 10})} \subset C_{({\bf 10})}$, the divisor $D$
becomes a collection of intersection points of $D$ and $\bar{c}_{({\bf 10})}$
(with the multiplicity of an intersection as the coefficient of the 
intersection point). The collection of these points defines a divisor 
on the matter curve $\bar{c}_{({\bf 10})}$. This definition of 
restriction is readily applied to the second component div($\xi+\tilde{a}_4$) 
of $\gamma$, but not to the first one div($\xi$), 
because the intersection point or the multiplicity of the intersection 
is not well-defined, when $D={\rm div}(\xi)$ is restricted onto 
$\bar{c}_{({\bf 10})}={\rm div}(\xi)$ itself.  
Thus, we replace $\gamma$ by another divisor of 
$C_{({\bf 10})}$ that is linearly equivalent to $\gamma$, and move the first 
component $4\lambda\bar{c}_{({\bf 10})}$ of $\gamma$ away 
from the matter curve $\bar{c}_{({\bf 10})}$. 
Let us take an arbitrary holomorphic section $\psi$ of 
${\cal O}(5K_S + \eta)$ that is different from $a_5$. 
Then $a_5/\psi$ is a rational function on $S$.
We define a divisor $\gamma'$ by\footnote{$\psi$ is not 
a rational function but a holomorphic section, and it is 
not conventional to use such a notation as ${\rm div} \; \psi$.
Its meaning will be clear, however. That is the zero locus of 
the section $\psi$. We will use this notation in this article.} 
\begin{equation}
 \gamma' = \gamma_{FMW} - 4 \lambda \; {\rm div}(a_5/\psi)
  = \lambda \left( - 5 \; {\rm div}(\xi + \tilde{a}_4)
                   + 4 \; {\rm div} \; \psi \right),
\end{equation}
and use this one instead of $\gamma$. Now 
\begin{equation}
 \gamma'|_{\bar{c}_{({\bf 10})}} = - 5 \lambda \; p_{E6} 
 + 4 \lambda \; {\rm div} \; \psi|_{\bar{c}_{({\bf 10})}}.
\end{equation}
Now all the components of the divisors on the matter curve 
$\bar{c}_{({\bf 10})}$ are understood.

The $E_6$-type singularity points $p_{E6}$ on $\bar{c}_{({\bf 10})}$ are 
characterized as the zero loci of $a_4$ on $\bar{c}_{({\bf 10})}$. 
Therefore, they give the divisor on the matter curve for the line bundle 
${\cal F}_{({\bf 10})}$ 
\begin{eqnarray}
 K_S|_{\bar{c}_{({\bf 10})}} + \frac{1}{2} p_{E6} 
  + \gamma_{FMW}|_{\bar{c}_{({\bf 10})}} & \sim & 
 K_S|_{\bar{c}_{({\bf 10})}} + \left( \frac{1}{2} - 5 \lambda \right) p_{E6} 
  + 4 \lambda \; {\rm div} \; \psi|_{\bar{c}_{({\bf 10})}}, \nonumber \\ 
 & = & \left(K_S + {\rm div} (a_4^{1/2 - 5 \lambda} \psi^{4 \lambda} )
       \right)|_{\bar{c}_{({\bf 10})}} 
   \equiv D_{({\bf 10})}|_{\bar{c}_{({\bf 10})}} , 
\end{eqnarray}
which % and ${\cal F}_{({\bf 10})}$ 
can be regarded as a divisor $D_{({\bf 10})}$ 
on $S$ restricted onto the matter curve $\bar{c}_{({\bf 10})}$. 
%
%\begin{equation}
% D_{({\bf 10})} \equiv K_S +
%  {\rm div} \left(a_4^{1/2 - 5 \lambda} \psi^{4\lambda} \right)
% \end{equation}
%

Since the sheaf cohomology group of our interest is of the form 
\begin{equation}
 H^0 \left( \bar{c}_{({\bf 10})}; {\cal F}_{({\bf 10})} \right) = 
 H^0 \left( \bar{c}_{({\bf 10})} ; {\cal O}_{\bar{c}_{({\bf 10})}} (
   D_{({\bf 10})}|_{\bar{c}_{({\bf 10})}} ) \right),
\end{equation}
and since $\bar{c}_{({\bf 10})}$ is a divisor of a complex surface $S$, 
we can use the short exact sequence 
\begin{equation}
0 \rightarrow {\cal O}_S(-\bar{c}_{({\bf 10})} + D_{({\bf 10})} ) 
  \rightarrow {\cal O}_S(D_{({\bf 10})} ) 
  \rightarrow {\cal O}_{\bar{c}_{({\bf 10})}} (D_{({\bf 10})}|_{\bar{c}_{({\bf 10})}} )
 \rightarrow 0
\label{eq:short-10}
\end{equation}
in calculating the cohomology group of our interest. 
The long exact sequence of their cohomology groups is 
\begin{equation}
\vcenter{\xymatrix@R=10pt@M=4pt@H+=22pt{
0 \ar[r] &
  H^0 \left( S; {\cal O}(D_{({\bf 10})}-\bar{c}_{({\bf 10})}) \right) \ar[r] &
  H^0 \left( S; {\cal O}(D_{({\bf 10})} ) \right) \ar[r] & % \ar@{->>}[r] & 
  H^0 \left( \bar{c}_{({\bf 10})} ; {\cal O}( D_{({\bf 10})}|_{\bar{c}_{ ({\bf 10})}} ) 
      \right)
 \ar`[rd]^<>(0.5){}`[l]`[dlll]`[d][dll] & 
\\
 &    
  H^1 \left( S; {\cal O}(D_{({\bf 10})}-\bar{c}_{({\bf 10})}) \right) \ar[r] & 
  H^1 \left( S; {\cal O}(D_{({\bf 10})} ) \right) \ar[r] & % \ar@{->>}[r] & 
  H^1 \left( \bar{c}_{({\bf 10})} ; {\cal O}( D_{({\bf 10})}|_{\bar{c}_{ ({\bf 10})}} ) 
      \right)  
  \ar`[rd]^<>(0.5){}`[l]`[dlll]`[d][dll] & 
 \\
 &    
  H^2 \left( S; {\cal O}(D_{({\bf 10})}-\bar{c}_{({\bf 10})}) \right) \ar[r] & 
  H^2 \left( S; {\cal O}(D_{({\bf 10})} ) \right) \ar[r] & % \ar@{->>}[r] & 
 0.&  
}}
\label{eq:long-10}
\end{equation}
Thus, the cohomology groups on the matter curve can be obtained from 
the line-bundle valued cohomology groups on $S$. 

Let us now take a specific value of $\lambda$. We will 
use $\lambda = -1/2$ in order to minimize the net chirality, 
though it does not give the realistic number of the generations  
(\cite{Curio})
\begin{equation}
 N_{\rm gen} = \gamma|_{\bar{c}_{({\bf 10})}} \cdot \bar{c}_{({\bf 10})} 
 = - \lambda \eta \cdot (5K_S + \eta) = - \lambda (17 H) \cdot (2H)
 \rightarrow 17. 
\label{eq:Ngen-10}
\end{equation}
Models with $N_{\rm gen}=17$ are certainly not ``realistic'', but 
we will keep carrying out this calculation, because 
this is only for illustration of the basic techniques, and the generalization 
or application to other cases will be straightforward.

When $\lambda = -1/2$, 
\begin{eqnarray}
 D_{({\bf 10})} & \sim & K_S + 3 (4K_S + \eta) -2 (5K_S+\eta) \sim 8H, \\
-\bar{c}_{({\bf 10})} + D_{({\bf 10})} & \sim & 6H.
\end{eqnarray}
Since $h^0(\P^2; {\cal O}(d H) ) = {}_{d+2} C_2$ (for $d \geq 0$), 
and $h^1(\P^2; {\cal O}(dH)) = h^2(\P^2; {\cal O}(dH))= 0$, 
we find that 
\begin{equation}
 H^1 \left( \bar{c}_{({\bf 10})}; {\cal F}_{({\bf 10})} \right) 
 =  H^1 \left( \bar{c}_{({\bf 10})}; 
            {\cal O}(D_{({\bf 10})}|_{\bar{c}_{({\bf 10})}}) 
     \right) \simeq 0,
\end{equation}
that is, there is no massless chiral multiplets in the 
$\overline{\bf 10}$ representation of $\SU(5)_{\rm GUT}$, and 
\begin{equation}
 H^0 \left( \bar{c}_{({\bf 10})}; 
            {\cal O}(D_{({\bf 10})}|_{\bar{c}_{({\bf 10})}}) \right) \simeq 
 {\rm Coker}\left[ 
  H^0 \left( S; {\cal O}(D_{({\bf 10})}-\bar{c}_{({\bf 10})}) \right) \rightarrow
  H^0 \left( S; {\cal O}(D_{({\bf 10})} ) \right)
\right].
\label{eq:10repr-ExVI}
\end{equation}
Since the map $H^0 \left( S; {\cal O}(D_{({\bf 10})}-\bar{c}_{({\bf 10})}) 
\right) \rightarrow H^0 \left( S; {\cal O}(D_{({\bf 10})} ) \right)$ 
on the right hand side of (\ref{eq:10repr-ExVI}) is 
an injective map from the ${}_8 C_2 = 28$-dimensional space 
to the ${}_{10} C_2 = 45$-dimensional space, 
the cokernel is a $17$-dimensional space, which is 
the same as we expected from (\ref{eq:Ngen-10}).

Our goal here in this section, however, is not just to 
obtain the number of independent massless fields in 
representations ${\bf 10}$ and $\overline{\bf 10}$ separately.
As long as we use the form of flux (\ref{eq:gamma-FMW}) on a 
spectral surface of an $E_8$ Higgs bundle globally defined on $S$, 
what we have done so far is not different from the corresponding 
analysis in Heterotic string compactifications on elliptically fibered 
Calabi--Yau 3-folds, and the necessary techniques have been well established.  
We proceed further, in this article, and determine (holomorphic) 
{\it wavefunctions} of the independent massless fields on their matter curves. 
In F-theory compactifications, a zero-mode wavefunction can be 
described on a complex curve (or on the complex surface $S$); we do not 
have to deal with a wavefunction on the complex 3-fold base $B_3$ 
like on a Calabi--Yau 3-fold in a Heterotic compactification,  
and this makes our task easier.

Let us describe $H^0(S; {\cal O}(D_{({\bf 10})}))$ and 
$H^0(S; {\cal O}(-\bar{c}_{({\bf 10})}+D_{({\bf 10})}))$ in the way 
we explained with $X = S$ in section \ref{ssec:review-hol-wave}. 
For $S = \P^2$, we choose an open covering of $S$ as 
$\{ U_{\check{S}}, U_{\check{T}}, U_{\check{U}} \}$, where 
$ U_{\check{S}} = \{ [S: T: U] \in \P^2 | S \neq 0 \}$, and 
$U_{\check{T}}$ and $U_{\check{U}}$ are defined similarly. 
A Cartier divisor description of $D_{({\bf 10})}$ is given by 
\begin{equation}
 \left(U_{\check{S}}, \frac{a_4^3}{STU\psi^2 }\frac{1}{S^{8}} \right), \quad 
 \left(U_{\check{T}}, \frac{a_4^3}{STU\psi^2 }\frac{1}{T^{8}} \right), \quad 
 \left(U_{\check{U}}, \frac{a_4^3}{STU\psi^2 }\frac{1}{U^{8}} \right), 
\end{equation}
whereas for a Cartier divisor description of 
$-\bar{c}_{({\bf 10})} + D_{({\bf 10})}$, 
a rational function for it on each patch is given 
by the rational function for $D_{({\bf 10})}$ shifted by $a_5$ 
on the same patch, because $\bar{c}_{({\bf 10})}$ is 
the zero locus of $a_5$, and thus 
\begin{equation}
\left(U_{\check{S}}, \frac{a_4^3}{STU a_5 \psi^2}\frac{1}{S^6} \right), \quad 
\left(U_{\check{T}}, \frac{a_4^3}{STU a_5 \psi^2}\frac{1}{T^6} \right), \quad 
\left(U_{\check{U}}, \frac{a_4^3}{STU a_5 \psi^2}\frac{1}{U^6} \right).
\end{equation}
Note that $a_4 \in \Gamma(S; {\cal O}(4K_S + \eta)) = 
\Gamma(\P^2; {\cal O}(5H))$ is regarded as a homogeneous function of degree 
five of the homogeneous coordinates $[S:T:U]$ of $\P^2$, 
and $a_5, \psi \in \Gamma(S; {\cal O}(5K_S + \eta))$ 
as homogeneous functions of degree two. 
The transition functions for the line bundle ${\cal O}_S(D_{({\bf 10})})$ 
are given by 
$g_{\check{S}\check{T}}=(T/S)^{8}$ on $U_{\check{S}}\cap{U}_{\check{T}}$, 
$g_{\check{T}\check{U}}=(U/T)^{8}$ on ${U}_{\check{T}}\cap{U}_{\check{U}}$, 
and $g_{\check{U}\check{S}}=(S/U)^{8}$ on ${U}_{\check{U}}\cap{U}_{\check{S}}$. 
For ${\cal O}_S(-\bar{c}_{({\bf 10})} + D_{({\bf 10})})$, they are given by 
$g_{\check{S}\check{T}}=(T/S)^{6}$ on ${U}_{\check{S}}\cap{U}_{\check{T}}$, 
$g_{\check{T}\check{U}}=(U/T)^{6}$ on ${U}_{\check{T}}\cap{U}_{\check{U}}$, 
and $g_{\check{U}\check{S}}=(S/U)^{6}$ on ${U}_{\check{U}}\cap{U}_{\check{S}}$. 

A global holomorphic section of a line bundle ${\cal O}(D)$ 
corresponds to a rational function $\tilde{f}$ on $S$ for the divisor $D$; 
for $-\bar{c}_{({\bf 10})} + D_{({\bf 10})}$ and 
$D_{({\bf 10})}$, their rational functions are  of the forms  
\begin{equation}
\tilde{f} = F^{(6)} a_5 \frac{STU \psi^2}{a_4^3} \quad {\rm and} \quad 
\tilde{f} = F^{(8)}\frac{STU \psi^2}{a_4^3}, 
\end{equation}
respectively, where $F^{(6)}$ and $F^{(8)}$ are homogeneous functions 
on $\P^2$ of degree 6 and 8, respectively. 
$H^0(S; {\cal O}(-\bar{c}_{({\bf 10})} + D_{({\bf 10})}))$ is regarded 
naturally as a subset of $H^0(S; {\cal O}(D_{({\bf 10})}))$, because 
$\bar{c}_{({\bf 10})}$ is effective; to be more explicit, $a_5 F^{(6)}$ 
can be regarded as a special form of $F^{(8)}$. 
% \kaku{Since a zero mode wave function on the matter curve 
% $\overline{c}_{(10)}$ is an element of 
% $H^0 \left( \bar{c}_{({\bf 10})}; 
% {\cal O}(D_{({\bf 10})}|_{\bar{c}_{({\bf 10})}}) \right)$, 
% using (\ref{eq:10repr-ExVI}), it can be identified with 
% $\tilde{f} = (F^{(8)} \; {\rm mod} \; a_5 F^{(6)})(STU \psi^2/a_4^3)$;} 
Since the zero mode matter fields are given by (\ref{eq:10repr-ExVI}), 
the zero mode wavefunctions are identified with 
$\tilde{f} = (F^{(8)} \; {\rm mod} \; a_5 F^{(6)})(STU \psi^2/a_4^3)$;  
this is quite reasonable, because this says that only $F^{(8)}$ on 
$\bar{c}_{({\bf 10})}$ -- on the locus $a_5 = 0$ -- is relevant. 
In the description using the trivialization patches, 
zero modes in the representation ${\bf 10}$ 
have wavefunctions on $\bar{c}_{({\bf 10})}$ 
given by 
% \kaku{A holomorphic section of the vector bundle 
% ${\cal O}(D_{({\bf 10})}|_{\bar{c}_{({\bf 10})}})$
% gives a zero mode in the representation ${\bf 10}$, and   
% as explained in section \ref{ssec:review-hol-wave}, 
% using one of the above rational functions and the trivialization patches, 
% it can be given on each of the patches 
% $U_{\check{S}} \cap \bar{c}_{({\bf 10})}$, 
% $U_{\check{T}} \cap \bar{c}_{({\bf 10})}$ 
% and $U_{\check{U}} \cap \bar{c}_{({\bf 10})}$ by  
% }
%
\begin{equation}
\tilde{f}_{\check{S}} = \frac{F^{(8)} \; {\rm mod} \; 
   a_5 F^{(6)}}{S^{8}}, \quad 
\tilde{f}_{\check{T}} = \frac{F^{(8)} \; {\rm mod} \; 
   a_5 F^{(6)}}{T^{8}}, \quad 
\tilde{f}_{\check{U}} = \frac{F^{(8)} \; {\rm mod} \; a_5 F^{(6)}}{U^{8}}.
\label{eq:10repr-wavefcn}
\end{equation} 
Classes of holomorphic functions on $U_{\check{S}}$, $U_{\check{T}}$ and 
$U_{\check{U}}$ above correspond to uniquely defined holomorphic functions 
on $U_{\check{S}} \cap \bar{c}_{({\bf 10})}$, 
$U_{\check{T}} \cap \bar{c}_{({\bf 10})}$ 
and $U_{\check{U}} \cap \bar{c}_{({\bf 10})}$.   
These wavefunctions do not depend on the choice of $\psi$, but on 
its divisor class. One can arbitrarily choose 
independent ones among $F^{(8)} \; {\rm mod} \; a_5 F^{(6)}$ to specify 
a basis in the $N_{\rm gen} = 45 - 28 = 17$ dimensional vector space of 
the zero modes in the representation ${\bf 10}$. 

%%%%%%%%%%%%%%%%%%%%%%%%%%%%%%%%%%%%%%%%%%%%%%%%%%%%%%%%%%
\subsubsection{Zero Mode Wavefunctions in the Representation 
$\bar{\bf 5}$--${\bf 5}$}
%%%%%%%%%%%%%%%%%%%%%%%%%%%%%%%%%%%%%%%%%%%%%%%%%%%%%%%%%%

Let us now move on to calculate the wavefunctions of zero modes 
in the representation $\bar{\bf 5}$--${\bf 5}$. 
The vector space of chiral multiplets in the representation 
$\bar{\bf 5}$ is \cite{Penn5, BMRW, Hayashi-1}
\begin{equation}
 H^0 \left( \tilde{\bar{c}}_{(\bar{\bf 5})}; 
                 \widetilde{\cal F}_{(\bar{\bf 5})}\right) = 
 H^0 \left( \tilde{\bar{c}}_{(\bar{\bf 5})}; 
  {\cal O} \left( 
        \tilde{\imath}^*_{(\bar{\bf 5})} K_S + \frac{1}{2} p_{A6} 
        + \gamma|_{\tilde{\bar{c}}_{({\bar{\bf 5})}}}
           \right) 
    \right);
\label{eq:H0-5bar-repr}
\end{equation}
Here, $\nu_{\bar{c}_{({\bar{\bf 5})}}}: \tilde{\bar{c}}_{(\bar{\bf 5})} 
\rightarrow \bar{c}_{({\bar{\bf 5})}}$ 
resolves 
% \kaku{the double points of all the $D_6$-type codimension-3 singularities 
% on the matter curve $\bar{c}_{({\bar{\bf 5})}}$,} 
all the double points of $\bar{c}_{({\bar{\bf 5})}}$ at the 
$D_6$-type codimension-3 singularities, and 
$\tilde{\imath}_{(\bar{\bf 5})}=i_{(\bar{\bf 5})}
\circ \nu_{\bar{c}_{(\bar{\bf 5})}}:
\tilde{\bar{c}}_{(\bar{\bf 5})} \rightarrow S$ pulls $K_S$ back from $S$ 
to the covering matter curve $\tilde{\bar{c}}_{(\bar{\bf 5})}$. 
$p_{A6}$ denotes the divisor consisting of all the $A_6$-type singularity 
points on $\bar{c}_{(\bar{\bf 5})}$, which is given by the zero locus of 
$\tilde{R}^{(5)}_{\rm mdfd}|_{\bar{c}_{(\bar{\bf 5})}}$ except the $D_6$-type 
singularity points (which are on the locus $a_5=0$) \cite{Hayashi-1}. 
Strictly speaking, $p_{A6}$ in the divisor 
$\tilde{\imath}^*_{(\bar{\bf 5})} K_S + \frac{1}{2} p_{A6} 
+\gamma|_{\tilde{\bar{c}}_{({\bar{\bf 5})}}}$ in (\ref{eq:H0-5bar-repr})
should be its pull back to the covering matter curve 
$\tilde{\bar{c}}_{(\bar{\bf 5})}$, not on the matter curve 
$\bar{c}_{(\bar{\bf 5})}$.
But the only difference between the two curves is only around the 
$D_6$-type singularity points, and this should not cause any problems. 
The last component ``$\gamma|_{\tilde{\bar{c}}_{({\bar{\bf 5})}}}$'' 
of the divisor is meant to be\footnote{
This is, by no means, something like ``$\gamma$ on $C_{({\bf 10})}$ 
restricted on 
$\tilde{\bar{c}}_{(\bar{\bf 5})}$''.
$\tilde{\bar{c}}_{(\bar{\bf 5})}$ is not a divisor of 
$C_{({\bf 10})}$. The notation should not be taken literally here.} 
the contribution coming from the 4-form flux $G^{(4)}$ 
in the Calabi--Yau 4-fold $X$. 
It is a 2-form on $\tilde{\bar{c}}_{(\bar{\bf 5})}$ and  
should be obtained by integrating the 4-form flux over 
the vanishing 2-cycle fibered over the 
$\tilde{\bar{c}}_{(\bar{\bf 5})}$ \cite{Hayashi-1}. 
Needless to say, one and the same $G^{(4)}$ on $X$ should be used 
for the calculation of wavefunctions both in the representation 
${\bf 10}$--$\overline{\bf 10}$  and in $\bar{\bf 5}$--${\bf 5}$, 
or otherwise, the net chiralities for the both representations are not 
guaranteed to be the same.\footnote{In perturbative Type IIB string 
compactifications with D7-branes in a Calabi--Yau 3-fold, 
one usually begins with specifying vector bundles 
(with a shift of $K_S^{1/2}$ due to Freed--Witten anomaly) separately 
on individual holomorphic 4-cycles (where D7-branes are wrapped). 
In this way, although all of these ``vector bundles'' are supposed 
to descend from one and the same 4-form flux $G^{(4)}$ in F-theory language, 
the common origin of these ``bundles'' is not guaranteed. 
This is why the Bianchi identities of \kaku{the} Ramond--Ramond fields 
(especially $d G^{(3)}$ and $d \tilde{F}^{(5)}$), or equivalently the 
Ramond--Ramond tadpole cancellation conditions, need to be imposed 
later on, in order to make sure the common origin of the ``bundles''.}
It seems formidable at this moment 
% \kaku{to derive 
% $\bar{c}_{({\bf 10})}$ and $\tilde{\bar{c}}_{(\bar{\bf 5})}$ 
% directly from a single 4-form flux $G^{(4)}$ on $X$.} 
to identify possible variety of $G^{(4)}$ in 4-fold $X$, and find 
out how it emerges on $\bar{c}_{({\bf 10})}$ and 
$\tilde{\bar{c}}_{(\bar{\bf 5})}$.} 

When an $E_8$ Higgs bundle is defined globally on the GUT divisor $S$, 
with a divisor $\gamma$ given on a spectral surface $C_{({\bf 10})}$ for 
the representation ${\bf 10}$, however, 
there is a simple prescription \cite{Hayashi-1}: 
\begin{equation}
 \gamma|_{\tilde{\bar{c}}_{({\bar{\bf 5})}}}
 =\tilde{\pi}_{D*} \left(\gamma|_{D} \right).
\label{eq:pi-gamma-D}
\end{equation}
Here, $D$ is a curve in $C_{({\bf 10})}$, and 
$\tilde{\pi}_{D}: D\to\tilde{\bar{c}}_{(\bar{\bf 5})}$.
See \cite{Penn5, BMRW, Hayashi-1} for more details. 
For example, when $\gamma$ is given in the form (\ref{eq:gamma-FMW}) 
as in \cite{FMW},\footnote{It has been shown explicitly in this case, 
by using purely the F-theory language (\ref{eq:H0-10-repr}, 
\ref{eq:H0-5bar-repr}, \ref{eq:gamma-FMW}, \ref{eq:pi-gamma-D}), 
that the net chirality in the $\bar{\bf 5}$--${\bf 5}$ sector is 
confirmed to be the same as the one in the 
${\bf 10}$--$\overline{\bf 10}$ sector \cite{Hayashi-1} (see 
also \cite{BMRW}).} the divisor
$\tilde{\pi}_{D*}\left(\gamma|_{D}\right)$  
on the covering matter curve $\tilde{\bar{c}}_{(\bar{\bf 5})}$ consists of 
all the $E_6$-type and resolved $D_6$-type singularity points on 
$\tilde{\bar{c}}_{(\bar{\bf 5})}$, because the support of $\gamma$ 
in (\ref{eq:gamma-FMW}) is only in the fiber of $\bar{c}_{({\bf 10})}$; 
each of the $E_6$-type singularity points contributes to the 
divisor with the coefficient $3\lambda$,\footnote{
Near a $E_6$-type singularity, $\gamma$ and $D$ intersect at 
$(\tilde{a}_4, \xi) = (0,0)$ in Figure~\ref{fig:divisorsATtypeA}. 
The dashed--dotted irreducible piece $\xi=0$ in $\gamma$ 
comes with the coefficient $4\lambda$, and intersects $D:\tilde{a}_4=0$ 
transversally, while the dotted piece $\xi+\tilde{a}_4=0$ in $\gamma$ 
has the coefficient $-\lambda$, and also intersects $D$ transversally. 
Thus, $\gamma|_D$ is a point at $(\tilde{a}_4, \xi)=(0,0)$ (locally) 
with the coefficient $4\lambda + (-\lambda) = 3\lambda$.} 
and so does each of the resolved $D_6$-type points with the coefficient 
$-2 \lambda$.\footnote{
The spectral surface $C_{({\bf 10})}$ is a 5-fold 
cover on $S$, and  at each point of $S$, there are five points 
$p_{i,j,k,l,m}$ on $C_{({\bf 10})}$. 
At a $D_6$-type singularity point in $S$, the five points 
are projected down to the $D_6$-type point by $\pi_{C_{({\bf 10})}}$. 
Among them, one of them, say, $p_m$ is in the zero section of $K_S$, 
while the four others satisfy $p_i \boxplus p_j = 0$ and 
$p_k \boxplus p_l = 0$, 
where $\boxplus$ means summation in the fiber vector space of 
$K_S \rightarrow S$. The curve $D$ passes through $p_{i,j,k,l}$, 
but not through $p_m$. The divisor $\gamma$ on $C_{({\bf 10})}$ 
intersects $D$ at all of $p_{i,j,k,l}$, because of the second term 
of (\ref{eq:gamma-FMW}). Thus, each of the four points 
$p_{i,j,k,l}$ at the intersection of $\gamma$ with $D$ 
contributes to the divisor $\gamma|_{D}$ 
with the coefficient $-\lambda$. Since the map $\tilde{\pi}_D$, 
however, sends $p_i$ and $p_j$ to one point 
and $p_k$ and $p_l$ to another in $\tilde{\bar{c}}_{(\bar{\bf 5})}$, 
each of the two points $\tilde{\pi}_D(p_i) = \tilde{\pi}_D(p_j)$ and 
$\tilde{\pi}_D(p_k) = \tilde{\pi}_D(p_l)$ in 
$\tilde{\bar{c}}_{(\bar{\bf 5})}$ 
contributes to $\tilde{\pi}_{D*}\left(\gamma|_D\right)$ 
with the coefficient $(-\lambda) + (-\lambda) = -2\lambda$.}
Note that the number of the resolved $D_6$-type points is 
twice as many as the one of the original $D_6$-type points.  

Now the divisor specifying the line bundle on 
$\tilde{\bar{c}}_{(\bar{\bf 5})}$ is completely identified with 
a linear combination of collection of points on the curve. 
Let us use this explicit form of $\tilde{\cal F}_{(\bar{\bf 5})}$ 
to calculate the zero-mode wavefunctions. 
For the matter curve $\overline{c}_{({\bf 10})}$,
the calculation of wavefunctions eventually became 
%a problem of 
finding holomorphic sections of the line bundle on the GUT divisor $S$, 
because the line bundle ${\cal F}_{({\bf 10})}$ on the matter curve 
$\bar{c}_{({\bf 10})} \subset S$ can be regarded as 
${\cal O}(D_{({\bf 10})}|_{\bar{c}_{({\bf 10})}})$, using the divisor 
$D_{({\bf 10})}$ on $S=\P^2$. 
%In the $\bar{\bf 5}$--${\bf 5}$ representation sector, 
For the matter curve $\tilde{\bar{c}}_{(\bar{\bf 5})}$, 
however, the situation is a little different.
The curve $\tilde{\bar{c}}_{(\bar{\bf 5})}$ is not a divisor of $S$, 
but is regarded as a divisor of the surface $\tilde{S}$ obtained 
by blowing up 
all the $D_6$-type points of $S$, because the covering matter curve 
$\tilde{\bar{c}}_{(\bar{\bf 5})}$ is obtained by resolving the double point 
singularities at all the $D_6$-type points on 
$\bar{c}_{(\bar{\bf 5})} \subset S$.  
\kaku{Thus, if the divisor 
$\tilde{\imath}^*_{(\bar{\bf 5})} K_S + \frac{1}{2} p_{A6}
+\gamma|_{\tilde{\bar{c}}_{({\bar{\bf 5})}}}$ 
on the curve $\tilde{\bar{c}}_{(\bar{\bf 5})}$} 
in (\ref{eq:H0-5bar-repr}) 
can be regarded as the restriction of 
a divisor of the ambient space $\tilde{S}$, then the same techniques 
as for the matter curve $\bar{c}_{({\bf 10})}$ can be used in calculating 
the wavefunctions of independent zero modes on the curve 
$\tilde{\bar{c}}_{(\bar{\bf 5})}$. 

Let us see that this is indeed possible. 
\kaku{The blow-up $\nu_S: \tilde{S} \rightarrow S$ gives 
the exceptional divisors $E_{D6} = \sum_P E_{D6; P}$ 
at all the $D_6$-type points, where $P$ labels 
$\# D_6$ different $D_6$-type points in $S$.} 
Since the $D_6$-type points of $S$ are zeros of order one of both 
$\tilde{R}^{(5)}_{\rm mdfd}$ and $a_5$, 
the divisors $\nu_S^* ({\rm div} \tilde{R}^{(5)}_{\rm mdfd})$ and 
$\nu_S^* ({\rm div} \; a_5)$ contain $E_{D6}$ in addition to the 
proper transform of ${\rm div} \tilde{R}^{(5)}_{\rm mdfd}$ and 
${\rm div} \; a_5$: 
\begin{equation}
 \nu_S^* ({\rm div} \tilde{R}^{(5)}_{\rm mdfd}) = 
   \overline{{\rm div} \tilde{R}^{(5)}_{\rm mdfd}} + E_{D6}, \qquad 
 \nu_S^* ({\rm div} \; a_5) = \overline{{\rm div}\; a_5} + E_{D6}.
\end{equation}
The proper transforms $\overline{{\rm div} \tilde{R}^{(5)}_{\rm mdfd}}$ 
\kaku{and} 
$\overline{{\rm div} \; a_5}$ do not intersect the covering matter curve 
$\tilde{\bar{c}}_{(\bar{\bf 5})}$ in $E_{D6}$ for a generic choice of 
the complex structure. Thus, 
\begin{eqnarray}
 \tilde{\imath}^*_{(\bar{\bf 5})} K_S 
  + \frac{1}{2}p_{A6} + \tilde{\pi}_{D*} \gamma_{FMW} 
 & = & (\nu_S^* (K_S))|_{\tilde{\bar{c}}_{(\bar{\bf 5})}}
  +\frac{1}{2} \; 
   \overline{{\rm div} \tilde{R}^{(5)}_{\rm mdfd}}|_{\tilde{\bar{c}}_{(\bar{\bf 5})}}
  + 3\lambda \; \overline{{\rm div}\; a_5}|_{\tilde{\bar{c}}_{(\bar{\bf 5})}} 
  -2\lambda \; E_{D6}|_{\tilde{\bar{c}}_{(\bar{\bf 5})}}  \nonumber \\
 & = & \left. \left( \nu^*_S (K_S) + \frac{1}{2} \; 
   \overline{{\rm div} \tilde{R}^{(5)}_{\rm mdfd}}
  + 3\lambda \; \overline{{\rm div} \; a_5}
  -2\lambda \; E_{D6}\right) \right|_{\tilde{\bar{c}}_{(\bar{\bf 5})}}.
\end{eqnarray}
Therefore, by using a divisor 
\begin{eqnarray}
D_{(\bar{\bf 5})} & = & \nu^*_S ( K_S) + \frac{1}{2} \; 
   \overline{{\rm div} \tilde{R}^{(5)}_{\rm mdfd}}
  + 3\lambda \; \overline{{\rm div} \; a_5}
  -2\lambda \; E_{D6} \nonumber \\
  & = & \nu^*_S \left(
  K_S + \frac{1}{2} \; {\rm div} \tilde{R}^{(5)}_{\rm mdfd}
  +  3 \lambda \; {\rm div} \; a_5 \right) - \left(\frac{1}{2} 
  + 5 \lambda \right) E_{D6} 
\end{eqnarray}
on the ambient space $\tilde{S}$, 
the line bundle $\tilde{\cal F}_{(\bar{\bf 5})}$ on the curve 
$\tilde{\bar{c}}_{(\bar{\bf 5})}$ can be regarded as 
the restriction of the line bundle ${\cal O}_{\tilde{S}}(D_{(\bar{\bf 5})})$ 
of $S=\P^2$ onto the covering matter curve. 

Just like we did before for the matter curve $\bar{c}_{({\bf 10})}$ 
in the last subsection, we can use the short exact sequence 
\begin{equation}
 0 \rightarrow 
 {\cal O}_{\tilde{S}} (- \tilde{\bar{c}}_{(\bar{\bf 5})} + D_{(\bar{\bf 5})})
  \rightarrow 
 {\cal O}_{\tilde{S}} (D_{(\bar{\bf 5})}) \rightarrow 
 {\cal O}_{\tilde{\bar{c}}_{(\bar{\bf 5})}} (D_{(\bar{\bf 5})}|_{\tilde{\bar{c}}_{(\bar{\bf 5})}})
  \rightarrow 0.
\label{eq:short-5bar}
\end{equation}
Its cohomology long exact sequence is 
\begin{equation}
\vcenter{\xymatrix@R=10pt@M=4pt@H+=22pt{
0 \ar[r] &
  H^0 \left( \tilde{S}; {\cal O}(D_{(\bar{\bf 5})}-\bar{c}_{(\bar{\bf 5})}) \right) 
        \ar[r] &
  H^0 \left( \tilde{S}; {\cal O}(D_{(\bar{\bf 5})} ) \right) \ar[r] & 
                                                        % \ar@{->>}[r] & 
  H^0 \left( \tilde{\bar{c}}_{(\bar{\bf 5})} ; 
     {\cal O}( D_{(\bar{\bf 5})}|_{\tilde{\bar{c}}_{ (\bar{\bf 5})}} ) 
      \right)
 \ar`[rd]^<>(0.5){}`[l]`[dlll]`[d][dll] & 
\\
 &    
  H^1 \left( \tilde{S}; 
   {\cal O}(D_{(\bar{\bf 5})}-\tilde{\bar{c}}_{(\bar{\bf 5})}) \right) \ar[r] & 
  H^1 \left( \tilde{S}; {\cal O}(D_{(\bar{\bf 5})} ) \right) \ar[r] & 
                                                   % \ar@{->>}[r] & 
  H^1 \left( \tilde{\bar{c}}_{(\bar{\bf 5})} ; 
                {\cal O}( D_{(\bar{\bf 5})}|_{\tilde{\bar{c}}_{ (\bar{\bf 5})}} ) 
      \right)  
  \ar`[rd]^<>(0.5){}`[l]`[dlll]`[d][dll] & 
 \\
 &    
  H^2 \left( \tilde{S}; 
         {\cal O}(D_{(\bar{\bf 5})}-\tilde{\bar{c}}_{(\bar{\bf 5})}) \right) \ar[r] & 
  H^2 \left( \tilde{S}; {\cal O}(D_{(\bar{\bf 5})} ) \right) \ar[r] & 
         % \ar@{->>}[r] & 
 0.&  
}}
\label{eq:long-5bar}
\end{equation}
A zero mode in the representation $\bar{\bf 5}$ is an element of 
the third term 
$H^0 \left( \tilde{\bar{c}}_{(\bar{\bf 5})} ; 
{\cal O}( D_{(\bar{\bf 5})}|_{\tilde{\bar{c}}_{(\bar{\bf 5})}})\right)$ 
in (\ref{eq:long-5bar}), which can be expressed by using 
the cohomology groups of 
${\cal O}_{\tilde{S}}(D_{(\bar{\bf 5})})$ and 
${\cal O}_{\tilde{S}}(-\tilde{\bar{c}}_{(\bar{\bf 5})}+D_{(\bar{\bf 5})})$ 
on $\tilde{S}$.

Let us focus on the specific case $\lambda = -1/2$ once again.
We will first determine the number of independent zero modes in 
the representations $\bar{\bf 5}$ and ${\bf 5}$ , respectively, before 
we begin to discuss their wavefunctions. For this purpose, 
just the topological data 
\begin{eqnarray}
 D_{(\bar{\bf 5})} & \sim & (-3H) + \frac{24}{2}H - \frac{3}{2}(2H) + 2 E_{D6} 
      = 6H + 2 E_{D6}, \\
 -\tilde{\bar{c}}_{(\bar{\bf 5})} + D_{(\bar{\bf 5})} & \sim & 
      -(21H - 2E_{D6}) +(6H+2E_{D6}) = - 15 H + 4E_{D6}, 
\end{eqnarray}
are sufficient, where $H$ is the pull back of the hyperplane class of 
$S = \P^2$ to $\tilde{S}$, to yield 
\begin{equation}
 h^0(\tilde{S};{\cal O}(6H+2E_{D6})) = 28,  \qquad 
 h^0(\tilde{S}; {\cal O}(- 15 H+4E_{D6})) = 0, 
\end{equation}
and, with the help of the Serre duality, 
\begin{eqnarray}
 h^2(\tilde{S};{\cal O}(6H+2E_{D6})) & = & 
   h^0(\tilde{S}; {\cal O}(-9H -E_{D6})) = 0, \\
 h^2(\tilde{S};{\cal O}(-15H+4E_{D6})) & = &
   h^0(\tilde{S};{\cal O}(12H-3E_{D6})) = 0, 
\end{eqnarray}
where the fact that $\# D_6 = 16$ in Example VII
was used,\footnote{
A section of ${\cal O}(12H - 3E_{D6})$ is a homogeneous function 
of degree 12 on $S=\P^2$ and has zeros of order three at the 
$\# D_6$ points to be blown up.
There are ${}_{14} C_2 = 91$ monomials of degree 12 
in the homogeneous coordinates. Since a section of 
${\cal O}(12H - 3E_{D6})$ has a zero of order three at each point of $E_{D6}$, 
$6(=1+2+3)$ constraints are imposed on the 91 coefficients of the
monomials for each one of type $D_6$ points. Since the total number 
$6\times\#{D6}=96$ of the constraints is greater than the number of 
coefficients of the monomials for $\#D6=16$, there is no non-trivial 
element in $H^0(\tilde{S}; {\cal O}(12 H - 3 E_{D6}))$.} 
as seen in Table \ref{tab:top-data}. 
Since the index theorem for a divisor $D$ 
\begin{eqnarray}
\chi(\tilde{S};{\cal O}(D)) & \equiv & h^0(\tilde{S};{\cal O}(D))
-h^1(\tilde{S};{\cal O}(D))+h^2(\tilde{S};{\cal O}(D)) \\
 & = & \int_{\tilde{S}} {\rm ch}(D) {\rm td} (T\tilde{S})
\end{eqnarray}
gives  
\begin{equation}
\chi(\tilde{S}; {\cal O}(6H+2E_{D6})) = 12,
\qquad
\chi(\tilde{S};{\cal O}(-15H+4E_{D6})) = -5,
\end{equation}
one finds that the remaining 1st cohomology groups can also be 
determined as follows:
\begin{equation}
h^1(\tilde{S}; {\cal O}(6H+2E_{D6})) = 16, \qquad 
h^1(\tilde{S}; {\cal O}(- 15 H+4E_{D6})) = 5.
\end{equation}

We therefore conclude in this example (with $\lambda = -1/2$) that 
the chiral multiplets in the representation $\bar{\bf 5}$ fits into 
an exact sequence 
\begin{equation}
0 \rightarrow H^0(\tilde{S}; {\cal O}_{\tilde{S}}(D_{(\bar{\bf 5})}))
  \rightarrow
     H^0(\tilde{\bar{c}}_{(\bar{\bf 5})}; \tilde{\cal F}_{(\bar{\bf 5})}) 
  \rightarrow 
     H^{1}(\tilde{S}; {\cal O}(
            D_{(\bar{\bf 5})}-\tilde{\bar{c}}_{(\bar{\bf 5})}))
  \rightarrow H^{1}(\tilde{S}; \mathcal{O}(D_{(\bar{\bf 5})})  )   ,
 \label{eq:5bar}
\end{equation}
while the anti-chiral multiplets in the representation $\bar{\bf 5}$ 
% of $\SU(5)_{\rm GUT}$
are identified with 
\begin{equation}
 H^1(\tilde{\bar{c}}_{(\bar{\bf 5})};\tilde{\cal F}_{(\bar{\bf 5})}) 
   \simeq 
 {\rm Coker} \left[ 
    H^1 \left(\tilde{S}; {\cal O}(
          D_{(\bar{\bf 5})}-\tilde{\bar{c}}_{(\bar{\bf 5})})\right) 
 \rightarrow 
    H^1(\tilde{S}; {\cal O}(D_{(\bar{\bf 5})})) 
              \right].
\label{eq:5hc-ExVI}
\end{equation}
There are $28 + N$ massless chiral multiplets 
in the representation $\bar{\bf 5}$, 
and $(16-(5-N))$ massless multiplets in the representation ${\bf 5}$; 
here $N$ is the dimension of the kernel of the map 
\begin{equation}
H^{1}(\tilde{S}; \mathcal{O}(
  D_{(\bar{\bf 5})}-\tilde{\bar{c}}_{(\bar{\bf 5})})) \rightarrow
  H^{1}(\tilde{S}; \mathcal{O}(D_{(\bar{\bf 5})})). 
\label{eq:map-H1-H1-5bar}
\end{equation}
In this particular example, the net chirality is 
$N_{\rm gen} = (28+N)-(11+N)=17$, the same as in the 
${\bf 10}$--$\overline{\bf 10}$ sector, and $11+N$ pairs 
of extra chiral multiplets in the 
representation ${\bf 5}$--$\bar{\bf 5}$ are in the low-energy spectrum. 

A part of the chiral multiplets in the representation $\bar{\bf 5}$, 
are given by the restriction of 
the independent wavefunctions 
%of the 28 independent elements 
%chiral multiplets in 
$H^0 (\tilde{S}; {\cal O}(D_{(\bar{\bf 5})}))$
% in the $\bar{\bf 5}$ representation are given by 
%
\begin{equation}
 \tilde{f} = \nu_S^* \left( F^{(6)} 
    \frac{STU \sqrt{a_5^3}}{\sqrt{\tilde{R}^{(5)}_{\rm mdfd}}} 
                    \right) \in 
  H^0 \left(\tilde{S}; {\cal O}_{\tilde{S}} (D_{(\bar{\bf 5})}) \right),
  \label{eq:5bar_hol}
\end{equation}
of $H^0 (\tilde{S}; {\cal O}(D_{(\bar{\bf 5})}))$ onto the curve 
$\tilde{\bar{c}}_{(\bar{\bf 5})}$, 
where $F^{(6)}$ is one of the 28 homogeneous monomials of degree 6 on $\P^2$ 
for each of the wavefunctions.
In fact, since $h^0(\tilde{S};{\cal O}
   (-\tilde{\bar{c}}_{(\bar{\bf 5})}+D_{(\bar{\bf 5})})) =0$ 
in this case, the rational function $\tilde{f}$ can simply 
be restricted on 
$\tilde{\bar{c}}_{(\bar{\bf 5})} 
\subset \tilde{S}$, without taking a quotient. 
%  \kaku{to give a wavefunction in 
%  $H^0(\tilde{\bar{c}}_{(\bar{\bf 5})}; \tilde{\cal F}_{(\bar{\bf 5})})$.} 
In order to cast this rational function on $\tilde{S}$ 
into a description in terms of the trivialization patches, we need a 
Cartier-divisor description of $D_{(\bar{\bf 5})}$. The rational functions 
$\varphi_a$ on the patches $U'_{\check{S}, \check{T}, 
\check{U}}$---$U_{\check{S}, \check{T}, \check{U}}$ with the punctured 
$D_6$ points---can be chosen as $\sqrt{\tilde{R}^{(5)}/a_5^3}/(STU)$ 
divided by $S^6$, $T^6$ and $U^6$, respectively, 
(and pulled back by $\nu_S^*$).
% , in a region away from the $D_6$ singularity points. 
Thus, the coefficient functions $\tilde{f}_a$ \kaku{of the line bundle} 
on these patches are simply given by 
\begin{equation}
\nu^*_S\left(\frac{F^{(6)}}{S^6} \right), \qquad 
\nu^*_S\left( \frac{F^{(6)}}{T^6} \right) \quad {\rm and} \quad 
\nu^*_S\left(\frac{F^{(6)}}{U^6} \right),
\label{eq:5barrepr-wavefcn}
\end{equation}
respectively. 
In a patch $U_{a}$ including the exceptional curves 
$E_{D6} = \sum_P E_{D6; P}$, however, $\varphi_a$ must have the exceptional 
curves as zeros of order two. Thus, the coefficient function 
$\tilde{f}_a = \varphi_a \tilde{f}$ in the patch $U_{a}$ becomes zero 
at all the points on the exceptional curves 
$E_{D6;P}|_{\tilde{\bar{c}}_{(\bar{\bf 5})}}$\kesu{.}\kaku{, since 
$\tilde{f}$ in (\ref{eq:5bar_hol}) has no poles.} 
It is the values of the wavefunctions at these points that are used 
in the calculation of down-type and charged lepton Yukawa 
couplings,\footnote{
The holomorphic wavefunction $\tilde{f}_a$ in patches 
$U'_{\check{S},\check{T}, \check{U}}$ (where type $D_6$ points 
are punctured out) approaches a finite non-zero value toward one of 
the $D_6$-type points, but the wavefunction $\tilde{f}_a$ 
in a patch containing the corresponding exceptional curve 
becomes zero on the exceptional curves. 
As we will explain in the next section, physics is better understood 
in terms of unitary-frame wavefunctions, and a unitary-frame wavefunction 
vanishes wherever its wavefunction $\tilde{f}_a$ in this section vanishes. 
Thus, the unitary frame wavefunction also vanishes on the exceptional curves 
$E_{D6}$ in this example; 
the behavior of the holomorphic wavefunction $\tilde{f}_a$  
in the patch $U'_{\check{S}, \check{T}, \check{U}}$ cannot be used to 
infer the value of unitary frame wavefunction on the exceptional curves, 
because $E_{D6}$ are not covered by these patches.} 
and in this example, we found that all the 28 independent chiral 
multiplets $\{\tilde{f}_a\}$ in the representation $\bar{\bf 5}$ vanish at these 
points.\footnote{It should be kept in mind, however, 
that even when $\tilde{f}_a$ on $\tilde{\bar{c}}_{(\bar{\bf 5})}$ 
vanishes at a type $D_6$ point, the wavefunctions $(\psi,\chi)$ 
may be non-zero at points on $S$ around the $D_6$-type point. 
Reference \cite{HV-Nov08} also pointed out that even wavefunctions 
that vanish at a $D_6$-type point contribute to the down-type/charged 
lepton Yukawa matrices, although the contributions are somewhat 
suppressed.  Given the fact that the bottom-quark and tau-lepton 
Yukawa couplings are not as large as the top-quark Yukawa couplings 
(assuming not extremely large $\tan \beta$ in supersymmetric Standard 
Models), it should be remembered that this may not be a phenomenological 
problem. }  

The remaining $N$ independent chiral multiplets in the representation 
$\bar{\bf 5}$ are characterized by the kernel of the map 
\begin{equation}
H^{1}(\tilde{S}; \mathcal{O}(
  D_{(\bar{\bf 5})}-\tilde{\bar{c}}_{(\bar{\bf 5})})) \rightarrow
  H^{1}(\tilde{S}; \mathcal{O}(D_{(\bar{\bf 5})})). 
\label{eq:map-H1-H1-5bar-2}
\end{equation}
The existence of these elements is reasonable, because not all 
of the global holomorphic sections 
on the curve can be extended to global holomorphic sections over 
the surface $\tilde{S}$. 
For a global holomorphic sections $(U_a, \tilde{f}_a)$ on the curve, 
one can always find a local holomorphic section $\tilde{f}'_a$ 
of the line bundle ${\cal O}_{\tilde{S}}(D_{(\bar{\bf 5})})$ 
in an open set (abusing notations, $U_a$) of $\tilde{S}$ so that 
$\tilde{f}'_a|_{\tilde{\bar{c}}_{(\bar{\bf 5})}} = \tilde{f}_a$. 
In $U_a \cap U_b$ with the transition function $g_{ab}$ of 
${\cal O}_{\tilde{S}}(D_{(\bar{\bf 5})})$, therefore, 
$\tilde{h}_{ab} \equiv \tilde{f}'_a - g_{ab} \tilde{f}'_b$ 
can be regarded as a section of line bundle 
${\cal O}(D_{(\bar{\bf 5})}- \tilde{\bar{c}}_{(\bar{\bf 5})})$, 
since $\tilde{h}_{ab}$ vanishes on the curve $\tilde{\bar{c}}_{(\bar{\bf 5})}$. 
From the definition of \v{C}ech cohomology, it is clear that a set of 
$\tilde{h}_{ab}$'s defines an element of 
$H^{1}(\tilde{S}; \mathcal{O}(
 D_{(\bar{\bf 5})}-\tilde{\bar{c}}_{(\bar{\bf 5})}))$. 
It is also evident that the 1-cochain $(U_a \cap U_b, \tilde{h}_{ab})$ 
is the image of the coboundary map of a 0-cochain $(U_a, \tilde{f}'_a)$ 
in the \v{C}ech cohomology of ${\cal O}_{\tilde{S}}(D_{(\bar{\bf 5})})$. 
Thus, $(U_a \cap U_b, \tilde{h}_{ab})$ is in the kernel 
of (\ref{eq:map-H1-H1-5bar-2}).

\v{C}ech cohomology can be calculated by brute-force, 
as explained in textbooks. One does not have to 
resort to numerical calculations. 
We cannot motivate ourselves, however, to carry out explicit
calculations of \v{C}ech cohomology for this example, for it 
is far from being ``realistic''.

$H^1(\tilde{\bar{c}}_{(\bar{\bf 5})}; \tilde{\cal F}_{(\bar{\bf 5})})$ 
in (\ref{eq:5hc-ExVI}) allows us to determined wavefunctions 
of zero modes that are somehow associated with 
{\it anti-chiral multiplets} in the representation $\bar{\bf 5}$. 
It is of more interest, however, to study holomorphic 
wavefunctions of {\it chiral multiplets} in the representation ${\bf
5}$, because that is what we want to use in calculating up-type Yukawa 
couplings $Q \bar{U} H_u \subset {\bf 10} \; {\bf 10} \; H({\bf 5})$. 
For this purpose, we define $D_{({\bf 5})}$ by replacing\footnote{
In Heterotic string language, when a vector bundle $V$ is constructed 
as the Fourier--Mukai transform 
$V = R^0 p_{1*} \left[ {\cal P}_B \otimes 
 p_2^* {\cal O}(r/2 + \gamma)\right]$, 
the dual bundle is given by 
$V^\times = R^0 p_{1*} \left[ {\cal P}_B^{-1} \otimes 
p_2^* {\cal O}(r/2 - \gamma)\right]$, using the same spectral surface 
$C_V$ for $V$ \cite{FMW-2, DI}. The difference between ${\cal P}_B$ and 
${\cal P}_B^{-1}$ is not relevant in 
%\kaku{${\cal F}_{({\bf 5})}$ and ${\cal F}_{(\overline{\bf 5})}$} 
${\cal F}_{({\bf 10})}$ and ${\cal F}_{(\overline{\bf 10})}$  
on the matter curve, but the difference of $\pm \gamma$ 
remain in the divisors defining ${\cal F}_{({\bf 10})}$ 
and ${\cal F}_{(\overline{\bf 10})}$, and gives rise to the net chirality.}  
$\gamma$ in $D_{(\bar{\bf 5})}$ by $-\gamma$, 
which practically corresponds to flipping the sign of $\lambda$, 
\begin{eqnarray}
 D_{({\bf 5})} & = & 
\nu^*_S \left( K_S + \frac{1}{2}{\rm div} \tilde{R}^{(5)}_{\rm mdfd}
  -  3 \lambda {\rm div} a_5 \right) - \left(\frac{1}{2} 
  - 5 \lambda \right) E_{D6}, \\ 
 & \sim & 12 H - 3 E_{D6},
\end{eqnarray}
where we used $\lambda = -1/2$ in the second line. Using a long 
exact sequence similar to (\ref{eq:long-10}, \ref{eq:long-5bar}), 
we obtain 
\begin{equation}
 H^0 \left( \tilde{\bar{c}}_{({\bf 5})} ;
                 {\cal O}(D_{({\bf 5})}|_{\tilde{\bar{c}}_{({\bf 5})}}) \right)
 \simeq {\rm Ker} \left[
   H^1 \left( \tilde{S}; 
        {\cal O}(D_{({\bf 5})} - \tilde{\bar{c}}_{(\bar{\bf 5})}) \right)
  \rightarrow H^{1}\left(\tilde{S}; \mathcal{O}(D_{(\bf 5)})\right) 
   \right]
\label{eq:5h-ExVI}
\end{equation}
which is precisely the Serre dual of (\ref{eq:5hc-ExVI}).

%%%%%%%%%%%%%%%%%%%%%%%%%%%%%%%%%%%%%%%%%%%%%%%%%%%%%%%%%%
\subsubsection{Matter Parity}
\label{sssec:matter-parity} 
%%%%%%%%%%%%%%%%%%%%%%%%%%%%%%%%%%%%%%%%%%%%%%%%%%%%%%%%%%%

%[remind readers of advantage of the $\Z_2$ scenario] $\quad$
Safely removing dimension-4 proton decay operators is the 
first step toward a theoretical framework for the flavor structure.  
To identify the right-handed neutrinos 
\kaku{among the various fluctuations in an F-theory compactification} 
is an issue closely related to this, 
because $\SU(5)_{\rm GUT}$-singlet right-handed neutrinos are 
some of fluctuations of moduli fields, and the vacuum expectation values of 
the moduli fields control the tri-linear couplings of massless 
modes \cite{TW-1, Tsuchiya}.
Imposing a $\Z_2$ symmetry (often called parity) is the most 
popular solution to the dimension-4 proton decay problem 
among phenomenologists. 
In F-theory compactifications, it requires 
a pair $(X, G^{(4)})$ of vacuum configurations 
to have a $\Z_2$ symmetry. The vector space 
of massless multiplets in a given representation splits into 
$+1$-eigenstates and $-1$-eigenstates of the generator $\sigma$ 
of the $\Z_2$ symmetry. That is, the $\Z_2$ parity can be assigned to 
the massless multiplets when a $\Z_2$ symmetric $(X, G^{(4)})$ is given. 
It has been discussed in section 4.1 of \cite{Tsuchiya} that 
a successful explanation of 
% \kaku{the mass scale of the Majorana right-handed neutrinos 
% in F-theory compactifications is possible without any conflict 
% with the $Z_2$ solution to the dimension-4 proton decay problem.} 
Majorana mass scale of right-handed neutrinos in flux 
compactification \cite{Tsuchiya} does not have a conflict with 
this $Z_2$ solution to the dimension-4 proton decay problem. 

%[top down v.s. bottom-up assignment of $\Z_2$] $\quad$ 
The $\Z_2$ symmetry solution itself is fine, but sometimes, 
it may not be the most economical way to start everything from 
a compact elliptic Calabi--Yau 4-fold $X$ and a flux $G^{(4)}$ on it.
One might sometimes be interested in deriving observable consequences 
by assuming existence of certain class of compactifications; providing 
a proof of existence of such compactification geometry is another 
business. One has to deal with a whole global compact Calabi--Yau 4-fold
$X$ for the latter purpose. But, local geometry of $X$ around a GUT divisor 
$S$ may sometimes be the only necessary assumption to get started 
for the former purpose. Thus, for this purpose, it will be convenient 
if we can impose such a $\Z_2$ symmetry and discuss its consequences 
(including parity assignment) in as bottom-up (local) manner as possible. 
That is what we present in the following, using the Example VII.

%\kaku{
%In this section, taking Example VII again, 
%we will demonstrate how the $\Z_2$ parity of the 
%low-energy spectrum is derived, imposing a $\Z_2$ symmetry only 
%on a local neighborhood of the GUT divisor $S$ in $X$ and on 
%the fluxes on the matter curves of $S$. 
%We will discuss a more precise meaning of it in terms of a spectral 
%surface $C$ and its line bundle ${\cal N}$ on it soon later. 
%Given a pair ($X, G^{(4)}$) with a $\Z_2$ symmetry defined globally, 
%the procedure we will demonstrate would be sufficient to find its 
%$\Z_2$ parity of the low-energy spectrum in the compactification. 
%}

\bigskip
\noindent
{\bf $\Z_2$ Symmetry on Geometry}

%[get started: $\P^2$] $\quad$ 
Let us assume that $S = \P^2$, as in Example VII, and consider 
a $\Z_2$ transformation acting on $S$ as the map
\begin{equation} 
\sigma: [S: T: U] \mapsto [-S: -T: U], \quad 
{\rm or\ equivalently,} \quad
[S: T: U] \mapsto [S: T: -U].
\label{eq:map-on-P2}
\end{equation}
There are two loci of fixed points of the transformation $\sigma$ 
in $S = \P^2$. One is a point $[0: 0: 1] \in \P^2$, and the other is 
$[S: T: 0] \simeq \P^1 \subset \P^2$. We should emphasize, however, 
that we will {\it not} take a quotient by this $\Z_2$ transformation, but 
we just assume a $\Z_2$-symmetric background configuration.
% We would like to consider how the transformation $\sigma$ 
% can be extended to a local geometry of $X$ and $G^{(4)}$. 
We would like to consider whether this $\sigma$ symmetry transformation 
can be extended to a local geometry of $X$ and $G^{(4)}$ there, and 
if it is possible, how. 

%[use of Higgs bundle, focus on spectral surface first] $\quad$
A local geometry of $X$ and its flux configuration can be captured 
by a Higgs bundle on $S$. Let us consider a case where the Higgs bundle 
is given an Abelianized description, or in other words, a spectral data 
consisting of a spectral surface $C$ and a line bundle ${\cal N}$ on it. 
We will extend the $\Z_2$ transformation $\sigma$ 
onto the spectral surface first, 
and then onto the line bundle ${\cal N}$ later. 

%[$\Z_2$ on $K_S$] $\quad$ 
A spectral surface $C$ for a Higgs bundle in a representation 
is a divisor of the total space $\mathbb{K}_S$ 
of the canonical bundle $K_S \simeq {\cal O}(-3H)$ on $S = \P^2$.
Let us first discuss the $\Z_2$ transformation $\sigma$ on $\mathbb{K}_S$.
Using the three open subsets $U_{\check{S}}$, $U_{\check{T}}$ and 
$U_{\check{U}}$ of $S = \P^2$ that we have already introduced, 
$K_S$ is given a local trivialization. We can take 
$(T/S, U/S) \equiv (t, u)$ as local coordinates of $U_{\check{S}}$, 
and let the fiber coordinate of $K_S|_{U_{\check{S}}}$ be $\xi_s$.
Similarly, local coordinates and fiber coordinates are introduced 
on the other trivialization patches; $(S/T, U/T) \equiv (s', u')$ and 
$\xi_t$ on $U_{\check{T}}$, and $(S/U, T/U) \equiv (s'', t'')$ and 
$\xi_u$ on $U_{\check{U}}$. The transition function of $K_S$ is given by 
$g_{\check{S}\check{U}} = (S/U)^3 = u^{-3} = (s'')^3$ in 
$U_{\check{S}} \cap U_{\check{U}}$ and similarly on the other 
intersections of the patches. 
We extend the transformation $\sigma$ on $S = \P^2$ to $K_S$ 
so that the two-form $\xi_sdt\wedge{du}$ is left invariant 
under the transformation $\sigma$; more explicitly, 
in the coordinates,
\begin{eqnarray}
 (t(\sigma(p)), u(\sigma(p)), \xi_s(\sigma(p))) & = 
  (t(p), - u(p), - \xi_s(p)) &
% \sigma: & (t,u, \xi_s) \mapsto  (t,-u,- \xi_s) &
  {\rm in~}U_{\check{S}},  \nonumber \\
 (s'(\sigma(p)), u'(\sigma(p)), \xi_t(\sigma(p))) & = 
   (s'(p), - u'(p), - \xi_t(p)) & 
% & (s',u', \xi_t) \mapsto  (s',-u',- \xi_t) &
   {\rm in~}U_{\check{T}}, \\
  (s''(\sigma(p)), t''(\sigma(p)), \xi_u(\sigma(p))) & = 
  (- s''(p), - t''(p), \xi_u(p)) & 
% \sigma: & (s'', t'', \xi_u) \mapsto (- s'', -t'', \xi_u) & 
 {\rm in~}U_{\check{U}},  \nonumber 
\end{eqnarray}
where $p \in \mathbb{K}_S$ is mapped to $\sigma(p) \in \mathbb{K}_S$. 
The map of $\sigma$ in $U_{\check{S}}$, $U_{\check{T}}$ and 
$U_{\check{U}}$ are consistent in the overlapping regions such 
as $U_{\check{S}} \cap U_{\check{U}}$; note that the transition
functions $g_{\check{S}\check{U}}$ and $g_{\check{T}\check{U}}$ 
are $\Z_2$-odd, and $g_{\check{S}\check{T}}$ even. 

This $\Z_2$ transformation induces an $\SU(3) \subset \U(3)$ rotation 
on the three complex coordinates of $\mathbb{K}_S$. Any SU(3) transformations 
act on spinors exactly the same way, because the $\SU(3)$ in 
$\SU(3) \times \U(1) \subset \SU(4) \simeq \SO(6)$ is embedded purely 
in the $3 \times 3$ block of the fundamental representation of $\SU(4)$ 
[spinor representation of $\SO(6)$]. Thus, the $\Z_2$ transformation 
$\sigma$ on the complex coordinates also generates a $\Z_2$ transformation 
on spinors (not a transformation of order 4), and hence it is a 
$\Z_2$ symmetry of the whole theory. 

$R$-parity and matter parity in supersymmetric Standard Models are 
different only by fermion parity, which always exists. Thus, they 
are equivalent, and we only look for a $\Z_2$ symmetry that becomes 
matter parity. This is why we imposed an $\SU(3)$ condition above, 
to find a symmetry that is essentially non-$R$. 

%
% At a point in $U_{\check{S}} \cap U_{\check{U}}$ corresponding to 
% $(t, u) = (t_*, u_*)$ in $U_{\check{S}}$
% [to $(s'', t'') = (1/u_*, t_*/u_*)$ in $U_{\check{U}}$]
%
%

%[invariance of spec surface in a single patch] $\quad$ 
If the transformation $\sigma$ on $\mathbb{K}_S$ is to be 
a symmetry of the system, then the (defining equation of the) 
spectral surfaces should be preserved by $\sigma$.  
Let us consider a 5-fold\footnote{It is always 
optional to drop higher order terms in the polynomial in $\xi$, when 
one focuses on a local geometry\kesu{,} \kaku{near $\xi\simeq0$,} 
as in (\ref{eq:spec-surf-E6}).}
spectral cover given by 
\begin{equation}
 a_0 \xi^5 + a_2 \xi^3 + a_3 \xi^2 + a_4 \xi + a_5 \simeq 0.
\label{eq:spec-surf}
\end{equation}
%
% \kaku{Recall that the coefficients $a_r$ ($r = 0,2,3,4,5$) take values in 
% ${\cal O}_S(rK_S+\eta)$. Since the transformation $\sigma$ is defined on 
% $K_s$ so that $\xi_s{dt}\wedge{du}$ is left invariant under $\sigma$, 
% the coefficient $a_r$ should also transform as $(K_S)^r$ under $\sigma$. 
% Furthermore, the defining equation (\ref{eq:spec-surf}) is homogeneous 
% in the coefficients $a_r$, and an overall scaling $a_r\to\lambda{a}_r$ 
% does not alter the spectral surface $C$ as a divisor. 
% Thus, for example, on the patch $U_{\check{S}}$, since the fiber coordinate 
% $\xi=\xi_s$ transforms under $\sigma$ as $\xi_s\to-\xi_s$, 
% the coefficient $a_r$ on $U_{\check{S}}$, then, transforms as 
% %
% \begin{equation}
%  a_{r} (\sigma(p))= e^{i\beta}(-)^r a_{r}(p),
% \label{eq:ar-inv}
% \end{equation}
% %
% where the parameter $\beta$ of the overall scaling $e^{i\beta}$ can 
% only take $0$ or $1$ as its value for the $\Z_2$ transformation $\sigma$. 
% In other words, the scaling $e^{i\beta}$ accounts for an extension of 
% the transformation $\sigma$ to sections of ${\cal O}_S(\eta)$ and/or the 
% normal bundle $N_{S|B}\simeq{\cal O}_S(6K_S+\eta)$. 
% The extension is not yet unique, and the choice of $\beta=0$ will be called 
% Case A and the other $\beta=\pi$ Case B. See Table \ref{tb:ai}}
%
For the time being, we focus on a single trivialization patch, 
say, $U_{\check{U}}$, and consider a map of $K_S$ within the patch.
Thus, $\xi$ (say $\xi_u$) is the holomorphic coordinate in the fiber
direction, and $a_r$ ($r = 0,2,3,4,5$) are holomorphic 
functions on the base coordinates (say, $(s'', t'')$).
Suppose that $\sigma: p \mapsto \sigma(p)$ in 
$U_{\check{U}} \subset \P^2$ and 
\begin{equation}
 \xi_u(\sigma(p)) = e^{i\alpha} \xi_u(p).
\end{equation}
Arbitrary points $p$ in the spectral surface is mapped by $\sigma$ 
to the spectral surface, if 
\begin{equation}
 a_{r} (\sigma(p)) = e^{i (\beta - (6-r) \alpha)} a_{r}(p)
\label{eq:ar-inv}
\end{equation}
for some phase $\beta$ in the $U_{\check{U}}$ patch. 
Both $\alpha$ and $(\beta - (6-r)\alpha)$ need to be $2\pi/N$ 
for $\sigma$ to be an element of order $N$ in the symmetry group. 
In the case of our interest, $N = 2$, and $\alpha_u = 0$ in the patch 
$U_{\check{U}}$. We still have two options, 
$\beta_u = 0$ and $\beta_u = \pi$ in $U_{\check{U}}$, and we call them 
case A and case B, respectively.  

%[$\sigma$-invariance of spec surf in other patches] $\quad$
The spectral surface needs to be invariant under the 
symmetry transformation $\sigma$ on $\mathbb{K}_S$ in other 
patches like $U_{\check{S}}$ and $U_{\check{T}}$ as well. 
The condition for invariance is (\ref{eq:ar-inv}) in the other 
patches as well, with $\alpha_s = \pi$ in $U_{\check{S}}$ and 
$\alpha_t = \pi$ in $U_{\check{T}}$. Since the ``$\sigma$-invariance'' 
of $a_r$ (\ref{eq:ar-inv}) needs to be consistent between two 
overlapping regions, the phase $\beta$ cannot be chosen independently 
in different patches. Suppose that $\sigma$ acts on a line bundle, 
and let the fiber coordinates be $s_a$ in trivialization patches $U_a$.
When $\sigma$ induces a map on the total space of the line bundle 
given by $s_a(\sigma(p)) = e^{i \alpha_a} s_a(p)$, then the consistency 
of the map in the common subset of two trivialization patches 
$U_a \cap U_b$ is 
\begin{equation}
 g_{ab}(\sigma(p)) = e^{i \alpha_a} g_{ab}(p) e^{-i \alpha_b}.
\label{eq:consis-glue}
\end{equation}
$\alpha_u = 0$ and $\alpha_{s,t} = \pi$ for $K_S$ on $S = \P^2$ 
satisfies this relation. Since $a_r$'s are sections of 
${\cal O}(rK_S +\eta) = {\cal O}(6K_S + \eta) \otimes K_S^{-(6-r)}$, 
$e^{i\beta}$'s for the trivialization patches is for the bundle 
$N_{S|B_3}$. For $N_{S|B} \simeq {\cal O}((d-3)H)$ 
[or $\bar{c}_{({\bf 10})} \in |dH|$], 
$e^{i \beta} = (-1)^{d-3}$ in $U_{\check{S}}$ and $U_{\check{T}}$ 
in case A, and 
$e^{i \beta} = - (-1)^{d-3}$ in $U_{\check{S}}$ and $U_{\check{T}}$ 
in case B. For a given $\Z_2$ transformation $\sigma$ on $S = \P^2$ 
and an $\SU(3)$ lift of $\sigma$ acting on $\mathbb{K}_S$, two 
different possibilities exist: case A and B.
See Table~\ref{tb:ai}.
%%%%%%%%%%%%%%%%%%%%%%%%%%%%%%%%%%%%%%%%%%%%%%%%%%%%
\begin{table}[tb] 
\begin{center}
\begin{tabular}{c|ccccc|ccccc}
& & $U_{\check{S}}$ & & $U_{\check{T}}$ & & & & $U_{\check{U}}$ & & \\
 \hline
& $a_5$ & $a_4$ & $a_3$ & $a_2$ & $a_0$ & $a_5$ & $a_4$ & $a_3$ &
 $a_2$ & $a_0$ \\
 \hline
case A & $+$ & $-$ & $+$ & $-$ & $-$ & $+$ & $+$ & $+$ & $+$ & $+$ \\ 
case B & $-$ & $+$ & $-$ & $+$ & $+$ & $-$ & $-$ & $-$ & $-$ & $-$ \\
 \hline
\end{tabular}
\caption{\label{tb:ai} 
The $\mathbb{Z}_{2}$ parities of the holomorphic sections $a_{r}$. 
The sections $a_r$ have to satisfy the condition (\ref{eq:ar-inv}) in 
order for the spectral surface to be $\Z_2$ invariant, with the phase 
factor $\pm 1$ (we call them parity) that may be different in 
different trivialization patches. In the case $S = \P^2$ with 
$\bar{c}_{({\bf 10})} \in |d H|$, the parity of $a_r$ is 
$(-1)^{d-3}(-1)^{6-r}$in $U_{\check{S}}$ and $U_{\check{T}}$ patches, 
and is $+$ in $U_{\check{U}}$ in Case A. The parity assignment of Case A 
is opposite to that of case B. In this table, the parity assignment for 
the $d = 2$ case (Example VII) is given.}
\end{center}
\end{table}
%%%%%%%%%%%%%%%%%%%%%%%%%%%%%%%%%%%%%%%%%%%%%%%%%%%%%%%%%%%%%%%%%%%%

%[symmetry and tuning] $\quad$ 
In order for the spectral surface to be invariant under 
the $\Z_2$ transformation, the holomorphic sections $a_r$ 
\kaku{must satisfy the relation (\ref{eq:ar-inv}).}
Thus, the $\Z_2$ symmetry solution to the dimension-4 proton decay 
problem \kaku{restricts the choices} 
of the complex structure parameters. 
\kaku{As in other flux compactifications, fluxes} 
will ultimately decide whether this is 
an ugly tuning by hand (that may possibly be justified by 
anthropics), or a prediction of the pure statistics of flux vacua. 
For now, we will take the phenomenological approach, just 
assuming that the complex structure moduli are chosen at such 
a $\Z_2$ symmetric point for some reasons that we do not know yet, 
and study all the remaining consequences. 

%[matter curve geometry I-a] $\quad$
In Case B, there is an interesting consequence. 
Since all the $a_r$ are odd under the $\Z_2$
transformation, all of them vanish at the $\Z_2$-fixed point 
$(s'',t'') = (0,0)$ in $U_{\check{U}}$. The singularity 
of the Calabi--Yau 4-fold $X$ is enhanced to $\tilde{E}_8$ at the point. 
A single branch of the matter curve $\bar{c}_{({\bf 10})}$ passes 
through this point, while three branches of $\bar{c}_{(\bar{\bf 5})}$ 
pass through this point, too. This is because all of the $a_r$'s in the 
patch $U_{\check{U}}$ start from terms linear in 
the local coordinates $(s'', t'')$\kaku{ near the origin $(s'',t'') = (0,0)$}, 
and then $P^{(5)}$ becomes cubic in the local coordinates. 
This point is like one $E_6$ type point and one $D_6$ type point 
merging into one. 
The $\Z_2$ symmetry ensures \kaku{without any tuning} 
that the two points are at the same place. 
This is interesting from a phenomenological point of view, 
because it provides a possible explanation why the pair of heaviest 
mass-eigenstates ($t,b$) of up-type and down-type quarks is 
almost\footnote{\kaku{Decomposing the Cabbibo-Kobayashi-Maskawa matrix 
of the real world into the $2\times2$ block of the first and second 
generations and the block of the third generation, the components 
in the off-diagonal blocks, e.g., $V_{ub}$ and $V_{cb}$, are quite tiny, 
compared to the other components. 
}} in the same left-handed quark doublet. 
Let us take a basis of independent wavefunctions $\{ f_j \}$ 
($j = 1,\cdots, N_{\rm gen}$) of quark doublets on the matter curve 
$\bar{c}_{({\bf 10})}$, and denote $q_j$ $(j = 1,\cdots, N_{\rm gen})$ 
the corresponding low-energy degrees of freedom. It is then 
the linear combination $\sum_j f_j(p) q_j$ that dominantly has 
the up-type Yukawa coupling at this $\tilde{E}_8$ singularity, 
and exactly the same linear combination has the dominant 
down-type Yukawa coupling there.\footnote{Here, we assume that 
there is no torsion component in ${\cal F}_{({\bf 10})}$ on
$\bar{c}_{({\bf 10})}$ at this point of $\tilde{E}_8$ singularity. 
Only under this assumption, do we have an intuitive picture of
wavefunctions that behave smoothly. We further assume that 
the up-type and down-type Yukawa matrices generated at an 
$\tilde{E}_8$ singularity are approximately rank-1.} 
If these Yukawa couplings are dominant over the contributions from 
the other codimension-3 singularity points, this linear combination 
of left-handed quark doublets would certainly give the pair of a top 
and bottom quark, and hence the structure of the CKM matrix of the 
real world follows.\label{page:tb-pair}  

%[matter curve geometry I-b] $\quad$ 
Note that the existence of such an $\tilde{E}_8$ singularity  
point in the $\Z_2$-symmetric configuration 
is not specific to the case with $S = \P^2$ or with a particular choice 
of the topological class of $\bar{c}_{({\bf 10})}$. Whenever there 
is an isolated $\Z_2$-fixed point on the GUT divisor, 
\kaku{the fiber coordinate $\xi$ of $K_S$ at the fixed point is left invariant 
under the $\Z_2$ transformation, because both of the two local coordinates 
of $S$ are flipped under the transformation at the isolated fixed point.} 
The holomorphic sections $a_r$ are then either all $\Z_2$-even 
(like in Case A) or all $\Z_2$-odd (like in Case B), because the fiber
coordinate $\xi$ does not change its sign. Thus, it is in the latter case 
that this isolated fixed point always gives the interesting possibility 
we described above. 

%[$\Z_2$ in local def eq of CY-4 $X$] $\quad$ 
Let us take a moment here to see how this $\Z_2$ transformation 
is further lifted to a $\Z_2$ transformation in 
the local geometry (\ref{eq:def-eq}) of the Calabi--Yau 4-fold $X$. 
%
% \kaku{Now the transformation $\sigma$ needs to be realized on the 
% coefficients in the local defining equation (\ref{eq:def-eq}). 
% Since $x$, $y$, and $z$ are sections of ${\cal O}_S(10K_S+2\eta)$, 
% ${\cal O}_S(15K_S+3\eta)$, and $N_{S|B_3}$, respectively, 
% they transform under $\sigma$ as
%
% \begin{equation}
% x \to x, \quad y \to -e^{i\beta}y, \quad z \to e^{i\beta}z.
% \end{equation} 
% %
% It means that the GUT divisor $S$ at $(x,y,z)=(0,0,0)$ is invariant 
% under the $\Z_2$ transformation. 
% Furthermore, the holomorphic sections $a_r$ and $a'_r$ 
% of ${\cal O}_S(rK_S+\eta)$ and ${\cal O}_S((r-6)K_S)$, respectively, 
% transform under the transformation $\sigma$ as
% %
% \begin{equation}
% a_r \to e^{i\beta}(-)^r a_r, \qquad a'_r \to (-)^{r} a'_r, 
% \end{equation}
% %
% for $r=0,2,3,4,5$. Thus, as in (\ref{eq:ar-inv}), 
% the $\Z_2$ invariant local geometry (\ref{eq:def-eq}) of $X$ must 
% satisfy
% \begin{equation}
% a_r(\sigma(p))=e^{i\beta}(-)^r a_r(p), \qquad 
% a'_r(\sigma(p))=(-)^{r} a'_r(p). 
% \end{equation}
% }
Now the generator $\sigma$ of the $\Z_2$ symmetry needs to 
be realized as a symmetry on a geometry defined by (\ref{eq:def-eq}). 
The GUT divisor $S$ is covered by a set of 
trivialization patches $U_a$'s for $N_{S|B_3}$ and $K_S$, and 
a set of coordinates $(x_a, y_a, z_a)$ is introduced for each 
trivialization patch $U_a$. Those coordinates are identified 
with one another up to appropriate transition functions in the 
overlapping regions $U_a \cap U_b$. In a given patch, $U_a$, 
for example, the map $\sigma$ needs to satisfy\footnote{Since the 
GUT divisor $S$ is at $(x,y,z) = (0,0,0)$ in $X$, 
the $(x,y,z) =(0,0,0)$ locus need to be mapped to $(x,y,z) = (0,0,0)$.}
\begin{equation}
 z_a(\sigma(p)) = e^{i \beta_a} z_a(p), \qquad 
 x_a(\sigma(p)) = e^{i2(\beta_a - \alpha_a)} x_a(p), \qquad 
 y_a(\sigma(p)) = e^{i3(\beta_a - \alpha_a)} y_a(p);
\end{equation} 
no freedom can be introduced in the process of lifting the map 
$\sigma$ on $\mathbb{K}_S$ and $a_r$'s to that of local geometry of 
$X$. The $\sigma$ map on the coordinates $(x_a, y_a, z_a)$ are 
consistently glued together in $U_a \cap U_b$; to see this, 
recall that $e^{i\alpha_a}$ and $e^{i\beta_a}$ are phases in 
the trivialization patches satisfying (\ref{eq:consis-glue}) 
for line bundles $K_S$ and $N_{S|B_3}$, respectively. Since 
$z_a$'s form a section of $N_{S|B_3}$, and $x_a$'s [resp. $y_a$'s]
that of $(K_{B_3})^{-2}|_{S} =  (N_{S|B_3} \otimes K_S^{-1})^{2}$ 
[resp. $(N_{S|B_3} \otimes K_S^{-1})^3$], the phase factors above 
are exactly the ones satisfying (\ref{eq:consis-glue}) for these
line bundles.\footnote{
$dx/y$ can be chosen as the differential in the elliptic fiber
direction. At a fixed point of this $\sigma$ transformation, 
holomorphic (2,0)-form $dz \wedge dx/y$ in the ALE direction 
is rotated by $e^{i\beta} \times e^{-i (\beta - \alpha)} = e^{i\alpha}$.
This is the same as the phase rotation of $d\xi$ in the fiber direction 
of $K_S$. Thus, an $\SU(3)$ rotation in $K_S$ corresponds to an $\SU(4)$
rotation in a local geometry of $X$, regardless of the choice of $\beta$.} 
One can also see that higher-order coefficients in the $z$-series 
expansion $a'_r$'s should satisfy 
\begin{equation}
 a'_r(\sigma(p)) = e^{i \alpha_a (r-6)} a'_r(p)
\label{eq:ar'-inv}
\end{equation}
in patch $U_a$.

\bigskip
\noindent
{\bf $\Z_2$ Symmetry on Bundles}

%[parity on fluxes, descend all the way down to ${\cal F}$] $\quad$
A $\Z_2$ symmetry for a matter parity has to be defined in a system 
of both a 4-fold $X$ and a four-form flux $G^{(4)}$ on it, not just in $X$. 
\kaku{The $\Z_2$ transformation must then induce its action also 
on the line bundles ${\cal N}_{({\bf 10})}$ and 
$\widetilde{\cal N}_{(\bar{\bf 5})}$ on the spectral surfaces $C_{({\bf 10})}$ 
of ${\bf 10}$ and $\widetilde{C}_{(\bar{\bf 5})}$ of $\bar{\bf 5}$, 
respectively, of $SU(5)_{\rm GUT}$.
The line bundle ${\cal F}_{({\bf 10})}$ in (\ref{eq:H0-10-repr}) on 
the matter curve $\bar{c}_{({\bf 10})}$ is given by the 
restriction of ${\cal N}_{({\bf 10})} \otimes \pi_{C_{({\bf 10})}}^* K_S$ 
onto $\bar{c}_{({\bf 10})}$, while $\widetilde{\cal F}_{(\bar{\bf 5})}$ 
in (\ref{eq:H0-5bar-repr}) on $\tilde{\bar{c}}_{(\bar{\bf 5})}$ by 
the one of $\widetilde{\cal N}_{(\bar{\bf 5})} \otimes
\tilde{\pi}_{C_{(\bar{\bf 5})}}^* K_S$ onto $\tilde{\bar{c}}_{(\bar{\bf 5})}$. 
Thus, these line bundles on the matter curves must also be left 
invariant under the $Z_2$ transformation.}

\kesu{The $\Z_2$ symmetry in such a compact set-up must then 
be reflected in the line bundle ${\cal N}_{({\bf 10})}$ on the spectral 
surface $C_{({\bf 10})}$ for ${\bf 10}$ 
of $\SU(5)_{\rm GUT}$, and $\widetilde{\cal N}_{(\bar{\bf 5})}$ 
on the normalized spectral surface $\widetilde{C}_{(\bar{\bf 5})}$ for 
$\bar{\bf 5}$. 
The line bundles ${\cal F}_{({\bf 10})}$ in (\ref{eq:H0-10-repr}) and 
$\widetilde{\cal F}_{(\bar{\bf 5})}$ in (\ref{eq:H0-5bar-repr}) are 
obtained by 
restricting ${\cal N}_{({\bf 10})} \otimes \pi_{C_{({\bf 10})}}^* K_S$ 
and $\widetilde{\cal N}_{(\bar{\bf 5})} \otimes
\tilde{\pi}_{C_{(\bar{\bf 5})}}^* K_S$ on the (covering) matter curves 
$\bar{c}_{({\bf 10})}$ and $\tilde{\bar{c}}_{(\bar{\bf 5})}$. Thus, 
the $\Z_2$ symmetry must also exist in these line bundles. }

%[economical, bottom-up, genericity] $\quad$
As we already explained at the beginning of this 
section \ref{sssec:matter-parity}, 
% \kaku{it seems formidable to derive ${\cal F}_{({\bf 10})}$ and 
% $\widetilde{\cal F}_{(\bar{\bf 5})}$ directly from a single 4-form 
% $G^{(4)}$, which would determine the $\Z_2$ transformation 
% on them uniquely. Instead, we consider all the possible actions of 
% the $\Z_2$ transformation $\sigma$ on ${\cal F}_{({\bf 10})}$ and 
% $\widetilde{\cal F}_{(\bar{\bf 5})}$. They in turn define a $\Z_2$ 
% transformation on zero mode chiral multiplets in the low-energy spectrum, 
% {\it i.e.}, a matter parity.} 
it is often economical to 
deal only with local geometry for the purpose of deriving 
phenomenological consequences by using string theory (not for 
the purpose of providing existence proof for a realistic string
vacuum). We can thus use the bundles ${\cal N}_{({\bf 10})}$ and 
$\widetilde{\cal N}_{(\bar{\bf 5})}$ instead of $G^{(4)}$ as a place 
to start discussing $\Z_2$ symmetry (matter parity) in a system 
including the effects of fluxes. 
It is even more economical, however, to construct a $\Z_2$ symmetry 
transformation at the level of ${\cal F}_{({\bf 10})}$ and
$\widetilde{\cal F}_{(\bar{\bf 5})}$. 
Because a zero mode chiral multiplet in the low-energy spectrum is 
a holomorphic section of these bundles, we only need 
to introduce a $\Z_2$ symmetry transformation in these bundles to 
derive matter-parity assignment on the zero modes; not necessarily 
at the level of ${\cal N}_{({\bf 10})}$ and 
$\widetilde{\cal N}_{(\bar{\bf 5})}$. We adopt this strategy in this 
article; this is not only economical, but also most bottom-up and 
generic way to discuss matter parity assignment.  
If $\Z_2$ transformations are introduced independently to 
${\cal F}_{({\bf 10})}$ and $\widetilde{\cal F}_{(\bar{\bf 5})}$, 
however, there may be an inconsistency, because the $\Z_2$
transformation on these line bundles should originate from a 
$\Z_2$ transformation on the common 4-form flux $G^{(4)}$. 
We will come back soon later to discuss the consistency.
% of the $\Z_2$ transformations on 
% ${\cal F}_{({\bf 10})}$ and $\widetilde{\cal F}_{(\bar{\bf 5})}$ 
% with the logical fact that they should be induced from the $\Z_2$ 
% transformation of a 4-form $G^{(4)}$. 

%[condition for the invariance]
The line bundles ${\cal F}_{({\bf 10})}$ and 
$\widetilde{\cal F}_{(\bar{\bf 5})}$ are both given by the divisors. 
A line bundle ${\cal O}(D)$ is invariant under a transformation 
$\sigma$, if and only if its divisor $D$ is invariant; 
$\sigma^* D = D$. The divisors 
for the line bundles ${\cal F}_{({\bf 10})}$ and 
$\widetilde{\cal F}_{(\bar{\bf 5})}$ consist of $i^* K_S + p_{E6/A6}$ 
and the flux dependent components.
The flux-independent part is given by the divisors 
$K_S$, ${\rm div} \; a_4$ and ${\rm div} \; \tilde{R}^{(5)}_{\rm mdfd}$, 
and they are invariant under $\sigma$, when the 
conditions (\ref{eq:ar-inv}, \ref{eq:ar'-inv}) are satisfied.
However, one needs to make sure that the flux dependent part 
$\gamma|_{\bar{c}_{({\bf 10})}}$ and 
``$\gamma|_{\tilde{\bar{c}}_{(\bar{\bf 5})}}$'' are also left 
invariant under the transformation $\sigma$, 
although the invariance of these $\gamma$'s should follow 
from invariance of four-form flux $G^{(4)}$ under the transformation $\sigma$. 

%[twin lifts to bundle isomorphism] $\quad$ 
A $\Z_2$-invariant line bundle is $\Z_2$-equivariant (see e.g. the
appendix A of \cite{equivariant}). 
\kaku{Thus, the action of the $\Z_2$ transformation $\sigma$ on 
the matter curves $\bar{c}_{({\bf 10})}$ and $\tilde{\bar{c}}_{(\bar{\bf 5})}$ 
can be promoted to its $\Z_2$ action $\phi_{\sigma}$ on the total spaces 
of the line bundles ${\cal F}_{({\bf 10})}$ and 
$\widetilde{\cal F}_{(\bar{\bf 5})}$.} 
It does not mean, however, that 
the bundle isomorphism $\phi_\sigma$ on ${\cal F}_{({\bf 10})}$ is 
determined uniquely for a given $\Z_2$ transformation $\sigma$.
For a bundle isomorphism 
$\phi_\sigma: {\cal F}_{({\bf 10})} \rightarrow 
\sigma^* {\cal F}_{({\bf 10})}$, there is another isomorphism 
$\phi'_g: {\cal F}_{({\bf 10})} \rightarrow \sigma^* {\cal F}_{({\bf
10})}$ that is different from $\phi_\sigma$ only by a multiplication 
of $(-1)$ in the rank-1 fiber. This ambiguity always exists for 
any $\Z_2$ action on a line bundle. 
Thus, there are two different ways to lift the transformation $\sigma$ 
to the $\Z_2$ transformation on ${\cal F}_{({\bf 10})}$. There are also 
two different ways to lift the $\sigma$ transformation to a $\Z_2$ 
bundle isomorphism of $\widetilde{\cal F}_{(\bar{\bf 5})}$, exactly 
for the same reason as above. 

%[cannot be chosen independently] $\quad$ 
We have already seen that the $\Z_2$ transformation $\sigma$ on $S$ 
can be lifted to two consistent $\Z_2$ transformations on $\mathbb{K}_S$ 
and configuration of spectral surfaces: Case A and Case B. 
Upon the further extension of them to the line bundles on the matter curves, 
for each of Case A and Case B, there apparently seem 
$2 \times 2$ different ways to lift the $\Z_2$ transformation acting 
on ${\cal F}_{({\bf 10})}$ and $\widetilde{\cal F}_{(\bar{\bf 5})}$.
This is not true, however. 
The zero modes on the matter curves are unified to 
the adjoint representation of the corresponding enhanced gauge group $G$ 
at a codimension-3 singularity point, along with the adjoint 
representation on the surface $S$ and the other. 
The action of the $\Z_2$ transformation on ${\cal F}_{({\bf 10})}$ 
induces the action on the zero modes on the curve $\bar{c}_{{\bf 10}}$, 
while the one on $\widetilde{\cal F}_{(\bar{\bf 5})}$ does on the zero 
modes on $\tilde{\bar{c}}_{(\bar{\bf 5})}$. However, the $\Z_2$ symmetry 
of the system $(X, G^{(4)})$ should become a symmetry of 
the gauge theory with the gauge group $G$. One should make sure 
that the $\Z_2$ symmetry is found not only in the multiplicity 
and wavefunctions of zero modes of individual irreducible 
representations of $G'' = \SU(5)_{\rm GUT}$, but also in 
the whole gauge theory including the interactions. 

% [descend from principal bundle, not generic, though] $\quad$
If we consider a case with an $\SU(5)_{\rm str}$-principal $K_S$-valued 
Higgs bundle globally defined on $S$, then we can consider a lift 
of $\Z_2$ transformation $\sigma$ to that on the principal bundle. 
Both the rank-5 Higgs bundle 
$(V_{({\bf 10})}, \varphi_{({\bf 10})})$
and rank-10 Higgs bundle 
$(\wedge^2 V_{({\bf 10})}, \rho_{\wedge^2 {\bf 5}}(\varphi_{({\bf
10})}))$ are given their $\Z_2$ transformation that descend from 
the $\Z_2$ transformation of the principal bundle. The $\Z_2$
transformation on ${\cal F}_{({\bf 10})}$ and 
$\widetilde{\cal F}_{(\bar{\bf 5})}$ obtained in this way preserves 
the commutation relation of $E_8$. We should not expect generically, 
however, that a globally defined $E_8$ Higgs bundle exists on $S$. 

% [use $D_6$ and $E_6$ instead] $\quad$ 
More bottom-up and generic way to determine the $\Z_2$ transformation 
on the line bundles is as follows. The two matter curves 
$\bar{c}_{({\bf 10})}$ and $\bar{c}_{(\bar{\bf 5})}$ meet at 
two different types of codimension-3 singularities; one is the 
$E_6$ type and the other is $D_6$ type. Field theory models at the 
$E_6$ type points are $E_6$ gauge theories, and those at the 
$D_6$ type points are $\SO(12)$ gauge theories. The $\Z_2$ symmetry 
of a whole system $(X, G^{(4)})$ should become a $\Z_2$ symmetry 
between the pairs of $E_6$ gauge theories mapped by $\sigma$.
Wiht this requirement on the $E_6$ gauge theories, we can uniquely 
determine the $\Z_2$ transformation on the sections of 
$\widetilde{\cal F}_{(\bar{\bf 5})}$. The same condition for a pair 
of $\SO(12)$ gauge theories uniquely determines the $\Z_2$ transformation 
on the bundle ${\cal F}_{({\bf 10})}$.

% [case A, use $U_{\check{S}}$ patch] $\quad$
For concreteness, let us take the case A lift of $\Z_2$ transformation 
$\sigma$ in Example VII as an example. We will describe the way the $\Z_2$
transformation is lifted to ones on the bundles ${\cal F}_{({\bf 10})}$ 
and $\widetilde{\cal F}_{(\bar{\bf 5})}$, and how the zero modes are 
classified into those with even parity and those with odd parity. 

% [specify our choice of $\Z_2$ on ${\cal F}_{({\bf 10})}$ and 
% $\widetilde{\cal F}_{(\bar{\bf 5})}$] $\quad$
As we have already mentioned above, there can be two different ways 
in lifting a $\Z_2$ transformation on a (covering) matter curve to 
one on the line bundle on it. In order to describe explicitly which 
one we use, it is convenient either to take one $\sigma$-invariant 
trivialization patch of the line bundle, or to take a pair of 
patches that are mapped by $\sigma$ to each other. 
In a given $\sigma$-invariant trivialization patch $U_a$, for example, 
sections of the line bundle are treated just as functions $\tilde{f}_a$, 
and the two different $\Z_2$ transformations of sections of the line 
bundles are described as 
$\tilde{f}(p)_a \rightarrow \tilde{f}(\sigma(p))_a$ or 
$\tilde{f}(p)_a \rightarrow - \tilde{f}(\sigma(p))_a$.
% In the Example VII, 
% the matter curve $\bar{c}_{({\bf 10})}$ is the zero locus of $a_5$, 
% and $a_5 \in \Gamma (\P^2; {\cal O}(2H))$ can be regarded as a 
% homogeneous function of degree two. In the case A, $a_5$ further has 
% to be of the form 
% %
% \begin{equation}
%  a_5 = a U^2 + (b S^2 + c ST + d T^2), 
% \end{equation}
% %
% as indicated in Table~\ref{tb:ai}. 
In the case A of Example VII, the matter curve 
$\bar{c}_{({\bf 10})}$ is covered by three trivialization patches, 
$\bar{c}_{({\bf 10})} \cap U_{\check{S}}$, 
$\bar{c}_{({\bf 10})} \cap U_{\check{T}}$, and 
$\bar{c}_{({\bf 10})} \cap U_{\check{U}}$ that are all 
$\sigma$-invariant. We find, in the appendix \ref{sec:Z2-assign}, 
that the $\Z_2$ transformation should turn 
$\tilde{f}_{({\bf 10});\check{S}}(p)$ 
into $- \tilde{f}_{({\bf 10}); \check{S}}(\sigma(p))$, 
in order to preserve the $\SO(12)$ commutation relations 
in a pair of $\SO(12)$ field theory local models that are 
mutually mapped to each other by $\sigma$; 
see Figure~\ref{fig:uspatch}.  
The $\Z_2$ transformation on $\tilde{f}_{\check{T}}$ and 
$\tilde{f}_{\check{U}}$ should be determined in a way consistent 
with the $\Z_2$ transformation on $\tilde{f}_{\check{S}}$, and 
this is possible.
This is sufficient information in uniquely specifying which one we
choose out of two possible $\Z_2$ transformations on ${\cal F}_{({\bf 10})}$. 
Similarly, the covering matter curve $\tilde{\bar{c}}_{(\bar{\bf 5})}$ 
is covered by trivialization patches including
$\tilde{\bar{c}}_{(\bar{\bf 5})} \cap U'_{\check{S}}$, 
$\tilde{\bar{c}}_{(\bar{\bf 5})} \cap U'_{\check{T}}$ and 
$\tilde{\bar{c}}_{(\bar{\bf 5})} \cap U'_{\check{U}}$, and all the three 
are $\Z_2$-invariant. We find in the appendix \ref{sec:Z2-assign} that 
the $\Z_2$ transformation on $\widetilde{\cal F}_{(\bar{\bf 5})}$ 
should take $\tilde{f}_{(\bar{\bf 5}); \check{S}}(p)$ to 
$+ \tilde{f}_{(\bar{\bf 5}); \check{S}}(\sigma(p))$, not the other one, 
this time, in order to preserve the $E_6$ commutation relations in a pair of 
$E_6$ field theory local models mapped to each other by $\sigma$.

% [classify ${\bf 10}$'s] $\quad$ 
We are now ready to work out the $\Z_2$ parity assignment for 
case A for the $N_{\rm gen} = 17$ independent zero modes in 
the ${\bf 10}$ representation in the Example VII. We have already seen 
that $\tilde{f}_{\check{S}}$'s are of the form given in 
(\ref{eq:10repr-wavefcn}). In the case A, we should take, as $a_5$,  
a homogeneous function of degree two in the form of 
\begin{equation}
 a_5 = a U^2 + (b S^2 + c ST + d T^2) \in \Gamma(\P^2; {\cal O}(2H)),  
\end{equation}
so that $a_5$ has the appropriate property under the case $A$ 
$\Z_2\vev{\sigma}$ transformation specified in Table~\ref{tb:ai}.
Assuming that $a \neq 0$, we can take 
\begin{equation}
 F^{(8)}(S, T, U)_- \equiv F^{(8)}(S, T) \quad {\rm and} \quad 
 F^{(8)}(S, T, U)_+ \equiv F^{(7)}(S, T) \; U
\end{equation}
as independent generators of (\ref{eq:10repr-ExVI}).
Since the coefficient functions 
$\tilde{f}_{({\bf 10}); \check{S}} = F^{(8)}(1,t)$ and 
$\tilde{f}_{({\bf 10}); \check{S}} = F^{(7)}(1,t) \; u$ in this
trivialization patch are even and odd under the $\Z_2$ transformation 
$\sigma: (t,u) \mapsto (t,-u)$, the former group of zero modes 
are odd under the $\Z_2$ transformation on ${\cal F}_{({\bf 10})}$, 
and the latter are even. Thus, we find 
\begin{equation}
h^{0}(\bar{c}_{({\bf 10})}; \mathcal{F}_{({\bf 10})})_{-}=9, \;\;\;\;\;
h^{0}(\bar{c}_{({\bf 10})}; \mathcal{F}_{({\bf 10})})_{+}=8.
\end{equation}
%

% [classification of $\bar{\bf 5}$'s] $\quad$ 
The $(28 + N)$ independent zero mode chiral multiplets in the 
$\bar{\bf 5}$ representation and $(11 + N)$ independent ones 
in the ${\bf 5}$ representation can also be classified by 
looking at whether their description as coefficient holomorphic 
functions $\tilde{f}_a$'s in the trivialization patch 
$\tilde{\bar{c}}_{(\bar{\bf 5})} \cap U'_{\check{S}}$. 
Coefficient holomorphic functions even under the $\Z_2$ transformation 
$\sigma: (t,u) \mapsto (t,-u)$ on 
$\tilde{\bar{c}}_{(\bar{\bf 5})} \cap U'_{\check{S}}$ are even under 
the $\Z_2$ acting on the bundle $\widetilde{\cal F}_{(\bar{\bf 5})}$, 
and the coefficient functions that are odd under the $\Z_2$ on 
$\tilde{\bar{c}}_{(\bar{\bf 5})} \cap U'_{\check{S}}$
are odd under the $\Z_2$ transformation acting on the line bundle.
Among the 28 zero mode chiral multiplets in the $\bar{{\bf 5}}$ 
representation in (\ref{eq:5bar_hol}, \ref{eq:5barrepr-wavefcn}), 
for example, $F^{(6)}(S; T; U)_+ \sim U^{6-2i}S^{j}T^{2i-j}$ leads to 
$\mathbb{Z}_{2}$-even coefficient functions 
$\tilde{f}_{(\bar{\bf 5});\check{S}}$ in the 
$\tilde{\bar{c}}_{(\bar{\bf 5})} \cap U'_{\check{S}}$ patch,  
and $F^{(6)}(S; T; U)_- \sim U^{5-2i}S^{p}T^{2i+1-p}$ to 
$\mathbb{Z}_{2}$-odd coefficient functions. 
Thus, we find in this model that 
\begin{eqnarray}
h^{0}(\tilde{\bar{c}}_{(\bar{{\bf 5}})}; 
      \tilde{\mathcal{F}}_{(\bar{{\bf 5}})})_{+} \geq
h^{0}(\tilde{S}; \mathcal{O}(D_{(\bar{{\bf 5}})}))_{+} &=& 16, \\
h^{0}(\tilde{\bar{c}}_{(\bar{{\bf 5}})}; 
      \tilde{\mathcal{F}}_{(\bar{{\bf 5}})})_{-} \geq
h^{0}(\tilde{S}; \mathcal{O}(D_{(\bar{{\bf 5}})}))_{-}&=& 12.
\end{eqnarray}
%

%%%%%%%%%%%%%%%%%%%%%%%%%%%%%%%%%%%%%%%%%%%%%%%%%%%%%%%%%%%%%%%%%%%

%%%%%%%%%%%%%%%%%%%%%%%%%%%%%%%%%%%%%%%%%%%%
\subsection{A Brief Comment on $\SU(5)_{\rm GUT}$ Symmetry Breaking}
\label{ssec:SU5-breaking}
%%%%%%%%%%%%%%%%%%%%%%%%%%%%%%%%%%%%%%%%%%%%%

For practical purpose, one will eventually be interested in 
vacua with the $\SU(5)_{\rm GUT}$ symmetry broken down to 
$\SU(3)_C \times \SU(2)_L \times \U(1)_Y$. 
${\bf 10}$-representation of Georgi--Glashow $\SU(5)_{\rm GUT}$ 
symmetry contains both quark doublets $Q$ and anti-up-type quarks 
$\bar{U}$. It is of theoretical interest to what extent the 
wavefunctions of quark doublets and anti-up-type quarks are 
mutually related, and to what extent they are different. 
In this section \ref{ssec:SU5-breaking}, we will use Example XVI
in Table~\ref{tab:top-data-B}, and illustrate the calculation 
of holomorphic wavefunctions in the presence of $\SU(5)_{\rm GUT}$ 
symmetry breaking. The technique of calculation itself, however, 
is essentially the same as the one in \ref{ssec:Ex-7} 

The $\SU(5)_{\rm GUT}$ symmetry can be broken down to the Standard 
Model gauge group by turning on non-trivial Wilson line in cases 
with $\pi_1 (S) \neq \{ 1 \}$ \cite{BHV-2}, or by turning on 
a line bundle on $S$ with non-zero first Chern class \cite{BHV-2, TW-2, DW-2}. 
We will assume the latter cases in the following calculation, although 
the $\SU(5)_{\rm GUT}$ symmetry breaking by Wilson lines remains a 
viable solution. The $\U(1)_Y$ gauge boson remains massless if 
the first Chern class of the line bundle is in the orthogonal complement 
of the image of $i_S^*: H^2(B_3) \rightarrow H^2(S)$~\cite{
Buican, BHV-2, DW-2}, 
or if there is a strongly coupled U(1) gauge boson coupled to the 
same 2-form in $H^2(B_3)$ as the $\U(1)_Y$ gauge boson 
does \cite{WY-IIB, loop, TW-2}. In the rest of the calculation of 
wavefunctions, it does not matter in which way the $\U(1)_Y$ gauge 
boson remains massless; the following calculation applies to both. 

The Example XVI in Table \ref{tab:top-data-B} takes a rational 
elliptic surface (denoted by $dP_9$) as the GUT divisor $S$, and 
the matter curve $\bar{c}_{\bf 10}$ for the ${\bf 10}$-representation 
fields of $\SU(5)_{\rm GUT}$ in a topological class 
\begin{equation}
 \bar{c}_{({\bf 10})} \in |3H - (E_1 + \cdots + E_7)|.
\label{eq:Ex17-class-10}
\end{equation}

Here, rational elliptic surfaces $S$ is regarded as $\P^2$ with 
nine points $p_i$ $(i=1, \cdots, 9)$ blown up: 
$\nu_{\P^2}: S \rightarrow \P^2$. $H$ is the pullback of a hyperplane 
divisor of $\P^2$ to $S$, and $E_i$ ($i = 1, \cdots, 9$) are the 
exceptional curve obtained by the blow up centered at $p_i$.
Eight points $p_i$ ($i=1, \cdots, 8$) can be chosen arbitrarily 
on $\P^2$, but the last point $p_9$ should be chosen at a 
special place in $\P^2$; any cubic curves on $\P^2$ passing 
through the eight points $p_i$ ($i=1,\cdots,8$) pass through 
another point in $\P^2$, and this special point is the last 
point $p_9$ to be blown up. 
Rational elliptic surface is also regarded as a subvariety of 
$\P^1 \times \P^2$ given by a homogeneous function of bi-degree $(1,3)$:
\begin{equation}
 V_0 F(S, T, U) + V_1 G(S, T, U) = 0,
\label{eq:1-3}
\end{equation}
where $[V_0: V_1]$ are the homogeneous coordinates of $\P^1$, 
and $[S: T: U]$ those of $\P^2$, and $F$ and $G$ are homogeneous 
functions on $\P^2$ of degree 3. The rational elliptic surface 
is regarded as an elliptic fibration over $\P^1$, where 
$\P^1$ with the coordinates $[V_0: V_1]$ becomes the base, and 
the fiber geometry is a cubic curve in $\P^2$ given by (\ref{eq:1-3}).
The topological class of the elliptic fiber is 
\begin{equation}
 [x] = 3H - (E_1 + \cdots + E_9), 
\end{equation}
because they are the cubic curves (that is, $3H$) passing through 
all the blown-up points $p_i$ ($i=1,\cdots, 9$).
For physicist-friendly reviews on rational elliptic surface, see 
\cite{dP9, Hayashi-1}.

Generic elements in the class (\ref{eq:Ex17-class-10}) are irreducible 
curve of genus 1, and regarded as cubic curves in $\P^2$ 
passing through the seven points $p_1$--$p_7$.
There are four type-$E_6$ points, and six type-$D_6$ points. 
In this example, if the complex structure is tuned as in 
(\ref{eq:E8-limit}), so that an $E_8$ Higgs bundle is defined 
globally, an intermediate-step description of flux given by 
(\ref{eq:gamma-FMW}) with $\lambda = -1/2$ results in a model 
with three generations.
\begin{equation}
 N_{\rm gen} = - \lambda \eta \cdot (5K_S + \eta) 
 = \frac{1}{2} (6 [x] + E_8 + E_9) \cdot ([x] + E_8 + E_9) = 6.
\label{eq:Ex-16-net-chi}
\end{equation}
%

%%%%%%%%%%%%%%%%%%%%%%%%%%%%%%%%%%%%%%%%%%%%%%%%%%%%%%%
\subsubsection{Introducing a Hypercharge Flux}
%%%%%%%%%%%%%%%%%%%%%%%%%%%%%%%%%%%%%%%%%%%%%%%%%%%%%%%%

We have assumed so far that the GUT divisor $S$ is a locus 
of $A_4$ singularity, and we continue to do so. 
The $\SU(5)_{\rm GUT}$ symmetry associated with this singularity 
locus can be broken down to the Standard Model symmetry group 
$\SU(3)_C \times \SU(2)_L \times \U(1)_Y$, by turning on a 
line bundle on $S$.
 
Various irreducible representations under the Standard Model gauge 
group have different $\U(1)_Y$ charges. Zero mode wavefunctions 
for an irreducible representation with a hypercharge $Y$ are sections 
of a line bundle that contains a factor ``$(L_Y)^{\otimes Y}$'' for some 
``line bundle $L_Y$''. Since $Y=1/6$ of left-handed quark doublets is 
the smallest absolute value of hypercharges in the Standard Model, 
all the $(L_Y)^{\otimes Y}$'s would be well-defined line bundles if 
$(L_Y)^{\otimes Y=1/6}$ were. But, this is not strictly necessary \cite{BHV-2}.
Various fields in the Standard Model are regarded as sections of 
bundles that are comprized of some representation of a rank-5 
$\SU(5)_{\rm str}$ Higgs bundle $V_5$ and a line bundle $(L_Y)^{\otimes Y}$, 
as shown in Table~\ref{tab:SU5-SU6}. 
%%%%%%%%%%%%%%%%%%%%%%%%%%%%%%%%%%%%%%%%%%%%%%%%%%%%%%
\begin{table}[tb]
\begin{center}
\caption{\label{tab:SU5-SU6} Various fields coming from roots of 
$E_8$ are regarded as sections of corresponding Higgs bundles 
shown in this table. The column I shows the representation under 
the $\SU(5)_{\rm str} \times \SU(5)_{\rm GUT} \subset E_8$, while 
the representations in the column II are those of 
$\SU(5)_{\rm str} \times \U(1)_Y \times \SU(3)_C \times \SU(2)_L$.
The hypercharges $Y$ multiplied by $6$ are shown on the shoulder 
of the representations under $\SU(5)_{\rm str}$. Massless fields 
in these representations are regarded as sections of ``bundles'' 
specified in the next column, if the symmetry $E_8$ is broken 
by an $\SU(5)_{\rm str}$ ``bundle'' $V_5$ and a line ``bundle'' $L_Y$ 
whose structure group is in the hypercharge direction of 
$\SU(5)_{\rm GUT}$. It is not that $V_5$ and $L_Y^{\pm 1/6}$ 
should be well-defined bundles, however. When $L_Y^{-5/6}$ 
is defined to be ${\cal L}_Y$, and 
$U_5 \equiv V_5 \otimes {\cal L}_Y^{-1/5}$, then all the ``bundles'' 
appear in combination of $U_5$ and ${\cal L}_Y$. It is $U_Y$ and 
${\cal L}_Y$ that needs to be well-defined bundles. 
If we are to introduce a notation $V_6 \equiv U_5 \oplus {\cal L}_Y$, 
then all the fields in this table are grouped into irreducible 
representations of $\SU(6)_{\rm str} \times \SU(3)_C \times \SU(2)_L$, 
as shown in the column III.
}
 \begin{tabular}{c|c|c|rl|c}
  & I & II & vector & bundle &  III \\
\hline 
$\bar{U}$ & & $({\bf 5}^{-4}, \bar{\bf 3}, {\bf 1})$ & 
 $V_5 \otimes L_Y^{-4/6}$ & $\simeq U_5 \otimes {\cal L}_Y$ & 
   $(\wedge^2 V_6, \bar{\bf 3}, {\bf 1})$ \\
$Q$ & $(V_5, {\bf 10})$ & $({\bf 5}^{+1}, {\bf 3}, {\bf 2})$ & 
 $V_5 \otimes L_Y^{+1/6}$ & $\simeq U_5$ & $(V_6, {\bf 3}, {\bf 2})$ \\
$\bar{E}$ & & $({\bf 5}^{+6}, {\bf 1}, {\bf 1})$ & 
 $V_5 \otimes L_Y^{+6/6}$ & $\simeq U_5 \otimes {\cal L}_Y^{-1}$ & 
 $({\bf adj.}, {\bf 1}, {\bf 1})$ \\
\hline
$\bar{D}$ & & $({\bf 10}^{+2},\bar{\bf 3},{\bf 1})$ &
  $(\wedge^2 V_5) \otimes L_Y^{+2/6}$ & $\simeq (\wedge^2 U_5)$ &
  $(\wedge^2 V_6, \bar{\bf 3}, {\bf 1})$ \\
$L, H_d$ & $(\wedge^2 V_5, \bar{\bf 5})$ &
  $({\bf 10}^{-3},{\bf 1},{\bf 2})$ &
  $(\wedge^2 V_5) \otimes L_Y^{-1/2}$ & $\simeq (\wedge^2 U_5) \otimes
  {\cal L}_Y$ & $(\wedge^3 V_6, {\bf 1}, {\bf 2})$ \\
\hline
$X$ & $({\bf 1}, {\bf adj.})$ & $({\bf 1}^{-5}, {\bf 3}, {\bf 2})$ & 
  $L_Y^{-5/6}$ & $\simeq {\cal L}_Y$ & $(V_6, {\bf 3}, {\bf 2})$
 \end{tabular}
\end{center}
\end{table}
%%%%%%%%%%%%%%%%%%%%%%%%%%%%%%%%%%%%%%%%%%%%%%%%%%%%%%
As explained 
in \cite{BHV-2, Blumenhagen-0908}, all the bundles appearing in 
Table~\ref{tab:SU5-SU6} are regarded as tensor products of 
some representations of ${\cal L}_Y \equiv (L_Y)^{-5/6}$ and 
$U_5 \equiv V_5 \otimes {\cal L}^{-1/5}$. Thus, we only need 
${\cal L}_Y$ and $U_5$ to be well-defined vector bundles; 
$(L_Y)^{1/6}$ or $V_5$ do not have to be. 

Chiral multiplets in the off-diagonal blocks of the $5 \times 5$ 
matrix of $\SU(5)_{\rm GUT}$ should not remain in the low-energy 
spectrum. Since the off-diaonal components have hypercharge 
$Y = \pm 5/6$, it would be a section of ${\cal L}_Y^{\pm 1}$, if there 
were a zero mode.
When the GUT divisor is $S = dP_k$ with $k = 0, \cdots, 9$, 
\begin{equation}
 {\cal L}_Y = {\cal O}_S(D_Y), \qquad D_Y = E_i - E_j \quad (i \neq j)
\end{equation}
is known to be a solution without an unnecessary massless modes 
from the off-diagonal components \cite{BHV-2}.
Since $ \pm (E_i - E_j)$ are not effective, $h^0(S; {\cal L}_Y^{\pm 1}) = 0$. 
Since $K_S \mp D_Y = - [x] \mp (E_i - E_j)$ are not effective, 
it also follows that $h^2(S; {\cal L}_Y^{\pm 1}) = 0$.
Finally, because
\begin{equation}
\chi(S; {\cal L}_Y^{\pm 1}) = {\rm td}_2(TS) 
  + \frac{[x]}{2} \cdot (\pm D_Y) + \frac{D_Y^2}{2} 
 = 1 + 0 + (-1) = 0,  
\end{equation}
$h^1(S ; {\cal L}_Y^{\pm 1}) = 0$. 

Now, let us construct a Higgs bundle $(U_5, \varphi)$. 
Since we want to preserve the gauge coupling unification, we keep 
the vacuum configuration of $\varphi$ to be strictly 
in $\SU(5)_{\rm str}$; we do not want to break the $\SU(5)_{\rm GUT}$ 
symmetry by the Higgs vev, or otherwise the branes for $\SU(3)_C$ 
and those for $\SU(2)_L$ would effectively be wrapped on different 
cycles, and their volumes are different, so that the gauge coupling 
unification is lost. The structure group of $U_5$, however, 
can be $\U(5)$, not $\SU(5)_{\rm str}$, and in fact, it should 
have non-zero first Chern class. Since we want ${\rm det} U_5 \simeq
{\cal L}_Y^{-1}$, we need to have $c_1(U_5) = - D_Y$. This is achieved 
if the traceless condition $\pi_{C*} \gamma = 0$ is replaced by 
\begin{equation}
 \pi_{C*} \gamma = - D_Y.
\label{eq:shifted-cond-gamma}
\end{equation}

In this article, we will not try to find the most generic way 
to find such $\gamma$. We restrict ourselves to the intermediate 
step approach explained in the previous section, and further assume 
the $E_8$ limit in the complex structure just for simplicity of 
the argument. We will be satisfied with only one example of $\gamma$ 
in this section, and use that choice of $\gamma$ to discuss 
holomorphic wavefunctions of zero modes of $\bar{U}$, $Q$ and $\bar{E}$.

Let us take 
\begin{equation}
 D_Y = E_8 - E_9.
\end{equation}
In order to construct $\gamma$ on the 5-fold spectral cover, we 
introduce an extra assumption. Let us suppose that the global
holomorphic section $a_5 \in \Gamma (S; {\cal O}(5K_S + \eta))$ 
is chosen so that the matter curve $\bar{c}_{({\bf 10})}$ 
($a_5 = 0$ locus) corresponds to a cubic curve in $\P^2$ that 
happens to pass through $p_8$ and $p_9$ as well, not just through 
$p_{1,\cdots, 7}$. In this case, the matter curve 
$\bar{c}_{({\bf 10})}$ consists of three irreducible pieces: one 
is the cubic curve described above, which belongs to the class 
$3H - (E_1 + \cdots + E_9) = [x]$, and the two others are 
$E_8$ and $E_9$. Since $\bar{c}_{({\bf 10})}$ is a divisor of 
the 5-fold spectral cover, and $E_8$ and $E_9$ are irreducible 
components of $\bar{c}_{({\bf 10})}$, we can choose a divisor 
$\gamma_0$ of the spectral cover as 
\begin{equation}
 \gamma_0 = E_9 - E_8.
\label{eq:gamma-0}
\end{equation}
It is now obvious that $\pi_{C*} \gamma_0 = - D_Y$.
This is basically along the line of idea in \cite{DW-3}; 
for non-generic choice of complex structure (of the spectral surface),
more choice for $\gamma$ is available in 
\begin{equation}
 H^2(C_{({\bf 10})}; \Z) \cap H^{1,1}(C_{({\bf 10})}; \R).
\end{equation}
Although it appears that an extra tuning of complex structure is 
necessary, in fact, the complex structure parameter that had to be 
tuned should become massive in the presence of fluxes like $\gamma_0$.
Vacua constructed in this way are just isolated from others by 
moduli potential. It is not that fine tuning is necessary \cite{DW-3}. 

Once one finds a divisor $\gamma_0$ on the spectral surface $C_{U_5}$ 
that satisfies (\ref{eq:shifted-cond-gamma}), then 
$\gamma = \gamma_0 + \gamma_{FMW}$ with $\gamma_{FMW}$ in the form 
of (\ref{eq:gamma-FMW}) also satisfy (\ref{eq:shifted-cond-gamma}) for 
any value of $\lambda$. If $\gamma_0 \cdot \bar{c}_{({\bf 10})} = 0$ 
in $C_{U_5}$ (which is the case for (\ref{eq:gamma-0}), as we see
shortly), then $\gamma_0$ does not contribute to the net chirality, 
and all the $\SU(3)_C \times \SU(2)_L \times \U(1)_Y$ irreducible 
components $\bar{U}, Q, \bar{E}$ within the ${\bf 10}$-representation 
of $\SU(5)_{\rm GUT}$ have the same net chirality.
The common net chirality comes from 
$\bar{c}_{({\bf 10})} \cdot \gamma_{FMW}$, which is then the same 
as (\ref{eq:Ex-16-net-chi}).

%%%%%%%%%%%%%%%%%%%%%%%%%%%%%%%%%%%%%%%%%%%%%%%%%%%%%
\subsubsection{Wavefunctions of $Q$, $\bar{U}$ and $\bar{E}$}
%%%%%%%%%%%%%%%%%%%%%%%%%%%%%%%%%%%%%%%%

Now that we have specified how the hypercharge flux 
is introduced, let us calculate the wavefunctions of 
independent matter fields in the presence of $\SU(5)_{\rm GUT}$ 
symmetry breaking. 
Zero modes are global holomorphic sections of line bundles on 
the (reducible) matter curve $\bar{c}_{({\bf 10})}$ given by 
\begin{equation}
 {\cal F}_{(\bar{U},Q,\bar{E})} = 
  {\cal O} \left( i^* K_S + \frac{1}{2}p_{E6} 
     + j^* (\gamma_{FMW} + \gamma_0) - \frac{6Y-1}{5} i^* D_Y\right).
\label{eq:curve-div-UQE}
\end{equation}
Note a subtle difference between pull-backs by 
$i: \bar{c}_{({\bf 10})} \hookrightarrow S$ and 
$j: \bar{c}_{({\bf 10})} \hookrightarrow C_{({\bf 10})}$.
Now the divisors $D_Y$ and $\gamma_0$ are used in describing 
the line bundles. 
In the last component, $Y=-2/3$, $Y=+1/6$ and $Y=+1$ should be used 
in $(6Y-1)/5$ for $\bar{U}$, $Q$ and $\bar{E}$. 

In the process of finding a candidate for $\gamma_0$, we have chosen 
a limit of complex structure so that the matter curve 
$\bar{c}_{({\bf 10})}$ consists of three irreducible components:
\begin{equation}
 \bar{c}_{({\bf 10})} = \bar{c}_{({\bf 10})0} + E_8 + E_9; 
\end{equation}
the latter two are isomorphic to the base $\P^1$ with the coordinates 
$[V_0: V_1]$, and the first one $\bar{c}_{({\bf 10})0}$ belongs to 
the fiber class $[x]$ (and hence genus one).
The ramification locus of the spectral cover on the matter curve 
$\bar{c}_{({\bf 10})}$ (which are also the $E_6$ type points) 
are also distributed to the three irreducible components. 
Since these points are characterized as zero locus of 
$a_4 \in \Gamma (S;{\cal O}(4K_S + \eta)) = 
 \Gamma(S; {\cal O}(2[x] + E_8 + E_9))$ on the matter curve, 
it turns out that the irreducible components 
$\bar{c}_{({\bf 10})0}$, $E_8$ and $E_9$, respectively, have 
\begin{equation}
 (2[x] + E_8 + E_9) \cdot [x] = 2, \qquad 
 (2[x] + E_8 + E_9) \cdot E_8 = 1, \qquad 
 (2[x] + E_8 + E_9) \cdot E_9 = 1
\end{equation}
of them. 

The first three components of the divisor for ${\cal F}_{\bar{U},Q,\bar{E}}$
remain the same as the divisor for ${\cal F}_{({\bf 10})}$ 
in (\ref{eq:H0-10-repr}). We have seen in section \ref{ssec:Ex-7}
that these components define a divisor on the matter curve as a linear 
combination of points on the curve $\bar{c}_{({\bf 10})}$ (mod linear 
equivalence), and that the divisor is obtained as a restriction of 
the divisor $D_{({\bf 10})}$ on $S$ restricted on 
$\bar{c}_{({\bf 10})}$. Nothing has to be changed in this story, even 
though the matter curve $\bar{c}_{({\bf 10})}$ is no longer
irreducible, and the four type-$E_6$ points (ramification points) 
are distributed to the three irreducible components in the way we have 
seen above.

We still have a little more work to do with regard to 
the latter two components in (\ref{eq:curve-div-UQE}), 
just like we did with $j^* \gamma_{FMW}$ in section \ref{ssec:Ex-7}.
Irreducible components of the support of $D_Y = E_8 - E_9$ in $S$ 
share in common with the matter curve $\bar{c}_{({\bf 10})}$. The 
divisor $\gamma_0 = - E_8 + E_9$ in the spectral surface 
$C_{({\bf 10})}$ also share irreducible components of its 
support with the matter curve. Thus, the divisors $D_Y$ and $\gamma_0$ 
have to be replaced by their linearly equivalent ones in $S$ and 
$C_{({\bf 10})}$, respectively, so that the replaced ones---$D_Y'$ 
and $\gamma_0'$---have transverse intersection with the matter 
curve in $S$ and $C_{({\bf 10})}$, respectively. 

Let us take an arbitrary line in $\P^2$ passing through $p_8$, and 
denote by $H_8$ a homogeneous function of degree one on $\P^2$
whose zero locus is the line we have chosen. Similarly, let $H_9$ 
be a homogeneous function of degree one on $\P^2$ whose zero locus 
passes through $p_9$. Then, 
\begin{equation}
 {\rm div} \; \nu_{\mathbb{P}^{2}}^* \left(\frac{H_8}{H_9}\right) = E_8 + L_8 - E_9 - L_9,
\end{equation}
where $L_8$ and $L_9$ are divisors of the rational elliptic surface $S$
and are the proper transforms of the lines of $\P^2$ we introduced
above. $L_8 \in |H - E_8|$ and $L_9 \in |H - E_9|$.
Let us introduce 
\begin{equation}
 D'_Y = D_Y - {\rm div} \; \nu_{\mathbb{P}^{2}}^* \left(\frac{H_8}{H_9}\right) 
  = - L_8 + L_9.
\end{equation}
This divisor $D'_Y$ in $S$ is linear equivalent to the original $D_Y$, 
and intersect transversely with all the irreducible components of 
the matter curve $\bar{c}_{({\bf 10})}$. Thus, 
$i^*D'_Y = D'_Y|_{\bar{c}_{({\bf 10})}}$ defines a collection 
of intersection points with integer coefficients, a divisor on 
the matter curve. The degree of the divisor on the three irreducible 
components are shown in Table \ref{tab:withYflux}.
%%%%%%%%%%%%%%%%%%%%%%%%%%%%%%%%%%%%%%%%%%%
\begin{table}[tb]
 \begin{center}
\caption{\label{tab:withYflux}  Degree of various divisors (1st--5th
column) on the three irreducible components
% , $\bar{c}_{({\bf 10})0}$, $E_8$ and $E_9$, 
of the matter curve $\bar{c}_{({\bf 10})}$. When all these 
degrees are summed up, one obtains the degree of the line bundle 
${\cal F}_{\bar{U},Q,\bar{E}}$ (6th--8th column) on the three 
irreducible components. $i^* D'_Y$ contributes with a coefficient 
$-(6Y-1)/5 = +1, 0, -1$ for $\bar{U}$, $Q$ and $\bar{E}$, respectively.}
  \begin{tabular}{c|ccc|c|c||c|c|c}
   & $i^* K_S$ & $i^* a_4^3$ & $i ^* \psi^{-2}$ & $j^* \gamma'_0$ &
 $i^* D'_Y$ & $\bar{U}$ & $Q$ & $\bar{E}$ \\
\hline
$\bar{c}_{({\bf 10})0}$ 
      & 0 & 6 & $-4$ & $(+2-2)$ & $(-2+2)$ & $+2$ & $+2$ & $+2$ \\
$E_8$ & $-1$ & 3 & 0 & $1+1$ & $-1+0$ & $3$ & $4$ & $5$ \\
$E_9$ & $-1$ & 3 & 0 & $(-1-1)$ & $0+1$ & $1$ & $0$ & $-1$  \\
\hline 
total & & & & 0 & 0 & 6 & 6 & 6
  \end{tabular}
 \end{center}
\end{table}
%%%%%%%%%%%%%%%%%%%%%%%%%%%%%%%%%%%%%%%%%%%%%%%

We are now done with the divisor $i^* D_Y$, and only $j^* \gamma_0$ 
remains. Let us introduce a divisor $\gamma'_0$ on the spectral 
surface that is linear equivalent to $\gamma_0$, as follows:
\begin{equation}
 \gamma'_0 = \gamma_0 + 
  {\rm div} \; \pi_{C_{({\bf 10})}}^* \nu_{\mathbb{P}^{2}}^* \left( \frac{H_8}{H_9}\right). 
\end{equation}
The divisor $\gamma'_0$ restricted on $\bar{c}_{({\bf 10})}$ 
contains three intersection points of $L_8$ and $\bar{c}_{({\bf 10})}$ 
in $S$, with all coefficients being $+1$, and three intersection points 
of $L_9$ and $\bar{c}_{({\bf 10})}$ in $S$, with all the coefficients
being $-1$. All these contributions as a whole is the same as 
$ - i^* D'_Y$. In addition to these contribution, $\gamma'_0$ has two 
more support points; one is the ramification point on $E_8$ with the 
coefficient $+1$, and the other is the ramification point on $E_9$ with 
the coefficient $-1$. All of this information is summarized in 
Table~\ref{tab:withYflux}.

Now we have succeeded in describing the ``divisor'' on the right-hand 
side of (\ref{eq:curve-div-UQE}) for $\bar{U}$, $Q$ and $\bar{E}$ 
truly as divisors on the matter curve $\bar{c}_{({\bf 10})}$---a
collection of points of $\bar{c}_{({\bf 10})}$ with integral
coefficients. If these divisors were obtained as restrictions of some 
divisors on the rational elliptic surface $S$, then the same technique 
as in section \ref{ssec:Ex-7} could be used, so that the calculation of 
the wavefunctions would be converted into calculation of cohomology 
of line bundles on $S$. The question is in the part $j^* \gamma'_0$, 
since all other components are in the form of 
$i^* D_{({\bf 10})} - (6Y-1)/5 i^* D'_Y$. So far, it does not seem 
to the authors that $j^* \gamma'_0$ can be obtained as a restriction 
of some divisor of $S$.

In the example that we have studied so far, however, $[V_0: V_1]$
coordinates of $\P^1$ can be used for the irreducible components $E_8$ 
and $E_9$, and the other component $\bar{c}_{({\bf 10})0}$ of the 
matter curve is described as a cubic curve in $\P^2$ with the
coordinates $[S: T: U]$. Thus, the wavefunctions can be calculated 
much like in a way we present in section \ref{sec:unit-wavefcn}, 
using the coordinates of $\P^1$ and $\P^2$. 
In other generic examples, one might have to somehow introduce 
open covering of the matter curve, and do the calculation of 
\v{C}ech cohomology, using the precise data of the divisor we obtained 
above (where information of complex structure is not lost).  

In the example we have been studying explicitly in this 
section \ref{ssec:SU5-breaking}, the line bundle 
${\cal F}_{\bar{U}}$ is a degree $+2$ line bundle, when it is 
restricted on the genus one $\bar{c}_{({\bf 10})0}$ component 
(see Table \ref{tab:withYflux}). 
Since $j^* \gamma'_0 - (6Y-1)/5 i^* D'_Y$ vanishes on 
$\bar{c}_{({\bf 10})0}$ for the anti-up-type quarks with $Y = -2/3$, 
the Wilson line part of this degree-two line bundle is determined 
from the $K_S + p_{E6}/2 + j^* \gamma_{FMW}$ part. The line bundle 
${\cal F}_{(\bar{U})}$ restricted on $E_8$ and $E_9$ are 
${\cal O}(+3)$ and ${\cal O}(+1)$, respectively. Thus, 
\begin{equation}
{\rm dim} \; H^0(\bar{c}_{({\bf 10})0}; {\cal
 F}_{\bar{U}}|_{\bar{c}_{({\bf 10})0}}) = 2,  \quad 
{\rm dim} \; H^0(E_8; {\cal O}(+3)) = 4, \quad 
{\rm dim} \; H^0(E_9; {\cal O}(+1)) = 2.
\end{equation}
Let the independent generators of the three vector spaces as 
$\tilde{f}_{0;i_0}$ $(i_0 = 1,2)$, $\tilde{f}_{8;i_8}$ ($i_8 = 1,\cdots,4$)
and $\tilde{f}_{9; i_9}$ ($i_9 = 1,2$), respectively. 
The wavefunction $\tilde{f}$ on the matter curve $\bar{c}_{({\bf 10})}$ 
can be an arbitrary linear combination 
$\sum_{i_0} c_{i_0} \tilde{f}_{0; i_0}$ on $\bar{c}_{({\bf 10})0}$, 
$\sum_{i_8} c_{i_8} \tilde{f}_{8; i_8}$ on $E_8$ and 
$\sum_{i_9} c_{i_9} \tilde{f}_{9; i_9}$ on $E_9$, and hence there 
are $2 + 4 + 2$ arbitrary complex parameters that can be chosen 
almost independently; the only constraint is that the values of 
$\tilde{f}$ on $\bar{c}_{({\bf 10})0}$ and on $E_8$ [resp. $E_9$] 
should be the same \cite{Tsuchiya}. This condition introduces 
two constraints, and 
six independent degrees of freedom are left. Since there are six 
massless $\bar{U}$'s in this example, and since we know that the net 
chirality is six, there should be no massless fields in the conjugate 
representation of $\bar{U}$ in the low-energy spectrum in this example. 

Similarly, the same number of $Q$-type and $\bar{E}$-type independent 
massless chiral multiplets are obtained in this example. 
The line bundle ${\cal F}_Q$ is degree $+2$, $+4$ and $0$, 
when restricted upon the irreducible components 
$\bar{c}_{({\bf 10})0}$, $E_8$ and $E_9$, respectively.
Thus, it follows that 
\begin{equation}
 h^0(\bar{c}_{({\bf 10})0}; {\cal F}_{Q}|_{\bar{c}_{({\bf 10})0}}) = 2,  
  \qquad 
 h^0(E_8; {\cal O}(+4)) = 5, \qquad 
 h^0(E_9; {\cal O}) = 1,
\end{equation}
and there are $(2+5+1) = 8$ linear-combination coefficients of 
the wavefunctions, but $(8-2)= 6$ remain free after requiring that 
$\tilde{f}$ on $\bar{c}_{({\bf 10})0}$ and that on $E_8$ [resp. $E_9$] 
should take the same value at the intersection point. 

The line bundle ${\cal F}_{\bar{E}}$ for the $\bar{E}$-type fields 
is degree $+2$, $+5$ and $-1$, respectively, on the irreducible 
components $\bar{c}_{({\bf 10})0}$, $E_8$ and $E_9$.
Thus, there is no holomorphic section on $E_9 \simeq \P^1$.
Consequently the holomorphic sections on $\bar{c}_{({\bf 10})0}$ 
should vanish at the point where $\bar{c}_{({\bf 10})0}$ and $E_9$ 
intersect. Thus, there is only one independent choice on
$\bar{c}_{({\bf 10})0}$. After requiring that $\tilde{f}$ on $E_8$ 
should take the same value at the intersection point as $\tilde{f}$ 
on $\bar{c}_{({\bf 10})0}$, only six independent wavefunctions are left.

In this example, six independent holomorphic wavefunctions are obtained 
for all of $\bar{U}$, $Q$ and $\bar{E}$, and no massless chiral
multiplets were predicted in the conjugate representations of 
$\bar{U}$, $Q$ or $\bar{E}$. The holomorphic wavefunctions are different 
for $\bar{U}$, $Q$ and $\bar{E}$. To see this, it will be sufficient 
to point out 
in this example that all the independent wavefunctions vanish on $E_9$ 
for the massless $\bar{E}$'s, all the independent wavefunctions are 
constant and (generically) non-zero over the entire $E_9 \simeq \P^1$ 
for the massless $Q$'s, and finally, all the wavefunctions for 
massless $\bar{U}$'s vary over $E_9 \simeq \P^1$.

%%%%%%%%%%%%%%%%%%%%%%%%%%%%%%%%%%%%%%%
\subsection{Recap}
%%%%%%%%%%%%%%%%%%%%%%%%%%%%%%%%%%%%%%%

Different complex structure of a Calabi--Yau 4-fold $X$ 
results in different complex structure of the matter curve
$\bar{c}_{(R)}$ for representation $R$ in the GUT divisor $S$, 
and in different divisor class describing the line bundle 
$\tilde{\cal F}_{(R)}$. Holomorphic wavefunctions of massless 
modes are the global holomorphic sections of these line bundles.
In order to calculate (or at least evaluate) contributions to 
Yukawa couplings from multiple points of codimension-3 singularity 
of $E_6$ type (for up-type Yukawa couplings) and of $D_6$ type 
(for down-type and charged lepton Yukawa couplings), we need 
to be able to calculate the zero mode wavefunctions. 

Although essence of the prescription of calculation already 
appear in the literature, but we used explicit examples and 
carried out the calculation for illustrative purpose in this 
section. The first step is to fix $\gamma$ for charged matter 
fields in multiple representations of the unbroken symmetry $G''$ 
(like $\SO(10)$, $\SU(5)_{\rm GUT}$ and 
$\SU(3)_C \times \SU(2)_L \times \U(1)_Y$), so that all the $\gamma$'s 
for multiple representations can descend from a common 4-form flux 
on $X$. $\gamma_{FMW}$ of (\ref{eq:gamma-FMW}) 
for ${\bf 10}$ representation of $\SU(5)_{\rm GUT}$ and 
$\tilde{\pi}_{D*} \gamma$ for the $\bar{\bf 5}$ representation 
in section \ref{ssec:Ex-7} [and its variation in section
\ref{ssec:SU5-breaking} that incorporates $\SU(5)_{\rm GUT}$ symmetry 
breaking using a flux in hypercharge] is one of possible techniques 
that guarantees the common origin. We should keep in mind, however, 
that this construction relying on existence of 5-fold spectral cover
globally defined on $S$ is not available for F-theory compactifications
with generic complex structure (without a limit (\ref{eq:E8-limit})). 
If one has plenty of technique to calculate contributions of a given 
$H^{2,2}(X)$ flux to 4-cycles given by vanishing 2-cycle fibered over 
the covering matter curve $\tilde{\bar{c}}_{(R)}$, then there is no 
need to start from intermediate step of using $\gamma$'s on spectral 
surfaces.

The second step is to express the divisor specifying $\tilde{\cal F}_{(R)}$ 
($R = {\bf 10}, \bar{\bf 5}, {\bf 5}$ when $G'' = \SU(5)_{\rm GUT}$) 
really as a divisor on the curve $\tilde{\bar{c}}_{(R)}$, a linear
combination of points on $\tilde{\bar{c}}_{(R)}$. Linear equivalence 
of divisors sometimes need to be exploited. 

If the divisor obtained in this way can be regarded as restriction 
of some divisor of a surface (like $S$ or $\tilde{S}$) where 
the curve $\tilde{\bar{c}}_{(R)}$ is embedded, then the calculation 
of global holomorphic sections on the curve $\tilde{\bar{c}}_{(R)}$  
can be carried out as calculation of cohomology groups of line bundles 
on the surface. See (\ref{eq:long-10}), (\ref{eq:long-5bar}) and 
(\ref{eq:5h-ExVI}). This was the case in the example we studied in 
section \ref{ssec:Ex-7}, and in fact, this is always the case whenever 
$\gamma$ on the spectral surface for the ${\bf 10}$-representation of 
$G'' = \SU(5)_{\rm GUT}$ is given in the form of (\ref{eq:gamma-FMW}). 
It is much easier to calculate the wavefunctions as cohomology group 
elements on the {\it surface} than those as cohomology on the 
{\it curves} given as a subvariety of the surface. 

The divisors describing the line bundles $\tilde{\cal F}_{(R)}$ may not 
always be regarded as restriction of some divisors in $S$ or its blow-up 
at $D_6$ points, $\tilde{S}$. Even in such cases, brute-force
calculation using \v{C}ech cohomology may still be possible. 
As long as full information of divisor class is maintained 
(not just the degree of the line bundle), all the necessary 
information for calculation of the zero mode wavefunctions 
is not lost. 

In section \ref{ssec:SU5-breaking}, we presented a possible 
choice of hypercharge flux for $\SU(5)_{\rm GUT}$ symmetry 
breaking. It is meant primarily to provide an explicit 
example, where we can carry out calculations of wavefunctions 
in the presence of $\SU(5)_{\rm GUT}$ symmetry breaking. 
It is an example for illustrative purpose, and is not meant 
to be the ``best'' or ``most generic''. We believe that there 
will be a wide room for improvement. Many questions associated 
with the two Higgs doublets of the supersymmetric Standard Model 
must be closely related to this $\SU(5)_{\rm GUT}$ symmetry breaking 
(and the doublet--triplet splitting problem). This is an interesting 
and important subject, but we leave it as a future problem in this 
article.

The main focus of this article is to discuss flavor structure 
in F-theory compactifications in general. The notion of 
flavor structure (masses and mixing angles) is a well-defined problem, 
as long as there are $N_{\rm gen} > 1$ zero-mode fields (chiral 
multiplets) in the same representation of $G''$ for a given 
compactification. Thus, we consider that it is not strictly necessary 
for the purpose of discussing flavor structure, to find a specific 
geometry with $N_{\rm gen} = 3$, and just two Higgs doublet.\footnote{
Needless to say, it {\it is} important for the purpose of providing an 
existence proof.} It is an assumption in this article that 
compactification geometries with just two Higgs doublet and 
$N_{\rm gen} = 3$ net chirality does not have very specific (and 
unexpected) natures that have any consequences in the flavor pattern. 

\section{$D$-term and $F$-term}
\label{sec:DandF}
%%%%%%%%%%%%%%%%%%%%%%%%%%%%%%%%%%%%%%%%%%%%%%%%%%%%%%%

%%%%%%%%%%%%%%%%%%%%%%%%%%%%%%%%%%%%%%%%%%%%%%
\subsection{Kinetic Terms from Dimensional Reduction}
%%%%%%%%%%%%%%%%%%%%%%%%%%%%%%%%%%%%%%%%%%%%%%

Masses and mixing angles of quarks and leptons are determined 
by their kinetic terms as well as Yukawa couplings in low-energy 
effective theory. 
\begin{eqnarray}
 \Delta {\cal L}_{4D, {\rm eff.}} & = &  
  K^{(q)}_{ji} \bar{q}_j i \bar{\sigma}^\mu D_\mu q_i + 
  K^{(u)}_{ji} \bar{u^c}_j i \bar{\sigma}^\mu D_\mu u^c_i + 
  K^{(d)}_{ji} \bar{d^c}_j i \bar{\sigma}^\mu D_\mu d^c_i \nonumber \\
 & & + \left[ \lambda^{(u)}_{ij} u^c_i q_j h 
            + \lambda^{(d)}_{kj} d^c_k q_j h^* \right] 
     + {\rm h.c.} + \cdots.
\label{eq:SM-quark-Yukawa}
\end{eqnarray}
Physical Yukawa eigenvalues of the up-type quarks are square roots of 
the eigenvalues of 
$[K^{(q)}]^{-1} \lambda^{(u) \dagger} K^{(u) -1 T} \lambda^{(u)}$, 
for example. Observed flavor structure such as hierarchy among Yukawa 
eigenvalues may be due to some structure in the Yukawa matrices 
$\lambda^{(u)}$ and $\lambda^{(d)}$, but the kinetic mixing matrices 
$K^{(q)}$, $K^{(u)}$ and $K^{(d)}$ may also play some role. 
Even if Yukawa matrices [or kinetic mixing matrices] have some 
structure, it is (in principle) possible that the structure is not 
reflected in physical observables because of cancellation against 
some structure in the kinetic mixing matrices [or Yukawa matrices]. 
Zero modes in a given representation of the Standard Model gauge group 
form a vector space, and one can freely choose a basis of the vector 
space in writing down the effective Lagrangian of the Standard Model. 
Yukawa matrices $\lambda^{(u,d)}$ and kinetic mixing matrices 
$K^{(q,u,d)}$ depend on the choice of the basis. It is the physical 
observables that do not depend on the choice of basis, and 
we need to know both to derive physical observables.  

In supersymmetric compactifications, the Yukawa couplings are 
in the superpotential, and is protected from radiative corrections. 
On the other hand, the bilinear kinetic terms are in the K\"{a}hler 
potential, and in general, are subject to various corrections in the 
absence of unbroken extended supersymmetry. We do not try to provide 
a perfect formula for the kinetic terms here, but provide a trial 
expression (corrections to which still remain out of control). 

Suppose in an F-theory compactification on a Calabi--Yau 4-fold 
$X$, where $\pi_X: X \rightarrow B_3$ is an elliptic fibration on 
a 3-fold $B_3$, that the discriminant locus $\Delta = 0$ has  
multiple irreducible components, and a non-Abelian gauge theory 
is supported on one of them. The component of the non-Abelian 
gauge theory (like $\SU(5)_{\rm GUT}$ GUT sector) is denoted by $S$.
Physics of charged matter fields on $S$ ($\SU(5)_{\rm GUT}$-charged 
fields) is captured by an effective field theory on 7+1 dimensional 
spacetime $\R^{3,1} \times S$ \cite{KV, DW-1, BHV-1}; the leading order 
terms of the action is given by 
\begin{eqnarray}
 {\cal L}_{8D}^{\rm LO-bos} & = & \frac{M_*^4}{4\pi} 
   \tr {}' \left[ 
    \frac{\omega \wedge \omega}{2} \left( 
       \frac{1}{2} D^2 - \frac{1}{4} F_{\mu \nu} F^{\mu\nu}
       - \frac{\theta}{8} F_{\mu\nu} F_{\kappa\lambda} 
         \epsilon^{\mu\nu\kappa\lambda}
                         \right)
      \right. \nonumber \\
  & & \quad \qquad \left. +  \omega \left(
        i {\cal G}^* \wedge {\cal G} - D F^{(1,1)} 
      - i F_{\mu}^{(1,0)} \wedge F^{(0,1) \mu}  
                        \right)  \right. \nonumber \\
  & & \quad \qquad \left. 
      -  \alpha^* \left( F^{(2,0)} \wedge \overline{\cal H} 
               + {\cal G}^* \wedge D' \overline{\varphi} \right)
      -  \alpha \left( {\cal H} \wedge F^{(0,2)} 
               + {\cal G} \wedge D'' \varphi \right)
             \right. \nonumber \\
  & & \quad \qquad \left. 
       + \frac{|\alpha|^2}{2} \left( {\cal H} \wedge \overline{\cal H} 
		       + [\varphi, \overline{\varphi} ] D 
                       - D_\mu \varphi D^\mu \overline{\varphi}
                 \right)          
             \right],
\label{eq:bosonic}
\end{eqnarray}
and 
\begin{eqnarray}
 {\cal L}_{8D}^{\rm LO-fer} & = & \frac{M_*^4}{4\pi} \tr {}' \left[
   \frac{\omega \wedge \omega}{2} i (D_\mu \eta) \sigma^\mu \bar{\eta}
  - \omega \wedge (D_\mu \bar{\psi}) \wedge \bar{\sigma}^\mu \psi
   + \frac{|\alpha|^2}{2} i (D_\mu \chi) \wedge \sigma^\mu \bar{\chi} 
 \right. \nonumber \\
  & & \qquad \quad +
   \left( \sqrt{2} i \omega \wedge D' \eta \wedge \psi
    +  \sqrt{2} i \omega \wedge \bar{\psi} \wedge D'' \bar{\eta}
   \right) \nonumber \\
  & & \qquad \quad + \frac{|\alpha|^2}{2} 
   \left( \sqrt{2} i [ \overline{\varphi}, \eta ] \wedge \chi 
        + \sqrt{2} i \bar{\chi} \wedge [ \varphi, \bar{\eta}]
   \right) \nonumber \\
  & & \left. \qquad \quad + 
    \alpha^* \left(\bar{\chi} \wedge D' \bar{\psi} 
         - \frac{i}{2} \bar{\psi}[ \overline{\varphi}, \bar{\psi}]
     \right) 
%   \nonumber \\
%   & & \left. \quad + 
   +  \alpha \left(\chi \wedge D'' \psi 
         - \frac{i}{2} \psi [ \varphi, \psi ] \right) \right].
\label{eq:fermi-bilin}
\end{eqnarray}
Here, $D'$ and $D''$ are (1,0) and (0,1) parts of the covariant 
derivative $D = d + i \rho_{U_I} (\vev{A})$, respectively, 
and $M_*^{-4}$ corresponds to $(2\pi)^4 (\alpha')^2 g_s$ of 
the Type IIB string theory. 
All the physical bosonic component fields $A_\mu(x,y)$, $A_{\bar{m}}$, 
$A_m$, $\varphi_{mn}$ and $\overline{\varphi}_{\bar{m}\bar{n}}$ 
have mass-dimension $+1$, and all the fermionic component fields 
$\eta_\alpha(x,y)$, $\bar{\eta}_{\dot{\alpha}}$, 
$\psi_{\bar{m}\alpha}$, $\bar{\psi}_{m \dot{\alpha}}$, 
$\chi_{mn \alpha}$ and $\bar{\chi}_{\bar{m\bar{n}} \dot{\alpha}}$ 
mass-dimension $+3/2$. The mass dimension is $+2$ for all the 
auxiliary component fields $D(x,y)$, ${\cal G}_{\bar{m}}$, 
${\cal G}^*_m$, ${\cal H}_{mn}$ and 
$\overline{\cal H}_{\bar{m}\bar{n}}$. 
See \cite{Hayashi-2} for all other the details of the convention. 
The bosonic and fermion-bilinear terms above are meant only to be 
the leading order terms in the $1/M_*^4$-expansion. One will 
generally expect that these terms are followed by higher order terms 
such as $\tr (F^4)$, with coefficients that are not known. 
When these higher order terms are ignored, the conditions for 
supersymmetric background become  
\begin{eqnarray}
 (D) & & \omega \wedge F^{(1,1)} - 
   \frac{|\alpha|^2}{2} \left[ \varphi, \overline{\varphi} \right] = 0, 
   \label{eq:BPS-D}\\
 ({\cal H}) & &  F^{(0,2)} = 0,  \label{eq:BPS-H} \\
 ({\cal G}) & &  D'' \varphi = 0. \label{eq:BPS-G}
\end{eqnarray}

Let us consider a local geometry of $X$ containing a patch $U$ of $S$ 
that is approximately an ALE fibration over $U$, and the ALE fiber is 
deformation of singularity of one of ADE types. Let this 
type be $G$, and the undeformed part of the singularity be $G''$. 
The commutant of $G''$ in $G$ is denoted by $G'$. Then low-energy
physics coming out of this local geometry of $X$ is described
by\footnote{This sentence itself is not logical; with a non-compact 
patch of $U \subset S$, one cannot carry out Kaluza--Klein
reduction/decomposition on $U$ to obtain a low-energy effective theory 
on $\R^{3,1}$. The GUT divisor $S$ needs to be covered by patches 
$U_\alpha$'s, and field theory local models on $U_\alpha \times
\R^{3,1}$ need to be ``glued together'' to obtain a low-energy 
effective theory on $\R^{3,1}$. See \cite{Hayashi-2, Tsuchiya} 
for more on this discussion.} 
a local field theory on $U \times \R^{3,1}$ whose action is given as above, 
with the gauge group $G$. All the field contents, $(A_\mu, \eta_\alpha)$
and $(A, \varphi, \psi, \chi)$, are in the ${\bf adj.}$ representation
of $\mathfrak{g}$. Under the subgroup $G' \times G''$, 
this ${\bf adj.}$ representation is decomposed into 
irreducible pieces.
\begin{equation}
 {\rm Res}^G_{G' \times G''} \, \mathfrak{g}\mbox{-}{\bf adj.} 
 = ({\bf adj.}, {\bf 1}) + ({\bf 1}, {\bf adj.}) \oplus_I (U_I, R_I).
\label{eq:decomp}
\end{equation}
Background field configuration $(\vev{A}, \vev{\varphi})$ should be 
chosen within the first component in (\ref{eq:decomp}). 
% The idea 
% is that this local field theory on $U \times \R^{3,1}$ can predict 
% the same low-energy physics as the original local geometry of $X$.

Zero mode wavefunctions are characterized, in the field theory language 
above,  as infinitesimal deformation to the background field configuration 
that still preserves the supersymmetric conditions 
(\ref{eq:BPS-D}--\ref{eq:BPS-G}). For infinitesimal deformation 
$(\delta A^{(0,1)}, \delta \varphi^{(2,0)}) \equiv (\psi, \chi)$ 
in $(U_I, R_I)$, the zero mode equations are 
\begin{eqnarray}
 \omega \wedge D' \psi + 
 \frac{|\alpha|^2}{2} \rho_{U_I}(\vev{\overline{\varphi}}) \chi & = & 0, 
     \label{eq:0-eq-D}\\
 D'' \psi & = & 0,  \label{eq:0-eq-H} \\
 D'' \chi - i \rho_{U_I}(\vev{\varphi}) \psi & = & 0. \label{eq:0-eq-G}
\end{eqnarray}

Zero modes, degrees of freedom that appear in the effective theory 
below the Kaluza--Klein scale, are characterized by global information 
on $S$, not by conditions on wavefunctions in a local patch $U$ of $S$.
Zero modes in a representation $R_I$ of the unbroken symmetry $G''$ 
are identified with a vector space of holomorphic sections 
of a line bundle $\tilde{\cal F}_{(R_I)}$ on a complex curve 
$\tilde{\bar{c}}_{(R_I)}$, as in (\ref{eq:H0-10-repr}) and 
(\ref{eq:H0-5bar-repr}); both are of the form 
 \cite{Curio, DI, DW-1, BHV-1, Hayashi-1}.
\begin{equation}
  H^0\left(\tilde{\bar{c}}_{(R_I)}; K_{\tilde{\bar{c}}_{(R_I)}}^{1/2} \otimes
   {\cal L}_G\right).
\label{eq:as-H0-A}
\end{equation}
Let us take a basis $\{ \tilde{f}_{(R_I); i} \}_{i = 1,2, \cdots, N_{R_I}}$ of 
the vector space of global holomorphic sections, where $N_{R_I}$ is 
the multiplicity of the fields ($\approx$ number of generations) 
in a representation $R_I$. $S$ is covered by local patches $U_\alpha$, 
and each one of basis elements $\tilde{f}_{(R_I); i}$ is assigned 
a set of zero-mode wavefunction 
$(\psi_{(R_I); i; \alpha}, \chi_{(R_I); i ; \alpha})$ in the field 
theory local model on $U_\alpha$. The relation between 
$\tilde{f}_{(R_I); i}$ and 
$(\psi_{(R_I);i; \alpha}, \chi_{(R_I); i ; \alpha})$ has already been 
outlined in \cite{Hayashi-2}, but we will elaborate more on it later in 
section \ref{ssec:uni.vs.hol} and the appendix \ref{sec:Hitchin}.

Super Yang--Mills fields on 7+1 dimensions in the field theory local
models are decomposed into infinite degrees of freedom on 3+1
dimensions, and the decomposition of-course contains the massless
degrees of freedom:
\begin{eqnarray}
 A^{(0,1)} (x,y)_{(R_I); \alpha} & = & \vev{A}^{(0,1)}(y)_{(R_I); \alpha} 
   + \sum_i \psi_{(R_I); i; \alpha}(y) \phi_{(R_I); i}(x) + \cdots, 
    \label{eq:KK1}\\
 \varphi^{(2,0)}(x,y)_{(R_I); \alpha} & = & 
    \vev{\varphi}^{(2,0)}(y)_{(R_I) ; \alpha} + \sum_i 
    \chi_{(R_I); i ; \alpha}(y) \phi_{(R_I); i}(x) + \cdots, \label{eq:KK2}\\
 \psi^{(0,1)} (x,y)_{(R_I); \alpha} & = & \qquad \qquad \qquad \qquad 
   \sum_i \psi_{(R_I) ; i ; \alpha}(y) \lambda_{(R_I); i}(x) + \cdots, 
    \label{eq:KK3}\\
 \chi^{(2,0)} (x,y)_{(R_I); \alpha} & = & \qquad \qquad \qquad \qquad  
   \sum_i \chi_{(R_I) ; i ; \alpha}(y) \lambda_{(R_I); i}(x) + \cdots,  
    \label{eq:KK4}
\end{eqnarray}
where ellipses correspond to Kaluza--Klein massive modes. 
Apart from the $({\bf adj.}, {\bf 1})$ component in (\ref{eq:decomp}) 
(that is, for moduli multiplets $R_I = {\bf 1}$), the vev part 
vanish in (\ref{eq:KK1}, \ref{eq:KK2}).
Although $\psi$ and $\chi$ are used for two different meanings, 
there should be no confusion.\footnote{$\psi$ and $\chi$ used 
in the left hand sides are fields on 7+1 dimensions, and depend 
both on $(x, y)$. $\psi$ and $\chi$ on the right-hand sides are 
zero mode wavefunctions, corresponding to a finite number of 
degrees of freedom in 3+1 dimensions. The wavefunctions depend 
only on the coordinates of the internal space.}
Complex scalar fields $\phi_{(R_I); i}(x)$ and Weyl fermions 
$\lambda_{(R_I); i}(x)$ on 3+1 dimensions form $N_{R_I}$ chiral 
multiplets 
\begin{equation}
 \Phi_{(R_I); i}(x, \theta, \bar{\theta}) = \phi_{(R_I); i}(x) + 
   \sqrt{2} \theta \lambda_{(R_I); i}(x) + \cdots 
\label{eq:def-chiral}
\end{equation}
in representation $R_I$ of the unbroken symmetry group $G''$.
The zero-mode wavefunctions 
$(\psi_{(R_I) ; i ; \alpha}, \chi_{(R_I) ; i; \alpha}) = 
 (\psi_{U_I; i; \alpha}, \chi_{U_I; i ; \alpha})$
in a field theory local model on $U_\alpha \subset S$ are in
representation $U_I$ of the structure group $G'$ in $U_\alpha$.
We assign mass dimension $+1$ [resp. $+3/2$] to the effective scalar 
[resp. fermionic] fields $\phi_{(R_I)}$ [resp. $\lambda_{(R_I) \alpha}$]
in 3+1 dimensions, as usual. Thus, the mass dimension of the zero-mode 
wavefunctions $(\psi_{\bar{m}}(y), \chi_{mn}(y))$ is zero. 

Yukawa couplings in the superpotential (in compactifications preserving 
${\cal N} = 1$ supersymmetry) are calculated from the last term of 
(\ref{eq:fermi-bilin}), if the Yukawa couplings involve at least one
charged matter field (that is, one from a non-$({\bf adj.}, {\bf 1})$
component in (\ref{eq:decomp})). Substituting the mode decomposition 
(\ref{eq:decomp}, \ref{eq:KK1}--\ref{eq:KK4}) and 
using (\ref{eq:def-chiral}), 
\begin{eqnarray}
 \Delta {\cal L}_{4D, {\rm eff.}} & = & \frac{M_*^4}{4\pi} (i \alpha)
  \sum_\alpha \int_{U_\alpha} 
   \tr {}' {}_{G\mbox{-}{\bf adj.}} \left( 
        \chi \left[ A, \psi \right] 
   - \frac{1}{2} \psi \left[ \varphi, \psi \right]
        \right), \label{eq:source-of-Yukawa} \\
 & = & - \frac{1}{3!} \sum_{R_I,i, R_J,j, R_K,k} 
    (\lambda_{ijk}^{(R_I,R_J,R_K)})^{a,b,c}  \nonumber \\
 & & \qquad 
     \left( \lambda_{(R_I); i}^a \lambda_{(R_J);j}^b \phi_{(R_K);k}^c 
          + \lambda_{(R_J); j}^b \lambda_{(R_K);k}^c \phi_{(R_I);i}^a 
          + \lambda_{(R_K); k}^c \lambda_{(R_I);i}^a \phi_{(R_J);j}^b 
     \right) \nonumber 
\end{eqnarray}
becomes a tri-linear interaction in the effective superpotential, 
\begin{equation}
 d^2 \theta \; \Delta W_{\rm eff.} = 
 d^2 \theta \;  \frac{1}{3!} \sum_{R_I,i, R_J,j, R_K,k} 
    (\lambda_{ijk}^{(R_I,R_J,R_K)})^{a,b,c} \; 
       \Phi_{(R_I);i}^a \Phi_{(R_J);j}^b \Phi_{(R_K);k}^c \;
 %   \rightarrow & &  
% %  \Delta W_{4D, {\rm eff}} \simeq 
% %   \left( \sum_\alpha \lambda^{(R_I,R_J,R_K)}_{i, j, k; \alpha} \right)
% %   \Phi_{(R_I); i} \Phi_{(R_J); j} \Phi_{(R_K); k}, 
% %    \\ 
\end{equation}
with the coupling given by 
\begin{eqnarray}
  (\lambda^{(R_I, R_J, R_K)}_{i, j, k})^{a,b,c} & = &
   \frac{M_*^4}{4\pi} f^{Aa, Bb, Cc} \alpha \sum_\alpha  \nonumber \\
   & & 
  \! \! \! \! \! \! \! \! \! \! \! \! \! \! \! \! \! \! \! \! \! \!\!\!\!
     \int_{U_\alpha} 
       \chi^A_{U_I;i;\alpha} \psi^B_{U_J;j;\alpha} \psi^C_{U_K;k;\alpha}
    +  \chi^B_{U_J;j;\alpha} \psi^C_{U_K;k;\alpha} \psi^A_{U_i;i;\alpha}
    +  \chi^C_{U_K;k;\alpha} \psi^A_{U_i;i;\alpha} \psi^B_{U_j;j;\alpha}.
%     \frac{(-i \alpha)}{2} \int_{U_\alpha} \tr {}' {}_{G'} \left(
%     \psi_{(R_i); i; \alpha} 
%     \left[ \chi_{(R_J); j; \alpha}, \psi_{(R_K); k; \alpha}
%			 \right]\right)  \nonumber \\
% & & \qquad \qquad   + ((R_I,i), (R_J, j), (R_K,k) {\rm ~cyclic}).
\label{eq:F-Yukawa}
\end{eqnarray}
Here, $a, b, c = 1, \cdots, {\rm dim} \; R_{I,J,K}$  
[resp. $A,B,C = 1, \cdots, {\rm dim} \; U_{I,J,K}$] label different 
weights of $R_{I,J,K}$ [resp. $U_{I,J,K}$] representations of $G''$ 
[resp. $G'$]. $f^{Aa,Bb,Cc}$ is the structure function of the Lie
algebra of $\mathfrak{g}$: $[t^{Bb}, t^{Cc}] = i f^{Aa,Bb,Cc} t^{Aa}$.
The expression (\ref{eq:F-Yukawa}) for the Yukawa couplings is 
dimensionless, as it should be. One can also see that it is symmetric 
under the exchange of $(Iia,Jjb,Kkc)$; both the structure constant 
and the overlap integration of the wavefunctions are anti-symmetric 
under the exchange of $JjBb \leftrightarrow KkCc$.
Yukawa matrices $\lambda^{(u)}_{ij}$ and $\lambda^{(d)}_{kj}$ of 
supersymmetric Standard Models are calculated in this way.\footnote{
To be more precise, the mode decompositions 
in (\ref{eq:KK1}--\ref{eq:KK4}) contain Kaluza--Klein excited states, 
the their tri-linear couplings (and mass terms) also descend 
from the last line of (\ref{eq:fermi-bilin}).
When the Kaluza--Klein states are integrated out, higher-order terms 
may be generated in the effective superpotential, but not the tri-linear 
terms (apart from the redefinition of the chiral multiplets). 
Possible corrections to this leading-order term may come from higher
order terms that are already omitted 
in (\ref{eq:bosonic}, \ref{eq:fermi-bilin}), but this issue has 
not been discussed in the literature so far. }

How can one determine the kinetic mixing matrix
(e.g. $K^{(R_I)}_{ji}$ ($R_I = q, u^c, d^c$) in (\ref{eq:SM-quark-Yukawa})) 
in low-energy effective theories from geometric data of 
F-theory compactifications? We need the kinetic mixing 
matrices as well as the superpotential Yukawa couplings in order 
to understand physical flavor structure. 
The most naive try is to substitute the same mode decomposition 
(\ref{eq:decomp}, \ref{eq:KK1}--\ref{eq:KK4}) into the action 
(\ref{eq:bosonic}, \ref{eq:fermi-bilin}) on 7+1 dimensions. 
Now, the (leading order terms of the) effective action\footnote{By
``effective'' action, we mean that any effects associated with 
``stringy'' excitations have already been integrated out, and are
included in coefficients of the (possibly higher order) terms 
of the field theory Lagrangian, whatever the ``stringy excitations'' may
be in F-theory.} on 7+1 dimensions is described in terms of infinite 
degrees of freedom in 3+1 dimensions. At energy scale way below the 
Kaluza--Klein scale, however, only finite degrees of freedom 
can be on-shell initial/final states. An effective field-theory 
description on 3+1 dimensions should be written only with the 
finite degrees of freedom, with all other heavy degrees of freedom 
integrated out. 
A practical way to make a first try (cheating) in this situation, however, 
is just to truncate heavy modes (ignore the effects from the heavy 
modes), instead of integrating them out to obtain a low-energy 
effective theory. We will stick to this prescription 
in the rest of this article. 
% This will not be very bad as a first attempt.

%%%%%%%%%%%%%%%%%%%%%%%%%%%%%%%%%%%%%%%%%%%%%%%%%%%%%%%%
\subsection{Unitary Frame and Holomorphic Frame}
\label{ssec:uni.vs.hol}
%%%%%%%%%%%%%%%%%%%%%%%%%%%%%%%%%%%%%%%%%%%%%%%%%%%%%%%%

Let us take a moment here to clarify a subtle yet important 
issue. We begin with reminding ourselves of known facts in 
mathematics. 

In Heterotic string compactification on a Calabi--Yau 
3-fold $Z$, supersymmetric conditions for the gauge field 
configuration on $Z$ are 
\begin{eqnarray}
  & & \omega \wedge \omega \wedge F \propto 
      \omega \wedge \omega \wedge \omega, \label{eq:BPS-Het-D} \\
  & & F^{(0,2)} = 0. \label{eq:BPS-Het-F}
\end{eqnarray}
The condition (\ref{eq:BPS-Het-F}) means that the (0, 1)-form 
part of the background gauge field configuration is integrable 
with respect to the anti-holomorphic derivative $\bar{\partial}$, 
and that $V$ can be regarded as a holomorphic vector bundle. 

Let us set up notations for the following discussion.
The background configuration of gauge field 
\begin{equation}
 A = A^{(1,0)} + A^{(0,1)} = A_m dz^m + A_{\bar{m}} d \bar{z}^{\bar{m}}
\end{equation}
takes its value in the structure group $\mathfrak{g}'$,  
but it does not have to be a well-defined $\mathfrak{g}'$ 
valued 1-form over the entire $Z$. 
One needs to have a well-defined $\mathfrak{g}'$-valued 
1-form $A_a$ on a open patch 
$U_a$ in $Z$, and $U_a$'s cover the entire $Z$.
$A_a$ on $U_a$ and $A_b$ on $U_b$ should be related by 
\begin{equation}
 (d + i A_a)^A_{\; B} = (g_{ab})^A_{\; C} (d + i A_b)^C_{\; D} (g_{ba})^D_{\; B}
\end{equation}
in $U_a \cap U_b$, where the transition functions $g_{ab}$ take their 
values in the structure group $G'$, and satisfy 
$g_{ab} \cdot g_{ba} = {\bf 1}$, and 
$g_{ab} \cdot g_{bc} \cdot g_{ca} = {\bf 1}$. 
The background gauge field configuration defines a vector bundle 
$\pi : V \rightarrow Z$, and $U_a$'s are trivialization patches. 
$\pi^{-1}(U_a) \subset V$ is isomorphic to $U_a \times \C^{{\rm rank}V}$.
By taking a trivialization patches, we have implicitly fixed a frame 
on $U_a$, a basis $\{ e_{A(a)}\}$ ($A = 1,\cdots,{\rm rank} V$) 
in the $({\rm rank} \; V)$-dimensional fiber vector space $\pi^{-1}(z)$ 
for each point $z \in U_a$. A section $s$ of $V$ 
is described in $U_a$ as $s = e_{A(a)} s^A(z)_a$, 
using ${\rm rank} \; V$ functions on $U_a$. In $U_a \cap U_b$, 
\begin{equation}
 s^A(z)_a = (g_{ab})^A_{\; B} s^B(z)_b, \qquad 
 e_{A(a)} (g_{ab})^A_{\; B} = e_{B(b)} \qquad ({}^\forall z \in U_a \cap U_b), 
\label{eq:trans-fcn}
\end{equation}
so that $s$ is a well-defined section over the entire manifold $Z$.

When (\ref{eq:BPS-Het-F}) is satisfied, one can find a $G^{'c}$-valued 
function ${\cal E}_a$ on $U_a$, and take a new frame 
$\{ \tilde{e}_{A(a)} \}$ ($A = 1, \cdots, {\rm rank} \; V$) satisfying 
\begin{equation}
 \tilde{e}_{B(a)}  =  e_{A(a)} \; ({\cal E}_a)^A_{\; B},
\end{equation}
so that the $(0, 1)$-part of the background gauge field vanishes. 
$G^{'c}$ is the complexification of $G'$, like SL($n, \C$) for 
$\SU(n)$. To be more explicit, the $(1,0)$ and $(0,1)$ parts 
of covariant derivative $\nabla \equiv d + iA$ are denoted by 
$D' \equiv \partial + i A^{(1,0)}$ and 
$D'' \equiv \bar{\partial} + i A^{(0,1)}$, respectively, and 
the covariant derivative in the frame using $\{ e_{A(a)} \}$ 
and the covariant derivative $\tilde{\nabla} = \tilde{D}' + \tilde{D}''$
in the $\{ \tilde{e}_{A(a)} \}$ frame are related by 
\begin{equation}
 D' = {\cal E}_a \tilde{D}' ({\cal E}_a)^{-1}, \qquad 
 D'' = {\cal E}_a \tilde{D}'' ({\cal E}_a)^{-1}. 
\end{equation}
Since the vanishing $(0,1)$ part of the connection 
in the $\{ \tilde{e}_{A)} \}$ frame means 
$\tilde{D}'' = \bar{\partial}$, the $G^{'c}$-valued matrix 
${\cal E}_a$ should be chosen so that 
\begin{equation}
 i(A^{(0,1)}_a)^A_{\; B} = 
  ({\cal E}_a)^A_{\; C} \bar{\partial} ({\cal E}_a^{\; -1})^C_{\; B}.
\label{eq:D}
\end{equation}
The condition (\ref{eq:BPS-Het-F}) guarantees that such 
${\cal E}_a$ can be found. 

In the original frame using $\{ e_{A(a)} \}$, the background 
configuration [fluctuations] of the connection 
takes its value in $\mathfrak{g}'$ [in $\mathfrak{g}$], and is represented as 
a Hermitian matrix (in unitary representations). This frame
is called unitary frame. On the contrary, in the frame using 
$\{ \tilde{e}_{A(a)} \}$, the connection 
$\tilde{A} = \tilde{A}^{(1,0)} + \tilde{A}^{(0,1)}$ 
is no longer Hermitian. $\tilde{A}^{(0,1)} = 0$, by definition,  
whereas  
\begin{eqnarray}
 i \tilde{A}^{(1,0)}_a & = & {\cal E}_a^{-1} (\partial + iA^{(1,0)}) {\cal E}_a  
 = {\cal E}_a^{-1} \partial {\cal E}_a 
  - {\cal E}_a^{-1} (i A^{(0,1)})^\dagger {\cal E}_a 
 = \cdots = H_a^{-1} \partial H_a, \label{eq:rel-A10-H}\\
(H_a)_{AB} & \equiv & ({\cal E}_a)^\dagger {\cal E}_a 
 = (({\cal E}_a)^C_{\; A})^* \delta_{CD} ({\cal E}_a)^D_{\; B}.   
\label{eq:def-Hform}
\end{eqnarray}
This frame is called holomorphic frame, because the 
transition functions $\tilde{g}_{ab}$ in their matrix 
representation are holomorphic in the holomorphic 
coordinates of $Z$. Sections $s = e_{A(a)} s^A(z)_a$ 
are also expressed in the holomorphic frame as 
\begin{equation}
 s = \tilde{e}_{A(a)} \tilde{s}^A(z)_a, \qquad 
  s^A(z)_a = ({\cal E}_a)^A_{\; B} \tilde{s}^B(z)_a, 
\end{equation}
and $(D'')^A_{\; B} s^B = 0$ now means that all $\tilde{s}^A(z)_a$ 
are holomorphic. 

% For holomorphic vector bundles, 
% it is known that the moduli space of gauge field configuration 
% satisfying (\ref{eq:BPS-Het-D}) is the same as that of stable vector 
% bundles [Narashimhan--Seshadri, Donaldson, ...]. 
% This is why stable vector bundles are dealt with in
% phenomenological study of Heterotic string theory, instead of 
% gauge field configurations satisfying 
% (\ref{eq:BPS-Het-D}, \ref{eq:BPS-Het-F}).

When a unitary frame description is provided for a vector bundle $V$, 
a natural metric is introduced in the space of sections of $V$.
Namely, we have a metric, when a $\mathfrak{g}'$-valued gauge-field 
background (in a unitary representation) is given to a vector bundle $V$, 
and all the transition functions $g_{ab}$ take their values in 
(the unitary representation of) the compact group $G'$, not in 
its complexification $G^{'c}$.  The Hermitian inner product of 
$s = e_{A(a)} s^A(z)_a$ and $s' = e_{A(a)} s^{'A}(z)_a$ is given by  
\begin{equation}
 \langle s' | s \rangle \equiv 
  \int_Z {\rm vol}.(Z) \; (s^{'A})^* \; \delta_{AB} \; s^B(z).
\label{eq:inn-pro-unit}
\end{equation}
Here, ${\rm vol}.(Z)$ is the volume form of $Z$. The integrand is
well-defined, that is, $(s^{'A})^* \; \delta_{AB} \; s^B(z)$ remain 
the same, regardless of which one of trivializations is used when 
a point $z$ is in $U_a \cap U_b$; this is because the transition 
functions $(g_{ab})^B_{\; C}$ are unitary. If this Hermitian inner
product  
is written in terms of component description in a holomorphic frame, 
that is 
\begin{equation}
 \langle s' | s \rangle = \int_Z ({\rm vol}.(Z)) \; 
   \tilde{s}^A(z)_a \; (H_a)_{AB} \; \tilde{s}^B(z)_a, 
% \qquad \qquad  
% (H_a)_{AB} \equiv 
%  ({\cal E}_a)^\dagger{}_A^{\; C} \delta_{CD} ({\cal E}_a)^D_{\; B}.
\end{equation}
using the Hermitian matrix $(H_a)_{AB}$ given in (\ref{eq:def-Hform}).
Once again, the integrand does not depend on the choice of
trivialization patch, and this Hermitian inner product 
$\langle s' | s \rangle$ is well-defined.
A norm $|| s ||$ can also be defined from this Hermitian inner product 
for a section $s$ of $V$. Note also that the covariant derivative 
$\tilde{\nabla}$ is already given, when a holomorphic vector bundle $V$
is endowed with a Hermitian inner product (or ${\cal E}_a$'s). This is because 
$\tilde{D}'' = \bar{\partial}$ by definition, and $i\tilde{A}^{(1,0)}$ 
is determined from $H_a$ by (\ref{eq:rel-A10-H}). 

That is all for a brief review of necessary mathematics, 
and let us get back to physics. 
In F-theory compactifications, Higgs bundles on the GUT divisor
$(n-1)$-fold $S$ is used instead of vector bundles on $n$-fold $Z$, 
in describing gauge-theory degrees of freedom. 
A $\mathfrak{g}'$-valued gauge field background 
$A$ and a $\mathfrak{g}' \otimes_\R \C$-valued $(2, 0)$-form 
field $\varphi$ on a local patch $U_\alpha$ of $S$ defines a canonical 
bundle valued Higgs bundle $(V, \varphi)$, when the supersymmetric conditions 
(\ref{eq:BPS-D}--\ref{eq:BPS-G}) are satisfied. The gauge field
background $A$ satisfying (\ref{eq:BPS-H}) defines a holomorphic 
vector bundle $V$ on $U_\alpha \subset S$, just as discussed above. 
Thus, in trivialization patches $U_a$'s in $U_\alpha \subset S$, 
both unitary frame and holomorphic frame description are possible. 
To be more precise, because of the decomposition (\ref{eq:decomp}), 
holomorphic vector bundles $\rho_{U_I}(V)$ have both unitary frame 
and holomorphic frame descriptions. 
The unitary frame and holomorphic frame are denoted by 
$\{ e_{A(a)} \}$ ($A=1, \cdots, {\rm dim} \; U_I$), and 
 $\{ \tilde{e}_{A(a)}\}$ ($A = 1, \cdots ,  {\rm dim} \; U_I$), 
respectively. They are related by a complexified gauge transformation 
${\cal E}_a$:
\begin{equation}
 e_{A(a)} \; (\rho_{U_I}({\cal E}_a))^A_{\; B} = \tilde{e}_{B(a)}.
\end{equation}

All the fields appearing in the effective-theory Lagrangian 
(\ref{eq:bosonic}, \ref{eq:fermi-bilin}) take their values 
in the ${\bf adj.}$ representation of $\mathfrak{g}$. 
This means that we have already chosen a set of trivialization 
patches $U_a$'s in $U_\alpha$. Furthermore, it is a description 
in a unitary frame. The gauge field $A_\mu dx^\mu$ and 
$A_m dz^m + A_{\bar{m}}d\bar{z}^{\bar{m}}$ with its value in 
$\mathfrak{g}$-{\bf adj} is Hermitian. The Hermitian conjugate 
of $A^{(1,0)} = A_m dz^m$ is $A^{(0,1)} = A_{\bar{m}}
d\bar{z}^{\bar{m}}$, and similarly, 
$\overline{\varphi} = \varphi^{(0,2)} = 
 \overline{\varphi}_{\bar{m}\bar{n}} 
 d\bar{z}^{\bar{m}}\wedge d\bar{z}^{\bar{n}}$ is the Hermitian conjugate of 
$\varphi = \varphi^{(2,0)}$.
``$\tr' $'' appearing in (\ref{eq:bosonic}, \ref{eq:fermi-bilin}) 
is the Killing metric of the Lie algebra $\mathfrak{g}$, which 
is a $G$-invariant metric of the fiber vector space 
$\mathfrak{g}$-{\bf adj.}. This corresponds to ``$\delta_{AB}$'' in 
(\ref{eq:inn-pro-unit}), which was used in defining a norm 
in a unitary frame description in space of sections.
%
% , but they can be 
% decomposed into irreducible pieces $(U_I, R_I)$ as 
% in (\ref{eq:decomp}). Those fields in a component $(U_I, R_I)$ 
% are regarded as scalar, 1-form or 2-form that take values 
% in a bundle $\rho_{U_I}(V)$. An implicit assumption is already 
% made in writing down the action (\ref{eq:bosonic},
% \ref{eq:fermi-bilin}); all the fields are dealt with in 
% component descriptions in a unitary frame, by taking some 
% basis $\{ e_{A(a)}\}$.  
%
% All the fields are in a single trace 
% $\tr {}'$ in the action (\ref{eq:bosonic}, \ref{eq:fermi-bilin}), 
% and that is the Killing metric of Lie algebra $\mathfrak{g}$. 
% This Killing metric in the fiber vector space of $\rho_{U_I}(V)$
% is regarded as the $\delta_{AB}$ in (\ref{eq:inn-pro-unit}); it is 
% invariant under adjoint action of group elements of $G$. 
% The supersymmetry conditions (\ref{eq:BPS-D}--\ref{eq:BPS-G}) for 
% the background $(A, \varphi)$ and the zero mode equations 
% (\ref{eq:0-eq-D}--\ref{eq:0-eq-G}) for the $(U_I, R_I)$ component 
% should be regarded as expressions in the component descriptions 
% in the unitary frame. $\varphi^\dagger$ in (\ref{eq:BPS-D}) and 
% (\ref{eq:0-eq-D}) are simply the complex conjugation of 2-forms 
% on the trivialization patch in $S$, when $\varphi$ is expanded 
% as a sum of products of independent Hermitian generators 
% of $\mathfrak{g}'$ and their coefficient complex-valued 2-forms.

We have already seen how the gauge field (connection) background 
in a holomorphic frame description is related to that in a unitary 
frame description. The $\varphi^{(2,0)}$ field background in the two 
descriptions are related by 
\begin{equation}
 \rho_{U_I}(\vev{\varphi})^A_{\; B}(z)_a =
    \rho_{U_I}({\cal E}_a)^A_{\; C}(z) \; 
    \rho_{U_I}(\widetilde{\vev{\varphi}})^C_{\; D}(z)_a \;
    \rho_{U_I}(({\cal E}_a)^{-1})^D_{\; B}, 
\label{eq:C}
\end{equation}
or $\varphi = {\cal E} \cdot \widetilde{\varphi} \cdot {\cal E}^{-1}$, 
for short.
The condition (\ref{eq:BPS-G}) means that $\widetilde{\varphi}^A_{\; B}(z)_a$
are holomorphic functions on $U_a$. The $(0,2)$ component of $\varphi$
field in the holomorphic frame, $\widetilde{\overline{\varphi}}$, 
is related to the one in the unitary frame through 
$\overline{\varphi} = {\cal E} \cdot
\widetilde{\overline{\varphi}} \cdot {\cal E}^{-1}$. 
Thus, 
\begin{equation}
 \widetilde{\overline{\varphi}} = 
 ({\cal E}_a)^{-1} \varphi^\dagger ({\cal E}_a) = 
 (H_a)^{-1} (\widetilde{\varphi})^\dagger (H_a).
\end{equation}
$\widetilde{\vev{\overline{\varphi}}}$ in the holomorphic frame 
is not a simple Hermitian conjugate of $\widetilde{\vev{\varphi}}$.
All other fluctuations of the fields, simply denoted by $s$, 
also have component descriptions in both frames, 
$e_{A(a)} \; s^A(z)_a$ and $\tilde{e}_{A(a)} \; \tilde{s}^A(z)_a$, 
and they are related by 
$s^A(z)_a = ({\cal E}_a)^A_{\; B} \tilde{s}^B(z)_a$.

% Apart from the $\varphi$ field background being a 2-form on 
% $U_\alpha \subset S$, it takes its value in $\mathfrak{g}' \otimes_R \C$, 
% and hence is regarded as a section of 
% ${\bf adj.} (\mathfrak{g}') \otimes_R \C \simeq V \otimes V^\times$ 
% vector bundle. Thus, $\varphi$ field background also has two 
% component description descriptions, both in matrix form; one 
% is $\varphi^A_{\; B}$ corresponding to $e_{A} \otimes (e^\times)^B$, 
% where $\{ (e^\times)^A \}$ is the dual basis of $\{ e_A \}$, and the 
% other is $\widetilde{\varphi}^A_{\; B}$ corresponding to 
% $\tilde{e}_A \otimes (\tilde{e}^\times)^B$. These two component 
% descriptions in matrix form are related by 

% The condition The $\varphi$ field can be regarded 
% both as a $\mathfrak{g}' \otimes_R \C$-valued (2,0)-form on $U_a$, and 
% as a section of $K_{U_\alpha} \otimes \mathfrak{g}'$. When we take the second
% picture, the condition (\ref{eq:BPS-G}) means that the background 
% configuration $\tilde{\varphi}$ is a holomorphic section of 
% $K_{U_\alpha} \otimes \mathfrak{g}'$.

The zero-mode equations (\ref{eq:0-eq-D}--\ref{eq:0-eq-G}) in a 
unitary frame description can be rewritten in a holomorphic frame.
\begin{eqnarray}
 \omega \wedge \tilde{D}' \tilde{\psi} + \frac{|\alpha|^2}{2} 
   \rho_{U_I} \left(\widetilde{\bar{\varphi}}\right) \; \tilde{\chi} 
 & = & 0, \label{eq:0-eq-D-hol}\\
 \bar{\partial} \widetilde{\psi} & = & 0, \label{eq:0-eq-H-hol} \\
 \bar{\partial} \tilde{\chi} - i \rho_{U_I}(\widetilde{\varphi}) \tilde{\psi}
  & = & 0. \label{eq:0-eq-G-hol}
\end{eqnarray}
Thus, the condition (\ref{eq:0-eq-H-hol}) implies that 
the $({\rm rank} \; U_I)$-component (0,1)-form $\tilde{\psi}$ 
in the holomorphic frame are expressed locally as 
\begin{equation}
 \tilde{\psi} = \bar{\partial} \; {}^\exists \tilde{\Lambda}, 
\label{eq:f2psi}
\end{equation}
where $\tilde{\Lambda}$ is a $({\rm rank} \; U_I)$-component function.
The condition (\ref{eq:0-eq-G-hol}) further implies that 
\begin{equation}
 \widetilde{\chi} = i \rho_{U_I}(\widetilde{\varphi}) \tilde{\Lambda} 
  + \tilde{f}_{(R_I); a},
\label{eq:f2chi}
\end{equation}
where $\tilde{f}$ is a $\rho_{U_I}(V)$-valued holomorphic (2,0)-form.
$\tilde{\Lambda}$ and $\tilde{f}$ cannot be chosen independently, because 
the condition (\ref{eq:0-eq-D-hol}) is imposed on them. But, one 
will notice that when a pair 
$(\tilde{\Lambda}, \tilde{f}_{(R_I); a})$ satisfies
(\ref{eq:0-eq-D-hol}), 
so does $(\tilde{\Lambda} - k, \tilde{f}_{(R_I); a} 
 + \rho_{U_I}(\widetilde{\varphi}) k )$ for 
arbitrary holomorphic sections $k$ of $\rho_{U_I}(V)$. 
The zero mode wavefunction $(\tilde{\psi}, \tilde{\chi})$, however, 
are not affected by this shifting by $k$; this is not physical. 
Thus, zero modes in representation $R_I$ of the unbroken symmetry 
$G''$ correspond to holomorphic sections $\tilde{f}_{(R_I); a}$ 
modulo $ + \rho_{U_I}(\widetilde{\varphi}) k$: 
\begin{equation}
 {\rm Coker} \left( H^0(U_a; \rho_{U_I}(V)) \rightarrow 
  H^0(U_a; \rho_{U_I}(V) \otimes K_S) \right).
\label{eq:coker}
\end{equation}

A spectral surface $C_{(R_I)}$ is defined as a subspace of
$\mathbb{K}_{U_\alpha}$, the total space of the canonical bundle 
$K_{U_\alpha}$ on $U_\alpha$, 
as the zero-locus of a characteristic equation\footnote{The definition of the 
spectral surface does not depend on whether the component description 
in unitary frame is used, or the one in holomorphic frame is used.} 
\begin{equation}
  {\rm det} \left( \xi {\rm Id} - 2\alpha \rho_{U_I} (\varphi_{12}) \right) 
= {\rm det} \left( \xi {\rm Id} - 2 \alpha \rho_{U_I} (\widetilde{\varphi}_{12})
	    \right)
= 0, 
\end{equation}
where $\varphi_{12}$ or $\widetilde{\varphi}_{12}$ are (2,0)-forms 
with $dz^1 \wedge dz^2$ stripped off, and $\xi$ is the coordinate 
of the fiber rank-1 vector space of $K_{U_\alpha}$, when $dz^1 \wedge dz^2$ 
is chosen as the local frame of $K_{U_\alpha}$
If the holomorphic vector bundle $\rho_{U_I}(V)$ (considered as a sheaf
on $U_\alpha$) is expressed as a pushforward of a sheaf 
$\widetilde{\cal N}_{(R_I)}$ on the spectral surface $\widetilde{C}_{(R_I)}$, 
\begin{equation}
 \rho_{U_I} (V) = \pi_{C_{(R_I)}*} \widetilde{\cal N}_{(R_I)}, 
\end{equation} 
where 
$\pi_{C_{R_I}}: \widetilde{C}_{(R_I)} \hookrightarrow 
\mathbb{K}_{U_\alpha} \rightarrow U_\alpha$, then 
$\tilde{f}_{(R_I); a}$ and $k$ can be regarded as holomorphic 
sections of $\widetilde{\cal N}_{(R_I)} \otimes K_{U_\alpha}$ and 
$\widetilde{\cal N}_{(R_I)}$ on the spectral surface, respectively. 
The covering matter curve $\tilde{\bar{c}}_{(R_I)}$ can be regarded 
as $\xi = 0$ locus in the spectral surface $\widetilde{C}_{(R_I)}$. 
Since only the cokernel really matters to charged matters 
in (\ref{eq:coker}), only the holomorphic sections 
$\tilde{f}_{(R_I); a}$ of $\widetilde{\cal N}_{(R_I)} \otimes K_{U_\alpha}$ 
restricted on the $\xi = 0$ subspace ($\tilde{\bar{c}}_{(R_I)}$) 
is relevant to the charged matter wavefunctions. 
We calculated holomorphic wavefunctions of massless charged matter 
fields in section \ref{sec:hol-wavefcn}. They are characterized 
as holomorphic sections $\tilde{f}$ of a line bundle 
\begin{equation}
 \tilde{\cal F}_{(R_I)} = 
   \left(\widetilde{\cal N}_{(R_I)} \otimes K_{U_\alpha} \right)
     |_{\tilde{\bar{c}}_{(R_I)}}
\end{equation}
on a curve $\tilde{\bar{c}}_{(R_I)}$. The holomorphic wavefunctions 
in the previous section and zero mode wavefunctions $(\psi, \chi)$ 
(or $(\tilde{\psi}, \tilde{\chi})$) are related in the way we have 
described so far. This argument has already appeared in
\cite{Hayashi-2}, but we clarified a bit in this article, by making 
a clear distinction between the descriptions in unitary and holomorphic 
frames. See the appendix \ref{sec:Hitchin} for more about the 
distinction between the two frames. 

%%%%%%%%%%%%%%%%%%%%%%%%%%%%%%%%%%%%%%%%%%%%%%%%%%%
\subsection{Kinetic Mixing Matrices as Inner Products}
%%%%%%%%%%%%%%%%%%%%%%%%%%%%%%%%%%%%%%%%%%%%%%%%%%%

Having clarified all these things above, we can now describe 
how the kinetic mixing matrix $K^{(R_I)}_{ji}$ are determined from 
geometric data of compactifications (where we allow ourselves to employ 
Kaluza--Klein truncation as a first try). Only the kinetic mixing 
of charged matter fields are discussed here; apart from right-handed 
neutrinos, all the charged matter fields in the Standard Model 
originate from off-diagonal components $(U_I, R_I)$.

Charged matter chiral multiplets in supersymmetric compactifications 
in $G''$-$R_I$ representation are identified with 
\begin{equation}
 H^0 \left( \tilde{\bar{c}}_{(R_I)} ; 
    (\widetilde{\cal N}_{(R_I)} \otimes K_{U_\alpha})
 |_{\tilde{\bar{c}}_{(R_I)}} \right);
\label{eq:as-H0-B}
\end{equation} 
the line bundle $\widetilde{\cal N}_{(R_I)}$ is decomposed as 
${\cal O}(r/2) \otimes {\cal L}_G$, where ${\cal L}_G$ is as in 
(\ref{eq:as-H0-A}), and $r$ is the ramification divisor 
on $\widetilde{C}_{(R_I)}$ associated with the projection 
$\tilde{\pi}_{\widetilde{C}_{(R_I)}}: 
 \widetilde{C}_{(R_I)} \rightarrow U_\alpha$.
It was shown in \cite{Curio, Hayashi-1, Hayashi-2} that 
\begin{equation}
 \left.  \left(K_S + \frac{1}{2}r \right) \right|_{\tilde{\bar{c}}_{(R_I)}} = 
%  \left.  \left(K_S + \frac{1}{2}(C_{(R_I)} - K_S) \right) 
%  \right|_{\tilde{\bar{c}}_{(R_I)}} = 
  \frac{1}{2} K_{\tilde{\bar{c}}_{(R_I)}}, 
\end{equation}
and hence the line bundle in (\ref{eq:as-H0-B}) is the same as 
(\ref{eq:as-H0-A}). 

Let us take a basis $\{ \tilde{f}_{(R_I); i} \}$ in the vector 
space (\ref{eq:as-H0-A}, \ref{eq:as-H0-B}) (or equivalently, in 
(\ref{eq:H0-10-repr}, \ref{eq:H0-5bar-repr})). Each one of basis elements 
$\tilde{f}_{(R_I); i}$ defines an element 
$\tilde{f}_{(R_I); i ; \alpha}$ in (\ref{eq:as-H0-B}), which is 
a holomorphic section of a line bundle on a curve 
$\tilde{\bar{c}}_{(R_I)} \cap U_\alpha$ within a single field-theory 
local model.
In a trivialization patch $U_a$ of a Higgs bundle on $U_\alpha$, 
$\tilde{f}_{(R_I); i; a}$ becomes a holomorphic function on the curve. 
A zero mode wavefunction $(\tilde{\psi}_{(R_I); i ;a}, 
\tilde{\chi}_{(R_I); i; a})$ on $U_a$ is assigned for 
$\tilde{f}_{(R_I); i; a}$ through the relation 
(\ref{eq:f2psi}, \ref{eq:f2chi}, \ref{eq:coker}).
The component description of the zero mode wavefunction 
in holomorphic frame obtained so far is related to that 
in unitary frame through 
\begin{equation}
\left(\psi^A, \chi^A \right)_{(R_I); i; a} = 
({\cal E}_a)^A_{\; B} \; (\tilde{\psi}^B, \tilde{\chi}^B)_{(R_I); i; a},  
\end{equation}
where $A, B = 1, \cdots, {\rm dim} \; U_I$. It is this zero-mode
wavefunction in unitary frame description that should be 
used in the mode decomposition (\ref{eq:KK1}--\ref{eq:KK4}) and 
be substituted to the action (\ref{eq:bosonic}, \ref{eq:fermi-bilin}).

Through a straightforward calculation starting from 
(\ref{eq:bosonic}, \ref{eq:fermi-bilin}), one can see that 
\begin{eqnarray}
 \Delta {\cal L}_{4D, {\rm eff}} & = & K_{i\bar{j}}^{(R_I)} 
  \left(
     - \left[ (D_\mu \phi^\dagger_j) (D^\mu \phi_i) \right]
     + i  \left[ (D_\mu \bar{\lambda}_{j \dot{\alpha}}) 
         \bar{\sigma}^{\mu \; \dot{\alpha} \alpha} \lambda_{i \alpha} \right]
  \right) \\
  & = & d^4 \theta \; K_{i\bar{j}}^{(R_I)} 
    \left[ \Phi_{(R_I); j}^\dagger \Phi_{(R_I); i}\right] 
   = d^4 \theta \; K(\Phi_{(R_I)}, \Phi_{(R_I)}^\dagger)
\end{eqnarray}
becomes the $D$-term of supersymmetric compactifications, 
with the kinetic mixing matrix given by 
\begin{eqnarray}
 K_{i\bar{j}}^{(R_I)} & = & \frac{c_{(U_I,R_I)} M_*^4}{4\pi} \int_S ({\rm vol}) 
 \left(
  2  \left[ (\psi^\dagger_j)_m (\psi_i)_{\bar{m}} \right] h^{\bar{m}m}
+ |2\alpha|^2 \left[ 
         (\chi^\dagger_j)_{\bar{m}\bar{n}} (\chi_i)_{mn} \right] 
    h^{\bar{m}m} h^{\bar{n}n}
  \right) \nonumber \\
&  &
  \!\!\!\!\!\!\!\!\!\!\!\!\!\!\!\!\!\!\!
 = \frac{ c_{(U_I,R_I)} M_*^4}{4\pi} \int_S ({\rm vol}) 
 \left(
  2 
     \left[ (\psi^\dagger_j)^{\bar{m}} (\psi_i)^{m} \right] h_{m\bar{m}}
+ |2\alpha|^2 \left[ 
         (\chi^\dagger_j)_{\bar{m}\bar{n}} (\chi_i)_{mn} \right] 
    h^{\bar{m}m} h^{\bar{n}n}
  \right),  \label{eq:K.F.-truncation}
\end{eqnarray}
where $c_{(U_I, R_I)}$ is a constant that may depend on the irreducible
component $(U_I, R_I)$ in (\ref{eq:decomp}). 
In the last line, we define the zero-mode wavefunctions 
$(\psi_i)^m$ and their complex conjugates $(\psi^\dagger_j)^{\bar{m}}$
by using the K\"{a}hler metric $h_{m\bar{m}}$ through 
$(\psi_i)^m \equiv h^{\bar{m}m} (\psi_i)_{\bar{m}}$, and 
$(\psi^\dagger_j)^{\bar{m}} \equiv h^{\bar{m}m} (\psi^\dagger_j)_m$.
% The complex scalar $\phi_i$ and Weyl fermion $\lambda_i$ in the 
% same chiral multiplet 
% $\Phi_i = \phi_i + \sqrt{2} \theta \lambda_i + \cdots$ share the same 
% kinetic term (as expected from unbroken ${\cal N} = 1$ supersymmetry). 
The expression for the kinetic mixing matrix $K^{(R_I)}_{i\bar{j}}$ 
above is dimensionless (as expected), since we have set all the 
component fields $\psi_{\bar{m}}$ and $\chi_{mn}$ of zero mode 
wavefunctions to be dimensionless. 
One will also notice that $K^{(R_I)}_{i\bar{j}}$ is obviously positive 
definite (as it should be); this kinetic term descends from 
the positive definite kinetic term 
in (\ref{eq:bosonic}, \ref{eq:fermi-bilin}), and hence that is 
not surprising at all.

To be more precise, we have obtained the expression above 
for $K^{(R_I)}_{i\bar{j}}$ only for a contribution from a 
single trivialization patch $U_a$. The wavefunctions 
$(\psi, \chi)_{(R_I);i; a}$ in patch $U_a$, however, are related 
to those in an adjacent patch $U_b$ by a transition function 
$(g_{ab})$ that takes its value in the representation $\rho_{U_I}$ 
of a compact group $G'$, and the metric $\tr {}'_{U_I}$ is invariant 
under this $G'$ transformation. Thus, the inner 
product (\ref{eq:K.F.-truncation}) is well-defined over the entire 
region $U_\alpha$ where a single field-theory local model is defined. 

When a field theory local model on $U_\alpha \subset S$ is glued 
to another field-theory local model on $U_\beta \subset S$ (by
definition, $U_\alpha \cap U_\beta \neq \phi$), the structure 
group $G'$ and the rank of representation $U_I$ may differ in 
$U_\alpha$ and $U_\beta$ in general. For example, a field-theory 
model with $G = E_6$ gauge group on $U_\alpha$ is glued to 
a model with $G = \SO(10)$ gauge group on $U_\beta$, 
the structure group $G' = \U(2)$ on $U_\alpha$ and $G' = \U(1)$ on 
$U_\beta$. ${\bf 10}$-representation fields of $G'' = \SU(5)_{\rm GUT}$
originates from the $(U,R) = ({\bf 2}, {\bf 10})$ component in the $E_6$
model, but it is from $(U, R) = (+2 , {\bf 10})$ in the $\SO(10)$ model.
See \cite{Hayashi-2} for various other examples.
Field-theory local models with different structure groups can be 
glued together, if size of some topological 2-cycles varies over $S$, 
and becomes small or large in different places in $S$. 
The wavefunctions in a doublet representation of $G' = \U(2)$ in
$U_\alpha$ can be identified with single component wavefunctions 
in $U_\beta$, if the doublet wavefunctions in one component are 
much smaller than in the other component in $U_\alpha \cap U_\beta$. 
At the level of this approximation in this identification (gluing 
process in $U_\alpha \cap U_\beta$), 
the expression in the form of inner product (\ref{eq:K.F.-truncation})
in the patch $U_\alpha$ is approximately the same as the one 
in the other patch $U_\beta$, and hence the integrand of 
(\ref{eq:K.F.-truncation}) is well-defined (approximately) over the 
entire $G''$-singularity surface $S$.

The wavefunctions $e_{A(a)} \psi^A = \tilde{e}_{A(a)} \tilde{\psi}^A$
and $e_{A(a)} \chi^A = \tilde{e}_{A(a)} \tilde{\chi}^A$ can be regarded 
certainly as $\rho_{U_I}(V)$-valued (0,1)-form and (2,0)-form, 
respectively, but they can be regarded also as sections of 
$\rho_{U_I}(V) \otimes TS$ and $\rho_{U_I}(V) \otimes K_S$,
respectively. In component description, they are 
\begin{equation}
 [e_{A(a)} \otimes (d\bar{z}^{\bar{m}} h_{m\bar{m}})] \; 
  \psi^{A m}(z,\bar{z})_a, \qquad 
 [e_{A(a)} \otimes (dz^m \wedge dz^n)] \; \chi^{A}_{mn}(z,\bar{z})_a.
\end{equation}
The expression for the kinetic mixing matrix $K^{(R_I)}_{i\bar{j}}$ 
(\ref{eq:K.F.-truncation}) is written in terms of this component 
description, and it is regarded as a sum of inner product of sections 
of holomorphic vector bundles $\rho_{U_I} \otimes TS$ (the first term) 
and $\rho_{U_I}(V) \otimes K_S$ (the second term). The metric 
in this inner product is constant and a canonical one 
for the $\rho_{U_I}(V)$ part; the basis $\{ e_{A(a)}\}$ provides 
a unitary frame description of $\rho_{U_I}(V)$, and the metric descends 
from the Killing metric of $\mathfrak{g}$). On the other hand, 
$h_{\bar{m}m}$ and $h^{\bar{m}m} h^{\bar{n}n}$ are used for the 
metric of the inner product in the first and second terms 
of (\ref{eq:K.F.-truncation}), respectively, for the $TS$ and $K_S$ 
part. This is because $(h_{m\bar{m}} d\bar{z}^{\bar{m}})$ and 
$(dz^m \wedge dz^n)$ are the holomorphic frames of the bundle 
$TS$ and $K_S$. We will call this component description as 
unitary--holomorphic frame. 
%%%%%%%%%%%%%%%%%%%%%%%%%%%%%%%%%%%%%%%%%%%%%%%%%%%%%%%%%%%%%%
\begin{table}[tb]
 \begin{center}
\caption{\label{tab:frame} A summary of three different choices of 
frame.}
  \begin{tabular}{c|c|c|c}
   & unitary--unitary & unitary--holomorphic & holomorphic--holomorphic
   \\
\hline 
frame & $e_{A(a)} \otimes e_{\bar{m}}^{\; M}$ &
        $e_{A(a)} \otimes d\bar{z}^{\bar{m}} h_{m\bar{m}}$ & 
        $\tilde{e}_{A(a)} \otimes d\bar{z}^{\bar{m}} h_{m\bar{m}}$ \\
coeff. & $\hat{\psi}^A_{M \; a}$ 
      & $\psi^{Am}_{\; a}$ & $\tilde{\psi}^{Am}_a$ \\
fib. metric & $\delta_{AB} \otimes \delta_{MN}$ & 
        $\delta_{AB} \otimes h_{m\bar{m}}$ & 
        $(H_a)_{AB} \otimes h_{m\bar{m}}$ \\
\hline
frame & $e_{A(a)} \otimes {\rm det} (e_m^{\; M}) (dz^1 \wedge dz^2)$ & 
        $e_{A(a)} \otimes (dz^m \wedge dz^n)$ & 
        $\tilde{e}_{A(a)} \otimes (dz^m \wedge dz^n)$ \\ 
coeff. & $\hat{\chi}^A_{\; a}$ 
    & $\chi^A_{mn\; a}$ & $\tilde{\chi}^A_{mn \; a}$ \\
fib. metric & $\delta_{AB}$ &
     $\delta_{AB} \otimes 2 h^{\bar{m}m} h^{\bar{n}n}$ & 
     $(H_a)_{AB} \otimes  2 h^{\bar{m}m} h^{\bar{n}n}$ \\
\hline  
  \end{tabular}
 \end{center}
\end{table}
%%%%%%%%%%%%%%%%%%%%%%%%%%%%%%%%%%%%%%%%%%%%%%%%%%%%%%%%%%%%%%

It is possible to express all the zero-mode wavefunctions by 
using the basis $\tilde{e}_{A(a)} \otimes (h_{m\bar{m}} d\bar{z}^{\bar{m}})$
and $\tilde{e}_{A(a)} \otimes (dz^m \wedge dz^n)$. We will call 
this holomorphic--holomorphic frame. The kinetic mixing function 
(\ref{eq:K.F.-truncation}) can be written as a sum of inner products 
of the two holomorphic vector bundles $\rho_{U_I} \otimes TS$ and 
$\rho_{U_I} \otimes K_S$, where $(H_a) \otimes h_{m\bar{m}}$ and 
$(H_a) \otimes h^{\bar{m}m} h^{\bar{n}n}$ are used as the metric 
of the inner product. 

Alternatively, one can think of expressing wavefunctions in a 
unitary--unitary frame. There is no rule that one always has 
to take $(dz^1 \wedge dz^2)$ as the basis of the rank-1 fiber 
vector space of $K_S$, for example. We define a vierbein for the 
K\"{a}hler metric in a field-theory local model by 
\begin{equation}
 h_{m\bar{m}} = e_m^{\; M} e_{\bar{m}}^{\; N} \delta_{MN}. \qquad \qquad 
 M, N = 1,2.
\end{equation}
A new frame $e_{A(a)} \otimes (e_M^{\; m} h_{m\bar{m}} d\bar{z}^{\bar{m}})$ 
can be chosen for the bundle $\rho_{U_I}(V) \otimes TS$, and 
$e_{A(a)} \otimes ({\rm det} (e_m^{\; M}) dz^1 \wedge dz^2)$ for 
the bundle $\rho_{U_I} (V)\otimes K_S$. Component fields 
$\hat{\psi}^A_{M}(z,\bar{z})_a$ and $\hat{\chi}^A(z, \bar{z})_a$ 
in this frame are not rescaled upon rescaling of the local 
coordinates $z^m, \bar{z}^{\bar{m}}$. Strictly speaking, 
the notion of (absolute value of) wavefunctions being large or small 
makes sense only in the component fields in this unitary--unitary frame; 
the transition functions of $\hat{\psi}^A_{M}(z,\bar{z})_a$ and 
$\hat{\chi}^A(z, \bar{z})_a$ are strictly unitary, and the norm of these
component fields at a point $z \in U_\alpha$ does not depend on the
choice of trivialization patches $U_a$ or $U_b$ in evaluating 
the norm. 
The fiber metric in the inner product in (\ref{eq:K.F.-truncation}) simply 
becomes $\tr {}'_{U_I} \otimes \delta_{MN}$ in this frame, and the inner 
product in the fiber vector space is simply integrated over $S$ 
after being multiplied by a volume form of $S$.  

The kinetic mixing matrix $K^{(R_I)}_{i\bar{j}}$ is now written simply 
by using the inner product of sections of 
$\rho_{U_I}(V) \otimes (TS \oplus K_S)$: 
\begin{equation}
 K^{(R_I)}_{i \bar{j}} = \frac{2 c_{(U_I, R_I)} M_*^4}{4\pi} 
  \langle (\hat{\psi}_j, \alpha \hat{\chi}_j) |
          (\hat{\psi}_i, \alpha \hat{\chi}_i) \rangle 
 = \frac{2 c_{(U_I, R_I)}}{\alpha_{\rm GUT}}  
   \frac{   
  \langle (\hat{\psi}_j, \alpha \hat{\chi}_j) |
          (\hat{\psi}_i, \alpha \hat{\chi}_i) \rangle }{{\rm vol} (S) }. 
\label{eq:K.F.-inn-prod}
\end{equation}
The inner product itself, however, does not depend on the choice of
frame. Reference \cite{Hitchin-1} already introduced this K\"{a}hler 
metric to the moduli space of Higgs bundle. In applications to physics, 
this K\"{a}hler metric of the vector space of zero modes is the 
one obtained by dimensional reduction of the 
action (\ref{eq:bosonic}, \ref{eq:fermi-bilin})
while simply truncating Kaluza--Klein modes. 

%%%%%%%%%%%%%%%%%%%%%%%%%%%%%%%%%%%%%%%%%%%%%%
\subsection{Kinetic Mixing Matrices Localized on Matter Curves }
\label{ssec:KMM-curve}
%%%%%%%%%%%%%%%%%%%%%%%%%%%%%%%%%%%%%%%%%%%%%

Although the kinetic mixing matrices $K^{(R_I)}_{i\bar{j}}$ and 
Yukawa matrices $\lambda^{(R_I, R_J, R_K)}_{i,j,k}$ are calculated 
by using the zero mode wavefunctions $(\psi,\chi)$ on $S$, all 
the information of zero mode wavefunctions should be contained 
in the original holomorphic sections $\tilde{f}_{(R_I); i}$ on the 
(covering) matter curves $\tilde{\bar{c}}_{(R_I)}$. Thus, it would 
be nice if it is possible to calculate Yukawa couplings and kinetic 
mixing matrices directly from the holomorphic sections, without solving 
$(\psi,\chi)$ (perhaps) numerically.\footnote{For a certain type of 
singlet-${\bf 5}$-$\bar{\bf 5}$ Yukawa couplings, a guess of such 
an expression has been proposed \cite{Tsuchiya}.}

Along a stretch of matter curve away from codimension-3 singularity
points, the zero-mode equations (\ref{eq:0-eq-D}--\ref{eq:0-eq-G}) 
becomes 
\begin{eqnarray}
 i \left(h_{t\bar{t}} \partial_n \psi_{\bar{n}} + 
         h_{n\bar{n}} \partial_t \psi_{\bar{t}} \right)
   - (F z_n)^* (2\alpha \chi_{tn}) & \simeq & 0, \\
 \bar{\partial}_{\bar{t}} \psi_{\bar{n}} - 
   \bar{\partial}_{\bar{n}} \psi_{\bar{t}} & \simeq & 0, \\
 \bar{\partial}_{\bar{t}} (2\alpha \chi_{tn}) 
    - i (F z_n) \psi_{\bar{t}} & \simeq & 0, \\
 \bar{\partial}_{\bar{n}} (2\alpha \chi_{tn}) 
    - i (F z_n) \psi_{\bar{n}} & \simeq & 0.
\end{eqnarray}
Here, we have chosen a field theory local model whose structure group 
$G'$ is $\U(1)$, and a charged matter field of interest is in
one-dimensional representation $U_I$ under the structure group. 
Local coordinates $(z_t, z_n)$ are chosen so that $z_n = 0$ is the 
local defining equation of the matter curve; subscripts $t$ and $n$ 
stand for tangential and normal directions of the curve, respectively.
We assumed that the $\varphi$ field background varies as 
\begin{equation}
  \alpha \rho_{U_I} (\vev{\varphi}) = 
  \alpha \rho_{U_I} (\vev{\varphi}_{tn}) (2 dz_t \wedge dz_n) \simeq 
 (F z_n) ( dz_t \wedge dz_n ).
\end{equation}
$F$ is a coefficient % (that may vary over $z_t$) 
of mass-dimension $+2$, and is expected to be of order $M_*^2$.
We ignored the gauge field background in the covariant
derivatives, because they are of order of ${\cal O}(1/R)$, 
whereas $\rho_{U_I}(\vev{\varphi}_{tn})$ becomes of order 
$M_*^2 \cdot (1/M_*) = M_*$ and is larger than $1/R$, 
even at a distance of order $1/M_*$ from the matter curve. 
% [comment of some refs? or in summary section?] 
An approximate solution to these equations is 
\begin{eqnarray}
  i\psi_{\bar{n}} & = & - \frac{1}{\sqrt{h_{t\bar{t}}}} e^{ -i {\rm Arg}(F)} 
     \exp \left[- |F| \frac{|z_n|^2}{\sqrt{h_{t\bar{t}}}} \right] 
          f_{(R_I); i; a}(z_t), \label{eq:approx-Gauss-psi}\\
  (2\alpha \chi_{tn}) & = & 
     \exp \left[- |F| \frac{|z_n|^2}{\sqrt{h_{t\bar{t}}}} \right] 
          f_{(R_I); i; a}(z_t),   \label{eq:approx-Gauss-chi}
\end{eqnarray}
and $\psi_{\bar{t}} = 0$. Just like we have already ignored 
the gauge field background, we have ignored possible variation 
of $h_{t\bar{t}}$, $|F|$ and $f_{(R_I); i; a}(z_t)$ along 
the coordinate $z_t$ on the matter curve.
$f = {\cal E}_a \cdot \tilde{f}$'s are the unitary frame version 
of the holomorphic wavefunctions on the matter curve. 

Along the matter curve, $\chi_{tn}$ looks like a section of 
$K_S|_{\bar{c}_{(R_I)}}$, while $\psi^n = \psi_{\bar{n}} h^{\bar{n}n}$
is regarded as a section of $N_{\bar{c}_{(R_I)}|S} = 
{\cal O}(\bar{c}_{(R_I)})$. These two bundles are not the same in general.
$f_{(R_I); i}$ are regarded as sections of a line bundle on the matter 
curve that includes a factor $K_{\bar{c}_{(R_I)}}^{1/2}$. 
Because of the adjunction formula, it is an ``average'' of the two:
\begin{equation}
 \frac{1}{2} K_{\bar{c}_{(R_I)}} = \frac{1}{2} 
     \left(K_S|_{\bar{c}_{(R_I)}} + \bar{c}_{(R_I)} \right).
\end{equation}
The difference between the two bundles is 
\begin{equation}
 \bar{c}_{(R_I)} - K_S|_{\bar{c}_{(R_I)}} = 
 \left(C_{(R_I)} - K_S\right)|_{\bar{c}_{R_I}} = r|_{\bar{c}_{R_I}},
\end{equation}
and hence all of $\chi_{tn}$, $\psi^n$ and $f$ are regarded 
as sections of the same line bundle 
$K_S|_{\bar{c}_{(R_I)}} = N_{\bar{c}_{(R_I)}|S} =
K_{\bar{c}_{(R_I)}}^{1/2}$ \cite{BHV-1}, if and only if 
the ramification divisor $r$ vanishes on the matter curve. 

Put differently, there is no difference among the three bundles 
$N_{\bar{c}_{(R_I)}|S}$, $K_S|_{\bar{c}_{(R_I)}}$ and 
$K_{\tilde{\bar{c}}_{(R_I)}}^{1/2}$ along the stretches of 
matter curves away from the ramification points; in F-theory 
compactifications with $G'' = A_4$ singularity along the 
GUT divisor $S$, the spectral surfaces is ramified 
at the $E_6$-type points on the curve $\bar{c}_{({\bf 10})}$, 
and at the $A_6$-type points on the curve $\bar{c}_{(\bar{\bf 5})}$, 
but nowhere else. Thus, along the segments of matter curves away 
from these codimension-3 singularity points, $\psi^m$, $\chi_{mn}$ and 
$f$ are virtually sections of the same bundle. 
Using the approximate solution  
(\ref{eq:approx-Gauss-psi}, \ref{eq:approx-Gauss-chi}), 
contributions to the kinetic mixing matrices $K^{(R_I)}_{i\bar{j}}$ 
(\ref{eq:K.F.-truncation}) from regions along the segments of 
matter curves are rewritten as follows:
\begin{eqnarray}
 K^{(R_I)}_{i\bar{j}} & \simeq & \frac{4 c_{(U_I, R_I)} M_*^4}{4\pi} 
  \int_S ({\rm vol}(S)) 
 \;  (f^*_{j} f_i )(z_t, \bar{z}_{\bar{t}}) \; (h^{\bar{n}n} h^{\bar{t}t}) 
 \exp \left[ - 2 |F| \frac{|z_n|^2}{\sqrt{h_{t\bar{t}}}} \right], \\
 & \simeq & \int_{\bar{c}_{(R_I)}} ({\rm vol}(\bar{c}_{(R_I)})) \; 
   \frac{c_{(U_I, R_I)} M_*^4}{2|F|}  (f^*_j \sqrt{h^{\bar{t}t}} f_i)(z_t,
   \bar{z}_{\bar{t}}).  
 \label{eq:K.F.-curve}
\end{eqnarray}
Integration in the normal direction $dz_n d\bar{z}_{\bar{n}}$ was 
carried out between the first and second lines; the physical width 
of the Gaussian profile in the normal direction is 
$[ h_{n\bar{n}}\sqrt{h_{t\bar{t}}} / 2|F|]^{1/2}$.
Interestingly, $\sqrt{h^{\bar{t}t}}$ appears in the last expression. 
This factor perfectly agrees with the Hermitian metric of the inner
product of sections of a line bundle containing 
$K_{\tilde{\bar{c}}_{(R_I)}}^{1/2}$ in the unitary--holomorphic frame.

Given the result above, it is tempting to claim that the kinetic mixing 
matrices are given by the inner product of sections $f$ on the covering 
matter {\em curves}.\footnote{In fact, \cite{BHV-2} has already made
this claim.} Certainly we have discussed the contributions
to (\ref{eq:K.F.-truncation}) only from regions of $S$ along segments 
of the matter curve away from $A_6$ or $E_6$ points; contributions 
from these points have not been evaluated or taken into account. 
However, at least in a limit where $M_* R \rightarrow \infty$, 
the contributions that we have already evaluated should dominate 
over those around ramification points; the former contributions 
effectively integrate over a region whose volume is of order 
$(R^2 / M_*^2)$, while the volume around ramification points\footnote{
The former contribution integrates over a volume $R^2$ along the curve 
and a volume $(1/M_*^2)$ in the normal directions of the curve. The
width of the Gaussian wavefunction is of order 
$1/\sqrt{|F|} \sim 1/M_*$. The zero mode wavefunctions do not fall off 
in a Gaussian profile in the normal direction near the ramification
points, but they still fall off as $e^{- (M_* |z_n|)^{3/2}}$. 
See \cite{Hayashi-2}, an appendix of \cite{Tsuchiya}, and the appendix 
\ref{sec:Hitchin} of this article.} 
is of order $(1/M_*^4)$. The integrand is not expected to be singular 
at/around the ramification points, since zero modes are holomorphic 
sections of a line bundle \cite{Hayashi-1}, and their component 
descriptions in trivialization patches should always be smooth. 
Smooth configuration of $f$ on the matter curve is very likely 
to correspond to smooth configuration of $(\psi, \chi)$ on $S$, 
as we study in detail in the appendix \ref{sec:Hitchin}.
Thus, the kinetic mixing matrices are calculated effectively 
by looking at contributions from segments away from the ramification 
points, and hence (\ref{eq:K.F.-curve}) is fine. 

Furthermore, there is an extra piece of evidence that supports 
a case for writing the kinetic mixing matrices in terms of the 
inner products on the covering matter curves. 
Certainly we have been unable to determined the zero mode
(hypermultiplet) wavefunctions in a region of complex surface around 
a $A_4$ or $E_6$ type point. But, at least we have succeeded in 
the appendix \ref{sec:Hitchin} in determining the wavefunction profile 
in a cross section of the matter curves that passes right through 
the $A_4$ or $E_6$ type point. 
Using the wavefunction profile instead of (\ref{eq:approx-Gauss-psi}, 
\ref{eq:approx-Gauss-chi}) and integrating (\ref{eq:K.F.-truncation}) 
in the normal directions of the matter curve, we obtained 
in the appendix \ref{sssec:reduced-rmfy-pt} 
an expression (\ref{eq:K.F.-rmfy-pt}) reduced to the matter curve 
at the $E_6$ or $A_6$ type points. This expression is also given 
by a inner product of holomorphic sections on the matter curve even 
at the ramification points, and the Hermitian metric there turns 
out to be $\sqrt{h^{t\bar{t}}}$, the right one for 
$K_{\tilde{\bar{c}}_{(R)}}^{1/2}$.

It still remains unclear what to think of $|F|$ 
in (\ref{eq:K.F.-curve}). Although we ignored the $z_t$-dependence 
of $|F|$ in the approximate solution 
(\ref{eq:approx-Gauss-psi}, \ref{eq:approx-Gauss-chi}), it does 
depend on $z_t$. Thus, at the next-to-leading order in 
$1/(M_*R)$ expansion, we can neither pull $|F|^{-1}$ out of
the integration over the matter curve, nor trust the form of 
the solution (\ref{eq:approx-Gauss-psi}, \ref{eq:approx-Gauss-chi}), 
which eventually led to (\ref{eq:K.F.-curve}). 
The value of $|F|$ goes to infinity near the $A_6$ points (for ${\bf 5}$ 
and $\bar{\bf 5}$ representation fields) and $E_6$ points (for ${\bf
10}$ and $\overline{\bf 10}$ representation fields). The factor 
$1/|F|$ in (\ref{eq:K.F.-curve}) will presumably be replaced by 
$H^{1/2}_{0*}(-A_1/A_0)_* / (|c|M_*^2)$ in (\ref{eq:K.F.-rmfy-pt}), 
but we have not figured out how this happens. 

To conclude, we have so far found that the kinetic mixing matrix 
$K^{(R_I)}_{i\bar{j}}$ can be given in the form of fiber-wise inner 
product of a pair of holomorphic sections on the covering matter curve, 
$\tilde{f}_j \sqrt{h^{t\bar{t}}} \tilde{f}_i$, integrated (with some 
weight) over the matter curve: (\ref{eq:K.F.-curve},
\ref{eq:K.F.-rmfy-pt}). There is not much we can say about 
the weight at this moment. We should also keep in mind that 
all the discussion above is based on simply truncation of Kaluza--Klein 
excited states. 

%%%%%%%%%%%%%%%%%%%%%%%%%%%%%%%%%%%%%%%%%%%%%%%%
\subsection{F-term Yukawa Couplings in Terms of Unitary Frame}
\label{ssec:F-Yukawa}
%%%%%%%%%%%%%%%%%%%%%%%%%%%%%%%%%%%%%%%%%%%%%%%%

An expression (\ref{eq:F-Yukawa}) for F-term Yukawa 
couplings is descried in terms of unitary--holomorphic frame 
wavefunctions $\psi$ and $\chi$. The overlap integral 
of zero-mode wavefunctions in (\ref{eq:F-Yukawa}), however, 
can also be written in terms of unitary--unitary frame 
wavefunctions. 
\begin{equation}
 \int_{U_\alpha} \chi^A_i \wedge \psi^B_j \wedge \psi^C_k 
  = 8 \int_{U_\alpha} {\rm vol}. \; 
   \hat{\chi}^A_i \hat{\psi}^B_{j;M} \hat{\psi}^C_{k;N} \; \epsilon^{MN}.
\label{eq:1}
\end{equation}
The values of component fields $\chi_{uv}$ and
$\psi_{\bar{m}}$ in the unitary--holomorphic frame are different 
for different choices of holomorphic coordinates, and the values 
are not physical. The values of the unitary--unitary frame 
wavefunctions, $\hat{\chi}$ and $\hat{\psi}_M$, are not affected 
by the coordinate choice, and hence their values are better suited 
for the purpose of discussing whether Yukawa couplings from a given
codimension-3 singularity is large or small. 

% Gauge indices are contracted 
% by using the structure constant of the Lie algebra. 
% The product of the zero-mode wavefunctions are integrated over $S$, 
% and since the integrand of (\ref{eq:F-Yukawa}) is already in the form 
% of differential form, it just has to be integrated. 

% The overlap integration for the F-term Yukawa couplings can also 
% be written in terms of holomorphic--holomorphic frame wavefunctions, 
% without changing anything but replacing $\psi$ and $\chi$ by 
% $\tilde{\psi}$ and $\tilde{\chi}$; this is because 
% the superpotential has a complexified symmetry $G^c$, 
% rather than just $G$. Once the holomorphic frame wavefunctions are 
% known, then the F-term Yukawa couplings can be calculated by 
% the overlap integration, without using the K\"{a}hler metric of 
% $S$ at all. 

If the expression (\ref{eq:1}) is rewritten in terms of 
wavefunctions $f$ on the matter curves, that would be more 
practically useful in obtaining estimate of low-energy 
Yukawa couplings. Although we do not dream to rewrite 
it rigorously in terms of the wavefunctions on the curves, 
approximate expressions are obtained in the following.

Let us begin with the down-type and charged lepton Yukawa couplings.
These Yukawa couplings are generated in all the regions around 
$D_6$-type points. The matter curve $\bar{c}_{({\bf 10})}$ for 
$Q, \bar{E} \subset {\bf 10}$ and two branches of the covering 
matter curve $\tilde{\bar{c}}_{(\bar{\bf 5})}$ for 
$\bar{D}, L \subset \bar{\bf 5}$ and $H_d \subset \bar{\bf 5}$
pass through any one of $D_6$-type points. 
For all the three pieces of curves in a region around a $D_6$-type
point, their spectral surfaces vary linearly in the normal 
coordinates of these pieces of the curves. Thus, the 
unitary--holomorphic frame wavefunctions are approximated by 
(\ref{eq:approx-Gauss-psi}, \ref{eq:approx-Gauss-chi}).
The unitary--unitary frame wavefunctions are 
\begin{equation}
 \alpha \hat{\chi} \sim  \frac{\alpha \chi}{\sqrt{{\rm det}
  h_{m\bar{m}}}} 
 \sim  
   \frac{f}{\sqrt{{\rm det} h_{m\bar{m}}}}   \qquad 
 \hat{\psi} \sim \frac{1}{\sqrt{h_{n\bar{n}}}} \psi_{\bar{n}} 
 \sim \frac{1}{\sqrt{h_{n\bar{n}}}}
 \frac{1}{\sqrt{h_{t\bar{t}}}} f 
\label{eq:uu-frame-psichi-f}
\end{equation}
on the matter curve; here, we do not pay much attention to complex 
phases. $f = f_{tn}$ is a coefficient function of a section for a 
frame $e \otimes (dt \wedge dn)$ of $\widetilde{\cal F} = 
(\widetilde{\cal N}\otimes K_S)|_{\tilde{\bar{c}}_{(R)}}$. 
Thus, the down-type Yukawa coupling from a given region around 
a $D_6$ type point is roughly 
\begin{equation}
 \Delta \lambda^{(d)}_{kj} \sim f_{(\bar{D});k} \times
  f_{(Q); j} \times f_{(H_d)} \times 
   \frac{1}{({\rm det} \; h_{m\bar{m}})^{3/2}}(M_*^4 \; {\rm vol}),
\label{eq:d-Yukawa-exprA}
\end{equation}
where ${\rm vol}$ above is the physical volume (measured by using metric
$h_{m\bar{m}}$) where the Gaussian wavefunctions are not exponentially 
suppressed. 

This expression is a little ugly, in that it contains explicit 
dependence on the choice of local coordinates. When the 
unitary frame wavefunction on the matter curve $\hat{f}$ is introduced 
as in 
\begin{equation}
 \hat{f} \sim \frac{M_*}{\sqrt{|F_{tnn}|}} (h^{t\bar{t}})^{1/4} f_{tn},
\label{eq:unitary-f-rk1}
\end{equation}
however, $\hat{f}$ does not depend on the choice of local coordinates 
any more. Here, the subscripts of $F_{tnn}$ reminds us of how $F$
changes under the coordinate re-parametrization in 
$\alpha \varphi_{tn} \sim F z_n$. The right-hand side has 
$(-1/2) - 2 \times (1/4) + 1 = 0$ covariant indices in the tangential 
direction, and $2 \times (-1/2) + 1 = 0$ covariant indices in the normal 
direction. Using this definition of the unitary frame wavefunction
$\hat{f}$ the down-type Yukawa matrix generated at a given $D_6$ type 
point (\ref{eq:d-Yukawa-exprA}) is rewritten approximately as 
\begin{equation}
 \Delta \lambda^{(d)}_{kj} \sim \hat{f}_{(\bar{D});k} \times
  \hat{f}_{(Q); j} \times \hat{f}_{(H_d)} \times 
   \left(\frac{\sqrt{|F_{\cdot \cdot \cdot}|}}
              { M_* h_{\cdot \cdot}^{3/4}} \right)^3 \; 
   (M_*^4 \; {\rm vol}). 
\label{eq:d-Yukawa-exprB}
\end{equation}
None of unitary frame wavefunctions or the physical volume depends on
the choice of coordinates, and the remaining factor $( \cdots )^3$ does 
not, either, as the factor within the parenthesis has $(3/2)$ covariant 
indices in the numerator and $2 \times (3/4)$ in the denominator. 
Thus, in the expression (\ref{eq:d-Yukawa-exprB}), the Yukawa coupling 
from a region around a type-$D_6$ point is given approximately by 
a product of physical (coordinate independent) quantities, unitary frame
wavefunctions $\hat{f}$ on the matter curves, physical volume where the 
$(\psi, \chi)$ wavefunctions are not exponentially suppressed and a
factor that is expected not be different so much from order one. 

The F-term Yukawa couplings should be independent of choice of 
K\"{a}hler metric, $h_{m\bar{m}}$, but it is not easy to see 
this in the expression (\ref{eq:d-Yukawa-exprB}). 
However, one only needs to remember that the Gaussian wavefunction 
(\ref{eq:approx-Gauss-psi}, \ref{eq:approx-Gauss-chi}) leave 
a physical volume 
$h_{n\bar{n}} |z_n|^2 \sim h_{n\bar{n}} \sqrt{h_{t\bar{t}}} / |F_{tnn}|$ 
in the normal directions. Thus, the physical volume of the region of 
unsuppressed wavefunctions is approximately 
\begin{equation}
 \left(\frac{1}{|F_{uvv}|} \sqrt{h_{u\bar{u}}} h_{v\bar{v}}\right) 
 \left(\frac{1}{|F_{vuu}|} \sqrt{h_{v\bar{v}}} h_{u\bar{u}}\right) 
 = \frac{1}{|F|^2} ({\rm det} h_{m\bar{m}})^{3/2}.
\label{eq:d-Yukawa-vol}
\end{equation}
Thus, one can see that the explicit metric dependence 
$({\rm det} h_{m\bar{m}})^{-3/2}$ in the F-term Yukawa coupling 
(\ref{eq:d-Yukawa-exprA}) is canceled by the metric dependence of 
the physical volume. 

The up-type quark Yukawa couplings in the superpotential 
are given by the following overlap integral
\begin{eqnarray}
 \Delta \lambda^{(u)}_{ij} & \sim & \frac{M_*^4}{4\pi} 
\alpha \epsilon_{AB} \int 
  \chi^A_{(\bar{U}); i} \wedge \psi^B_{(Q); j} \wedge \psi_{(H_u)}
- \chi^B_{(Q); j} \wedge \psi^A_{(\bar{U}); i} \wedge \psi_{(H_u)}, \\
  & \simeq & \frac{M_*^4}{4\pi} \alpha \int
  \left( \chi_{\uparrow; (\bar{U});i} \wedge \psi_{\downarrow; (Q); j} 
       + \chi_{\uparrow; (Q);j} \wedge \psi_{\downarrow; (\bar{U}); i}
  \right) \wedge \psi_{(H_u)}.
\label{eq:u-Yukawa-overlap}
\end{eqnarray}
Here, $A, B$ label two weights $\uparrow, \downarrow$ of the doublet 
representation $U_I = {\bf 2}$ of the $\U(2)$ structure group 
in field theory local models of 
$E_6 \rightarrow \vev{\U(2)} \times \SU(5)_{\rm GUT}$ symmetry
breaking. The structure constant of the $E_6$ Lie algebra 
is of the form $\epsilon_{AB} \epsilon^{abcde}$, with the totally 
anti-symmetric contraction of $\SU(5)_{\rm GUT}$ indices. 
The $\epsilon^{abcde}$ part becomes $\epsilon^{abc}$ contraction 
of $\SU(3)_C$ and $\epsilon^{de}$ of $\SU(2)_L$ in the Standard Model, 
which we omitted from the expression above. 
The $\chi_{(H_u)} \wedge \psi_{(Q)} \wedge \psi_{(\bar{U})}$ term 
is omitted from the expression above. That is because it almost vanishes, 
since both $\psi_{(Q)}$ and $\psi_{(\bar{U})}$ have 
their values primarily in the normal direction of the matter curve 
$\bar{c}_{({\bf 10})}$. Furthermore, we have shown in the 
appendix \ref{sec:Hitchin} that the ${\bf 10}$--$\overline{\bf 10}$ 
hypermultiplet wavefunction is non-zero in $\chi_{A=\uparrow}$ and 
$\psi_{A = \downarrow}$, not in the components 
$\chi_{A=\downarrow}$ or $\psi_{A=\uparrow}$. Hence we arrived at the 
expression above. 

At the intuitive level, we think that the Yukawa couplings are 
localized at/around the $E_6$-type codimension-3 singularity points.
Charged matter fields are M2-branes wrapped on 2-cycles whose volume 
vanishes on the matter curve $\bar{c}_{({\bf 10})}$ and
$\bar{c}_{(\bar{\bf 5})}$. All the relevant 2-cycles have vanishing 
volume at the $E_6$-type codimension-3 singularity points, and hence 
the M2-branes can reconnect there without blowing themselves to be 
large. In the field theory language, however, we have not managed 
to show that this is indeed the case. Let us take a set of local 
coordinates $(u,v)$, so that $v = 0$ is the matter curve 
$\bar{c}_{({\bf 10})}$, and $u = 0$ the matter curve 
$\bar{c}_{(\bar{\bf 5})}$. It is known that the $\psi_{(H_u)}$ 
wavefunction is in the Gaussian profile in the $u$ direction, and 
is localized along $u=0$ ($\bar{c}_{(\bar{\bf 5})}$). 
The wavefunctions of $(\psi, \chi)$ for the fields in ${\bf 10}$ 
representation of $\SU(5)_{\rm GUT}$ have not been determined 
in regions around $E_6$-type points. An approximate solution on the 
$u = 0$ slice is found in the appendix \ref{sec:Hitchin}, and found 
to be localized around $v = 0$ (almost the same result was already 
obtained in \cite{Hayashi-2}). The wavefunction along $u \neq 0$ slice, 
however, is not known yet. Thus, strictly speaking, it has not been 
shown yet that the overlap integral (\ref{eq:u-Yukawa-overlap}) is
localized also in the $v$ direction. 

It is still hard to doubt, however, that the overlap 
integral (\ref{eq:u-Yukawa-overlap}) is localized around $E_6$-type 
points. We have seen in \cite{Hayashi-2} that zero-mode equation 
for the $({\bf 2}, {\bf 10})$ component around the $E_6 \rightarrow
\SU(5)_{\rm GUT}$ deformation is the same as that for 
the $({\bf 2}, \bar{N})$ component around $A_{N+1} \rightarrow A_{N-1}$ 
deformation. Thus, the $\SU(5)_{\rm GUT}$-${\bf 10}+\overline{\bf 10}$ 
hypermultiplet must have the same wavefunction profile as the 
$\SU(N)$-$\bar{N}+N$ hypermultiplet. The local geometry of 
the $A_{N+1} \rightarrow A_{N-1}$ deformation with generic complex
structure has a Type IIB D7-brane interpretation. $N$ D7-branes are 
at $\xi = 0$ in Figure~\ref{fig:divisorsATtypeA}~(b), and the spectral 
surface in the figure is interpreted as another D7-brane intersecting with 
the $N$ D7-branes. Since this is a simple intersecting D7--D7 system, 
bifundamental open strings are believed to be localized around the 
intersection curve. Thus, even in the field theory language, the 
hypermultiplet wavefunctions of $({\bf 2}, \bar{N}) + {\rm h.c.}$ and 
those of $({\bf 2}, {\bf 10}) + {\rm h.c.}$ are also believed to 
be localized along the matter curve.\footnote{Reference \cite{HV-Nov08} 
derived hierarchical pattern of Yukawa eigenvalues contributed by 
individual $E_6$ / $D_6$ type points, assuming Gaussian
profile of hypermultiplet wavefunctions and triple intersection of
matter curves. The hierarchical pattern per se, however, does not 
strongly depend on the precise form of the hypermultiplet
wavefunctions. Under an assumption that the hypermultiplet wavefunctions
$\chi$ and $\psi$ of the $({\bf 2}, {\bf 10}) + {\rm h.c.}$ component 
fall off very quickly (say, exponentially), a discussion similar to the
one in \cite{HV-Nov08} leads to a prediction that the up-type Yukawa 
eigenvalues from a given $E_6$ type point is in the hierarchy 
$\epsilon^8 :  \epsilon^4 : 1$. (The up-type Yukawa matrix of 
the low-energy effective theory, however, is obtained by summing up 
such contributions from all the $E_6$ type points.) } 
The overlap integration (\ref{eq:u-Yukawa-overlap}), then, is also 
localized in both $u$ and $v$ directions, and hence is localized 
in local regions around $E_6$-type points.

Let us accept the assumption that the overlap integral 
(\ref{eq:u-Yukawa-overlap}) is localized around $E_6$ type points, 
and move on. Then the Yukawa couplings from a given $E_6$ type point 
is approximately products of three unitary--unitary frame wavefunctions 
multiplied by $(M_*^4 {\rm vol})$; ${\rm vol}$ is the volume where none 
of the wavefunctions are exponentially suppressed. Let us rewrite 
the overlap integral (\ref{eq:u-Yukawa-overlap}) into a form 
that is more useful in reading out the leading order contribution 
from a given $E_6$ type point. To do this, note first that 
the unitary--holomorphic frame $\chi_{\uparrow; (\bar{U}/Q)}$ 
and $\psi_{\downarrow; (\bar{U}/Q)}$ are given by the
holomorphic--holomorphic frame wavefunctions as 
\begin{equation}
 \chi_{\uparrow; ({\bf 10})} = \underline{\cal E}^{-1}
  \tilde{\chi}_{\uparrow; ({\bf 10})}, \qquad 
 \psi_{\downarrow; ({\bf 10})} = \underline{\cal E} 
  \tilde{\psi}_{\downarrow; ({\bf 10})}.
\end{equation}
See the appendix \ref{sec:Hitchin} for the definition of 
$\underline{\cal E}$ and its profile. Thus, 
$\underline{\cal E}^{-1}$ and $\underline{\cal E}$ cancel in 
(\ref{eq:u-Yukawa-overlap}), and $\chi_{\uparrow; ({\bf 10})}$ and 
$\psi_{\downarrow; ({\bf 10})}$ in (\ref{eq:u-Yukawa-overlap}) can be 
replaced by their holomorphic--holomorphic frame counterpart.\footnote{
This is a reflection of the fact that superpotential has $G^c$ symmetry, 
not just $G$.} Using (\ref{eq:uu-frame-psichi-f}) for $\psi_{(H_u)}$, 
and analogous relation for the ${\bf 10}$-representation fields 
based on (\ref{eq:metric-dep}) instead 
of (\ref{eq:approx-Gauss-psi}, \ref{eq:approx-Gauss-chi}), we obtain 
\begin{equation}
 \Delta \lambda^{(u)}_{ij} \sim \hat{f}_{(\bar{U}); i} \times
  \hat{f}_{(Q); j} \times \hat{f}_{(H_u)} \times 
%    \left(\frac{|c|^2}{h_{u\bar{u}}}\right)^{\frac{1}{6}} 
   \left[\frac{\sqrt{|F_{vuu}|}}{M_*} \frac{|c|^{7/3}}{c} 
         h_{u\bar{u}}^{-11/12} h_{v\bar{v}}^{-3/2} H_{0*}^{-1/2} \right]
   \;  (M_*^4 {\rm vol}.).
\label{eq:u-Yukawa-exprA}
\end{equation}
The unitary frame wavefunction $\hat{f}_{(H_u)}$ on the 
${\bf 5}$ representation matter curve is defined by
(\ref{eq:unitary-f-rk1}), while the unitary frame wavefunctions 
$\hat{f}_{(\bar{U})}$  and $\hat{f}_{(Q)}$ on the ${\bf 10}$
representation matter curve are given by 
\begin{equation}
 \hat{f} = \left[ \left( - \frac{A_1}{A_0}\right)_* H_{0*}^{1/2} 
    \frac{(h^{v\bar{v}})^{1/4} (h^{u\bar{u}})^{1/2}}{|c|} \right]^{1/2} 
   \tilde{f}_{uv}.
\end{equation}
The unitary frame wavefunctions $\hat{f}$ in this definition do 
not depend on reparametrization of local coordinates. See the appendix 
\ref{sec:Hitchin} for the definition of $H_{0*}$, $(A_1/A_0)_*$, $|c|$ 
and all other necessary details. 
The expression (\ref{eq:u-Yukawa-exprA}) is given by a product of 
physical unitary frame wavefunctions on the matter curves and 
the physical volume of unsuppressed wavefunctions and a dimensionless 
factor $[ \cdots ]$ that is expected to be of order unity. 

% Although it may not be obvious at first sight, this contribution to 
% the F-term up-type Yukawa couplings does not depend on the metric 
% $h_{m\bar{m}}$; this can be seen by an argument similar to the one
% employed around (\ref{eq:d-Yukawa-vol}).

Let us suppose that the unitary frame wavefunctions $\hat{f}$ 
on the matter curve do not have a particular structure and take 
values of order one. Then the F-term Yukawa couplings at type $D_6$ 
points and type $E_6$ points are of order unity, as the physical 
volume with unsuppressed wavefunctions are expected to be of order 
$1/M_*^4$. On the other hand, the kinetic mixing matrix in the D-term 
(\ref{eq:K.F.-inn-prod}) is of order $(M_* R_{\rm GUT})^2 \sim
\alpha_{\rm GUT}^{-1/2}$. Therefore, the overall normalization\footnote{
We did not count the power of $\pi$'s.} of physical Yukawa couplings 
in 3+1 dimensions scales as $\alpha_{\rm GUT}^{3/4} \sim g_{\rm GUT}^{3/2}$. 

If the volume of $\bar{\bf 5}$ representation curve is larger than 
that of ${\bf 10}$ representation curve, then the physical down-type 
and charged lepton Yukawa couplings tend to have smaller overall 
normalization than the physical up-type Yukawa couplings. Since 
the matter curve $\bar{c}_{(\bar{\bf 5})}$ tends to have a higher 
``degree'' than the matter curve $\bar{c}_{({\bf 10})}$, the assumption 
is likely to be true. The phenomenological outcome is in nice agreement 
with smaller mass-eigenvalues of bottom and tau than top in the real world. 

%%%%%%%%%%%%%%%%%%%%%%%%%%%%%%%%%%%%%%%%%%%%%%%%%%%
\section{Unitary Frame Wavefunctions on Matter Curves}
\label{sec:unit-wavefcn}
%%%%%%%%%%%%%%%%%%%%%%%%%%%%%%%%%%%%%%%%%%%%%%%%%%%

%%%%%%%%%%%%%%%%%%%%%%%%%%%%%%%%%%%%%%%%%%%%%%%%%%%%%
\subsection{Genus 0 Matter Curve}
%%%%%%%%%%%%%%%%%%%%%%%%%%%%%%%%%%%%%%%%%%%%%%%%%%%%%

Line bundles on a genus 0 curve $\P^1$ are characterized by 
their degree, $N$. Let the metric on $\P^1 \simeq S^2$ be 
$ds^2 = d \theta^2 + \sin^2 \theta d\phi^2$, where 
$(\theta, \phi)$ are coordinates on $S^2$ ($\theta \in [0, \pi]$ 
and $\phi \in [0, 2\pi]$). Unitary connection appearing in 
covariant derivative $D = d + i A$ are given by 
\begin{eqnarray}
     A = - N \frac{\cos \theta - 1}{2} d\phi 
     & &{\rm on~the~north~patch~}U_n~ (\theta \neq \pi), 
\\
     A = - N \frac{\cos \theta + 1}{2} d \phi
     & &{\rm on~the~south~patch~}U_s~ (\theta \neq 0).
\end{eqnarray}
The transition function is $g_{sn} = e^{i N \phi}$ between the two
patches $U_n$ $(\theta \neq \pi)$ and $U_s$ ($\theta \neq 0$). 
A section of a line bundle specified by an integer $N$ is 
described by a basis vector $e_{n/s}$ of the rank-1 fiber and 
its coefficient function $\psi_{n/s}$ in each one of the trivialization 
patches $U_n$ and $U_s$; they define the same section in the 
overlapping region $U_n \cap U_s$, if $e_{n/s}$, $\psi_{n/s}$ and
$g_{sn}$ satisfy the relations 
\begin{eqnarray}
 e_n(z) = g_{sn}(z)\, e_s(z), \qquad 
 \psi_s(z)  = g_{sn}(z)\, \psi_n(z) \qquad ({}^\forall z \in U_a \cap U_b). 
\end{eqnarray}

Holomorphic coordinates on $\P^1$ are introduced through 
the stereographic projection: 
\begin{equation}
 \zeta = \tan \left(\frac{\theta}{2} \right)
  e^{-i\phi}  \quad {\rm in~}U_n , \qquad 
  z = \cot \left(\frac{\theta}{2}\right) 
  e^{i \phi}  \quad {\rm in~}U_s, 
\end{equation}
with the relation $z = 1/\zeta$ in the intersection $U_n \cap U_s$. 
The transition function is $g_{sn} = e^{i N {\rm Arg}(z)}$. 
Using these coordinates, the connection is written as 
\begin{eqnarray}
 i A_n & = & - \frac{N}{2} \frac{|\zeta|^2}{|\zeta|^2 + 1} 
   \left(\frac{d\zeta}{\zeta} - \frac{d\bar{\zeta}}{\bar{z}}\right)
  \qquad {\rm in~}U_n, \\
 i A_s & = & - \frac{N}{2} \frac{|z|^2}{1 + |z|^2}
   \left(\frac{dz}{z} - \frac{d\bar{z}}{\bar{z}}\right)
  \qquad {\rm in~}U_s.
\end{eqnarray}

Let us choose complexified gauge transformations ${\cal E}_n$ and 
${\cal E}_s$ taking their values in $\U(1)^c = \C^\times$ as 
\begin{equation}
 {\cal E}_n = \left(1 + |\zeta|^2 \right)^{- \frac{N}{2}}, \qquad 
 {\cal E}_s = \left(1 + |z|^2 \right)^{- \frac{N}{2}}.
\end{equation}
The basis vectors $\tilde{e}_{n/s}$ in the fiber vector space 
in the trivialization patches $U_{n/s}$ are related to 
$e_{n/s}$ by 
\begin{eqnarray}
\tilde{e}_n={\cal E}_n e_n,
\qquad
\tilde{e}_s={\cal E}_s e_s.
\end{eqnarray}
Then, the connection in the new frame are 
\begin{eqnarray}
 i \tilde{A}_n & = & i A_n + {\cal E}_n^{-1} d {\cal E}_n 
  = - N \frac{\bar{\zeta} d \zeta}{1 + |\zeta|^2}, \\
 i \tilde{A}_s & = & i A_s + {\cal E}_s^{-1} d {\cal E}_s
 = - N \frac{\bar{z} dz}{1 + |z|^2}. 
\end{eqnarray}
Now the (0,1) part of the connection has been ``gauged-away'', 
at the cost of allowing $\tilde{A}_{n/s}$ to be non-Hermitian.
The transition function is given by 
\begin{equation}
 \tilde{g}_{sn} = {\cal E}_s^{-1} g_{sn} {\cal E}_n = z^N,
\end{equation}
which depends holomorphically on the local coordinate $z$ 
in $U_n \cap U_s$.

For line bundles with non-negative degree $N \geq 0$, there 
are $N+1$ independent holomorphic sections. They form an 
$(N+1)$-dimensional vector space, and a basis of the vector 
space can be chosen arbitrarily. We can take a basis, for example, 
as 
\begin{equation}
 \{ (\tilde{f}_{j;n}, \tilde{f}_{j;s}) = (\zeta^{N-j}, z^j) \} \qquad 
  (j = 0, \cdots, N).
\label{eq:hol-sec-P1}
\end{equation}
These are the descriptions of the holomorphic sections in the 
trivialization patches in the holomorphic frame. 
In the unitary frame, these holomorphic sections become 
\begin{equation}
 (f_{j;n}, f_{j;s})  = \left(
  e^{i (j-N) \phi } \cos^j \left( \frac{\theta}{2} \right) 
                    \sin^{N-j} \left( \frac{\theta}{2} \right),  \;\; 
  e^{i j \phi } \cos^j \left( \frac{\theta}{2} \right) 
                    \sin^{N-j} \left( \frac{\theta}{2} \right) 
\right). 
\label{eq:unit-0-P1}
\end{equation}
Note that $f_{j;s} = g_{sn} f_{j; n} = e^{iN\phi} f_{j; n}$, and 
$|f_{j; s}| = |f_{j; n}|$ in the unitary frame description.

If the kinetic mixing matrix is to be calculated by the inner product 
of the sections in the unitary frame description on the curve, then 
the $(N+1)$ independent zero modes of the ${\cal O}(N)$ bundle 
on $\P^1$ has a kinetic mixing matrix 
\begin{equation}
 K_{i\bar{j}} = \langle f_j | f_i \rangle 
 \propto \frac{1}{4\pi} \int^{2\pi}_{0} d \phi \int^{1}_{-1} d (\cos \theta) 
  (f_{j; s/n}^* f_{i;s/n}) = \delta_{ij} 
  \int_0^1 dx x^j (1-x)^{N-j} = \frac{\delta_{ij} }{{}_N C_j \; (N+1)}, 
\end{equation}
where the change of the variables $x=(\cos \theta + 1)/2$ was done,  
and the volume measure of $\P^1$ is proportional to $dx$. 
The basis of the vector space of holomorphic sections (\ref{eq:hol-sec-P1}, 
\ref{eq:unit-0-P1}) form an orthogonal basis. 

\begin{figure}[tb]
\begin{center}
 \begin{tabular}{ccc}
   \includegraphics[width=.4\linewidth]{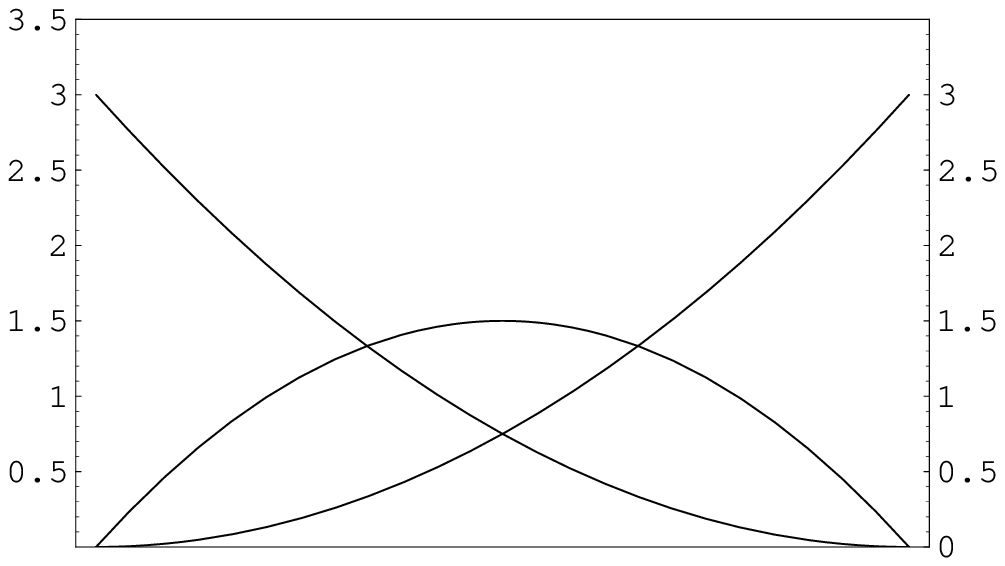}  
   & $\qquad \qquad $&
   \includegraphics[width=.4\linewidth]{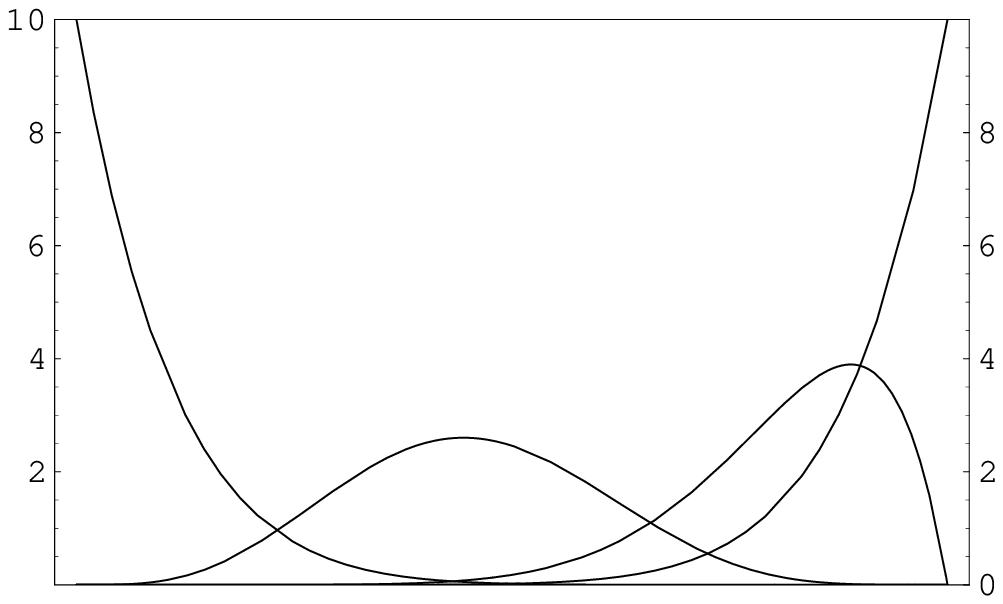}  
\\
  (a) & & (b)
 \end{tabular}
\caption{\label{fig:g=0} Behavior of zero mode wavefunctions coming from
 ${\cal O}(N)$ line bundle on a $g = 0$ curve $\P^1$. (a) is for the
 $N=2$ case, and (b) for $N=9$. $|f_{j;n/s}|^2 dx$ is presented (after
 normalized properly) for all the three zero modes in (a), whereas 
only those for $j = 0,1,5,9$ are presented in (b).}
\end{center}
\end{figure}

Figure~\ref{fig:g=0} shows the behavior of wavefunctions of 
the independent zero modes on a genus 0 curve. 
%%%%%%%%%%%%%%%%%%%%%%%%%%%%%%%%%%%%%%%%%%%%%%%%%%%%
%%%%%%%%%%%%%%%%%%%%%%%%%%%%%%%%%%%%%%%%%%%%%%%%%%%%%%
For line bundles with any degree $N$, independent canonically normalized
zero modes (i.e., forming an orthonormal frame with respect to the inner
product) are localized in different places in the $\cos \theta$ axis. 
Their wavefunctions do not have nodes, but they are all orthogonal 
because of their different $\phi$ dependence. 
For large $N$, the wavefunctions behave as 
\begin{equation}
 |f_{j;n/s}|^2 \propto 
  \exp \left[- \frac{N}{2}\frac{1}{x_{0,j}(1-x_{0,j})} (x-x_{0,j})^2 
     + {\cal O}((x-x_{0,j})^3) \right], \qquad 
  x_{0,j} = \frac{j}{N}.
\end{equation}
The width of the localization band in the $\cos \theta$ axis decreases 
as $\propto 1/\sqrt{N}$.

%%%%%%%%%%%%%%%%%%%%%%%%%%%%%%%%%%%%%%%%%%%%%%%%%%
\subsection{Genus 1 Matter Curve}
\label{ssec:torus}
%%%%%%%%%%%%%%%%%%%%%%%%%%%%%%%%%%%%%%%%%%%%%%%%%

Let us assume a flat metric $ds^2 = dx^2 + dy^2$ on a genus 1 curve 
$T^2$ with the local coordinates $(x,y)$ being dimensionless. 
Two independent periods of $T^2$ are given by  
\begin{equation}
 (x,y) \sim (x+1,y), \qquad (x + \tau_1, y+\tau_2).
\end{equation}
Let us consider the wavefunctions of the zero modes for the 
hermitian connection 
\begin{equation}
 A = \xi_1 \left(dx - \frac{\tau_1}{\tau_2} dy \right) + 
 \frac{1}{\tau_2} \left( \xi_2 
                        - 2\pi N \left(x - \frac{\tau_1}{\tau_2}y
		                 \right)
                  \right) dy
\label{eq:gauge-T2-bg}
\end{equation} 
in the unitary frame description, where $(\xi_1, \xi_2)$ 
correspond to two independent degrees of freedom of Wilson line on $T^2$. 
This connection defines a degree $N$ line bundle on $T^2$, 
because $dA = - (2\pi N)dx \wedge dy/\tau_2$. 
The zero mode solutions for this connection should satisfy the 
twisted periodicity conditions
\begin{equation}
 f(x+1,y) = e^{2\pi i N \frac{y}{\tau_2}} f(x,y), \qquad 
 f(x+\tau_1, y+\tau_2) = f(x,y).
\label{eq:T2-periodicity-A}
\end{equation}

A holomorphic coordinate $z \equiv x + i y$ is introduced on $T^2$
with the complex structure $\tau\equiv\tau_1+i\tau_2$ on it.
The zero mode solutions for the connection (\ref{eq:gauge-T2-bg})
are the sections of the line bundle satisfying
\begin{equation}
\bar{D}\, f \equiv \frac{1}{2} (D_x + i D_y) \, f = 
\frac{1}{2} \left((\partial_x + i A_x) + i (\partial_y + i A_y)\right) 
  f = 0
\end{equation}
with the periodicity conditions (\ref{eq:T2-periodicity-A}).
There are $N$ independent modes $\{ f_j \}$ ($j \in \Z / N \Z$) 
satisfying those conditions: 
\begin{eqnarray}
 f_j & = & \sum_{m \in \Z} e^{- 2\pi i \frac{m}{N}j} 
  e^{-i\left(x-\frac{\tau_1}{\tau_2}y\right) \xi_1} 
  e^{2\pi i N 
      \left(x - \frac{\tau_1}{\tau_2}y - \frac{\xi_2}{2\pi N}\right)
      \left( \frac{y}{\tau_2} + \frac{m}{N} 
             + \frac{\xi_1}{2\pi N}\right)
    }
  e^{\pi i N \tau \left(\frac{y}{\tau_2} + 
                        \frac{m}{N}+\frac{\xi_1}{2\pi N}
                  \right)^2}, \\
 & = & e^{-i \left(x - \frac{\tau_1}{\tau_2} y \right) \xi_1} 
       e^{- \pi i \frac{N}{\tau} 
           \left(\frac{\bar{\tau}\tilde{z} - \tau \bar{\tilde{z}}}
                      {\bar{\tau} - \tau} 
           \right)^2 }
       e^{\pi i \frac{N}{\tau} \tilde{z}^2}
       \vartheta_{0,0}\left(\tilde{z}- \frac{j}{N}; \frac{\tau}{N} \right).
\label{eq:0-mode-T2-A}
\end{eqnarray}
with 
\begin{eqnarray}
 \tilde{z} & = & z - \frac{\xi_2 - \tau \xi_1}{2\pi N},
\end{eqnarray}
where the theta function 
$\vartheta_{0,0}\left(u;\tilde\tau\right)$ is defined by
\begin{eqnarray}
 \vartheta_{0,0}(u; \tilde{\tau}) & = & 
  \sum_{m \in \Z} e^{\pi i m^2 \tilde{\tau}} e^{2\pi i m u}.
\label{eq:theta-def}
\end{eqnarray}
The kinetic mixing matrix of these zero modes are given by 
\begin{equation}
 K_{k\bar{l}} \propto 
    \frac{1}{\tau_2} \int_{T^2} dx dy (f_l^* f_k )
  = \delta_{k,l} \sqrt{\frac{N}{2\tau_2}}.
\end{equation}
Thus, the $N$ independent zero modes do not have mixing, and 
share the same normalization. That is, the K\"{a}hler potential 
does not yield any flavor structure at all in this case. 
Therefore, our focus will move to the profiles and the overlapping of 
the zero mode solutions to give rise to any flavor structure 
in the Yukawa couplings.

Let suppose that the complex structure parameter $\tau$ of 
the genus 1 matter curve is such that $\tau' \equiv -1/\tau$
satisfies ${\rm Im}\tau' \gg 1$. This is the case when 
\begin{equation}
\tau_1  < \tau_2 \ll 1.  
\label{eq:cpx-str-4reduction}
\end{equation}
Under the condition ${\rm Im}\, \tau' \equiv \tau'_2 \gg 1$, 
modular transformation of the theta function 
\begin{equation}
 \vartheta_{0,0}(u; \tilde{\tau}) = i \sqrt{\frac{i}{\tilde{\tau}}}
  e^{- \frac{\pi i}{\tilde{\tau}} u^2 } 
 \vartheta_{0,0}\left(\frac{u}{\tilde{\tau}}; - \frac{1}{\tilde{\tau}}\right) 
\end{equation}
enables us to find the behavior of the wavefunction easily.
The series-expansion expression of the theta function (\ref{eq:theta-def})
receives the dominant contribution only from $m=0$ term, when 
the imaginary part of $\tilde{\tau} = - N/\tau = N \tau'$ (that is, 
$N \tau'_2$) is much larger than 1. The wavefunctions behave as 
\begin{equation} 
 f_j \simeq \exp \left(- \pi N \tau'_2 
   \left( x - \frac{\tau_1}{\tau_2} y - \frac{\xi_2}{2\pi N} 
          - \frac{j}{N}\right)^2
                \right) e^{\pi i N \delta(x,y)_j }, 
\end{equation}
where $\delta(x,y)_j$ are phases that varies only mildly over $T^2$, 
even when $\tau'_2 \gg 1$. They are approximately in Gaussian profile 
along the one of the two real directions within the matter curve 
$T^2$, with the peak located at equal distance at 
\begin{equation}
\left(x - \frac{\tau_1}{\tau_2} y \right)_j = 
   \frac{\xi_2}{2\pi N} + \frac{j}{N},
\label{eq:peak-pos-inX}
\end{equation}
and the width $d$ of the Gaussians are all the same, and it is 
indeed given by 
\begin{equation}
 d^2 = \frac{1}{2\pi N \tau'_2}.
\end{equation}
%
%%%%%%%%%%%%%%%%%%%%%%%%%%%%%%%%%%%%%%%%%%%%%%%%%%%%%%%%%
\begin{figure}[tb]
 \begin{center}
     \includegraphics[width=.4\linewidth]{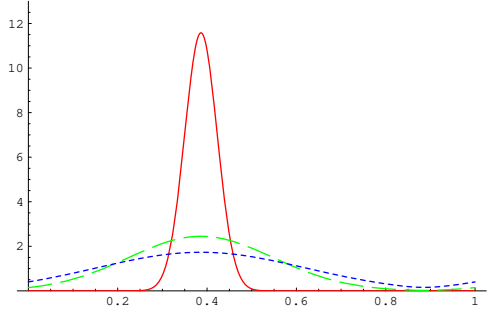}  
\caption{\label{fig:Gaussian-tau2} (color online)
Profile of zero mode wavefunctions depend on the complex structure 
parameter of $T^2$. In this figure, $|f_{j=1}(s)|$ is plotted 
for $N = 3$ over an interval $s \in [0, 1]$ on a $t\simeq 0.5$ slice, 
for three different values of $\tau_2$. 
The red curve (in solid line) is for $\tau_2 = 0.02$, 
green curve (dashed) for $\tau_2 = 0.5$, and 
blue one (dotted) for $\tau_2 = 1.0$. For all the three cases, 
we used $\tau_1 = 0.01$. The zero mode wavefunction has a localized
profile when the condition (\ref{eq:cpx-str-4reduction}) is satisfied.}
 \end{center}
\end{figure}
%%%%%%%%%%%%%%%%%%%%%%%%%%%%%%%%%%%%%%%%%%%%%%%%%%%%%%%%%
See Figure~\ref{fig:Gaussian-tau2}. 
The Gaussian width $d$ can be smaller than the periodicity 
of $T^2$, $x \rightarrow x+1$, or even smaller than the distance 
between the adjacent Gaussian wavefunctions $1/N$,  
if the complex structure parameter $\tau'_2$ is much larger than 1.
Contrary to the genus 0 curve $\P^1$, the genus 1 curve $T^2$ has 
the complex structure parameter which qualitatively changes 
the profile of zero mode wavefunctions. By tuning this parameter, 
the zero mode wavefunctions cam be made localized within the matter 
curve, and the wavefunctions have exponentially small tail outside 
the localization centers \cite{HSW}. This is important in generating 
small Yukawa couplings.   

Note that the zero modes with the Gaussian wavefunctions are 
localized at the coordinate (\ref{eq:peak-pos-inX}), where 
the value of the second term in (\ref{eq:gauge-T2-bg}) vanishes 
mod Kaluza--Klein momentum in the $y$ direction. 
%%%%%%%%%%%%%%%%%%%%%%%%%%%%%%%%%%%%%%%%%%%%%%%%%%%%%%%%%
\begin{figure}[tb]
 \begin{center}
     \includegraphics[width=.4\linewidth]{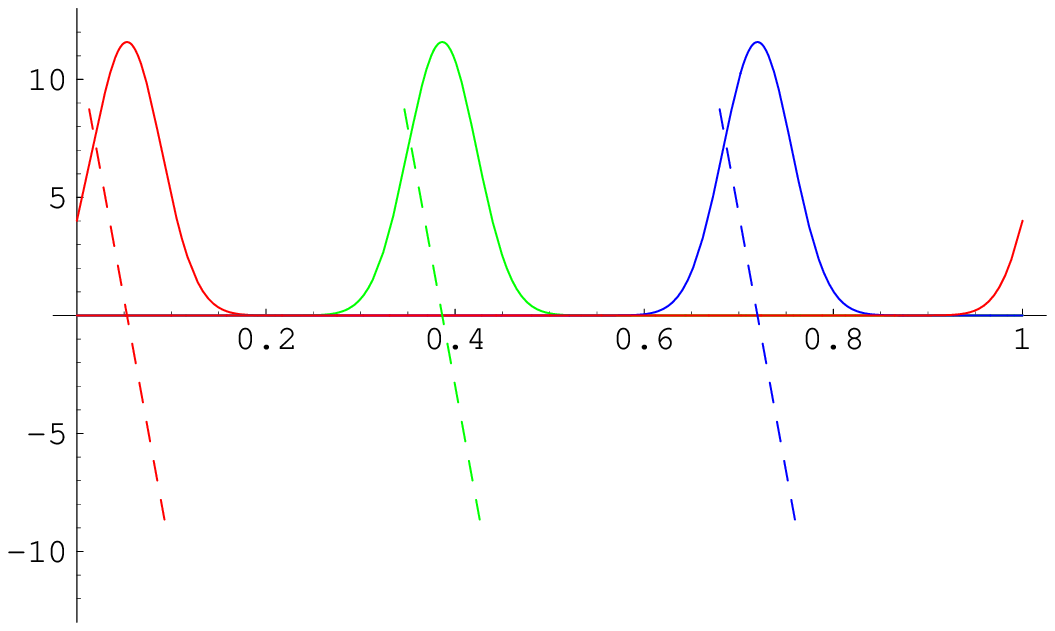}  
\caption{\label{fig:domain-wall} 
Relation between the profile of $N$ independent zero modes 
$|f_j(s)|$ and background ``scalar field'' configuration 
$\phi(s) + 2\pi \Z$. Gaussian-profile zero modes are localized 
around the locus of $\phi(s) + 2\pi \Z = 0$. We used $N = 3$, 
$\tau_2 = 0.02$ and $\tau_1 = 0.01$ for this figure.}
 \end{center}
\end{figure}
%%%%%%%%%%%%%%%%%%%%%%%%%%%%%%%%%%%%%%%%%%%%%%%%%%%%%%%%%
See Figure~\ref{fig:domain-wall}
Here, as we take the limit (\ref{eq:cpx-str-4reduction}), 
the torus $T^2$ becomes much shorter in the $dy$ direction than 
in the $(dx - (\tau_1/\tau_2)dy)$ direction, 
and we can study zero modes in a picture that is dimensionally 
reduced in the $dy$ direction. 
The second term of (\ref{eq:gauge-T2-bg}) 
then appears as a mass parameter or a Higgs field varying 
over the coordinate $(x - (\tau_1/\tau_2)y)$. 
It would be more obvious if you change the coordinates $(x,y)$ to $(s,t)$ as 
\begin{eqnarray}
s=x-{\tau_1\over\tau_2}y, 
\qquad 
t={y\over\tau_2}.
\end{eqnarray}
Then, the connection (\ref{eq:gauge-T2-bg}) may be rewritten as
\begin{eqnarray}
A=\xi_1 ds + \left(\xi_2-2\pi{N} s \right) dt=a_s+\phi\wedge{dt},
\end{eqnarray}
and the dimensional reduction along the $t$ direction gives the 
scalar field background $\phi(s)=\xi_2-2\pi{N} s$ in the lower
dimensions. Fermion zero modes (and their bosonic partners, 
too, in compactifications preserving supersymmetry) are localized 
at the place the mass parameter or the VEV of the Higgs field 
vanishes; this is the domain wall fermion, which has long been used in 
the phenomenology community for various models of flavor structure 
(e.g. \cite{AS}). 
The gauge field background (\ref{eq:gauge-T2-bg}) generates the 
net chirality of $N$ generations of ``the domain wall fermions''.
The zero mode wavefunctions become Gaussian if one 
makes only on assumption (\ref{eq:cpx-str-4reduction}) in the complex 
structure of the matter curve. We also know how the relative peak 
positions of the Gaussian wavefunctions are determined. 
At least within the case we studied, where the metric is assumed 
implicitly to be flat and the gauge field strength is constant 
over $T^2$, that the peak positions of the $N$ independent zero 
modes are equally separated in the direction of $T^2$ that remains 
long in the limit of (\ref{eq:cpx-str-4reduction}).

A qualitatively similar result follows also when $\tau_2 \gg 1$. 
To see this, it is better to take a new basis $\{ f'_k \}$ 
($k \in \Z / N \Z$) of the vector space of holomorphic sections as 
\begin{equation}
 f'_k = \sum_{m \in k + N \Z} 
  e^{-i\left(x-\frac{\tau_1}{\tau_2}y\right) \xi_1} 
  e^{2\pi i N 
      \left(x - \frac{\tau_1}{\tau_2}y - \frac{\xi_2}{2\pi N}\right)
      \left( \frac{y}{\tau_2} + \frac{m}{N} 
             + \frac{\xi_1}{2\pi N}\right)
    }
  e^{\pi i N \tau \left(\frac{y}{\tau_2} + 
                        \frac{m}{N}+\frac{\xi_1}{2\pi N}
                  \right)^2}. 
\label{eq:0-mode-T2-B}
\end{equation}
When $\tau_2 = {\rm Im} \, \tau \gg 1$, each term in $m$ 
summation has a Gaussian profile coming from the last factor, 
with the peak of the Gaussian profile located at 
\begin{equation}
 \frac{y}{\tau_2} \in - \frac{\xi_1}{2\pi N} - \frac{k}{N} - \Z.
\label{eq:peak-pos-inY}
\end{equation}
The width $d$ for $(y/\tau_2)$ is given by $d^2 = 1/(2\pi N \tau_2)$, 
and is much smaller than the periodicity of $(y/\tau_2)$ 
(that is, 1), if $\tau_2 \gg 1$. Only one term in the 
$m \in k + N \Z$ summation effectively contribute to the wavefunction 
in the fundamental domain of $T^2$ ($(y/\tau_2) \in [0,1]$), and 
that is a Gaussian profile with the peak position given by 
(\ref{eq:peak-pos-inY}) with the $- \Z$ part chosen appropriately. 
The $N$ independent Gaussian zero modes are placed at a equal distance 
in the $(y/\tau_2)$ axis, with the distance given by $1/N$ of the 
periodicity.

The origin of exponential behavior of the zero mode wavefunctions 
may be understood as follows.\footnote{Closely related discussion is 
found in \cite{BHV-2} in an attempt to generate hierarchically small 
trilinear couplings involving $\SU(5)_{\rm GUT}$ singlet fields. 
It is not that the authors perfectly understood the argument 
of \cite{BHV-2}, but we are certainly benefited from it.  }
The zero modes are defined by the condition 
$\bar{D} f = (\bar{\partial} + i A^{(0,1)}) f = 0$. Thus, 
roughly speaking, $f$ is like 
\begin{equation}
 f \sim \exp \left[ - i \int^{z'} A^{(0,1)}_{\bar{z}} d\bar{z} \right].
\label{eq:trial}
\end{equation}
Of course, this argument is too naive, and in particular, there is 
clearly the freedom to add any holomorphic functions to the exponent 
locally.\footnote{Globally though, such a freedom will not exist. 
This problem will essentially be to find global holomorphic sections.} 
But the bottom line is that $A^{(0,1)}$ already contains linear 
dependence on the coordinate, and it is further integrated to 
be the exponent. Thus, the zero mode wavefunction contains 
a quadratic dependence on the coordinate in the exponent. 
Unless the exponent is purely imaginary, so that the zero mode 
$f$ only varies in its complex phase, it exhibits the Gaussian 
behavior somewhere on the matter curve.\footnote{The peak positions 
of such localized zero mode wavefunctions also depend on 
the linear term in the exponent.} 
The gauge field background (\ref{eq:gauge-T2-bg}) becomes 
\begin{equation}
 A = \frac{1}{2 i \tau_2} \left[
 \left\{(\xi_2 - \tau \xi_1) - 2\pi N 
         \left( \frac{\tau \bar{z} - \bar{\tau} z}{2 i \tau_2} \right) 
 \right\} dz -
 \left\{(\xi_2 - \bar{\tau} \xi_1) - 2\pi N 
         \left( \frac{\tau \bar{z} - \bar{\tau} z}{2 i \tau_2} \right) 
 \right\} d\bar{z} 
\right]
\end{equation}
when it is expressed in terms of complex coordinate. Since there are 
so many parameters with complex phases already involved, it is 
impossible to keep the exponent pure imaginary, and that is why 
we obtain the Gaussian behavior in the zero mode wavefunctions. 

This argument also allows us to make an estimate of the degree 
of exponential hierarchy appearing in the zero mode wavefunctions. 
In the expression (\ref{eq:trial}), let us naively treat $A_{\bar{z}}$ 
as $F_{z\bar{z}} z$. Then, the expression (\ref{eq:trial}) becomes 
\begin{equation}
 f \sim \exp \left[ F_{z\bar{z}} z \bar{z} \right], 
\end{equation}
where we ignored linear terms in the exponent and any overall phase 
or sign of the exponent. Thus, the value of the real part of 
the exponent may be approximately described on the matter curve by 
\begin{equation}
 |F| ({\rm distance})^2 \sim (2\pi N) 
   \frac{({\rm distance})^2}{{\rm vol}},
\end{equation}
where $N$ is the degree of the line bundle with the gauge field 
background. The square of the distance between two points in a complex curve 
(in the above numerator) ranges from zero to some finite value. 
The maximum value of the numerator is of the same order as 
the volume of the curve (in the denominator) for generic complex
structure $\tau = \tau_1 + i \tau_2$ (c.f., \cite{Ibanez}). 
The ratio, however, roughly becomes $\tau'_2/4$ when\footnote{
Note that the maximum distance between two points in $T^2$ is 
not $\tau'_2$ but $\tau'_2/2$.} $\tau'_2 \gg 1$ 
($\tau_2/4$ when $\tau_2 \gg 1$) in the case of genus 1 curve. 
This intuitive argument already captures the essence of the Gaussian 
profile obtained by our rigorous calculations for 
the genus 1 curves.

The degree $N = N_{\rm gen} + g - 1$ of a line bundle on a genus $g$
curve cannot be chosen arbitrarily for practical applications. 
For curves with small genus, $N \sim N_{\rm gen}$. Thus, not much 
exponential hierarchy can be expected for generic complex structure. 
Quantitatively, 
\begin{equation}
  \frac{{\rm min}|f_j|}{ {\rm max} |f_j|}  \sim 
 \exp \left[ - (2\pi N_{\rm gen}) \frac{\tau^{(')}_2}{4} \right] 
 \sim 10^{- 0.68  N_{\rm gen} \tau^{(')}_2}
 = 10^{- 2.0 (N_{\rm gen}/3) \tau^{(')}_2}.
\end{equation}
Even for generic choice of complex structure parameter $\tau^{(')}_2$, 
some combinations of small numbers like $10^{-2}$ from the 
left-/right-handed quarks/leptons could be sufficient in generating 
the eigenvalues of quarks and leptons much smaller than unity. 
However, the electron Yukawa coupling and up-quark Yukawa coupling 
in the Standard Model are of order $10^{-6}$ and $10^{-5}$, respectively. 
It looks a little too difficult to obtain such small numbers from
overlap of wavefunctions that are not smaller than $10^{-2}$.
A boost in the exponential profile by the complex structure parameters 
such as $\tau^{(')}_2$ of the genus 1 curve, however, makes it easier 
to obtain the hierarchical arrangement of the Yukawa eigenvalues, 
as we discuss in section \ref{sec:flavor}. 

Before ending this section \ref{ssec:torus}, let us briefly 
comment on how the wavefunctions 
(\ref{eq:0-mode-T2-A}, \ref{eq:0-mode-T2-B}) are related to 
holomorphic wavefunctions like those obtained 
in section \ref{sec:hol-wavefcn}. To this end, we need to 
work out the relation between the holomorphic frame 
wavefunctions and unitary frame wavefunctions explicitly 
in the case of genus 1 curve, because the wavefunctions 
(\ref{eq:0-mode-T2-A}, \ref{eq:0-mode-T2-B}) describe coefficient
functions in the unitary frame. In fact, the transition function appearing 
in (\ref{eq:T2-periodicity-A}) is U(1) valued.

\begin{figure}[tb]
 \begin{center}
     \includegraphics[width=.3\linewidth]{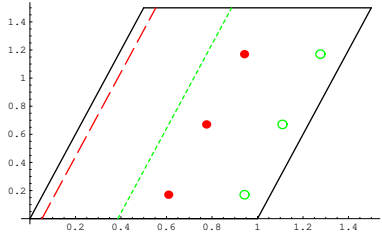} 
\caption{\label{fig:open-cover-T2} (color online)
Red filled dots in the fundamental domain of $T^2$ are the 
zero points of $f_{j=0}$, and green open circles the zero points 
of $f_{j=1}$. Here, we use $N = 3$. The open patches $U_j$ ($j = 0,1$) 
are $T^2$ without the zero points of $f_j$'s. Thus, the entire 
region of $T^2$ is covered by $U_0$ and $U_1$. These zero points 
are right at the ``opposite phase'' of the peak positions 
(\ref{eq:peak-pos-inX}). The peak position of $f_0$ and 
that of $f_1$ in the $s$ axis in the $\tau'_2 \gg 1$ limit 
are shown by the red dashed line and green dotted line, 
respectively in the figure.} 
 \end{center}
\end{figure}

Holomorphic frame wavefunctions---holomorphic sections of 
a line bundle---are described by an open covering $\{ U_a \}$ 
of $T^2$, transition functions $g_{ab}$ on $U_a \cap U_b$, and the 
holomorphic functions $\tilde{f}_{j; a}$ on $U_a$ glued together 
by the transition functions $g_{ab}$ on $U_a \cap U_b$.
To specify the open covering, note that the zero of the theta functions 
$\vartheta_{0,0}(u; \tilde{\tau})$ are at  
\begin{equation}
 u \equiv \frac{1+\tilde{\tau}}{2} \qquad {\rm mod ~} \Z + \tilde{\tau} \Z.
\end{equation}
Therefore, on $T^2$, each of the wavefunctions $f_j$ 
in (\ref{eq:0-mode-T2-A}) has $N$ zero points $z=z_n$ $(n=1,\cdots,N-1)$ 
\begin{equation}
 z_n\equiv \frac{\xi_2 - \tau \xi_1}{2\pi N} + 
  \frac{j}{N} + \frac{N + \tau}{2N} + \frac{n}{N} \tau \qquad 
{\rm mod~} \Z + \tau \Z.
\end{equation}
So, let us define an open subset $U_a$ of $T^2$ as $T^2$ without 
the zero points of $f_a$. Since the zero points of $f_j$'s 
are different for different $j \in \Z / N \Z$, two of the 
open sets, say, $U_0$ and $U_1$ are sufficient to 
cover the entire $T^2$. See Figure\ref{fig:open-cover-T2}.
%%%%%%%%%%%%%%%%%%%%%%%%%%%%%%%%%%%%%%%%%%%%%%%%%%%%%%%%%
%%%%%%%%%%%%%%%%%%%%%%%%%%%%%%%%%%%%%%%%%%%%%%%%%%%%%%%%%

To the zero mode wavefunction $f_j$ in (\ref{eq:0-mode-T2-A}), 
let us assign a holomorphic function $f_{j; a}$ on $U_a$ as
\begin{equation}
 \tilde{f}_{j;a} = \frac{f_j}{f_a} = 
  \frac{\vartheta_{0,0}\left(\tilde{z} - \frac{j}{N}; \frac{\tau}{N}\right)}
       {\vartheta_{0,0}\left(\tilde{z} - \frac{a}{N}; \frac{\tau}{N}\right)}.
\label{eq:0-mode-T2-C}
\end{equation}
This is holomorphic on $U_a$, because the zero points of $f_a$ 
have been removed from $U_a$, and all the non-holomorphic factors 
in (\ref{eq:0-mode-T2-A}) cancel out. 
Thus, this should be the holomorphic frame description of the unitary 
frame wavefunction $f_j$. The transition function on $U_a \cap U_b$ 
is given by $g_{ab} = f_b/f_a$, which is also holomorphic in $z$ on 
$U_a \cap U_b$. The ratio of the theta functions 
like (\ref{eq:0-mode-T2-C}) and the transition functions are 
expressed by rational functions of $\wp$ and its derivative on 
the genus 1 curve. Thus, the holomorphic frame wavefunctions 
are expressed in terms of holomorphic coordinates of the genus 1 curve, too. 

%%%%%%%%%%%%%%%%%%%%%%%%%%%%%%%%%%%%%%%%%%%%%%%%%%%%%%%%%%%%%%
\subsection{Higher Genus Matter Curves}
%%%%%%%%%%%%%%%%%%%%%%%%%%%%%%%%%%%%%%%%%%%%%%%%%%%%%%%%%%%%%%

Let us also take a look at zero mode wavefunctions on higher genus 
($g > 1$) curves for certain line bundles. We do not try to be 
exhaustive, but a limited case study still may still illustrate 
how the wavefunction profiles depend on complex structure of 
the curves. 

Before entering into detailed discussion, however, it is useful 
to recap some aspects of the theta function. 
The theta function (\ref{eq:theta-def}) has the periodicity 
conditions
\begin{equation}
 \vartheta_{0,0}(u + 1; \tilde{\tau}) = \vartheta_{0,0}(u; \tilde{\tau}), 
\qquad 
 \vartheta_{0,0} (u+\tilde{\tau};\tilde{\tau}) 
  = e^{- \pi i \tilde{\tau}} e^{-2\pi i u} \vartheta_{0,0}(u; \tilde{\tau}) 
\label{eq:periodicity-theta}
\end{equation}
along $a$-cycle $u\to{u+1}$ and $b$-cycle $u\to{u}+\tilde{\tau}$, 
respectively, 
on a genus 1 curve identified with $T^2 = \C / (\Z + \Z \tilde{\tau})$.
Since this is not perfectly periodic on $T^2$, it is regarded 
as a holomorphic section of a certain line bundle, rather than a 
holomorphic function on $T^2$. The transition function, however, 
takes its value in $\C^\times$, not in $\U(1)$.
From the theta function, a unitary frame description of the same section
of the line bundle is obtained easily:
\begin{equation}
 f(u; \tilde{\tau}) = e^{\pi i \tilde{\tau} \left(\frac{y}{\tilde{\tau}_2}\right)^2} 
   \vartheta_{0,0}(u;\tilde{\tau}). 
\label{eq:temp3}
\end{equation}
It obeys the periodicity conditions
\begin{equation}
 f(u + 1; \tilde{\tau}) = f(u; \tilde{\tau}), \qquad 
 f(u + \tilde{\tau} ; \tilde{\tau}) = 
   e^{- 2 \pi i \left( x - \frac{\tilde{\tau}_1}{\tilde{\tau}_2}y \right)} f(u; \tilde{\tau}).
\end{equation}
The transition function 
$e^{-2\pi i \left( x - \frac{\tilde{\tau}_1}{\tilde{\tau}_2}y \right)}$ takes 
its value strictly in $\U(1)$.

This is the essence of what we have done in section \ref{ssec:torus} 
to obtain holomorphic sections of a line bundle and to relate them  
to their unitary frame descriptions, although $N$ is not yet introduced 
here. The expression (\ref{eq:0-mode-T2-A}) might appear more
complicated than what it is, but the essential point is to consider 
$\vartheta(u;\tilde{\tau}) = 
e^{\pi i u^2/\tilde{\tau}} \vartheta_{0,0}(u; \tilde{\tau})$
instead of $\vartheta_{0,0}(u; \tilde{\tau})$ so that it satisfies the 
periodicity condition
\begin{equation}
 \vartheta(u+1; \tilde{\tau}) = e^{- \pi i (-1/\tilde{\tau})} 
    e^{- 2 \pi i (-u/\tilde{\tau})} \vartheta(u; \tilde{\tau}), \qquad 
 \vartheta (u+\tilde{\tau}; \tilde{\tau}) = \vartheta(u; \tilde{\tau}).
\end{equation}
Thus, $\vartheta(u+1; \tilde{\tau})$ is periodic along the $b$-cycle 
$u \rightarrow u + \tilde{\tau}$, but is not periodic along the $a$-cycle 
$u\to{u}+1$, as opposed to $\vartheta_{0,0}(u; \tilde{\tau})$.
It is just a holomorphic change in the basis of rank-1 fiber, and 
is one of the $\C^\times = \U(1)^c$ gauge transformations, and 
$\vartheta$ is regarded as a holomorphic section of the same line 
bundle on $T^2$ as $\vartheta_{0,0}$. 
A unitary frame description of $\vartheta$ is given by 
\begin{equation}
 g(u; \tilde{\tau}) = 
  e^{ \pi i (-1/\tilde{\tau})
      \left(x - \frac{\tilde{\tau}_1}{\tilde{\tau}_2} y \right)^2 } 
  \vartheta(u; \tilde{\tau})
 = e^{- \pi i \frac{1}{\tilde{\tau}} 
      \left(\frac{\bar{\tilde{\tau}} u - \tilde{\tau}\bar{u}}
                 {\bar{\tilde{\tau}} - \tilde{\tau}}
      \right)^2
     } 
   e^{\pi i \frac{u^2}{\tilde{\tau}}} \vartheta_{0,0}(u; \tilde{\tau}),
\label{eq:temp2}
\end{equation}
and satisfies the periodicity conditions 
\begin{equation}
g(u+1; \tilde{\tau}) = e^{2 \pi i \frac{y}{\tilde{\tau}_2} }g(u; \tilde{\tau}),
\qquad 
g(u;\tilde{\tau}) = g(u;\tilde{\tau}),
\end{equation}
with the transition function $e^{2 \pi i (y/\tilde{\tau}_2)}$
taking its value in $\U(1)$.
Thus, the first factor of the right hand side of (\ref{eq:temp2}) 
is to convert $\vartheta$ into its unitary frame description, and 
the second factor is the $\U(1)^c$ gauge transformation that 
switch the direction of non-periodicity. 
The formulation in section \ref{ssec:torus} was based on the description 
(\ref{eq:temp2}), along with a shift due to the Wilson lines $\xi_{1,2}$ 
and the change in multiplicity $N$, but it could have used an 
appropriate modification of (\ref{eq:temp3}). We chose to use a 
description along the line of (\ref{eq:temp2}) and the gauge choice 
(\ref{eq:gauge-T2-bg}) so that the connection with the domain wall 
fermion is seen easily.
Because of the simplicity of the expression of (\ref{eq:temp3}), 
we will use the description like (\ref{eq:temp3}) for higher genus 
matter curves in the rest of this section.

A Jacobian $J(\Sigma_g)$, a complex $g$-dimensional manifold, is 
defined for a genus $g$ curve $\Sigma_g$ as follows. 
An integral basis 
$\{ \alpha_i, \beta_i \}_{i=1,\cdots,g}$ of $H_1(\Sigma_g; \Z)$ can be
chosen so that $\alpha_i \cdot \alpha_j = 0$, 
$\beta_i \cdot \beta_j = 0$ and $\alpha_i \cdot \beta_j = \delta_{ij}$.
A normalized basis of a complex vector space $H^0(\Sigma_g; \Omega^1)$, 
$\{ \omega_i \}_{i=1,\cdots, g}$ are such that 
$\int_{\alpha_i} \omega_j = \delta_{ij}$. 
Depending on the complex structure of the curve $\Sigma_g$, 
a period matrix $\tau_{ij}$ is defined by 
\begin{equation}
 \tau_{ij} = \int_{\beta_j} \omega_i.
\end{equation}
The Riemann bilinear relations in \cite{GH} pp. 231--232 mean 
that $\tau_{ij} = \tau_{ji}$. The Jacobian $J(\Sigma_g)$ is given by 
\begin{equation}
  \C^g / (\vec{m}+\tau \cdot \vec{n}) = \C^g / ((m_i, \tau_{ij} n_j)_{i=1,\cdots,g}).
\end{equation} 
The $2g$ periods correspond in complex coordinates $u_i$ 
($i=1,\cdots,g$) of $\C^g$ to $u_i \rightarrow u_i + \delta_{il}$ 
and $u_i \rightarrow u_i + \tau_{il}$.
The Jacobian can be identified with the original curve when $g=1$, 
but otherwise, due to the Abel theorem in \cite{GH}, p. 232, 
a genus $g$ complex curve can be embedded into its complex $g$ 
dimensional Jacobian. 

% \kesu{When the theta function (\ref{eq:theta-def}) is regarded as 
% a holomorphic section of a line bundle on the Jacobian of 
% genus one curve, the theta function can be generalized for 
% higher genus cases. Complex coordinates $u_i$ ($i=1,\cdots,g$) are 
% introduced on $\C^g$, and the theta function} 
The theta function on $\C^g$ is defined by 
\begin{equation}
 \vartheta(u,\tau) = \sum_{\vec{m} \in \Z^g} 
   e^{\pi i \vec{m}^T \tau \vec{m} }
   e^{2 \pi i \vec{m}^T \vec{u} } 
  = \sum_{m_i \in \Z} e^{\pi i m_i \tau_{ij} m_j } 
   e^{2 \pi i m_i u_i }.
\end{equation}
It satisfies 
\begin{equation}
 \vartheta(u_i + \delta_{ik} ; \tau) = \vartheta(u_i, \tau), \qquad 
 \vartheta(u_i + \tau_{ik} ; \tau) = 
    e^{- \pi i \tau_{kk} - 2 \pi i u_k} \vartheta(u_i ; \tau).
\end{equation}
Since it is not perfectly periodic in the translation by 
$\delta u_i = \tau_{ik}$, it is not a function on the Jacobian, 
but it is regarded as a holomorphic section of a line bundle 
on the Jacobian.

Just like we considered $\vartheta_{0,0}(z; \tau/N)$ for the genus 1 
case in section \ref{ssec:torus}, we would also like to introduce 
something analogous to $N$ of the genus one case into the higher genus case.
Let us consider another complex torus $\C^g / \Lambda'$, whose periods 
correspond to $z_i \rightarrow z_i + N_i \delta_{il}$, and 
$z_i \rightarrow z_i + \tau'_{il}$ ($l = 1,\cdots, g$). 
The integers $N_i$ ($i=1,\cdots, g$) are non-zero,  
but for our practical purpose, we will take $N_i = 1$ for $i = 1,\cdots,g-1$, 
and $N_g = N_{\rm gen}$. The relation between $J(\Sigma_g) = \C^g/\Lambda$ 
and $\C^g / \Lambda'$ is specified shortly. 
On the complex torus $\C^g/\Lambda'$, we introduce the following 
``functions'' for $(k_i) \in ( \Z / N_i \Z)$:
\begin{equation}
  \vartheta_k(z) = \sum_{m_i \in k_i + N_i \Z} 
  e^{\pi i \sum_{i,j} \frac{m_i}{N_i} {\tau'}_{ij} \frac{m_j}{N_j}}
   e^{2 \pi i \sum_{i} \frac{m_i}{N_i} z_i}.
 \label{eq:theta-def-multi}
\end{equation}
These functions satisfy 
\begin{equation}
 \vartheta_k(z_i + N_l \delta_{il}) = \vartheta_k(z_i), \qquad 
 \vartheta_k(z_i + \tau'_{il}) = e^{- \pi i \tau'_{ll} - 2 \pi i z_l} \vartheta_k(z_i).
\end{equation}
Thus, they are sections of a line bundle on $\C^g/\Lambda'$.
When we take $N_i = 1$ for all $i=1,\cdots, g-1$ but $N_g = N_{\rm gen}$, 
there are $| \prod_{i=1}^g  (\Z / N_i \Z) | = N_{\rm gen}$ global sections. 

It is not difficult to obtain a unitary frame description of 
these holomorphic sections. 
Let us introduce $2g$ real coordinates $x_i$, $y_i$ ($i=1,\cdots,g$) 
on $\C^g$, so that $z_i = N_i x_i + \sum_{j}\tau'_{ij} y_{j}$. 
We also rewrite $\tau'_{ij}$ in terms of real-valued $X_{ij}$ and 
$Y_{ij}$ as $\tau'_{ij} = X_{ij} + i Y_{ij}$.
With these notations, we find 
\begin{equation}
  f'_k(x_i,y_i) \equiv e^{- \pi \sum_{i,j} y_{i} Y_{ij} y_{j}} 
  \vartheta_k(z_i)
\end{equation} 
obeying the periodicity conditions 
\begin{equation}
 f'_k(x_i + \delta_{ij}, y_i) = f'_k(x_i,y_i), 
 \qquad 
 f'_k(x_i, y_{i} + \delta_{ij}) = 
   e^{-\pi i X_{jj}} e^{-2\pi i (N_j x_j + \sum_{k}X_{jk}y_{k})} 
   f'_k(x_i, y_i).
\end{equation}
Thus, $f'_k$ is a global section of a U(1) bundle on $\C^g/\Lambda'$.

Let us suppose that ${\rm Im} \tau'_{gg} = Y_{gg} \gg 1$. 
Then, only one $m_g$ in the series expansion 
of (\ref{eq:theta-def-multi}) contributes dominantly, because 
$e^{- \pi (Y_{gg} /N_g^2) m_g^2}$ depends strongly on $m_g$. Let 
$m_g$ giving the largest contribution be $k_g$. Then, 
\begin{equation}
 f'_k \sim \exp 
  \left[ - \pi Y_{gg} y_{g}^2 - 2 \pi Y_{gg} \frac{k_g}{N_g} y_{g} \right]
 \propto 
 \exp \left[ - \pi Y_{gg} \left(y_{g} + \frac{k_g}{N_g} \right)^2 \right].
\end{equation} 
The $N_g = N_{\rm gen}$ global sections $f'_k$ ($k=1,\cdots, N_g$) 
in the unitary frame are in Gaussian profile in the $y_{g}$ direction, 
with the width parameter $d^2 = 1/(2\pi Y_{gg})$ possibly 
much smaller than the periodicity $1$ in this direction. 
They are localized, and the center positions of localization are 
all different from one another, when $Y_{gg} \gg 1$.

We are interested in line bundles and their global sections 
on a genus $g$ curve $\Sigma_g$, not in those on 
$J(\Sigma_g) = \C^g / \Lambda$ nor on $\C^g/\Lambda'$. 
However, once a map from $\Sigma_g$ to $J(\Sigma_g)$ and another from 
$\C^g/\Lambda$ to $\C^g/\Lambda'$ are given, then the line bundle and 
its global sections on $\C^g/\Lambda'$ can be pulled back to the genus 
$g$ curve $\Sigma_g$. 

The first step, from $\Sigma_g$ to its Jacobian, is sometimes 
called the Abel map in the literature. A point $p_0$ is 
arbitrarily chosen from $\Sigma_g$ and is called a base point. 
For any point $p \in \Sigma_g$, 
\begin{equation}
 u_i = \int^p_{p_0} \omega_i
\end{equation}
determines a point in $\C^g$ modulo $ \Lambda = \Z + \tau \Z$ 
by using a normalized basis of holomorphic one-forms 
$\omega_i$ ($i=1,\cdots,g$) on $\Sigma_g$. 
Thus, the image in $J(\Sigma_g) = \C^g/\Lambda$ is well-defined. 
In the case of $g=1$ curve, this map is trivial. In the $g > 1$ cases, 
this map gives a holomorphic embedding of $\Sigma_g$ into $J(\Sigma_g)$.

In the $g=1$ case, the period of the complex torus 
$\C^{g=1}/\Lambda'$ is chosen to be $(N, \tau') = (N, N \tau)$, and 
the map from $J(\Sigma_{g=1}) = \C^{g=1}/\Lambda$ to $\C^{g=1}/\Lambda'$ 
was given by $z = N u$. With this map, it will not be difficult to see 
that a linear combination $\sum_k e^{- 2\pi i k j/N} \vartheta_k$  
of $\vartheta_k$ in (\ref{eq:theta-def-multi}) becomes 
$\vartheta_{0,0}(u - j/N; \tau/N)$ that appears in (\ref{eq:0-mode-T2-A}).  
Without taking any linear combinations,  
$f'_k$ just above does correspond to $f'_k$'s in (\ref{eq:0-mode-T2-B}). 

For $g > 1$ cases, 
% it is a bit trickier to identify the map between $J(\Sigma_g)$ 
% and $\C^g/\Lambda'$, because a naive generalization, 
% $\tau'_{ij} = N_i \tau_{ij}$ is not symmetric under the exchange of 
% $i$ and $j$. Instead, we set $\tau' = \tau$, and we use a map 
% $z_i = N_i u_i$. This map is not one to one, but the image of 
% $J(\Sigma_g)$ wraps $\prod_i N_i = N_g$ times around $\C^g/\Lambda'$. 
let us take a limit $\tau_{ig} =0$ for $i = 1,\cdots, g-1$, 
and take $\tau'_{ij} = \tau_{ij}$ for $i,j, = 1,\cdots, g-1$ and 
$\tau'_{gg} = N_g \tau_{gg}$. The map from $J(\Sigma_g)$ to 
$\C^g/\Lambda'$ is given by $z_i = N_i u_i$.
Combining this map with the Abel map, the $g > 1$ curves can also be 
embedded into the complex torus $\C^g/\Lambda'$. Thus, the holomorphic 
line bundles and their global sections $f'_k$'s can be pulled back also 
to the curve $\Sigma_g$. 
 
The line bundle on $\C^g/\Lambda'$ pulled backed to $\Sigma_g$ determines 
a line bundle on $\Sigma_g$. The line bundle on $\Sigma_g$ can 
be characterized by its degree. It is calculated by following 
the discussion \cite{GH}, pp.334--335, and by modifying it a little
bit. It turns out that the degree is 
\begin{equation}
 \sum_i N_i = (g-1) + N_g = \frac{1}{2} {\rm deg} K_{\Sigma_g} + N_g.
\end{equation}
Pulling back the $N_g$ global holomorphic sections to $\Sigma_g$, 
we obtain $N_g$ independent elements of $H^0$ of the line bundle 
on $\Sigma_g$. On the other hand, the Riemann--Roch theorem tells us that 
$h^0 - h^1 = N_g$. Thus, if $h^1 = 0$, $f'_k$ obtained by pulling back 
the theta functions on $\C^g /\Lambda'$ to $\Sigma_g$ are all the 
independent elements of $H^0$ of a degree $N_g$ line bundle on a genus 
$g$ curve $\Sigma_g$.

We have already seen that the global section $f'_k$ have their  
localized profiles in $\C^g/\Lambda'$ with the center of localized positions 
at different places for different values of $k$, if 
\begin{equation}
Y_{gg} = {\rm Im} \tau'_{gg} = {\rm Im} \tau_{gg} \gg 1.
\end{equation}
When these wavefunctions are pulled back to the matter curve 
$\Sigma_g$, the $N_g$ independent modes must be localized in 
different places on $\Sigma_g$ along the direction of 
the gradient of $y_g$ on $\Sigma_g$. 
We have so far explicitly shown that independent zero mode wavefunctions 
have localized profiles on a higher genus curve, at least when 
$(g-1)$ complex structure parameters $\tau_{ig}$ for $i = 1,\cdots, g-1$ 
are set to zero and one more $\tau_{gg}$ is tuned as above. 
It remains unclear to us to what extent this condition on 
tuning of complex structure parameters can be relaxed while maintaining 
the localized profile of zero mode wavefunctions. 
If none of the complex structure parameters are tuned, 
on the other hand, then there is not many reasons to believe that 
the independent zero modes are particularly localized in $\Sigma_g$. 
Therefore, even in the case of higher genus curves, 
independent zero mode wavefunctions may or may not be 
localized within the matter curve, depending on the 
complex structure parameters of the curve $\Sigma_g$.

%%%%%%%%%%%%%%%%%%%%%%%%%%%%%%%%%%%%%%%%%%%%%%%%%%
\section{Realistic Flavor Structure from Localized Wavefunctions} 
\label{sec:flavor}
%%%%%%%%%%%%%%%%%%%%%%%%%%%%%%%%%%%%%%%%%%%%%%%%%%

Field theory local models for all the $E_6$ type 
[resp. $D_6$ type] codimension-3 singularity points 
have their own contributions to the $N_{\rm gen} \times N_{\rm gen}$ 
Yukawa matrices for the up-type quarks [resp. down-type quarks and 
charged leptons]. Since the contributions from a single field 
theory local model is approximately rank-1 for both 
up-type \cite{Hayashi-2} and down-type \cite{BHV-2} Yukawa matrices, 
it is a natural expectation that the up-type [resp. down-type] 
Yukawa matrices of the effective theory has a rank given by 
${\rm min}(N_{\rm gen}, \# E_6)$ [resp. ${\rm min}(N_{\rm gen}, \# D_6)$].
This is not consistent with the reality (where only one quark in the 
each sector is much heavier than the others in the same sector), 
unless there is only one $E_6$-type and one $D_6$-type codimension-3 
singularity points in the entire GUT divisor $S$.
We have seen in section \ref{sec:inv}, however, that such assumption 
is never satisfied in F-theory compactifications. 

Two possible solutions to this problem have already been mentioned.
One is to assume a globally defined $E_8$ Higgs bundle on the GUT 
divisor, and take a factorization limit of the spectral 
surface \cite{Caltech-0906}. 
This is partially motivated as a solution to the dimension-4 
proton decay problem \cite{TW-1, Tsuchiya}, but factorization for 
this purpose does not always predict that $\# E_6 = \# D_6 = 1$ 
in the ``relevant'' irreducible pieces of the matter curves. 
Thus, this is to introduce 
an extra condition to obtain approximately rank-1 up-type and down-type 
Yukawa matrices. The other, as we discussed in section \ref{ssec:F-Yukawa},  
is to assume in a $\Z_2$-symmetry solution to the dimension-4 proton 
decay problem that there are precisely two $E_6$ type points and two 
$D_6$ type points on the GUT divisor $S$, and they form pairs under 
the $\Z_2$ symmetry transformation. 

In this section, we propose another scenario that leads to realistic 
pattern of flavor observables. The alternative scenario does not need 
to assume $\# E_6 = \# D_6 = 2$ on the GUT divisor; it works for 
any cases, unless $\# E_6 = 0$ or $\# D_6 = 0$. We will exploit the 
fact that zero mode wavefunctions can be localized (as we have seen 
in the previous section) for a moderate tuning of complex structure 
parameter. None of complex structure parameters has to be tuned 
exactly to zero in this scenario.  

%%%%%%%%%%%%%%%%%%%%%%%%%%%%%%%%%%%%%%%%%%%%%%%
\subsection{A Short Review}
\label{ssec:flavor-review}
%%%%%%%%%%%%%%%%%%%%%%%%%%%%%%%%%%%%%%%%%%%%%%

It will be useful to provide a brief summary of observed flavor pattern 
in the real world, before we begin to discuss how to reproduce it 
in F-theory compactifications. One might think that it is a known 
old thing. A whole new understanding is required, however, after the 
discovery of neutrino oscillations with large mixing angles, and 
furthermore, conventional understanding on the ``flavor structure of 
the real world'' is often tainted too much by model building in 3+1 
dimensions based on flavor symmetries. For these reasons, we insert 
a short review here in section \ref{ssec:flavor-review}. 

Measured values of flavor observables keep the most detailed information 
of moduli values of compactifications. The precise values or detailed 
information, however, is not always the same as what we understand 
as an essence. One extracts phenomenological features from the measured 
values under an intention to make it easier to pin down the essence of 
underlying (and presumably microscopic) theory of flavor. 
The ``features'' to be extracted are therefore influenced inevitably 
by what one imagines as a framework of the microscopic theory. 
The ``features'' summarized below largely follows the observation 
in \cite{HSW}; this is because it shares much the same picture for 
the microscopic theory of flavor as in this article. 

\begin{itemize}
 \item {\bf Hierarchical Structure}: This is about the $N_{\rm gen} = 3$ 
mass {\bf eigenvalues} in a given sector. 
Mass terms of fermions in the Standard Model are classified into four 
sectors, based on the representations (and charges) of QCD and QED: 
up-type quarks ($u$, $c$, $t$), down-type quarks 
($d$, $s$, $b$), charged leptons ($e$, $\mu$, $\tau$), 
and neutrinos($\nu_{1,2,3}$). Each sector has three mass eigenvalues. 
The three mass eigenvalues in a given sector are all different by orders 
of magnitude (see the comment below for the neutrino sector, however), 
and this is called the hierarchical structure of the mass eigenvalues. 
Since the hierarchy among the mass eigenvalues is the same as the 
hierarchy among the Yukawa eigenvalues for the up-type, down-type 
quark and charged lepton  sectors, we can also call it the hierarchical 
structure of the Yukawa eigenvalues. 
\begin{itemize}
 \item The hierarchy among the Yukawa eigenvalues of the up-type quarks 
 seem to be larger than those among the eigenvalues of the down-type quark 
and charged lepton sectors. 
 \item As for the neutrino mass eigenvalues, only the difference between 
mass-squared, $\Delta m^2_{ij} = m^2_i - m^2_j$, have been measured 
experimentally. 
Since we do not know the values of the mass-eigenvalues themselves, 
it is not possible to argue for sure at this moment what kind of 
hierarchical pattern neutrino mass eigenvalues show.
If we assume the simplest scenario called ``normal hierarchy'', 
then the largest eigenvalue $m_3$ is inferred from $\Delta m^2$ of 
atmospheric neutrino oscillation, $m_3 \simeq \sqrt{|\Delta m^2_{\oplus}|}$, 
and the second largest eigenvalue from the one of solar neutrino
oscillation, $m_2 \simeq \sqrt{|\Delta m^2_{\odot}|}$. There is not
enough experimental data to infer the value of $m_1$, except that
$m_1 \ll m_2$ from assumption of the normal hierarchy itself. 
In this scenario, the hierarchy between $m_3$ and $m_2$, 
\begin{equation}
 \frac{\sqrt{\Delta m^2_\odot}}{\sqrt{\Delta m^2_\oplus}} \simeq 0.2,  
\end{equation}
is not more than the typical hierarchy among the adjacent 
Yukawa eigenvalues in the down-type / charged lepton sector.  
\end{itemize}
 \item {\bf Pairing Structure} and {\bf Generation Structure}: 
This is about the CKM {\bf mixing angles} among the 
{\bf left-handed quarks}. Left-handed quark doublets of the 
Standard Model form a $N_{\rm gen} = 3$-dimensional vector space.
The up-type Yukawa couplings $\lambda^{(u)}_{ij} u^c_i q_j h$ are 
``diagonalized'' by using unitary matrices $U^{(uL)}$ and $U^{(uR)}$ 
that make $(U^{(uR)})_{ik} \lambda^{(u)}_{ij} (U^{(uL)})_{jl} 
= \hat{\lambda}^{(u)}_{kl}$ diagonal. The mass-eigenbasis $\{ \hat{u}_{lL} \}$
($l = 1,2,3$) of the left-handed up-type quarks are related to 
the left-haded up-type quarks $u_{jL}$ in the original basis $q_j$ 
by $u_{jL} = (U^{(uL)})_{jl} \hat{u}_{lL}$. Similarly, the down-type 
Yukawa matrix $\lambda^{(d)}_{kj}$ is ``diagonalized'', and the 
mass eigenbasis $\{ \hat{d}_{lL} \}$ of the left-handed down-type 
quarks are related to the left-handed down-type quarks $d_{jL}$ 
in the original basis $q_j$ by $d_{jL} = (U^{(dL)})_{jl} \hat{d}_{lL}$.
The CKM mixing matrix is $V_{mn} = [(U^{(dL)})_{jm}]^* (U^{(uL)})_{jn}$, 
describing the coefficient of 
$\hat{\bar{d}}_{mL} \bar{\sigma}^{\mu} \hat{u}_{nL}$ in the W-boson current.  
The CKM matrix is the difference between $U^{(uL)}$ and $U^{(dL)}$, and 
it has nothing to do with other unitary matrices $U^{(uR)}$ or $U^{(dR)}$.
Thus, any features appearing in the CKM matrix tell us about the 
left-handed quark doublets (and possibly about the Higgs boson(s)), but 
no so much about other fields of the Standard Model. \\
The measured value of the CKM matrix has a distinct feature called 
the pairing structure, which means that there are only three entries 
of order unity in the unitary CKM matrix. 
Top quarks decay dominantly to bottom quarks, and charm quarks decay 
to strange quarks. 
The left-handed up-type quarks in the mass-eigenbasis are all paired up 
with their own partner left-handed down-type quarks in the mass-eigenbasis
in the electroweak current. That is the pairing structure.
Furthermore, the CKM matrix in the real world has a special pairing, 
called generation structure (or perfectly ordered pairing structure). 
The heaviest up-type quark ($t$) is paired with the heaviest down-type quark 
($b$) in the W-boson current, the middle up-type quark ($c$) is the 
partner of the middle down-type quark ($s$), and the lightest 
among the up-type and down-type quarks ($u$ and $d$) form the last 
pair. This is not a random pairing, but is a perfectly ordered pairing. 
The generation structure is one of special cases of the pairing
structure. The generation structure may appear in the real world 
just accidentally, by one chance in $N_{\rm gen}!$, but one may take this 
perfectly ordered pairing as an indication of an essence of microscopic 
theory of flavor. \\
The pairing structure and generation structure strongly suggests 
that there is a well-motivated basis 
$\{ q_j = (u_{jl}, d_{jL}) \}$ in the $N_{\rm gen}$ dimensional vector space 
in microscopic theory of flavor, so that $u_{jL}$ and $d_{jL}$ almost 
serve as the mass-eigenbasis'.
 \item {\bf Anarchy} \cite{anarchy}:
This is about the {\bf mixing angles} among the {\bf left-handed leptons}.
A unitary matrix called lepton flavor mixing matrix (also known as 
PMNS matrix) is defined similarly to the CKM matrix. It describes the 
mismatch between the mass-eigenbasis of the left-handed charged leptons 
and that of the left-handed neutrinos. Any features of the 
lepton flavor mixing matrix tell us about the lepton doublets 
(and possibly the Higgs boson(s)), but not so much about other fields. \\
It turns out after neutrino oscillation experiments that at least two 
out of three mixing angles in the lepton flavor mixing matrix are large. 
Thus, there is no pairing structure in the lepton flavor mixing matrix, 
and it is not even possible to argue whether there is a generation structure 
(perfectly ordered pairing structure) or not. 
\end{itemize}
It is important to note that all the phenomenological features 
above have implications different and independent from one anther. 
In particular, one should clearly keep in mind that the hierarchical 
structure of mass eigenvalues and the pairing/generation structure 
of the CKM mixing angles are totally independent. It is true that the both 
features are strongly correlated in models based on a U(1) flavor  
symmetry on 3+1 dimensions, but not necessarily in other theoretical 
frameworks of flavor structure.\footnote{This is why we intentionally avoid 
referring to the Wolfenstein parametrization of the CKM mixing matrix 
in this review. See e.g. \cite{HSW, HSW-2} for frameworks where the CKM 
mixing angles are predicted somewhat differently from the conventional 
Froggatt--Nielsen framework.} 

%%%%%%%%%%%%%%%%%%%%%%%%%%%%%%%%%%%%%%%%%%%%%%%%%%%%%%%
\subsection{Hierarchical Structure, Pairing Structure and Anarchy}
%%%%%%%%%%%%%%%%%%%%%%%%%%%%%%%%%%%%%%%%%%%%%%%%%%%%%%%

All the field-theory local models around $E_6$ and $D_6$ type
points have their own contributions to the up-type and down-type 
Yuakwa matrices in low-energy effective theory. In order to 
study the flavor structure in the effective theory, we need to 
be able to evaluate how important contributions individual 
field-theory local models make relatively. The relative importance 
is controlled by the value at a given codimension-3 singularity of the
unitary--unitary frame zero mode wavefunctions on the (covering) 
matter curves, as we saw in section \ref{ssec:F-Yukawa}. Behavior of 
the unitary--unitary frame wavefunctions was studied 
in section \ref{sec:unit-wavefcn}. Thus, we are now ready to take on 
the problem of flavor structure in the low-energy effective theory. 

{\bf Hierarchical Structure}

The up-type Yukawa matrix of the effective theory is given 
approximately by\footnote{We are careless about the overall factor 
that is irrelevant to the flavor structure here.} 
\begin{equation}
 \lambda^{(u)}_{ij} = \sum_A 
   \hat{f}_{(u^c);i}(A) \hat{f}_{(q);j}(A) \hat{f}_{(h_u)}(A),
\label{eq:up-fff}
\end{equation}
where $A$ labels all the ($\Z_2$-symmetric pair of) $E_6$ type 
singularity points in the GUT divisor $S$. Contribution from 
a given point $A$ is certainly approximately rank-1 \cite{Hayashi-2}, 
but generically there are more than one ($\Z_2$-symmetric pair of) 
$E_6$ type points. The up-type Yukawa matrix $\lambda^{(u)}$ would 
have a rank ${\rm min}(N_{\rm gen}, \# E_6)$, if none of the zero 
mode wavefunctions $\hat{f}_{(q)}$, $\hat{f}_{(u^c)}$ or 
$\hat{f}_{(h_u)}$ has some structure. 

\begin{figure}[tb]
 \begin{center}
\begin{tabular}{ccc}
     \includegraphics[width=.35\linewidth]{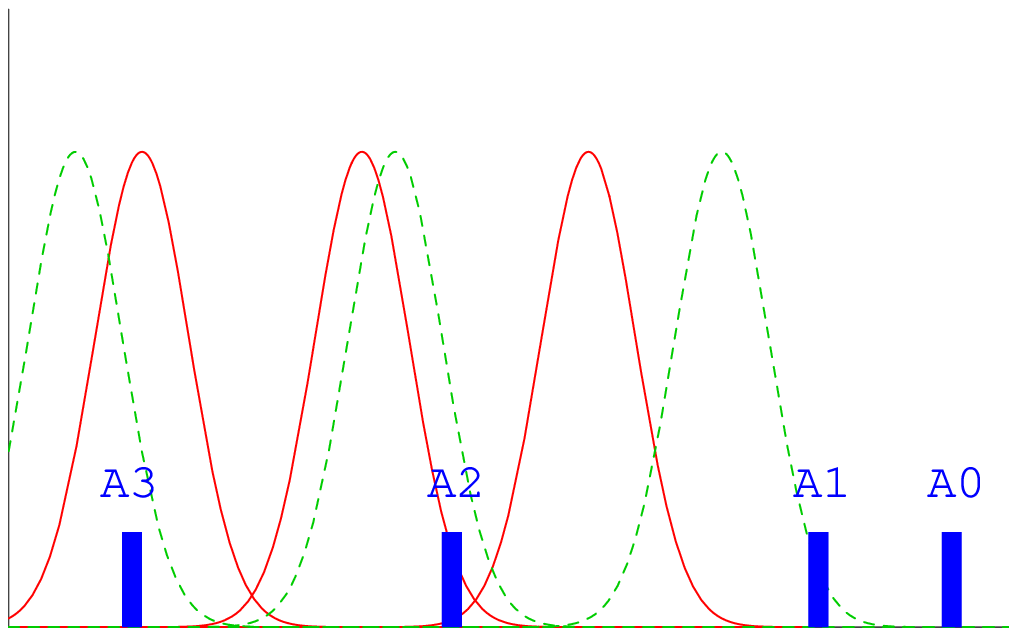}    & &
     \includegraphics[width=.35\linewidth]{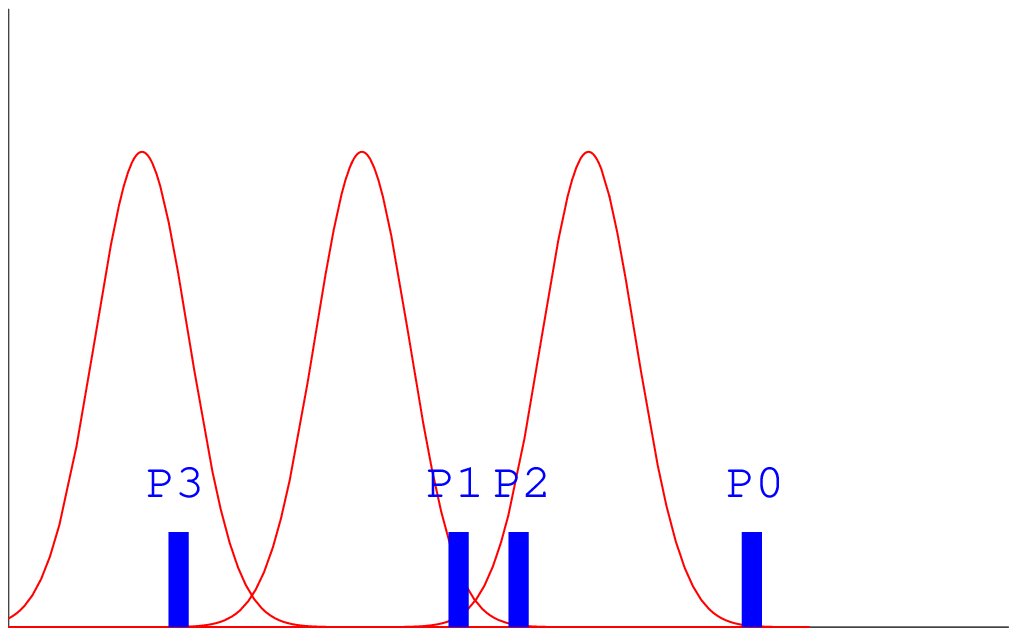}    \\
  (a): up-type Yukawa & & (b): down-type Yukawa
\end{tabular}
\caption{\label{fig:overlap} When a complex structure parameter of the 
matter curve $\bar{c}_{({\bf 10})}$ is tuned (which is possible when the
  genus of this curve is not zero), the zero mode wavefunctions on this 
matter curve are localized as schematically shown in this figure. The
  wavefunctions of $q_j$'s (solid/red) and those of $u^c_i$'s
  (dotted/green) are not necessarily localized at the same places 
on $\bar{c}_{({\bf 10})}$ in the presence of $\SU(5)_{\rm GUT}$ symmetry 
breaking gauge field background on $S$. The horizontal axis in this
  figure schematically represents the matter curve $\bar{c}_{({\bf 10})}$.
In the panel (a), $E_6$ type points labeled by $A$ in
  eq. (\ref{eq:up-fff}) are shown as blue boxes with labels $A_i$'s, 
while the $D_6$ type points labeled by $P$ in eq. (\ref{eq:down-fff}) 
as $P_i$'s in the panel (b). 
In a configuration of zero mode wavefunctions and the codimension-3 
singularity points on $\bar{c}_{({\bf 10})}$, localized quark doublet 
zero modes $q_3$, $q_2$ and $q_1$ from left to right almost correspond 
to mass-eigenstates $\hat{u}_{L3}$, $\hat{u}_{L2}$ and $\hat{u}_{L1}$,
and to mass-eigenstates $\hat{d}_{L3}$, $\hat{d}_{L1}$ and
  $\hat{d}_{L2}$. 
Thus, the pairing structure 
$q_3 \sim (\hat{u}_{L3}, \hat{d}_{L3})$, 
$q_2 \sim (\hat{u}_{L2}, \hat{d}_{L1})$ and 
$q_1 \sim (\hat{u}_{L1}, \hat{d}_{L2})$ follows. 
This is not a perfectly ordered pairing, though. }
 \end{center}
\end{figure}

Suppose that a complex structure parameter of the matter curve 
$\bar{c}_{({\bf 10})}$ is tuned so that the unitary--unitary frame 
zero mode wavefunctions in $Q, \bar{U}, \bar{E} \subset {\bf 10}$
are localized within the curve $\bar{c}_{({\bf 10})}$. We saw 
in section \ref{sec:unit-wavefcn} that this is possible, 
unless the genus of the curve $\bar{c}_{({\bf 10})}$ is zero. 
$N_{\rm gen}$ independent wavefunctions $\hat{f}_{(q);j}$ have 
exponentially localized profile with their center positions separated 
from one another for such a choice of one complex structure parameter 
of the matter curve. The zero mode wavefunctions $\hat{f}_{(u^c);i}$
are sections of a line bundle different from that of $\hat{f}_{(q);j}$; 
a brief and explicit discussion was presented in section \ref{ssec:SU5-breaking}
Thus, $\hat{f}_{(u^c); i}$'s are expected to be different
from $\hat{f}_{(q); j}$'s. But the $N_{\rm gen}$ independent
wavefunctions $\hat{f}_{(u^c);i}$ are also expected to have   
exponentially localized profiles, similarly to $\hat{f}_{(q);j}$'s, 
because the two group of fields $u^c_i$'s and $q_j$'s share 
the same matter curve $\bar{c}_{({\bf 10})}$.
See Figure~\ref{fig:overlap}.
%%%%%%%%%%%%%%%%%%%%%%%%%%%%%%%%%%%%%%%%%%%%%%%%%%%%%%%%%
%%%%%%%%%%%%%%%%%%%%%%%%%%%%%%%%%%%%%%%%%%%%%%%%%%%%%%%%%

For a given $E_6$ point $A$, the contribution from the local model 
to the up-type Yukawa matrix is approximately rank-1, with the 
largest eigenvalue proportional approximately to 
\begin{equation}
 \left({\rm max}_i | \hat{f}_{(u^c);i}(A) | \right) \times 
 \left({\rm max}_j | \hat{f}_{(q);j}(A) | \right) \equiv  
|\hat{f}_{(u^c);i_A}(A)| \times |\hat{f}_{(q);j_A}(A)|.
\label{eq:up-ij}
\end{equation}
Furthermore, the contribution $\Delta\lambda^{(u)}_{ij}(A)$ 
around $A$ is an $N_{\rm gen} \times N_{\rm gen}$ whose $(i_A,j_A)$ 
entry is almost the same as the largest eigenvalue, and all other 
entries are generically exponentially suppressed relatively to the 
$(i_A, j_A)$ entry. The ``eigenvalues'' of the up-type Yukawa 
matrix (\ref{eq:up-fff}) are approximately the $N_{\rm gen}$ largest 
of (\ref{eq:up-ij}) among the possible $A = 1, \cdots, \# E_6$, if 
$N_{\rm gen} \leq \# E_6$. It thus follows that there is exponentially 
hierarchy among the eigenvalues of the low-energy up-type Yukawa
matrix. When $\# E_6 < N_{\rm gen}$, higher-order contributions 
in the matrix form of (\ref{eq:up-ij}) and in the derivative 
expansion in \cite{HV-Nov08} should also be taken into account 
in order to determine the smallest eigenvalue of the up-type Yukawa 
matrix. By tuning the complex structure parameter of the 
matter curve $\bar{c}_{({\bf 10})}$, the unitary--unitary frame 
zero mode wavefunctions of $q_j$'s and $u^c_i$'s can have exponential 
profile, which eventually introduces exponential hierarchy among 
the $\#E_6$ contributions to the low-energy up-type Yukawa matrix. 
This is how the hierarchical structure of the up-type Yukawa eigenvalues 
can be reproduced. 

Reference \cite{HSW} or discussion in section \ref{ssec:torus} 
may be useful in getting the feeling of how much the complex structure 
parameter $\tau_2$ should be tuned to obtain the right ammount of 
hierarchical structure in the up-sector eigenvalues. To what 
extent the ${\bf 5}$-representation matter curve (or a type $E_6$ point)
looks like a point on the matter curve $\bar{c}_{({\bf 10})}$ is an important 
parameter in determining the required ammount of tuning. 

Once the complex structure parameter of the matter curve 
$\bar{c}_{({\bf 10})}$ is tuned to obtain exponential profiles 
in the zero mode wavefunctions of $(q,u^c,e^c) = {\bf 10}$, the 
hierarchical structure of the Yukawa eigenvalues in the 
down-type / charged lepton sector is derived. 
To see this, note that the corresponding Yukawa matrices of the 
effective theory are given by the sum of local contributions:
\begin{equation}
 \lambda^{(d)}_{kj} \simeq \sum_P 
   \hat{f}_{(d^c);k}(P) \; \hat{f}_{(q);j}(P) \; \hat{f}_{(H_d)}(P), \qquad 
 \lambda^{(e)}_{kj} \simeq \sum_P 
   \hat{f}_{(\ell);k}(P) \; \hat{f}_{(e^c);j}(P) \; \hat{f}_{(H_d)}(P), 
\label{eq:down-fff}
\end{equation}
where $P$ labels the type $D_6$ singularities and local models around
them. Since the quark doublet wavefunctions $\hat{f}_{(q);j}$ now have 
exponential profile on the matter curve $\bar{c}_{({\bf 10})}$, there 
is an exponential hierarchy among the values of 
\begin{equation}
 {\rm max}_{j \in \{ 1,\cdots, N_{\rm gen}\}} |\hat{f}_{(q);j}(P)|
\label{eq:down-j}
\end{equation}
for $P = 1, \cdots , \# D_6$. The $N_{\rm gen}$ largest among them 
roughly becomes the eigenvalues of the down-type / charged lepton 
Yukawa matrices in the effective theory. When $\# D_6 < N_{\rm gen}$, 
higher order corrections are also important.

We have assumed a tuning of only one parameter (or maybe more than one 
for higher genus case), the complex structure parameter of the 
curve $\bar{c}_{({\bf 10})}$. This assumption was 
introduced to obtain the hierarchical structure in the up-type Yukawa 
matrix. The hierarchical structure in the down sector and the charged 
lepton sector follows from this assumption. Furthermore, the
hierarchical structure tends to be stronger in the up-type sector 
the in the other two, because i) the hierarchy in the up sector comes 
from exponential profile of both $q_j$'s and $u^c_i$'s, while the 
hierarchy in the down / charged lepton sector comes purely from 
that of $q_j$ / $e^c_j$'s, and also because ii) there tends to be 
more $D_6$ type points than $E_6$ type points, as we see in all 
the examples in Table~\ref{tab:top-data} and \ref{tab:top-data-B}.
This observation originates from \cite{anarchy}, and is along the 
line of generalization in \cite{HSW, HSW-2}, but now adapted here 
with some new ingredients in a form suited to F-theory compactifications. 
Analysis in \cite{HSW} provides a sense of feeling of predicted pattern 
of the hierarchical Yukawa eigenvalues in the up and down/charged-lepton 
sectors in F-theory compactifications (with one tuning of the complex 
parameter), but the set-up there is not exactly the same. Thus, 
the prediction there should not be taken literally in this context. 

Neutrino Yukawa couplings are generated all along the curve 
$\tilde{\bar{c}}_{({\bf 5})}$ in the matter parity solution to 
the dimension-4 proton decay problem, and the Majorana masses 
of right-handed neutrinos from the entire bulk of a compact 
Calabi--Yau 4-fold for an F-theory compactification \cite{Tsuchiya}.
Without introducing extra assumptions, both the neutrino Yukawa matrix 
and Majorana mass terms in the effective theory are not expected to 
have any particular structure, except that there may still be some 
weak hierarchy originating from overlap integrals. Such hierarchical 
structures in the neutrino Yukawa couplings and Majorana mass terms 
tend to add up in the small mass terms generated by the see-saw
mechanism in such framework \cite{HSW}, in contrast against the
cancellation of hierarchical structure in the neutrino Yukawa matrix and
the Majorana mass terms in flavor models based on a U(1) flavor symmetry. 
As observed in \cite{HSW-2}, however, mass eigenvalues generated 
through the see-saw mechanism are not expected to have a very large 
hierarchy, when there are many right-handed neutrinos. Since all the 
complex structure moduli qualifies to be the right-handed neutrinos 
with Majorana masses in F-theory compactifications, there are indeed 
many right-handed neutrinos. Thus, it is a prediction that there is 
not a large hierarchy in the see-saw generated neutrino masses. 
This is in nice agreement with the measured value of $\Delta m^2_{ij}$
in the atmospheric and solar neutrino oscillation experiments. 

{\bf Pairing Structure}

When the complex structure parameter of $\bar{c}_{({\bf 10})}$ is 
tuned so that the zero mode wavefunctions have localized profile 
within the matter curve $\bar{c}_{({\bf 10})}$, there is a
well-motivated choice of basis $\{ q_j \}_{j = 1, \cdots, N_{\rm gen}}$ 
in the $N_{\rm gen}$-dimensional vector space of quark doublets.
Individual basis elements $q_j$ have localized wavefunctions
$\hat{f}_{(q); j}$ as shown schematically as solid curves in
Figure~\ref{fig:overlap}~(a) and (b). Because of the way we obtain 
hierarchical Yukawa eigenvalues in the down-type quark Yukawa matrix, 
the elements of the mass-eigenbasis of left-handed down-type quarks 
$\{ \hat{d}_{j L} \}_{j = 1,\cdots, N_{\rm gen}}$ are almost 
in one-to-one correspondence with the down-type components 
$d_{\sigma_d(j) L}$ of the original basis $q_{\sigma_d(j)}$ 
for some $\sigma_d \in \mathfrak{S}_{N_{\rm gen}}$. Suppose 
that the $(1 + N_{\rm gen} - l)$-the largest value of (\ref{eq:down-j}) 
for $l = 1, \cdots, N_{\rm gen}$ is the one at $P = P_{N+1-l}$ for 
$j = \sigma_d(l)$. This is how a permutation 
$\sigma_d \in \mathfrak{S}_{N_{\rm gen}}$ is determined. 
See Figure~\ref{fig:overlap}~(b).

Similarly, the elements of the mass-eigenbasis of left-handed up-type 
quarks $\{ \hat{u}_{l L}\}_{l = 1, \cdots, N_{\rm gen}}$ are generically
almost in one to one correspondence with the up-type components 
$u_{\sigma_u(l) L}$ of the original basis $q_{\sigma_u(l)}$ for some 
$\sigma_u \in \mathfrak{S}_{N_{\rm gen}}$. The permutation element 
$\sigma_u$ is determined by finding the $N_{\rm gen}$ largest values 
of (\ref{eq:up-ij}). The permutation will be determined by the 
$N_{\rm gen}$ largest values of the quark-doublet wavefunctions 
${\rm max}_{j} |\hat{f}_{(q);j}(A)|$ alone for most cases (statistically
in landscape of F-theory vacua), although there is a chance that 
exponentially small ${\rm max}_j |\hat{f}_{(q);j}(A_l)|$ at some $A_l$
is compensated by non-exponentially suppressed 
${\rm max}_i |\hat{f}_{(u^c);i}(A_l)|$. 

It thus follows that the up-type and down-type quark components 
of the original basis elements $q_j$ ($j = 1, \cdots, N_{\rm gen}$) 
are approximately the elements of some mass-eigenbasis of up-type 
and down-type quarks, in such a way as 
\begin{equation}
 q_j = (u_{jL}, d_{jL}) \approx \left(
   \hat{u}_{\sigma_u^{-1}(j)L}, \hat{d}_{\sigma_d^{-1}(j)L} \right).
\end{equation}
Therefore, the pairing structure is predicted, once the complex
structure parameter is tuned to reproduce the hierarchical structure 
of the up-type Yukawa eigenvalues. 

The CKM matrix approximately becomes a representation of 
$\sigma_u^{-1} \cdot \sigma_d$. Only $N_{\rm gen}$ entries 
are of order unity, and all others are small 
in the $N_{\rm gen} \times N_{\rm gen}$ unitary matrix. 
That is the manifestation of the pairing structure. 
On the other hand, one can also see that the generation structure 
(perfectly ordered pairing structure) does not follow
immediately.\footnote{
In the situation schematically shown in Figure~\ref{fig:overlap}, 
for example, the CKM matrix is approximately 
\begin{equation}
 V_{\rm CKM} \approx \left( \begin{array}{ccc}
		      & 1 & \\ 1 & & \\ & & 1
			    \end{array}\right).
\end{equation}
} 

We can deny a possibility that the perfectly ordered pairing 
in the CKM matrix in our vacuum may be realized just by one chance 
in $N_{\rm gen} ! = 6$; landscape of F-theory vacua containing ours 
may have only the pairing structure, but not the generation structure.
It is still possible that the $\Z_2$ matter parity may play some role, 
as we have already seen a hint in section \ref{sssec:matter-parity}. 
We have so far simply assumed that the $D_6$ type points and $E_6$ type 
points are scattered randomly on the matter curve 
$\bar{c}_{({\bf 10})}$. Within the class of vacua with a matter parity, 
however, some $E_6$ points and $D_6$ points are forced to come on top 
of each other, to form $\tilde{E}_8$ singularity points. In such cases, 
both $\sigma_u$ and $\sigma_d$ are determined by largest ones among 
${\rm max}_j | \hat{f}_{(q);j} (P = A)|$ at such $\tilde{E}_8$ points. 

{\bf Anarchy}

We have tuned the complex structure parameter of the matter curve 
$\bar{c}_{({\bf 10})}$ to fit the observed hierarchical structure in 
the up-type Yukawa couplings, and we have seen so far that it is 
the minimal tuning. There is no extra need to tune complex structure 
parameters of the other matter curve $\bar{c}_{(\bar{\bf 5})}$, 
so nature will not. None of the zero mode wavefunctions of 
$H_u \subset H({\bf 5})$ or lepton doublets $l_i$ is expected to have 
a localized profile on $\tilde{\bar{c}}_{(\bar{\bf 5})}$. Complex
structure moduli of a Calabi--Yau 4-fold $X$ are identified with 
the right-handed neutrinos in the matter parity scenario
\cite{Tsuchiya}; their wavefunctions are not localized at all.
Without a localized wavefunction profile, no structure is generated 
in the lepton flavor mixing matrix in the effective theory; that 
is the anarchy. 

The sharp contrast between the small mixing in the quark sector 
and large mixing in the lepton sector has remained a big 
theoretical puzzle for the last decade. A new word ``anarchy'' was 
coined in \cite{anarchy}, along with a new picture; there is a 
hierarchy among elements of a well-motivated basis of ${\bf 10}$'s, 
whereas there is just anarchy among the elements of a basis of 
$\bar{\bf 5}$'s (equivalently, there is not even a microscopically 
motivated choice of basis). The hierarchical structure (and anarchy), 
however, was still phrased in the conventional language of charge 
assignment of a U(1) flavor symmetry. A question remained who imposed 
this flavor symmetry, and how the U(1) charge assignment was
determined. The key idea of \cite{HSW, HSW-2} was that the U(1) flavor
symmetry can be replaced by wavefunctions with localized profiles 
while keeping successful phenomenology. There is no need to impose 
U(1) flavor symmetry, or to assign the U(1) charges to basis elements 
in order to reproduce the data successfully. The only necessary 
assumptions in \cite{HSW, HSW-2} were i) that the {\bf 10} fields 
(quark doublets) have localized wavefunctions in some internal 
directions, ii) that the Higgs boson also have wavefunctions localized 
in the same directions, and iii) that the $\bar{\bf 5}$ matter fields 
(lepton doublets) do not have localized wavefunctions. 

In F-theory compactifications, a complex structure parameter of 
the matter curve $\bar{c}_{({\bf 10})}$ can be tuned (if $g > 0$) 
in order to achieve the assumption i) above. Without an extra
assumption, iii) is automatically satisfied. Since the Higgs
wavefunction is localized along the curve $\bar{c}_{(\bar{\bf 5})}$, 
and since this curve intersects $\bar{c}_{({\bf 10})}$ transversely, 
the assumption ii) follows automatically in F-theory. Thus, we found 
that F-theory compactifications do have one parameter whose tuning 
realize the phenomenological contrast between the small mixing in 
the quark sector and the large mixing in the lepton sector. The tuning 
of the parameter now remains the only necessary assumption to be made 
phenomenologically.\footnote{If one allows oneself to resort to
anthropic arguments, then this tuning may be justified, because the 
tuning makes the lightest quarks and charged lepton light enough. } 

{\bf Remark}

We have a few more remarks on predictability before closing 
this section. 
There are two kinds of predictions; one is to tell the value 
of an observable that is expected to be measured in future 
experiments, and the other is to use some known facts in a 
new theoretical framework to derive some other known facts. 
The second kind of prediction is still better than just 
a hindsight explanation of a known fact, in that relations 
between seemingly independent facts are clarified. As we have 
already seen, the F-theory compactifications with a tuned 
complex structure parameter is a theoretical framework that has 
some predictions of the second kind in flavor physics. 

All the predictions of the second kind we obtained so far, 
however, are all qualitative, not quantitative. It is thus  
natural to think if there is any chance to make a 
quantitative prediction. It is easy to imagine that a plenty 
of complex structure moduli parameters are relevant in the 
process starting from calculating holomorphic wavefunctions. 
Physical observables also depend on K\"{a}hler moduli parameters. 
Since it is very unlikely that only one choice of flux 
configuration on a Calabi--Yau 4-fold is singled out as 
a consistent solution, we should expect that there are 
many choices in the value of moduli parameters. Thus, 
at least very naively, we do not expect that a precise 
quantitative relation between multiple flavor observables 
can be derived as a prediction of a reasonably broad class 
of F-theory compactifications. Qualitative predictions we have 
already discussed can still be extracted for generic vacua 
in F-theory compactifications with a tuned complex structure 
parameter of the matter curve $\bar{c}_{({\bf 10})}$. This 
attitude is similar to the one in \cite{HSW, HSW-2} in spirit.

We have not made all possible efforts to derive quantitative 
predictions, however.  Let us list up a couple of possible 
directions that might be pursued further in the following. 
\begin{itemize}
 \item The wavefunctions of anti up-type quarks are not the same 
as those of left-handed quark doublets, because of the symmetry breaking
from $\SU(5)_{\rm GUT}$ to $\SU(3)_C \times \SU(2)_L \times \U(1)_Y$. 
At this moment, it is not clear whether the wavefunctions of $u^c_i$'s 
can be totally independent from those of $q_j$, although we did not
assume any kinds of precise relations between them in the discussion 
in this section. In the case study of section \ref{ssec:torus}, 
for example, the line bundles for $u^c$ and $q$ should have the same 
degree on a genus 1 matter curve $\bar{c}_{({\bf 10})}$. The only 
difference between the two line bundles should thus be in the Wilson 
line (Jacobian). The width of the Gaussian-profile wavefunctions is 
completely common to all the $2 \times N_{\rm gen}$ independent zero 
modes of $u^c$'s and $q$'s. The center positions of the localized 
wavefunctions are determined only by $(\xi_1 + i \xi_2)$ for $q$ 
and that of $u^c$. It is not that the zero modes are localized 
at random places in the matter curve (c.f. \cite{HSW}). 
Thus, the real question will be to what extent such constraints 
among the zero mode wavefunctions can be generalized. 
 \item We did not use any particular properties of wavefunctions of 
the Higgs doublet(s). If it turns out that the wavefunctions of the (two) 
Higgs doublet(s) of the (supersymmetric) Standard Models have 
wavefunctions localized in the matter curve $\bar{c}_{(\bar{\bf 5})}$, 
for example, there will be more chance to say something more
quantitative. We have not fully exploited consequences coming from 
a $\Z_2$ symmetry introduced as a solution to the dimension-4 proton 
decay problem, although a hint is shown in section \ref{sssec:matter-parity}.
\end{itemize}
%

%%%%%%%%%%%%%%%%%%%%%%%%%%%%%%%%%%%%%%%%%%%%%%%%%%%
\section{Summary and Discussion}
\label{sec:discuss}
%%%%%%%%%%%%%%%%%%%%%%%%%%%%%%%%%%%%%%%%%%%%%%%%%%%

F-theory compactification is one of three large classes of string vacua 
that are able to generated the up-type Yukawa couplings (\ref{eq:1}) 
of $\SU(5)_{\rm GUT}$ unified theories. $\SU(5)_{\rm GUT}$-charged 
matter fields in low-energy effective theory have wavefunctions localized 
in internal space in F-theory compactification. This enables us to 
estimate flavor structure of low-energy effective theory in an 
intuitive way, for a large number of vacua all at once. In this article, 
we discussed flavor structure in F-theory compactifications in general. 

One could think of picking up one vacuum of string theory after another, 
studying flavor structure for each individual vacuum. Given the enormous 
number of semi-realistic vacua expected in string theory, however, such
an approach may no longer be an effective strategy to use string 
theory to learn something more about phenomenology. Thus, we try
to identify elements of geometric data of compactifications that 
{\it directly} control the flavor structure of low-energy effective 
theory. It is known that algebra of topological 2-cycles controls 
whether or not Yukawa couplings and dimension-4 proton decay operators 
are generated \cite{TW-1} (see also \cite{Tsuchiya}). We know ask 
what controls hierarchy among mass eigenvalues of quarks and leptons, 
and how the mixing angles depend on geometric data. With this approach, 
we try to cover as many semi-realistic vacua as possible from F-theory 
compactifications. 

This article therefore does not talk about specific choice of topology 
of a complex three fold $B_3$ and of a complex surface $S$ (called GUT 
divisor) in $B_3$ where $G'' = \SU(5)_{\rm GUT}$ gauge fields are localized.
It is true that the topology of $B_3$ and $S$ is crucial information 
in determining multiplicities of massless fields in various representations.  
We do not have a reason to believe, however, that the number of generations 
$N_{\rm gen}$ being three is a crucial element\footnote{Origin of Higgs boson, 
however, may play an important role in the flavor structure of quarks 
and leptons.} in the hierarchical structure of Yukawa eigenvalues and 
small mixing angles in the quark sector in the real world. We thus leave 
it as a separate issue to identify topological data of $B_3$, $S$ and 
fluxes that lead precisely to $N_{\rm gen} = 3$.

\begin{center}
..........................................................................
\end{center}

Yukawa couplings among three charged matter fields are generated 
in areas in the complex surface $S$ around isolated singularity 
points in F-theory compactifications.
The up-type Yukawa couplings are generated at type $E_6$ points, while 
the Yukawa couplings of down-type quarks and charged leptons at type 
$D_6$ points. See Figure~\ref{fig:schem-curve} and \ref{fig:realistic-curve}. 
Zero mode wavefunctions of fields in a given representation are obtained 
by working out holomorphic sections of a line bundle on a complex curve
\cite{Hayashi-1} (called (covering) matter curves); explicit examples 
in section \ref{sec:hol-wavefcn} in this article will give a rough 
picture of how to do this in practice.\footnote{
Sections \ref{sssec:matter-parity} and \ref{ssec:SU5-breaking} 
present explicit examples of how to implement matter parity and fluxes 
for $\SU(5)_{\rm GUT}$ symmetry breaking. 
That will be of interest for those who try to geometrically 
engineer supersymmetric Standard Models.} 
The zero mode wavefunctions determined on matter curves are 
fed into field-theory local models around the $E_6$ or $D_6$ points, and 
contributions to the low-energy Yukawa matrix from individual local models 
are calculated by overlap integration using the field theory introduced 
in \cite{BHV-1, DW-1}.

Let us begin with a summary of technical developments achieved 
in this article. 
It has been known to some extent in the literature \cite{BHV-1, 
Hayashi-1} how to feed the wavefunctions on the matter curve into 
the field-theory local models. There has still remained, however, a couple 
of theoretical issues to be clarified; i) clear and explicit enough 
distinction has not been made between descriptions of wavefunctions 
in holomorphic frame and unitary frame in the existing literature; 
ii) there has been a room to improve description of background field 
configuration and wavefunctions of fluctuations in field-theory local 
models around type $E_6$ points and type $A_6$ points, where spectral surface 
shows more complicated behavior than those considered in \cite{KV}. 
One cannot talk about physical flavor observables without facing 
these two issues. We addressed them both in section \ref{sec:DandF} and 
the appendix \ref{sec:Hitchin} in this article. 

We showed in the appendix \ref{sec:Hitchin} that branch 
loci in field-theory local models associated with Weyl group twists 
\cite{Hayashi-2} have an alternative description, where the branch locus 
is replaced by a smooth soliton solution. Thus, no singularity needs to 
be introduced in the field-theory description; no assumption\footnote{
e.g., whether something like a ``twisted sector'' needs to be introduced 
or not.} has to be made about how to deal with singularity. 
Pure field theory can provide a self-consistent description 
at the level of effective theory in 7+1 dimensions. We further used this 
reliable framework to study behavior of zero mode wavefunctions 
around the $E_6$ type points. Precise relation between the wavefunctions 
on matter curves and those in the $E_6$ field-theory local models 
is also clarified. Our results go beyond the contents of \cite{Hayashi-2}, 
and lay foundation for detailed study of up-type Yukawa matrix arising 
from a given point of $E_6$ type.

Clear distinction between holomorphic frame and unitary frame 
is introduced in section \ref{sec:DandF}. One cannot avoid this 
in establishing a precise relation between geometry for
compactifications and both the kinetic terms ($D$-term) and Yukawa 
couplings in the $F$-term in the effective theory on 3+1 dimensions. 
One could be careless about this distinction, if one is concerned 
only about the F-term Yukawa couplings. Observable masses and mixing 
angles, however, depend also on the $D$-term. Discussion even of 
{\it qualitative} pattern on flavor observables is made possible 
only after the clear distinction is made and everything is sorted out 
in section \ref{sec:DandF}. 

Kinetic mixing matrices---coefficients in the effective K\"{a}hler
potential---have not been discussed in the literature so far.\footnote{ 
An exception is \cite{BHV-2}, which has an expression for the kinetic 
mixing matrix (without justification) for complex structure without 
ramification of spectral surfaces.}
We found that they are determined by (\ref{eq:K.F.-truncation},
\ref{eq:K.F.-inn-prod}, \ref{eq:K.F.-curve}, \ref{eq:K.F.-rmfy-pt}). 
It should be noted, however, that we made assumptions---simply 
truncating Kaluza--Klein modes, and ignoring higher-order 
corrections---in deriving (\ref{eq:K.F.-truncation}, \ref{eq:K.F.-inn-prod}). 
An alternative expression for the kinetic mixing matrices is suggested 
by (\ref{eq:K.F.-curve}, \ref{eq:K.F.-rmfy-pt}), where only
zero mode wavefunctions on the matter curves are necessary. 
This alternative expression will be useful and convenient, 
but we should keep in mind that this suggestion is not firmly 
justified yet, as we discussed toward the end of section \ref{ssec:KMM-curve}.

\begin{center}
..........................................................................
\end{center}

Let us now move on to our results on the flavor structure in generic 
F-theory compactifications. The up-type and down-type Yukawa matrices 
in the low-energy effective theory consists of contributions from 
various points of $E_6$ type and $D_6$ type. Contribution from a given 
type-$E_6$ point is known to be approximately rank-1 \cite{Hayashi-2}. 
Local geometry around a type $D_6$ has a Type IIB (D7 and O7) interpretation, 
and contribution from a given type $D_6$ point is also approximately 
rank-1. Therefore, the up-type and down-type Yukawa matrices in 
the effective theory are expected generically to have rank 
${\rm min} (N_{\rm gen}, \# E_6)$ and ${\rm min} (N_{\rm gen}, \# D_6)$, 
respectively, 
$\# E_6$ and $\# D_6$ are the number of type $E_6$ points and that 
of type $D_6$ points on the complex surface of $\SU(5)_{\rm GUT}$ gauge 
fields $S$.
$\# E_6$ and $\# D_6$ are topological invariants of individual F-theory
compactifications.\footnote{References \cite{BHV-1, BHV-2} emphasized 
that three matter curves pass through type $E_6$ points generically 
in F-theory compactifications, but that is clearly against a known fact  
in Heterotic string compactifications (see e.g., \cite{Penn5};
$\bar{c}_{({\bf 10})} = \bar{c}_V$ is smooth and does not have 
a double point generically.) and duality between Heterotic string 
and F-theory. 
As explained in \cite{Hayashi-1, Hayashi-2}, only two matter curves 
pass through a type $E_6$ point generically. 
See the schematic figure \ref{fig:schem-curve}.
On the other hand, three matter curves always pass 
through type $D_6$ points, as explained in \cite{Hayashi-1, Hayashi-2}. 
This is because a Type IIB interpretation with D7-branes and an O7-plane 
is available for the local geometry around a type $D_6$ point; a
D7-brane always has its mirror image on the other side of an O7-plane, and 
this fact explains why three curves can meet at a point in a surface
without fine-tuning. See Figure 6 of \cite{Hayashi-2}. 
We thus disagree with a logic in \cite{Randall}.}
These topological invariants are calculated for some examples 
in Table 2 of \cite{Hayashi-1}.\footnote{The type $E_6$ points are 
called type (a) points in \cite{Hayashi-1}. The type $D_6$ points 
correspond to type (d) points there, and type $A_6$ points to type (c) 
points.} More examples are found in Table \ref{tab:top-data} and 
\ref{tab:top-data-B} of this article. Generically one cannot expect 
that $\# E_6 = 1$ or $\# D_6 = 1$. Thus, the up-type and down-type 
Yukawa matrices in the effective theory are {\it not} predicted to be 
approximately rank-1 generically in F-theory compactifications.  
We call it hierarchical structure problem.

Reference \cite{HV-Nov08-rev} proposed to focus on F-theory
compactifications where $\# E_6 = 1$ and $\# D_6 = 1$, so that 
the up-type and down-type Yukawa matrices are approximately 
rank-1, just like in the real world. According to \cite{HV-Nov08-rev}, 
this is a ``minimal and generic'' choice. But it is clear from 
Table \ref{tab:top-data} and \ref{tab:top-data-B} that 
$\# E_6 = \# D_6 = 1$ is not generic; none of the examples in 
the Table satisfies these conditions. Matter curves intersect 
one another not as a consequence of tuning of parameters, but 
the intersection number is determined by topology. Thus, there 
is no such notion as a ``minimal'' number of intersection. 
Focusing on models with $\# E_6 = \# D_6 = 1$ should be 
regarded as discrete tuning (choice of specific topology) 
to reproduce hierarchical Yukawa eigenvalues observed in the reality. 

We found in section \ref{sec:inv}, however, that this discrete 
tuning is not even possible. In a simple calculation (21)
% (\ref{eq:}), [refer by label] 
we showed that $\# E_6$ is always even in F-theory compactifications 
for $\SU(5)_{\rm GUT}$ unification. If $\# E_6 = 0$, the up-type 
Yukawa couplings are not generated. If $\# E_6 = 2$ or larger, then 
we cannot expect that the charm quark Yukawa eigenvalue is much 
smaller than that of top quark. 

Therefore, an extra assumption or tuning is necessary in order to 
obtain the hierarchical structure among Yukawa eigenvalues in 
effective theory of F-theory compactifications. We proposed two 
solutions this problem in this article, whose summary is given 
in the following. 

Small mixing structure in the CKM matrix is in principle 
independent from the hierarchical structure of Yukawa eigenvalues.
It is true that small mixing angles automatically follow (that is, 
predicted) from hierarchical Yukawa eigenvalues in flavor models 
in field theory on 3+1 dimensions based on a U(1) (Froggatt--Nielsen) 
flavor symmetry. But there are many other frameworks of flavor models 
where the structure of the CKM matrix does not follow immediately 
from hierarchical structure of the up-type and down-type Yukawa matrices. 
In flavor models discussed in \cite{HSW, HSW-2}, for example, the hierarchical 
eigenvalues originate from overlap of localized wavefunctions just as 
in F-theory compactifications, and the mechanism of generating 
the hierarchical eigenvalues allows a Froggatt--Nielsen interpretation. 
Structure in the mixing angles in the flavor models of \cite{HSW, HSW-2}, 
however, is different from the prediction of Froggatt--Nielsen U(1) symmetry.
The framework proposed in \cite{HV-Nov08-rev} (assuming that a geometry 
with $\# E_6 = 1$ and $\# D_6 = 1$ exists) is also one of such cases, 
as noted already in various literature (e.g., \cite{Randall, Caltech-0906}). 

We thus make it clear for the two solutions whether the structure of 
the CKM matrix is automatically predicted properly, or an extra tuning 
is necessary. Since another solution to the hierarchical structure problem 
was already pointed out in the literature \cite{Caltech-0906}, 
we will compare the solution of \cite{Caltech-0906} with the two in 
this article. 

Before we begin to describe the solutions to the hierarchical structure 
problem, we should have one remark. It is utterly pointless to talk about 
flavor structure of supersymmetric effective theory while ignoring 
the dimension-4 proton decay problem; the problem is closely associated 
with the question of what are the Higgs fields, and what are the 
right-handed neutrinos.
Questions of practical interest in the present context will be to 
discuss the flavor structure in vacua where one of solutions to 
the problem is already implemented.  
It is a separate theoretical challenge to try to derive in a more 
top-down manner why one of solutions to the dimension-4 proton decay 
problem is implemented. For now, we allow ourselves to take a
phenomenological approach; we just assume that one of the solutions 
is realized for some reason we do not know, focus on vacua where 
the solution is implemented, and move on to work out the rest of the 
phenomenological consequences such as flavor structure. 
Reference \cite{Tsuchiya} explains a couple of different ways to solve 
the dimension-4 proton decay problem. All of these solutions require 
tuning of continuous parameters, but we accept these tuning as 
a phenomenological approach. 
We will discuss in the following whether an extra tuning is required 
to reproduce the flavor structure of real world.  

\begin{center}
..........................
\end{center}

Our first solution to the hierarchical structure problem relies on 
a matter parity solution to the dimension-4 proton decay problem. 
F-theory compactifications with a matter parity \cite{Tsuchiya} will 
be arguably the most natural first try for those familiar with SUSY 
model building. There is no top-down reason for now (to our knowledge) 
to believe that a pair of Calabi--Yau 4-fold $X$ and a 4-form flux 
$G^{(4)}$ on it for F-theory compactification has a $\Z_2$ symmetry, 
although one could speculate that $\Z_2$ symmetry enhanced points may 
be somehow special in landscape of F-theory flux vacua. 
Section \ref{sssec:matter-parity} of this article conveys rough feeling 
of what it really takes to impose a $\Z_2$ symmetry that corresponds 
to matter parity. That is a combination of tuning some of continuous 
moduli parameters to zero, and discrete tuning of topological aspects 
of line bundles on the matter curve. 
We just accept this set of tuning without asking a reason, following 
the philosophy above. 

Under the $\Z_2$ symmetry, type $D_6$ points on the GUT divisor $S$ 
are either $\Z_2$ invariant, or form $\Z_2$ pairs. The baryon-number 
violating trilinear couplings $\bar{\bf 5} \; {\bf 10} \; \bar{\bf 5}$ 
vanish at the $\Z_2$ invariant type $D_6$ points, and have 
contributions opposite in sign from a $\Z_2$-pair of $D_6$ points; 
this is how the dimension-4 proton decay problem is solved. 
The $E_6$ type points on the GUT divisor $S$ are similarly classified 
into orbits of the $\Z_2$ symmetry action. 

If the type $E_6$ points form a single $\Z_2$-orbit, then the 
two $E_6$ points forming a $\Z_2$ pair give rise to exactly 
the same contribution (including sign) to the up-type Yukawa matrix 
of the effective theory, because of the symmetry. Although we know 
that $\# E_6 = 2$ or larger, if $\# E_6 = 2$ and the two $E_6$ points form 
a $\Z_2$-symmetric pair, then there is effectively only one $E_6$ 
field-theory local model. The up-type Yukawa matrix of the low-energy 
effective theory becomes approximately rank-1, for a reason just like in
\cite{Hayashi-2}. Smaller eigenvalues in the up-type Yukawa couplings 
are derived in a discussion similar to those in \cite{HV-Nov08-rev}; 
see also section \ref{ssec:F-Yukawa} of this article for a minor modification 
to the argument in \cite{HV-Nov08-rev}.
In order to reproduce hierarchical structure in the down-type Yukawa 
eigenvalues, we also need to assume that the type $D_6$ points form 
a single $\Z_2$-orbit. Predictions on the hierarchy of the down-type 
and charged lepton sector are the same as in \cite{HV-Nov08}.

The topological conditions on $\# E_6$ and $\# D_6$ set a very tight 
constraint on the choice of topology of the GUT divisor $S$ and its 
normal bundle within the base 3-fold $B_3$. We have not found an existence 
proof or proof of non-existence, but there is a room to make further 
effort to find out whether there is an example.

Small mixing structure of the CKM matrix does not follow in this 
solution to the hierarchical structure problem. The situation is 
just like in \cite{HV-Nov08-rev}. The mixing angles become small, 
when the $\Z_2$-orbit of $E_6$ points is placed close to the $\Z_2$ 
orbit of $D_6$ points. This is nothing but a continuous tuning 
of moduli parameters. 

To summarize, we need to introduce extra topological conditions 
on the $\Z_2$-orbits of $E_6$ points and $D_6$ points (in addition to 
assumptions for the existence of the $\Z_2$ symmetry itself) in order 
to avoid the rank-$N_{\rm gen}$ Yukawa matrices in the effective 
theory. An extra continuous tuning of moduli parameters is required 
to avoid fully generic CKM matrix. 
Bonus of this package of solutions to various problems (we will use a word
``framework'') is the prediction of the detailed hierarchy pattern of 
up-type, down-type and charged lepton Yukawa eigenvalues presented 
in \cite{HV-Nov08-rev} (augmented by the discussion in 
section \ref{ssec:F-Yukawa} of this article).
We will discuss later on in this section whether this bonus prediction 
fits well with the reality. 
We will come back to flavor structure in the neutrino masses and 
lepton mixing in this framework later on. 

\begin{center}
..........................
\end{center}

The second solution to the hierarchical structure problem was 
discussed in section \ref{sec:flavor}. It is based on an 
observation in section \ref{sec:unit-wavefcn}, which essentially 
dates back to \cite{HSW}, that zero mode
wavefunctions in unitary frame have localized profile {\it on} matter 
curves with {\it exponentially} suppressed tails, if some complex structure 
parameters of the curve are tuned to be large. 
See Figure~\ref{fig:Gaussian-tau2} and \ref{fig:overlap}. 
There are such parameters to be tuned, as long as 
the genus of a matter curve is not zero. 
% The tuning of this continuous moduli parameter itself has to be 
% done for the purpose of avoiding the rank-$N_{\rm gen}$ Yukawa matrices 
% in the effective theory. 
We find that we need to tune some complex structure parameters  
controlling the matter curve for $\SU(5)_{\rm GUT}$-{\bf 10} representation, 
in order to avoid the approximately rank-$N_{\rm gen}$ up-type Yukawa matrix 
in the effective theory. $N_{\rm gen}$ independent zero modes of 
quark doublets (and anti up-type quarks) have isolated and 
localized wavefunctions (see Figure~\ref{fig:overlap}). 
When the genus of $\bar{c}_{({\bf 10})}$ is 1, we know that just one 
parameter is fine.
Although individual $E_6$ field-theory local models give rise to 
approximately rank-1 contributions to the up-type Yukawa matrix 
in the effective theory, even the largest entry of the matrix from a 
given local model is often exponentially suppressed because of 
the exponential profile of the wavefunctions. Although the up-type 
Yukawa matrix in the effective theory consists of $\# E_6$ contributions 
from the $E_6$ local models, hierarchy is generated among them, 
and hence the hierarchical structure is obtained in the eigenvalues 
of the effective up-type Yukawa matrix.

Some continuous parameter can be tuned to fit the overall hierarchy 
in the up-type Yukawa eigenvalues. More detailed hierarchy pattern 
of the $N_{\rm gen}$ eigenvalues within the up-type sector will be 
predicted by an analysis similar to (but refined from) the one 
in \cite{HSW}. Although we did not present such an analysis in this 
article, the result will not be qualitatively different from the one 
in \cite{HSW}, and $N_{\rm gen}$ eigenvalues will be scattered 
in logarithmic axis, just like in the real world. It is important 
that the zero mode wavefunctions can have exponential profile, 
not just localized, in generating the hierarchy.

The exponential profile of the wavefunctions of the quark doublets 
(and anti-charged-leptons $e^c$) in the ${\bf 10}$ representation 
of $\SU(5)_{\rm GUT}$ also introduces exponential hierarchy among the 
$\# D_6$ contributions to the down-type (charged lepton) Yukawa 
matrix of the low-energy effective theory. As a result, the Yukawa 
eigenvalues in the down-type quark (charged lepton) sector also 
have a hierarchical structure. The hierarchy in these sectors 
is predicted to be weaker than that in the up-type sector, 
which is in good agreement with the real world.\footnote{The 
tuning of complex structure parameter of the matter curve 
for the ${\bf 10}$ representation fields of $\SU(5)_{\rm GUT}$ 
creates the contrast between the hierarchy and anarchy. } 

To be more precise, there are two different kinds of contributions 
for smaller Yukawa eigenvalues. 
The first kind of contribution is the mechanism discussed 
in \cite{HV-Nov08}, where derivative expansion (see also \cite{HSW}) and 
flux distortion of wavefunctions give rise to non-rank-1 contribution 
in a given local model. 
The smaller Yukawa eigenvalues in this mechanism are suppressed in 
powers\footnote{The value of 
unified gauge $\alpha_{\rm GUT}$ coupling constant is directly related 
to the ratio of the fundamental energy scale $M_*$ and the Kaluza--Klein
(GUT) scale $1/R$; $(M_* R)^4 = 1/\alpha_{\rm GUT}$. This is why the
parameter $\sqrt{\alpha_{\rm GUT}} \sim (M_* R)^{-2}$ is relevant to 
the profile of zero mode wavefunctions, and eventually sets the scale 
of hierarchical structure. Although simple power of $(M_* R)$ introduces
a scale in the wavefunction profile in the direction along the matter
curves, wavefunctions have exponential profile in the transverse
direction, and hence the parameter $\alpha_{\rm GUT}$ also enters into 
Yukawa couplings in the form of $e^{ - (M_* R)^\nu}$ for some power of 
$\nu$ (e.g., $\nu = 2, 3/2$). See \cite{Hayashi-2, Tsuchiya}. We assume 
implicitly in the main text that this exponentially suppressed 
contributions are smaller than the power suppressed ones.} 
of $\sqrt{\alpha_{\rm GUT}}$. A refined theoretical analysis 
in \cite{Tsuchiya} suggests that $\sqrt{\alpha_{\rm GUT}/\pi}$ should 
be the one. If this observation is taken at face value, then the 
predicted hierarchy in this mechanism tends to be too 
strong compared with the one in the real world \cite{Randall}. 
The prediction of \cite{HV-Nov08-rev} was that the up-type sector 
has $1: \epsilon^4: \epsilon^8$ hierarchy among the eigenvalues with 
$\epsilon = \sqrt{\alpha_{\rm GUT}} \simeq 0.2$, which results in 
$1: 1.6 \times 10^{-3} : 2.6 \times 10^{-6}$ ($\epsilon = 0.2/\sqrt{\pi}$ 
would make the hierarchy even more stronger).\footnote{The value 
of the unified coupling $\alpha_{\rm GUT} \sim 1/25$ assumes that 
there are no $\SU(5)_{\rm GUT}$-charged particles far below the 
GUT scale. In the presence of messenger sector fields in the 
gauge mediated supersymmetry breaking, this assumption is not 
satisfied any more. Larger value of $\alpha_{\rm GUT}$ would also imply 
that the Kaluza--Klein radius would not be much different from 
the ``string length''.}\raisebox{5pt}{,}\footnote{
This comment is applied also to the first solution to the hierarchical 
structure problem, and also to the solution relying upon factorization 
limit of spectral surfaces that we describe later.} 
This may be regarded as a hint that an extra contribution is needed. 
The other kind of contribution comes from multiple $E_6$ (and $D_6$) 
points with exponentially suppressed wavefunctions of the fields 
in the ${\bf 10}$ representation of $\SU(5)_{\rm GUT}$. The two different 
contributions compete in the smaller eigenvalues of Yukawa matrices. 
When the complex structure parameter of the ${\bf 10}$-representation 
curve is tuned very much, and the zero mode wavefunctions show 
extremely localized profiles on the matter curve, then the first type 
of contributions will eventually determine the hierarchical pattern, 
and the order of magnitude of the overall hierarchy. If the complex
structure parameter is tuned only minimally, the second contribution 
is at least just as important as the first one, and may even be more 
important.\footnote{It is an important observation that the value of 
$\alpha_{\rm GUT}$ sets the largest possible hierarchical structure 
in the Yukawa eigenvalues. Regardless of whether the second contribution 
dominates or not, there always exists contribution of the first kind. 
Except in the unusual case of accidental cancellation between the first 
and second kind of contributions, smaller eigenvalues cannot be smaller 
than those predicted from the first type of contributions alone. This is
one of very robust prediction on flavor physics in generic F-theory 
compactifications.}

In this solution to the hierarchical structure problem, 
we find in section \ref{sec:flavor} that the structure of 
the CKM matrix is predicted ``partially''. The CKM matrix 
in the real world is a unitary matrix with $N_{\rm gen}$ entries 
of order one and all other entries are small (called paring 
structure), and furthermore, the $N_{\rm gen}$ entries of 
order unity are aligned in the diagonal part (generation structure) 
\cite{HSW} (see also a review in section \ref{ssec:flavor-review}). 
The pairing structure is automatically predicted, though 
the generation structure is not. The generation structure is 
obtained by one chance in $1/N_{\rm gen}!$ from the pairing structure.

This second solution is available for the first two different solutions 
to the dimension-4 proton decay problem outlined in Introduction: matter 
parity, and factorization of spectral surfaces. 

In the matter parity solution to the dimension-4 proton decay problem, 
there may be a chance to derive even the generation structure, 
not just the pairing structure. The $\Z_2$ symmetry for matter parity 
acts on the $\SU(5)_{\rm GUT}$ locus $S$. If the $\Z_2$ transformation 
has a fixed point on $S$, and if the $\Z_3$ transformation is the 
case B type of Table~\ref{tb:ai} rather than case A, then 
the singularity in the transverse direction of $S$ is enhanced from 
$A_4 \simeq \SU(5)_{\rm GUT}$ to $\tilde{E}_8$. This point is like a 
$E_6$ point and $D_6$ point merged into one. The up-type and down-type 
Yukawa matrices generated in the local model around this point 
are approximately rank-1 with the up-type and down-type 
component of one and the same linear combination of the $N_{\rm gen}$ 
massless left-handed quark doublets. Thus, the generation structure 
may at least be reduced to a $1/(N_{\rm gen}-1)!$ discrete tuning problem 
from $1/N_{\rm gen}!$ discrete tuning problem, when there is one such 
$\Z_2$ fixed point. 

In the matter parity solution to the dimension-4 proton decay 
problem, Majorana masses are generated for right-handed neutrinos 
in flux compactifications of F-theory. The energy scale of the 
Majorana mass is predicted from flux compactifications, and is 
compatible with the upper bound of the Majorana mass inferred from 
the measured value of $\Delta m^2$ of atmospheric neutrino 
oscillation \cite{Tsuchiya}. The see-saw mechanism generates 
very small masses of left-handed neutrinos. Since right-handed neutrinos 
are identified with complex structure moduli of Calabi--Yau 4-fold $X$, 
there are generically enormous number of them \cite{Tsuchiya}. 
Thus, three eigenvalues of left-handed neutrinos are predicted not 
to have a large hierarchy, because of the same reason as in \cite{HSW-2}.
This is in nice agreement with the fact that the ratio 
$\sqrt{|\Delta m^2_{\odot}|}/ \sqrt{|\Delta m^2_\oplus|} \sim 0.2$ 
is not smaller than typical hierarchy among charged fermions. 
The anarchy (large mixing angles) is predicted\footnote{One might 
be interested in considering a limit where the matter curve (to 
be more precise, the spectral surface) for the $\bar{\bf 5}$ 
representation of $\SU(5)_{\rm GUT}$ is factorized, in the context 
of doublet--triplet splitting problem (massless $L$--$H_u$ and 
massless $H_d$--$H_u$), or dimension-5 proton decay problem. 
Although we do not find factorization for these purposes strictly 
necessary, even in such a factorization limit, the anarchy remains 
the prediction. When the matter curve for the $\bar{\bf 5}$ representation 
is factorized into two pieces, the intersection number of the two pieces 
is usually much larger than one. } for the lepton 
mixing matrix without an extra tuning of parameters, which is also 
what we see in the real world. 

\begin{center}
..........................
\end{center}

It must be fair to mention that one solution to the hierarchical 
structure problem has been proposed in the literature \cite{Caltech-0906}.
It relies on the factorized spectral surface solution 
\cite{TW-1, Tsuchiya, Caltech-0906} to the dimension-4 proton decay 
problem. First, a limit of complex structure (\ref{eq:E8-limit}) 
is taken (for existence of globally defined $E_8$ bundle), and then 
the coefficients of (\ref{eq:spec-surf}) 
are tuned so that the left-hand side of (\ref{eq:spec-surf}) is factorized. 
We have so far no idea how this limit could follow from top-down 
principle, though one could argue that this limit may be special 
in landscape of F-theory flux vacua, because a U(1) symmetry is 
enhanced at this limit. We will take a phenomenological approach 
and accept this tuning, because it is pointless to consider flavor 
physics in vacua with fast dimension-4 proton decay. 

The dimension-4 proton decay operators are absent because of 
an unbroken U(1) symmetry in this solution. This U(1) symmetry 
also forbids the Majorana masses of right-handed neutrinos, 
at least when the $\SU(5)_{\rm GUT}$ symmetry is broken by 
a Wilson line in $\pi_1(S) \neq \{ 1\}$, or by a hypercharge flux 
on $S$ while keeping the vector field massless in the mechanism of 
\cite{Buican, BHV-2}. 
Once this U(1) symmetry is broken spontaneously carelessly, 
then this is not a solution to the proton decay problem any more. 
Taking a $\Z_2$ quotient of this set-up also ruins this solution, 
because the spectral surface is no longer factorized in the 
quotient theory. We have thus nothing more to comment on 
with respect to neutrino masses and mixing angles in the lepton 
sector in this context. 
 
The heart of the idea of the solution to the hierarchical structure 
problem in \cite{Caltech-0906} is to consider a factorization 
limit of spectra surface that has a following property. When the 
spectral surface is factorized, the matter curve $\bar{c}_{({\bf 10})}$ 
for ${\bf 10}(+\overline{\bf 10})$ representation fields and 
$\bar{c}_{(\bar{\bf 5})}$ for $\bar{\bf 5}+{\bf 5}$ fields are also 
factorized into $\bar{c}_{({\bf 10})} = \sum_i \bar{c}_{({\bf 10})i}$, and 
$\bar{c}_{(\bar{\bf 5})} = \sum_j \bar{c}_{(\bar{\bf 5})j}$. Only one of 
$\bar{c}_{({\bf 10})i}$'s is identified with the matter curve of 
``our {\bf 10}'s''. Similarly, the matter curves for the Higgs fields 
and the matter $\bar{\bf 5}=(\bar{D},L)$ are identified with 
one of $\bar{c}_{(\bar{\bf 5})j}$'s separately. The topological intersection 
number $\# E_6$ and $\# D_6$ of ``our components'' of the matter curves 
need to be both one. That is the heart of the idea of \cite{Caltech-0906}, 
as we understand it. At least a discrete tuning is necessary beyond 
the factorization limit that is necessary purely for the solution 
to the dimension-4 proton decay problem.

Consequences in the hierarchical pattern and mixing angles are the 
same as in the first solution we discussed above. 

\begin{center}
..........................
\end{center}

It is true that the factorized spectral surface solution to the 
dimension-4 proton decay problem is totally screwed up when the U(1) 
symmetry is broken carelessly. There is a way to break it safely, however, 
so that the Majorana mass of right-handed neutrinos, while the 
dimension-4 proton decay operators are completely absent. 
That is the spontaneous R-parity violation solution in \cite{TW-1, KNW} 
and section 4.4 of \cite{Tsuchiya}. Flavor structure in vacua 
with this solution is also worth investigation separately. 
For that purpose, however, totally different approach is necessary, 
and we have nothing to comment on in this article. 

%%%%%%%%%%%%%%%%%%%%%%%%%%%%%%%%%%%%%%%%%%%%%%%%%%%%%%%%%

%%%%%%%%%%%%%%%%%%%%%%%%%%%%%%%%%%%%%%%%%%%%%%%
% \section*{Acknowledgements} % preferred spelling in BrE
  \section*{Acknowledgements}  % more usual in AmE

We would like to thank Alexey Bondal, Kentaro Hori, Kyoji Saito, and 
Martijn Wijnholt for useful comments and communications. 
TK would like to thank the organizers of Summer Institute 2009 at Fuji-Yoshida 
for hospitality, and TW thanks Aspen Center for Physics for hospitality, 
where TK and TW stayed during the last stage of this project.
This work was supported in part 
by JSPS Research Fellowships for Young Scientists (HH), 
by a Grant-in-Aid \#19540268 from the MEXT of Japan (TK), 
by Global COE Program ``the Physical Sciences Frontier'', MEXT, Japan (YT), 
and by WPI Initiative, MEXT, Japan (TW).

%%%%%%%%%%%%%%%%%%%%%%%%%%%%%%%%%%%%%%%%%%%%%%%%%%%

%%%%%%%%%%%%%%%%%%%%%%%%%%%%%%%%%%%%%%%%
\appendix

%%%%%%%%%%%%%%%%%%%%%%%%%%%%%%%%%%%%%%%%%%%%%%%%
\section{A Brief Note on Heterotic--F Duality Map}
\label{sec:Het-F}
%%%%%%%%%%%%%%%%%%%%%%%%%%%%%%%%%%%%%%%%%%%%%%%%

Duality map of moduli parameters of Heterotic and F-theory 
compactifications has often been described only in 
a specific limit. When the volume of the elliptic fiber 
is large compared with $\alpha'$ on the Heterotic side, 
the elliptic fibered K3 manifold in the dual F-theory description is 
in the stable degeneration limit. In this limit, the geometry of 
``half of the K3 manifold'' can be described by an elliptic 
rational surface $dP_9$. Moduli of Wilson lines in the 
elliptic fiber in one of the two $E_8$'s of Heterotic 
compactifications are mapped to moduli of complex structures
of the rational elliptic surface in F-theory compactifications. 
In this appendix, we write down a moduli map without taking a stable 
degeneration limit, so that both visible and hidden 
sector moduli (in the Heterotic language) are mapped into 
complex structure moduli of an elliptic fibered K3 manifold. 

Wilson lines in $T^2$ compactifications of Heterotic string 
are described by a spectral cover. One should note, however, 
that this picture relies on the supergravity approximation, 
where the volume of $T^2$ is larger than $\alpha'$. It is only 
when an effective field theory picture holds below the string 
scale that the notion of ``Wilson lines'' of the $E_8 \times E_8$ 
gauge theory is well-defined. More generally, in the stringy regime 
of Heterotic string compactifications on $T^2$,  
the classical distinctions among the ``Wilson lines'' moduli, 
the complex structure and K\"{a}hler moduli of $T^2$ 
cease to be clear in the whole moduli space 
$\Gamma \backslash \SO(2,18) / \SO(2) \times \SO(18)$. 
It is also possible, however, to extend the notion of ``Wilson lines'' 
into stringy regime of the moduli space of the Heterotic string 
compactifications, by mapping the spectral surface moduli once into 
F-theory moduli, and extending its definition into a region that is not 
in the stable degeneration limit. That is what we do in the following.

Suppose that the spectral cover for the visible and 
hidden $E_8$ are given, respectively, by  
\begin{eqnarray}
s_{vis} & = & a_5 xy + a_4 x^2 + a_3 y + a_2 x + a_0 = 0, 
\label{eq:spectral-cover-vis}\\
s_{hid} & = & a''_5 xy + a''_4 x^2 + a''_3 y + a''_2 x + a''_0 = 0, 
\label{eq:spectral-cover-hid}
\end{eqnarray}
where we assume that the respective structure groups of the visible 
and hidden sector vector bundles are at most $\SU(5) \subset E_8$, 
for simplicity.
The spectral surfaces (\ref{eq:spectral-cover-vis},\ref{eq:spectral-cover-hid})
are divisors on a Calabi--Yau 3-fold $Z$, and $Z$ is an elliptic fibration on 
a complex surface $S$ with its fibration given by a Weierstrass model 
\begin{equation}
 y^2 = x^3 + a'_2 x + a'_0.
\end{equation} 
All of $a_{5,4,3,2,0}$, $a''_{5,4,3,2,0}$ and $a'_{2,0}$ are 
sections of line bundles on $S$:
\begin{eqnarray}
a_{r} \in \Gamma \left(S; {\cal O}(r K_S + \eta_{\rm vis})\right), & \quad &
a''_{r} \in \Gamma \left(S; {\cal O}(r K_S + \eta_{\rm hid})\right), \\
a'_2 \in \Gamma \left( S; {\cal O}(-4 K_S) \right), & \quad & 
a'_0 \in \Gamma \left( S; {\cal O}(-6 K_S) \right).
\end{eqnarray}
In order to satisfy a consistency condition for Heterotic string 
compactifications, the divisors $\eta_{\rm vis}$ and $\eta_{\rm hid}$ 
on $S$ need to satisfy\footnote{This constraint comes from the Bianchi 
identity of 
the $B$-field in Heterotic string compactifications. When M5-branes 
are wrapped on curves in $S$ (the zero section of the elliptic fibration
$Z$ is identified with the base space $S$), there is an extra
contribution to this constraint equation. We do not try to generalize
the moduli map in such cases in this article. } 
$\eta_{\rm vis} + \eta_{\rm hid} = -12 K_S$.

The dual Calabi--Yau 4-fold $X$ on the F-theory side is given by 
\begin{eqnarray}
y^2 & = &   
       (a_5 xy + a_4 z x^2 + a_3 z^2 y + a_2 z^3 x + a_0 z^5) \nonumber \\
&&+  (x^3 + a'_2 z^4 x + a'_0 z^6) \label{eq:K3-fib} \\
&&+  z^2 (a''_5 xy + a''_4 z x^2 + a''_3 z^2 y + a''_2 z^3 x + a''_0 z^5).
  \nonumber 
\end{eqnarray}
By using the same sections ($a_{5,4,3,2,0}$, $a'_{2,0}$ and 
$a''_{5,4,3,2,0}$) on $S$, moduli parameters of the Heterotic string 
compactification are mapped to those of the F-theory compactification. 
$X$ is a K3 fibration over $S$, and the K3 fiber itself is an elliptic 
fibration over $\P^1$. $(x,y)$ are the coordinates of the elliptic 
fiber, and $z$ an inhomogeneous coordinate of the base $\P^1$.
 
The way that the data $a''_{5,4,3,2,0}$ of the spectral cover 
(\ref{eq:spectral-cover-hid}) in the hidden sector are used 
in (\ref{eq:K3-fib}) on the F-theory side can be justified as follows. 
The $\P^1$ base of the elliptic fibered K3  
is covered by two open patches, one using the coordinate $z$ above, 
and the other using $\zeta = 1/z$. 
% The coordinates $(x,y)$ of the elliptic fiber 
% on the Heterotic side are sections of line bundles, 
% and $(x, y)$ in (\ref{eq:K3-fib}) are regarded 
% as the ones of the elliptic fiber used in the patch of $\P^1$ 
% with the $z$-coordinate. 
The coordinates of the elliptic fiber are sections 
of line bundles ${\cal O}(4) = {\cal O}(-2 K_{\P^1})$ and 
${\cal O}(6) = {\cal O}(-3 K_{\P^1})$. 
$(x, y)$ in (\ref{eq:K3-fib}) are the coefficient functions in the 
trivialization patch of $\P^1$ with the $z$-coordinate. 
Let us denote the corresponding coordinate in the $\zeta$-patch  
as $(\tilde{x},\tilde{y}$). Then 
% Since $(x,y)$ and $(\tilde{x},\tilde{y})$
% are sections of ${\cal O}(4)\oplus{\cal O}(6)$, 
% where ${\cal O}(4)$ and ${\cal O}(6)$ are 
% ${\cal O}(-2K_{\P^1})$ and ${\cal O}(-3K_{\P^1})$, respectively, 
%
\begin{equation}
 \tilde{x} = \zeta^4 x, \qquad \tilde{y} = \zeta^6 y.
\end{equation}
Since the defining equation (\ref{eq:K3-fib}) is a section of 
${\cal O}(-6K_{\P^1}) = {\cal O}(12)$, the defining equation 
in the $\zeta$ patch becomes 
\begin{eqnarray}
\tilde{y}^2 & = &   
       (a''_5 \tilde{x}\tilde{y} + a''_4 \zeta \tilde{x}^2 
      + a''_3 \zeta^2 \tilde{y} + a''_2 \zeta^3 \tilde{x} + a''_0 \zeta^5) 
  \nonumber \\
    & + & (\tilde{x}^3 + a'_2 \zeta^4 \tilde{x} + a'_0 \zeta^6) 
  \label{eq:K3-fib-2} \\
    & +  & \zeta^2 (a_5 \tilde{x}\tilde{y} + a_4 \zeta \tilde{x}^2 
     + a_3 \zeta^2 \tilde{y} + a_2 \zeta^3 \tilde{x} + a_0 \zeta^5).
  \nonumber 
\end{eqnarray}
One can see that the description in the $z$-patch and in the 
$\zeta$-patch on the F-theory side treat the visible and hidden 
sectors of the Heterotic $E_8 \times E_8$ string on the same footing. 

The defining equation (\ref{eq:K3-fib}) can be regarded as a 
special case of (\ref{eq:def-eq}). The defining equation 
(\ref{eq:def-eq}) is meant to describe a local area around $S$, 
and can be used for any F-theory compactifications. 
$z$ is a normal coordinate of $S$ in $B_3$, and the coefficients 
of the monomials in $x$ and $y$ in (\ref{eq:def-eq}) 
are expressed in the $z$-series expansion. 
In an F-theory compactification with its Heterotic dual, however, 
there is a well-motivated choice of the normal coordinate $z$, 
all the $z$ series expansions of the coefficients stop at 
the next-to-next-to-leading order of $z$, and the geometry of 
the entire $X$ except the $\zeta = 0$ locus is given by the 
equation (\ref{eq:K3-fib}). 

Comparing (\ref{eq:K3-fib}) and (\ref{eq:def-eq}), we
further notice that $a'_{5,4,3} = 0$ in an F-theory 
compactification with its Heterotic dual.  When $a'_{5,4,3} = 0$, 
all the correction terms of $R^{(5)}$---those in the second line 
of (\ref{eq:R5})---vanish in such F-theory compactifications.
It is not that those corrections vanish only in the stable degeneration limit, 
but they vanish everywhere in the intersection of 
the moduli spaces of an F-theory compactification and of its Heterotic dual. 

Let us rewrite the defining equation (\ref{eq:K3-fib}) as 
the Weierstrass form
\begin{equation}
y^2 = x^3 + f x + g, \qquad f = \sum_{i=-4}^4 f_i z^{4-i}, \quad 
g = \sum_{i=-6}^{6} g_i z^{6-i}, 
\end{equation}
where 
\begin{eqnarray}
f_i & \in & 
    \Gamma \left( S; {\cal O}(-4 K_S + i (6K_S + \eta_{\rm vis}))\right)
  = \Gamma \left( S; {\cal O}(-4 K_S - i (6K_S + \eta_{\rm hid}))\right), \\
g_i & \in & 
    \Gamma \left( S; {\cal O}(-6 K_S + i (6K_S + \eta_{\rm vis}))\right)
  = \Gamma \left( S; {\cal O}(-4 K_S - i (6K_S + \eta_{\rm hid}))\right). 
\end{eqnarray}
The polynomials $f_i$'s and $g_i$'s are given as follows:
\begin{eqnarray}
f_4 & = & - \frac{1}{48} a_5^4, \\
f_3 & = & - \frac{1}{6} a_4 a_5^2, \\ 
f_2 & = & - \frac{1}{3} a_4^2 + \frac{1}{2} a_3 a_5 
          - \frac{1}{12} a_5^3 a''_5, \\
f_1 & = & a_2  - \frac{1}{6} a''_4 a_5^2 - \frac{1}{3} a_4 a_5 a''_5, 
\end{eqnarray}
and 
\begin{eqnarray}
g_6 & = & \frac{1}{864} a_5^6, \\
g_5 & = & \frac{1}{72} a_4 a_5^4, \\
g_4 & = & \frac{1}{18} a_4^2 a_5^2 - \frac{1}{24}a_3 a_5^3
        + \frac{1}{144}a_5^5 a''_5, \\
g_3 & = & \frac{2}{27} a_4^3 - \frac{1}{6}a_3 a_4 a_5 
        - \frac{1}{12}a_2 a_5^2 + \frac{1}{72} a''_4 a_5^4  
        + \frac{1}{18} a_4 a_5^3 a''_5, \\
g_2 & = & \frac{a_3^2}{4} - \frac{a_2 a_4}{3} - \frac{a'_2 a_5^2}{12} 
        + \frac{a_4 a''_4 a_5^2}{9}  - \frac{a''_3 a_5^3}{24} 
        + \frac{a_4^2 a_5 a''_5}{9} - \frac{ a_3 a_5^2 a''_5}{8} 
 + \frac{5 a_5^4 (a''_5)^2}{288}, \\
g_1 & = & a_0  - \frac{a'_2 a_4}{3} 
       + \frac{2 a_4^2 a''_4}{9} 
       - \frac{a''_3 a_4 a_5}{6} 
       - \frac{a_3 a''_4 a_5}{6} \nonumber \\
& & - \frac{a''_2 a_5^2}{12} - \frac{a_3 a_4 a''_5}{6} 
    - \frac{a_2 a_5 a''_5}{6} + \frac{a''_4 a_5^3 a''_5}{18} 
    + \frac{a_4 a_5^2 (a''_5)^2}{12}, 
\end{eqnarray}
and 
\begin{eqnarray}
f_0 & = & a'_2 - \frac{2 a_4 a''_4}{3}
        + \frac{a''_3 a_5 + a_3 a''_5}{2} - \frac{a_5^2 (a''_5)^2}{8}, \\
g_0 & = & a'_0 
        + \frac{a_3 a''_3}{2} 
        - \frac{a''_2 a_4 + a_2 a''_4}{3} 
        + \frac{(a''_4)^2 a_5^2 + a_4^2 (a''_5)^2}{18} 
        - \frac{a'_2 a_5 a''_5}{6}  \nonumber \\
 & & + \frac{2 a_4 a''_4 a_5 a''_5}{9}
     - \frac{a''_3 a''_5 a_5^2  + a_3 a_5 (a''_5)^2 }{8}
     + \frac{5 a_5^3 (a''_5)^3}{216}.
\end{eqnarray}
The polynomials $f_{i}$ ($i=-1,\cdots,-4$) and $g_{i}$ ($i=-1,\cdots,-6$) 
are also expressed in the same way as $f_{i}$ ($i=1,\cdots,4$) and 
$g_{i}$ ($i=1,\cdots,6$), respectively, by exchanging $a_r$ and $a''_r$ 
for each $r$.

These expressions are almost the same as (240--251) in \cite{Hayashi-1}, 
but now the hidden sector moduli parameters $a''_r$ are included. 
When the defining equation of a Calabi--Yau 4-fold $X$ is written in 
a Weierstrass form, our traditional understanding was that $f_0$ and 
$g_0$ on the F-theory side correspond to $a'_{2,0}$ on the Heterotic
side, $f_i$ and $g_i$ with $i > 0$ to the vector bundle moduli $a_r$ 
of the visible sector, and $f_i$ and $g_i$ with $i < 0$ to those $a''_r$ 
of the hidden sector. This statement largely remains true, but 
more precisely, more complicated mixings between the three sectors 
$a$'s, $a'$'s, and $a''$'s is observed above. 

Despite all these mixings, however, the location of an  
$A_6$-type codimension-3 singularity points  
in the visible sector does not receive corrections in any range of 
the complex structure moduli, even in a region not in the stable 
degeneration limit, as we have already mentioned above. 
On the Heterotic string side, the notion of a spectral surface is 
based purely on the supergravity approximation. 
It is only in the region of the moduli space where the volume of 
the elliptic fiber of $Z$ is parametrically larger than $\alpha'$ that 
the $E_8 \times E_8$ gauge theory can be studied separately from 
supergravity in 10-dimensions, or from stringy corrections. Thus, 
by writing down the defining equation of the dual geometry in the 
F-theory compactification, which can be applied for any region of 
the complex structure moduli space, we have effectively 
extended the definition of $a_r$ and $a''_r$ on the Heterotic side 
into a region where the supergravity approximation 
in Heterotic string theory is not valid. When the definition of 
``Wilson lines'' is extended in that way, the locus of an $A_6$ 
singularity is expressed in terms of these ``Wilson lines'' without 
corrections. 

%%%%%%%%%%%%%%%%%%%%%%%%%%%%%%%%%%%%%%%%%%%%%%%%%%%%%%%%%%
\section{Topological Invariants in Case of  Factorized Spectral Surface}
\label{sec:4+1}
%%%%%%%%%%%%%%%%%%%%%%%%%%%%%%%%%%%%%%%%%%%%%%%%%%%%%%%%%%

Suppose that the $E_8$ limit (\ref{eq:E8-limit}) is taken, and 
there is a 5-fold cover defined globally on the GUT divisor $S$. 
\begin{equation}
 a_0 \xi^5 + a_2 \xi^3 + a_3 \xi^2 + a_4 \xi + a_5 = 0.
\end{equation}
$a_r$'s are global holomorphic sections of line bundles 
${\cal O}(r K_S + \eta)$ for some divisor $\eta$ on $S$.
This 5-fold cover spectral surface $C_{({\bf 10})}$ splits 
into two irreducible pieces for certain limit of $a_r$'s. 

The 4+1 factorization in its most generic form \cite{Caltech-0906} 
is the limit where 
\begin{eqnarray}
& & a_0 = p_0 q_0, \quad 
 a_1 = p_1 q_0 + p_0 q_1 \equiv 0, \quad  a_5 = p_1 q_4, \\ 
& & a_2 = p_1 q_1 + p_0 q_2, \quad 
 a_3 = p_1 q_2 + p_0 q_3, \quad 
 a_4 = p_1 q_3 + p_0 q_4.
\end{eqnarray}
Here, $p_i \in \Gamma(S; {\cal O}(i K_S + \eta'))$, and 
$q_i \in \Gamma(S; {\cal O}(i K_S + \eta''))$, with divisors 
on $S$ satisfying $\eta' + \eta'' = \eta$. In this limit, 
the defining equation of the spectral surface factorizes as 
\begin{equation}
 (p_0 \xi + p_1) (q_0 \xi^4 + q_1 \xi^3 + q_2 \xi^2 + q_3 \xi + q_4) = 0.
\end{equation}
Because the spectral surface is a divisor of a non-compact space 
$\mathbb{K}_S$ in F-theory, not of a compact elliptic fibered space 
in Heterotic string compactification, the defining equation does not 
have to be written in terms of the coordinate of the elliptic fiber 
$(x, y)$. This is why the term $q_1$ is allowed.\footnote{The 4+1 
factorization in \cite{TW-1, KNW, Tsuchiya} corresponds to 
$q_1 = p_1 = 0$.} The spectral surface, 
and hence the corresponding Higgs field vev can take its value in 
S[$\U(4) \times \U(1)$]; the rank-4 part and rank-1 part do not have 
to be traceless separately. 

In this most generic 4+1 factorization limit, the matter curve 
$\bar{c}_{({\bf 10})}$ splits into two irreducible pieces, 
$\bar{c}_{({\bf 10})1} + {\bar{c}_{({\bf 10})4}}$,  
\begin{equation}
 \bar{c}_{({\bf 10})1} = \{ p_1 = 0 \} \in | K_S + \eta' |, \qquad 
 \bar{c}_{({\bf 10})4} = \{ q_4 = 0 \} \in | 4K_S + \eta''|.
\end{equation}
In case $\eta' = - K_S$, for example, the curve $\bar{c}_{({\bf 10})1}$ 
is void, and $\bar{c}_{({\bf 10})4} \in |5K_S + \eta|$.

The matter curve for the ${\bf 5}$ representation fields 
$\bar{c}_{({\bf 5})}$, given by 
$P^{(5)} = a_0 a_5^2 - a_2 a_5 a_3 + a_4 a_3^2 = 0$, also becomes
reducible. 
\begin{equation}
 P^{(5)} = (q_2 p_1^2 + q_3 p_1 p_0 + q_4 p_0^2) 
           (- q_4 q_1 p_1 + q_3 q_2 p_1 + q_3^2 p_0).
\end{equation}
Thus, the matter curve $\bar{c}_{({\bf 5})}$ consists of two pieces, 
$\bar{c}_{({\bf 5})4} + \bar{c}_{({\bf 5})6}$, 
\begin{eqnarray}
 \bar{c}_{({\bf 5})4} & = & \{ P^{(5)}_4 \equiv 
   (q_2 p_1^2 + q_3 p_1 p_0 + q_4 p_0^2)=0 \} 
   \in |4K_S + 2 \eta' + \eta'' |, \\
 \bar{c}_{({\bf 5})6} & = & \{ P^{(5)}_6 \equiv 
   (- q_4 q_1 p_1 + q_3 q_2 p_1 + q_3^2 p_0)=0 \}
   \in |6K_S + \eta' + 2\eta'' |.
\end{eqnarray}

There are multiple kinds of codimension-3 singularity points in this 
4+1 factorization limit. There are five types of such points on the 
matter curves $\bar{c}_{({\bf 10})1}+ \bar{c}_{({\bf 10})4}$:
\begin{itemize}
 \item [type 1] $p_1 = q_4 = 0$. Three curves $\bar{c}_{({\bf 10})1}$, 
       $\bar{c}_{({\bf 10})4}$ and $\bar{c}_{({\bf 5})4}$ pass through this 
       type 1 points. $a_5 = a_4 = 0$. The singularity is enhanced to $E_6$.
 \item [type 2] $p_1 = p_0 = 0$. The curves $\bar{c}_{({\bf 10})1}$ and 
       $\bar{c}_{({\bf 5})6}$ pass through the type 2 points, while 
       $\bar{c}_{({\bf 5})4}$ forms a double point there. $a_{0,2,3,4,5}
       = 0$. The singularity is enhanced to $\tilde{E}_8$.
 \item [type 3] $p_1 = q_3 = 0$. The curve $\bar{c}_{({\bf 10})1}$ 
       pass through the type 3 points, while $\bar{c}_{({\bf 5})6}$ 
       form a double point at each type 3 point. This is because the
       defining equation of $\bar{c}_{({\bf 5})6}$ is quadratic in 
       $q_3$ and $(p_1 \sim q_1)$. $a_5 = a_3 = 0$. The singularity 
       is enhanced to $D_6$.
 \item [type 4] $q_4 = (q_2 p_1 + q_3 p_0) = 0$. Type 4 points are zero 
       of order one of $\bar{c}_{({\bf 10})4}$, $\bar{c}_{(\bf 5)4}$ and 
       $\bar{c}_{({\bf 5})6}$. $a_5 = a_3 = 0$. The singularity is
       enhanced to $D_6$.
 \item [type 5] $q_4 = q_3 = 0$. The two curves $\bar{c}_{({\bf 10})4}$ 
       and $\bar{c}_{({\bf 5})4}$ intersect transversely at the type 5
       points. $a_5 = a_4 = 0$. The singularity is enhanced to $E_6$.
\end{itemize}
What used to be ``type $E_6$ points'' $a_5 = a_4 = 0$ for generic choice 
of $a_{0,2,3,4,5}$ are now distributed as 
\begin{eqnarray}
 \# E_6 & = & (5K_S + \eta) \cdot (4K_S + \eta)  \nonumber \\
&  & % \!\!\!\!\!\!\!\!\!\!\!\!\!\!\!\!\!\!\!\!\!\!\!\!\!\!\!\!
   \rightarrow 2 (K_S + \eta') \cdot (4K_S + \eta'')
 + (3K_S + \eta'') \cdot (4K_S + \eta'')
 + (K_S + \eta') \cdot \eta' \nonumber \\
& & =  2 \# 1 + \# 5 + \# 2,
\end{eqnarray}
and what use to be ``type $D_6$ points'' $a_5 = a_3 = 0$ are now 
distributed as 
\begin{eqnarray}
 \# D_6 & = & (5K_S + \eta) \cdot (3K_S + \eta) \nonumber \\
  & & \rightarrow 
      (K_S + \eta') \cdot \eta'
    + (K_S + \eta') \cdot (3K_S + \eta'')
    + (4K_S + \eta'') \cdot (3K_S + \eta' + \eta'') \nonumber \\
  & & = \# 2 + \# 3 + \# 4.
\end{eqnarray}
It will not be difficult to understand the multiplicity for each type 
of points from local behavior of $a_{3,4,5}$ around the codimension-3 
singularity points of type 1--5.

There are three other type of codimension-3 singularity points 
that are not on the matter curve 
$\bar{c}_{({\bf 10})1} + \bar{c}_{({\bf 10})4}$ in the 4+1 
factorization limit. They are all on the matter curve,
$\bar{c}_{({\bf 5})4}+\bar{c}_{({\bf 5})6}$.
To see the nature of these singularity points, note that the 
defining equation of the 10-fold spectral surface of the 
$\bar{\bf 5}$ representation factorizes into 
\begin{eqnarray}
C_{(\bar{\bf 5})4}: & & q_0 p_0^2 \xi^4 
+ (q_1 p_0^2 + 4 q_0 p_1 p_0) \xi^3
+ (q_2 p_0^2 + 3 q_1 p_1 p_0 + 6 q_0 p_1^2) \xi^2
% + (q_3 p_0^2 + 2 q_2 p_1 p_0 - q_1 p_1^2) \xi
+ R^{(5)}_4 \xi 
% + (q_4 p_0^2 + q_3 p_1 p_0 + q_2 p_1^2)
+ P^{(5)}_4 = 0, \label{eq:41-spec-surf-5bar-41}\\ 
C_{(\bar{\bf 5})6}: & & q_0^2 p_0 \xi^6 
+ 3 q_1 q_0 p_0 \xi^5 
+ (3 q_1^2 + 2 q_2 q_0) p_0 \xi^4
+ (-q_1^2 p_1 + 4 q_1 q_2 p_0) \xi^3 \nonumber \\
& & \quad + (- 4 q_4 q_0 p_0 - 2 q_2 q_1 p_1 + q_2^2 p_0 + q_3 q_1 p_0) \xi^2
+ R^{(5)}_6 \xi + P^{(5)}_6 = 0, 
\label{eq:41-spec-surf-5bar-44}
\end{eqnarray}
where 
\begin{eqnarray}
 R^{(5)}_4 & = & (q_3 p_0^2 + 2 q_2 p_1 p_0 - q_1 p_1^2), \\
 R^{(5)}_6 & = & - (4 q_4 q_1 p_0 + q_2^2 p_1 + q_3 q_1 p_1). 
\end{eqnarray}
All the terms in (\ref{eq:41-spec-surf-5bar-41}) are sections of 
${\cal O}(4K_S + 2\eta' + \eta'')$, and those 
in (\ref{eq:41-spec-surf-5bar-44}) are sections of 
${\cal O}(6K_S + \eta' + 2 \eta'')$. 
$R^{(5)}_4 \in \Gamma(S; {\cal O}(3K_S + 2\eta' + \eta''))$ and 
$R^{(5)}_6 \in \Gamma(S; {\cal O}(5K_S + \eta' + 2 \eta''))$ indicates 
where the spectral surfaces $C_{(\bar{\bf 5})4}$ and 
$C_{(\bar{\bf 5})6}$ are ramified over $S$. One can see through 
explicit computation that 
\begin{equation}
 \tilde{R}^{(5)}_{\rm mdfd} = P^{(5)}_4 R^{(5)}_6 + R^{(5)}_4 P^{(5)}_6.
\end{equation}
Both sides of this equation are sections of ${\cal O}(9K_S + 3\eta)$.
After preparing the language above, the three different types of 
codimension-3 singularities off the ${\bf 10}$-representation matter
curve are characterized as follows:
\begin{itemize}
 \item [type 6] points where $\bar{c}_{({\bf 5})4}$ and $\bar{c}_{({\bf
       5})6}$ intersect transversely, but not the type 2 points or type
       4 points on $\bar{c}_{({\bf 10})1} + \bar{c}_{({\bf 10})4}$.
 \item [type $A_6(4)$] zero points of $R^{(5)}_4$ on $\bar{c}_{({\bf 5})4}$ 
       that are not any one of type 1--6 points. 
 \item [type $A_6(6)$] zero points of $R^{(5)}_6$ on $\bar{c}_{({\bf 5})6}$ 
       that are not any one of type 1--6 points.
\end{itemize}

The number of these type 6, $A_6(4)$ and $A_6(6)$ points are worked out 
as follows. To begin with let us remind ourselves that there is a
relation 
\begin{equation}
 {\rm deg} K_{\bar{c}_{({\bf 5})}} = 2 K_S \cdot \bar{c}_{({\bf 5})} +
  {\rm deg} \tilde{R}^{(5)}_{\rm mdfd}
\label{eq:temp-P}
\end{equation}
for generic complex structure $a_{0,2,3,4,5}$. 
This identity is easily verified because 
\begin{eqnarray}
 {\rm deg} K_{\bar{c}_{({\bf 5})}} & = &
    (11K_S + 3\eta) \cdot (10 K_S + 3 \eta), \\
 {\rm deg} \tilde{R}^{(5)}_{\rm mdfd} & = & 
    (9 K_S + 3\eta) \cdot (10 K_S + 3 \eta).
\end{eqnarray}
At the 4+1 factorization limit, the matter curve $\bar{c}_{({\bf 5})}$ 
becomes reducible, and ${\rm deg} K_{\bar{c}_{({\bf 5})}}$ becomes 
\begin{equation}
 {\rm deg} K_{\bar{c}_{({\bf 5})}}  \rightarrow  
   {\rm deg} K_{\bar{c}_{({\bf 5})4}}
 + {\rm deg} K_{\bar{c}_{({\bf 5})6}}
 + 2 \bar{c}_{({\bf 5})4} \cdot \bar{c}_{({\bf 5})6}, 
 \label{eq:temp-U}
\end{equation}
which is easily seen by splitting the topological class of 
$\bar{c}_{({\bf 5})}$, $(10K_S + 3\eta)$, into 
$(4K_S + 2\eta' + \eta'')$ and $(6K_S + \eta' + 2 \eta'')$.
All the three contributions in the right-hand side are given 
by intersection numbers:
\begin{eqnarray}
 {\rm deg} K_{\bar{c}_{({\bf 5})4}} & = &
  (5 K_S + 2\eta' + \eta'') \cdot (4K_S + 2 \eta' + \eta''), \\
 {\rm deg} K_{\bar{c}_{({\bf 5})6}} & = & 
  (7 K_S + \eta' + 2 \eta'') \cdot (6K_S + \eta' + 2 \eta'')  \\
  \bar{c}_{({\bf 5})4} \cdot \bar{c}_{({\bf 5})6} & = & 
   (4K_S + 2 \eta' + \eta'') \cdot (6K_S + \eta' + 2 \eta'').  
\end{eqnarray}
In the meanwhile, we can study local behavior of $R^{(5)}_4$ and 
$R^{(5)}_6$ around various types of codimension-3 singularities, and 
find that 
\begin{eqnarray}
 {\rm deg} K_{\bar{c}_{({\bf 5})4}} & = & 
  2 K_S \cdot \bar{c}_{({\bf 5})4} + {\rm deg} R^{(5)}_4 
  =  2 K_S \cdot  \bar{c}_{({\bf 5})4} + \# A_6(4) + 4 \# 2,
   \label{eq:temp-R}\\
 {\rm deg} K_{\bar{c}_{({\bf 5})6}} & = & 
  2 K_S \cdot \bar{c}_{({\bf 5})6} + {\rm deg} R^{(5)}_6 
  =  2 K_S \cdot  \bar{c}_{({\bf 5})6} + \# A_6(6) + 2 \# 3 + \# 2. 
  \label{eq:temp-S}
\end{eqnarray}
One can also see from geometric intuition that 
\begin{equation}
 \bar{c}_{({\bf 5})4} \cdot \bar{c}_{({\bf 5})6} = 
  2 \# 2 + \# 4 + \# 6.
   \label{eq:temp-T}
\end{equation}
Thus, the number of all of type 6, $A_6(4)$ and $A_6(6)$ points 
are determined by the intersection numbers of various divisors. 
It serves as a consistency check of all the derivation above to 
note, by examining local behavior of $\tilde{R}^{(5)}_{\rm mdfd}$ 
around various types of codimension-3 singularities, that 
\begin{equation}
 {\rm deg} \tilde{R}^{(5)}_{\rm mdfd} = 9 \# 2 + 2 \# 3 + 2 \# 4 + 2 \#
  6
 + \# A_6(4) + \# A_6(6)
\label{eq:temp-Q}
\end{equation}
at the 4+1 factorization limit. One can see that 
(\ref{eq:temp-P}, \ref{eq:temp-Q}) for ${\rm deg} K_{\bar{c}_{({\bf
5})}}$ in (\ref{eq:temp-U}) and (\ref{eq:temp-R}--\ref{eq:temp-T}) 
in (\ref{eq:temp-U}) are perfectly consistent. 

Let us finally present the number of all the different types of 
codimension-3 singularities in the 4+1 factorization.
\begin{eqnarray}
 \# 1 & = & (K_S + \eta') \cdot (4K_S + \eta''), \\
 \# 2 & = & (K_S + \eta') \cdot \eta', \\
 \# 3 & = & (K_S + \eta') \cdot (3K_S + \eta''), \\
 \# 4 & = & (4K_S + \eta'') \cdot (3K_S + \eta' + \eta''), \\
 \# 5 & = & (4K_S + \eta'') \cdot (3K_S + \eta''), \\
 \# 6 & = & (4K_S + 2 \eta' + \eta'') \cdot (6K_S + \eta' + 2 \eta'')
               - 2 \# 2 - \# 4, \label{eq:nbr-6} \\
 \# A_6(4) & = & (4K_S + 2\eta' + \eta'') \cdot (3K_S + 2\eta' + \eta'')
     - 4 \# 2, \\
 \# A_6(6) & = & (6K_S + \eta' + 2 \eta'') \cdot (5K_S + \eta' + 2
  \eta'') - \# 2 - 2 \# 3.
\end{eqnarray}

Now, we are done with all the necessary mathematical preparation. 
As a physics application, we can assume this 4+1 factorization. 
Suppose that the Standard Model ${\bf 10}$-representation fields originate 
from $C_{({\bf 10})4}$, the Standard Model $\bar{\bf 5}$ matter fields 
from $C_{(\bar{\bf 5})4}$ and the Higgs fields $H({\bf 5})$ and 
$\bar{H}(\bar{\bf 5})$ from $C_{(\bar{\bf 5})6}$. Then an unbroken 
U(1) symmetry remains unbroken, and dimension-4 proton decay operators 
are forbidden by the symmetry. 

We have seen in section \ref{sec:inv} that $\#E_6$ is even in any 
F-theory compactifications and cannot be 1. If one further wants to 
realize the approximately rank-1 Yukawa matrices of the real world 
through this 4+1 factorization, then one has to arrange the divisors 
$\eta'$ and $\eta"$ so that $\# 5 = 1$ (instead of $\#E_6 = 1$) and 
$\# 4 = 1$ (instead of $\# D_6 = 1$). For this purpose, 
$\bar{c}_{({\bf 10})1}$ cannot be void. 

The Higgs curve $\bar{c}_{(\bar{\bf 5})6}$ and the matter curve 
$\bar{c}_{(\bar{\bf 5})4}$ intersect at type 6 points, and they are 
where neutrino Yukawa couplings are generated in this class of 
models. The number of type 6 points in the GUT divisor $S$ is 
given by (\ref{eq:nbr-6}). Since there is no specific reason to 
believe that the number (\ref{eq:nbr-6}) is one, generically 
neutrino Yukawa matrix of low-energy effective theory receives 
contributions from multiple type 6 points. Without an extra 
symmetry, there is no reason to believe that they are aligned. 
Thus, a contribution from a given type 6 point may have some 
structure, but it will be lost in the neutrino Yukawa matrix 
in the effective theory.

%%%%%%%%%%%%%%%%%%%%%%%%%%%%%%%%%%%%%%%%%%%%%%%%%%%%%%%%%%%
\section{Hitchin Equation at Branch Locus}
\label{sec:Hitchin}
%%%%%%%%%%%%%%%%%%%%%%%%%%%%%%%%%%%%%%%%%%%%%%%%%%%%%%%%%%%

Deformations of an $A$-$D$-$E$ type singularity are parametrized 
by $\mathfrak{h} \otimes \C/W$, where $\mathfrak{h}$ is a Cartan 
subalgebra of a Lie algebra $\mathfrak{g}$ corresponding to one of 
the $A$-$D$-$E$ type singularity, and $W$ is its Weyl group \cite{KM}. 
Thus, the field-theory description of the gauge-theory 
sector of F-theory compactifications must be in terms 
of a super Yang--Mills theory with a gauge group $G$ of the Lie 
algebra $\mathfrak{g}$. The field theory description may have 
branch loci, around which fields in the ${\bf adj.}$ representation 
are twisted by $W$.

In F-theory compactifications down to 3+1 dimensions, 
matter curves and branch loci are both codimension-1 
in the discriminant locus (which is a complex surface) $S$, 
and hence they intersect as many times as the intersection number 
of their divisors in $S$. 
Therefore, the branch locus inevitably comes into a description of 
charged matter fields in F-theory compactifications down to 3+1 
dimensions \cite{Hayashi-2}. 

In this appendix, we show that ``the branch locus''
can be described as a soliton in the super Yang--Mills 
theory on $S$, with no singularity in the soliton 
configuration. In \cite{Hayashi-2}, we discussed that we need 
three ingredients to give some of Yukawa couplings;  
a super Yang--Mills theory on $S$ with branch cuts, 
the Weyl-group twists along the branch cuts 
and a singular field configuration along the branch loci. 
This statement is not wrong, but 
an alternative description is given in this appendix. 
The new description is the super Yang--Mills theory on $S$ 
in a soliton background; neither branch cuts, branch loci nor 
singular behavior of background field configuration are necessary.
This new description (the new soliton, in particular) may 
be compared to the 't Hooft--Polyakov monopole in 
its relation to the Dirac monopole. The singularity in the previous 
description is replaced by the smooth soliton configuration in the 
field theory as a UV completion. 
Since a microscopic formulation of F-theory is still missing,  
it is good news that an UV safe description is given to the ``branch loci''.

In studying the soliton solution, we found that various 
techniques in studying vortex solutions of an Abelian Higgs model
are very useful. Thus, this appendix begins with a warming up, 
which may help ourselves to recall those useful techniques. 

%%%%%%%%%%%%%%%%%%%%%%%%%%%%%%%%%%%%%%%%%%%%%%%%%%%
\subsection{BPS Vortex in Abelian Higgs Model}
\label{ssec:vortex}
%%%%%%%%%%%%%%%%%%%%%%%%%%%%%%%%%%%%%%%%%%%%%%%%%%%

In an Abelian Higgs model in $d + 1$ dimensions, 
vortices are time-independent solutions that extend in $d-2$ 
spatial dimensions. It carries its energy density 
whose mass dimension is $(d-1)$. 

The energy of the Abelian Higgs model in $2+1$ dimensions is given by 
\begin{equation}
 E = \int_{\R^2} d^2 x \; 
  \left[ 
   (D_m \phi)^\dagger (D_m \phi) + \frac{1}{4 g^2} F_{mn}F_{mn} 
   + \frac{\lambda^2}{2} (|\phi|^2 - V)^2 
  \right], 
\end{equation}
where the Higgs field $\phi$ and the gauge field $A_m$ ($m=1,2$) have 
mass dimensions 1/2 and 1, respectively, and the parameters 
$g^2$, $\lambda^2$ and $V$ have all mass dimension 1. When $\lambda =
g$, a vortex solution satisfying the Bogomol'nyi equation 
\begin{equation}
 D'' \phi = 0, \qquad 
 - F_{12} = g^2 (V - |\phi|^2)
\label{eq:Bogomolnyi-eq}
\end{equation}
becomes a BPS soliton, with the energy $E = 2 \pi |N| V$, 
where the covariant derivative $D''=d\bar{v}\,D_{\bar{v}}$ 
is the anti-holomorphic derivative with respect to the local holomorphic 
coordinate $v = x_1 + i x_2$, along with the $(0,1)$ part 
$A_{\bar{v}}=\left(A_1+iA_2\right)/2$ of the gauge field.
We will restrict ourselves to the cases with 
$\int F_{12} dx_1 dx_2 = 2 \pi N < 0$.\footnote{
The BPS equation becomes $D' \phi = 0$ and $F_{12} = g^2 (V - |\phi|^2)$ 
in the cases with $N > 0$.} 

Let us find the BPS vortex solution more explicitly for 
the $N=1$ case. Without loss of generality, we assume that 
the vortex is placed at $v=0$. In order to see the behavior of 
the solution near the center of the vortex, we use a trial form 
\begin{eqnarray}
 \phi(v,\bar{v}) & = & \sqrt{V} \sum_{m=0}^\infty c_m v^{m+1} \bar{v}^m, 
\label{eq:Bogomolnyi-vortex-series-phi}
\\
 i A_{\bar{v}} & = & \sum_{m=0}^\infty a_{m+1} v^{m+1} \bar{v}^m.
\label{eq:Bogomolnyi-vortex-series}
\end{eqnarray}
Because the gauge field $A_m dx_m$ is Hermitian, 
its $(1, 0)$ part is given by 
$i A_v = - \sum_{m=0}^\infty a_{m+1}^* \bar{v}^{m+1} v^m$.
The equation $D'' \phi = 0$ gives the recursion relations 
\begin{equation}
  c_1 + a_1 c_0 = 0, \qquad 
  2 c_2 + (a_2 c_0 + a_1 c_1) = 0, \qquad 
  m c_m + \sum_{l=0}^{m-1} a_{m-l} c_l = 0,
\label{eq:vortex-rec-c}
\end{equation}
and hence for $m \geq 1$, $c_m/c_0$ is given by a homogeneous function of  
$a_l$ ($l=1,\cdots,m$) of weight $m$, where the weight of $a_l$ 
is given by $l$.

\begin{figure}[tb]
 \begin{center}
  \begin{tabular}{ccc}
   \includegraphics[width=.4\linewidth]{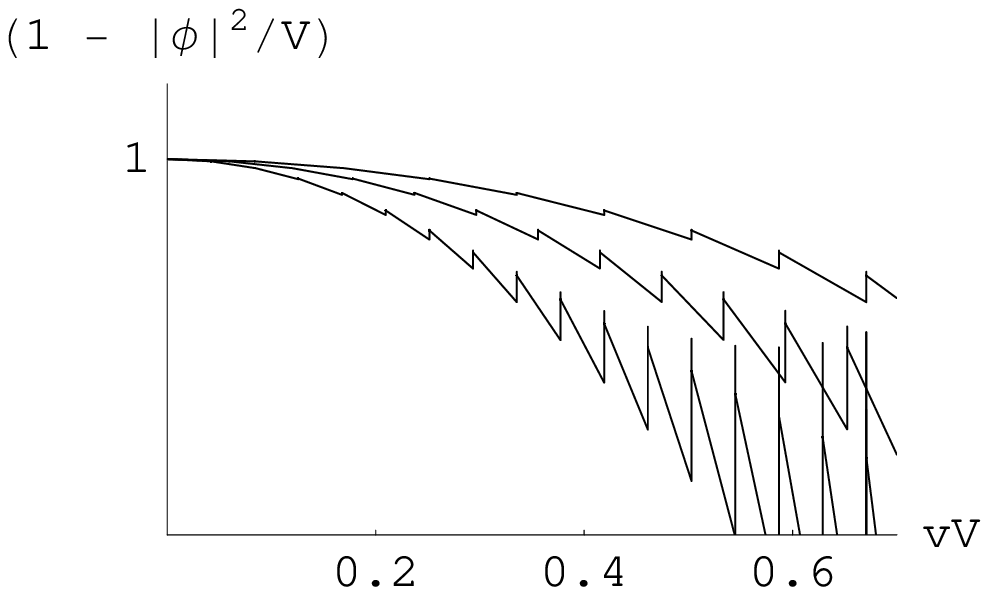}     
& & 
   \includegraphics[width=.4\linewidth]{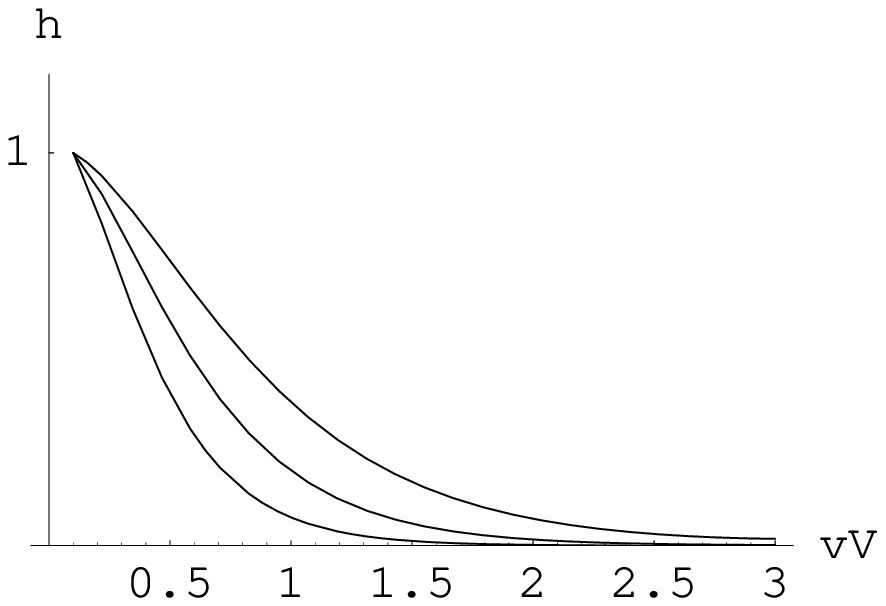}  \\
(a) & & (b)
  \end{tabular}
\caption{\label{fig:abelian-vortex} The figure (a) 
shows the behavior of $(V - |\phi|^2)$ of the BPS vortex solution of 
the Abelian Higgs model, for $g^2/V = 1$, $2$ and $4$ from right 
to left. Despite the zigzag behavior due to truncating the series 
expansion solution at finite order, we can clearly see that 
$|\phi|^2$ approaches $V$ at large $|v|$. 
The function $h(|v|^2)$---a numerical solution 
to (\ref{eq:h-diffeq-vortex})---is shown in the figure (b) for the same
values of $g^2/V$, which equals $1$, $2$ and $4$ from right to left.
Although the series expansion solution has a finite radius of 
convergence, the function $h$ can be used to describe the asymptotic 
behavior of the solution outside the convergence radius. 
In both (a) and (b), the dimensionless combination $v V$ is used 
for the horizontal axis.}
 \end{center}
\end{figure}

The other equation 
\begin{equation}
 2 i F_{v\bar{v}} = g^2 (V- |\phi|^2)
\label{eq:vortexBPS-mdfd}
\end{equation}
further requires that 
\begin{equation}
 4 a_1 = g^2 V, \qquad 
 4 \times 2 a_2 = - g^2 V |c_0|^2, \qquad 
 4 (m+1) a_{m+1} = - g^2 V |c_0|^2 \sum_{l=0}^{m-1} (c_l/c_0)(c_{m-1-l}/c_0)^*. 
\label{eq:vortex-rec-a}
\end{equation}
Here, we assumed on the left-hand side of each of the 
equations (\ref{eq:vortex-rec-a}) that all the $a_{m+1}$ are real. 
In other words, the gauge field $A_m dx_m$ only has its angular 
component $A_{\theta}$ in the circular coordinates $(r,\theta)$ of 
$v=re^{i\theta}$. Using (\ref{eq:vortex-rec-c}) we can determine 
$(c_l/c_0)$ in terms of $a_m$'s with $m \leq l$, 
and further all the $a_m$ are also determined in terms of  
$c_0$, $g^2$ and $V$, by systematically using (\ref{eq:vortex-rec-a}). 
The remaining step is to determine the value of $c_0$ by requiring 
that $|\phi|^2$ approaches $V$ at large $|v|$. Numerically,  
we found that the appropriate value is 
$c_0/V \simeq 0.92$ for $g^2/V = 1$, $c_0/V \simeq 1.30$ 
for $g^2/V = 2$
and $c_0/V \simeq 1.84$ for $g^2/V = 4$, for example. 
Figure~\ref{fig:abelian-vortex}~(a) shows how $|\phi|^2$ approaches 
$V$ at large $|v|$ for these values of $c_0$.
%%%%%%%%%%%%%%%%%%%%%%%%%%%%%%%%%%%%%%%%%%%%%%%%%%%%%%%%%%
%%%%%%%%%%%%%%%%%%%%%%%%%%%%%%%%%%%%%%%%%%%%%%%%%%%%%%%%%%

The $\phi$ field can be regarded as a section of a line bundle 
determined by the U(1) gauge field $A_mdx^m$. The gauge field 
is Hermitian, and the simple form $|\phi|^2$ appearing in the BPS 
condition or the kinetic term $|D_m \phi|^2$ can be regarded 
as the Hermitian inner product of $\phi$ or $D_m \phi$  
with an obvious choice of metric in the unitary frame description. 
Its holomorphic frame description $\tilde{\phi}= V^{3/2} v$ 
is related to the unitary one $\phi$ by 
$\phi = {\cal E} \tilde{\phi}$, where 
\begin{equation}
\qquad {\cal E} = V^{-1} \sum_{m=0}^{\infty} c_m |v|^{2m}.
\end{equation}

In order to find out the asymptotic behavior of the BPS vortex solution 
at large $|v|$, however, the series 
expansions (\ref{eq:Bogomolnyi-vortex-series-phi}, 
\ref{eq:Bogomolnyi-vortex-series}) are not useful. 
To study the asymptotic behavior, it is more useful to introduce 
a function $h(w)$ of $w = |v|^2$. 
The function $h(w)$ approaches zero at infinity, and 
$h(w) \rightarrow 1$ as $w \rightarrow 0$. 
The configuration  
\begin{equation}
 A = i \frac{1 - h(|v|^2)}{2} 
   \left( \frac{dv}{v} - \frac{d\bar{v}}{\bar{v}} \right), \qquad 
 \phi = \sqrt{V} \frac{v}{|v|} 
  \exp \left[ - \int^\infty_{|v|^2} \frac{dw}{2w} h(w) \right]
\label{eq:h-ansatz-vortex}
\end{equation}
is smooth at $v=0$, and the asymptotic forms of the fields at 
large $|v|$ are $A \sim - d \theta$ and $\phi \sim \sqrt{V} v/|v|$.
One can further see that the condition $D'' \phi = 0$ is 
satisfied. The remaining BPS condition (\ref{eq:vortexBPS-mdfd}) 
can be regarded as an equation that $h$ satisfies. 
Substituting (\ref{eq:h-ansatz-vortex}) into (\ref{eq:vortexBPS-mdfd}) 
and taking the derivative of both sides of the resulting equation 
with respect to $w = |v|^2$, one obtains 
\begin{equation}
 2 w h'' = (g^2 V + 2 h') h, 
\label{eq:h-diffeq-vortex} 
\end{equation}
where $'$ denotes a derivative with respect to $w = |v|^2$.
$h$ is supposed to be small at large $|v|$, and the last term 
can be ignored because it is the square of $h$. 
Then the equation (\ref{eq:h-diffeq-vortex}) is satisfied 
at large $w$ by 
\begin{equation}
 h \sim \exp \left[ - \sqrt{2 g^2 V} w^{\frac{1}{2}} \right]
  = e^{- \sqrt{2 g^2 V} |v|}.
\label{eq:asympt-h-vortex}
\end{equation}
The field strength $i F_{v\bar{v}} = - h'$ and $(V-|\phi|^2)$
are both exponentially small for large $|v|$, just like $h(|v|^2)$  is.
Figure~\ref{fig:abelian-vortex}~(b) shows the profile of 
$h(|v|^2)$ obtained by solving the differential equation 
numerically for $g^2/V = 1,2,4$.

The dimensionless parameter $\epsilon \equiv g^2/V$ controls 
the behavior of the BPS vortex configuration. 
On closer examination, one finds that the vortex solution 
is described in terms of $\epsilon$ by 
\begin{equation}
 \phi(v) = \sqrt{V} \phi_*(\sqrt{\epsilon} vV), \qquad 
 A_{\bar{v}}(v) = \sqrt{\epsilon} V a_{*}(\sqrt{\epsilon} vV), 
\end{equation}
where the pair ($\sqrt{V} \phi_*$, $V a_*$) is a solution to the BPS 
conditions (\ref{eq:Bogomolnyi-eq}) with $\epsilon = 1$.
The characteristic radius is given by  
$1/(\sqrt{\epsilon} V) = 1/\sqrt{g^2 V}$, 
which perfectly agrees with the form (\ref{eq:asympt-h-vortex}) of $h$. 
When the field configuration $\phi(v)$ is expressed in 
the series expansion as in (\ref{eq:Bogomolnyi-vortex-series-phi}), 
the leading order coefficient $c_0$ should scale as 
$c_0 = c_{0*} \sqrt{\epsilon} V$; we have already found the appropriate 
value of $c_0/V$ numerically for three different values of $\epsilon$, 
and indeed the results satisfy 
$c_0/V \simeq 0.92 \times \sqrt{\epsilon}$.

%%%%%%%%%%%%%%%%%%%%%%%%%%%%%%%%%%%%%%%%%%%%%%%%%%%
\subsection{Ramified Spectral Cover in Hitchin Equation}
\label{ssec:ramified-bg}
%%%%%%%%%%%%%%%%%%%%%%%%%%%%%%%%%%%%%%%%%%%%%%%%%%%

It is of theoretical as well as phenomenological 
interest to deepen our understanding of the field theory description  
of the codimension-3 singularities. 
$D_6$-type codimension-3 singularities in $\SU(5)_{\rm GUT}$ 
models are not particularly difficult. 
The field-theory local models have $\U(1)$ structure groups,  
not non-Abelian, and all the fluctuations are in single component 
representations of the $\U(1)$ Higgs bundle. 
The field theory descriptions of an enhanced $E_6$ singularity 
and an enhanced $A_6$ singularity, however, still need to be developed.

The structure group is a non-Abelian $\U(2)$ group for both, 
and the irreducible decompositions of the adjoint representation  
of $E_6$ and $A_6$ are as follows.
\begin{eqnarray}
 {\rm Res}^{E_6}_{\vev{\U(2)} \times \SU(5)_{\rm GUT}} {\bf adj.} & = & 
  ({\bf adj.}, {\bf 1}) +({\bf 1},{\bf adj.}) + ({\bf 1},{\bf 1}) \nonumber \\
 & & +
 \left[ ({\bf 2},{\bf 10}) + (\wedge^2 {\bf 2}, \bar{\bf 5}) \right] 
  + {\rm h.c.}, \label{eq:E6toA4} \\ 
 {\rm Res}^{\SU(7)}_{\vev{U(2)} \times \SU(5)_{\rm GUT}} {\bf adj.} & = & 
  ({\bf adj.}, {\bf 1}) +({\bf 1},{\bf adj.}) + ({\bf 1},{\bf 1}) \nonumber \\
 & + & ({\bf 2},\bar{\bf 5}) + {\rm h.c.}. \label{eq:A6toA4}
\end{eqnarray}
The $({\bf 2},{\bf 10})$ component in the decomposition of 
$E_6$ and the $({\bf 2},\bar{\bf 5})$ component in the decomposition 
of $A_6$ are both the doublets of the background Higgs 
bundle with the structure group $\U(2)$, and furthermore, 
the local behaviors of the spectral surfaces of these two components 
turned out to be exactly the same around each of the codimension-3 
singularities \cite{Hayashi-2}.
Thus, it is sufficient to work on a codimension-3 singularity 
for the enhancement of an $A_{N-1}$ singularity to $A_{N+1}$. 

Reference \cite{Hayashi-2} already worked out in great detail 
the field theory configuration around these types of codimension-3 
singularities. There is still a room for improvement, however, 
because the analysis in \cite{Hayashi-2} had two shortcomings. 
The first one is that the distinction between the unitary 
frame and the holomorphic frame was not clearly made there. 
The other one is the treatment of the branch locus itself. 
The codimension-3 singularity of the enhancement of $A_{N-1}$ to $A_{N+1}$  
is at the point where a matter curve and a branch locus meet.
Thus, there is no freedom to get around this problem. 
This section will provide a prescription to deal with these issues. 

Before we get started, a bit more explanation may be necessary 
on what the issue is in the treatment of the branch locus. 
In order to specify a field-theory local model, a (BPS) 
background in the field theory has to be determined so that 
the model corresponds to the local geometry of a Calabi--Yau 4-fold $X$. 
The principle is that the Hitchin map \cite{Hitchin-2} 
of $\varphi$ % \kesu{(or $\widetilde{\varphi}$)} 
has to reproduce some of the coefficients of the 
local defining equation of the complex 4-fold $X$. 
See \cite{KV, DW-1, Hayashi-2, DW-3, Tsuchiya} for more details. 
Since the Hitchin map specifies only eigenvalues of $\rho_{U_I}(\varphi)$ 
% \kesu{(or of $\rho_{U_I}(\widetilde{\varphi})$)}, 
the coefficients of the local geometry can determine only 
all the eigenvalues up to ordering, 
but not necessarily the entire matrix valued $\varphi$. 
% \kesu{(or $\widetilde{\varphi}$)}. 
A subtle problem remains on a locus of $S$ where two of 
the eigenvalues degenerate, that is, the branch locus.  
A $2 \times 2$ matrix with degenerate eigenvalues $\lambda$ 
may be either  
\begin{equation}
 \left( \begin{array}{cc} \lambda & \\
         1 & \lambda \end{array} \right) 
 \quad {\rm or} \quad 
  \left( \begin{array}{cc} \lambda & \\
           & \lambda \end{array} \right) 
\label{eq:Jordan-or-split}
\end{equation}
in the Jordan normal form. The above principle alone 
cannot determine which one we should take for the 
field-theory model of the local geometry. 
This issue has not been clarified yet in physics 
literature so far.

A spectral surface of a rank-2 Higgs bundle is always given 
locally in the form
\begin{equation}
 \xi^2 + 2 s_1(u,v) \xi + s_2(u,v) = 0,
\label{eq:spec-surf-rk2}
\end{equation}
where $(u,v) = (u_1,u_2)$ is a set of local coordinates 
on a complex surface $U_\alpha \subset S$, and $s_1$ and $s_2$ are 
locally functions of the local coordinates. 
This corresponds, in a deformation of $A_{N+1}$ to $A_{N-1}$,  
to a local geometry determined by 
\begin{equation}
  y^2 \simeq x^2 + z^N(z^2 + 2 s'_1 z + s'_2).
\label{eq:local-geometry-AN}
\end{equation}
The $K_{U_\alpha}$-valued Higgs-bundle in the doublet 
representation $U = {\bf 2}$ [resp. $U = \bar{\bf 2}$] 
determines the local behavior of zero mode wavefunctions 
in the $(U,R) = ({\bf 2}, \bar{N})$ [resp. $(U,R) = \bar{\bf 2}, N)$] 
representation. Here, we assume that all the coordinates $y, x, z$ 
are somehow made dimensionless, and so are the values of the local functions 
$s'_{1,2}$, by normalizing them with respect to some ``unit length'' 
$l_*$. The local functions $s_i$ ($i=1,2$) in (\ref{eq:spec-surf-rk2}) 
and $s'_i$ in the local defining equation (\ref{eq:local-geometry-AN})
are related \cite{Tsuchiya} by
\begin{equation}
 s_i = \left( \frac{l_*}{4\pi \alpha'} \right)^i s'_i,
\end{equation}
in order for $s_i$'s to have their proper mass dimensions $i$; 
this relation was determined by relying on the D7-brane 
interpretation of the deformation of an $A_{N+1}$ singularity to 
$A_{N-1}$, but the precise meaning of $\alpha'$ and $l_*$ remains 
unclear in the context of generic F-theory compactifications. 

The matter curve for the $\bar{N}+N$ representation of unbroken 
the $\SU(N)$ symmetry is $s_2 = 0$ in a local model with 
$\SU(N+2)$ gauge group.  The $s_2 = 0$ matter curve is for 
for the ${\bf 10}+\overline{\bf 10}$ representation in a $E_6$ local 
model with $\SU(5)_{\rm GUT}$ unbroken symmetry, and for the 
$\bar{\bf 5} + {\bf 5}$ representation in an $A_6$ local model, 
as we already saw in (\ref{eq:E6toA4}, \ref{eq:A6toA4}). 
The branch locus $s_1^2 - s_2 = 0$ in $U_\alpha \subset S$ is where 
the two eigenvalues in the $\varphi$ field {\it i.e.}, the two roots 
of $\xi$ degenerate, and the spectral surface is ramified over
$U_\alpha$. For simplicity, we will only consider the cases 
where $s_2 = - c^{2} M_*^3 v$ and $s_1 = c' M_*^2 u$ with some 
dimensionless constants $c$ and $c'$. 
This does not loose generality much, because 
the codimension-3 singularity of our interest is placed at the 
point $(s_1,s_2)=(0,0)$, and we can take the origin of the 
local coordinates $(u,v)$ at the codimension-3 singularity point. 
We assume a generic complex structure, so that $(u,v)= (0,0)$ 
is a simple zero locus of order one for both $s_1$ and $s_2$.

%%%%%%%%%%%%%%%%%%%%%%%%%%%%%%%%%%%%%%%%%%%%%%%%%%%%%%%%%%%%%%
\subsubsection{Ramified Spectral Curve with $s_1=0$}
\label{sssec:vortex-s-zero}
%%%%%%%%%%%%%%%%%%%%%%%%%%%%%%%%%%%%%%%%%%%%%%%%%%%%%%%%%%%%%%

Let us first study a field theory description of 
a simpler local geometry than the generic one 
(\ref{eq:local-geometry-AN}) by setting $s_1$ to zero. 
This may be regarded as assuming that $s_1$ varies very slowly 
along the $u$-direction, {\it i.e.}, with small $c'$, and focusing 
on the region $u \approx 0$ in such cases. 
It reduces the problem of finding solutions to the BPS equations 
on a complex surface to the one on a complex curve.
 
The BPS equations (\ref{eq:BPS-D},\ref{eq:BPS-G}) for the Higgs 
bundle on a complex surface yield  
\begin{equation}
 D'' \varphi^{(1)} = 0, \qquad 
  i F + [\varphi^{(1)}, \overline{\varphi}^{(1)}] = 0
\label{eq:Hitchin}
\end{equation}
on the complex curve.  
The $(1,0)$-form $\varphi^{(1)}$ on the curve $u=0$ in $U_\alpha$ 
is related to the $(2,0)$-form $\varphi$ on $U_\alpha \subset S$, by 
\begin{equation}
  \alpha \varphi = 2 \alpha \varphi_{uv} du \wedge dv 
 = \left[ \sqrt{h_{u\bar{u}}} du \right] \wedge \varphi^{(1)}.
\end{equation}
In fact, the equations (\ref{eq:Hitchin}) are derived from 
(\ref{eq:BPS-D}, \ref{eq:BPS-G}) by factoring out 
$h_{u\bar{u}}du \wedge d\bar{u}$, and they are identical to the 
original Hitchin equations.\footnote{In this article, we adopt 
a convention that the covariant derivative is $\nabla = d + i \rho(A)$. 
Both the 1-form $A$ and $\varphi^{(1)}+\overline{\varphi}^{(1)}$ are 
Hermitian, not anti-Hermitian.}

Although a Higgs bundle is described in differential geometry by 
the solution to the Hitchin equations, it is regarded in algebraic 
geometry as a pair $(V, \tilde{\varphi})$ of a holomorphic vector 
bundle $V$ on $U_\alpha$ and a holomorphic $\mathfrak{g}'$-valued 
$(2,0)$-form $\tilde{\varphi}$. When the vector bundle $V$ on 
$U_\alpha$ is given by the pushforward $V = \pi_{C*} {\cal N}$ 
of a sheaf ${\cal N}$ on a spectral cover $C$ 
($\pi_C: C \rightarrow U_\alpha \subset S$ is the projection), then 
the action of $\varphi$ (or $\tilde{\varphi}$ depending on the frame)
on ${\cal N}$ should be the same as that of $\xi$ restricted upon $C$, 
because $\xi - 2\alpha \varphi_{12}$ is the defining equation of 
the spectral surface; it should trivially act on any sheaves on 
$\mathbb{K}_{U_\alpha}$ that is given by a pushforward of a sheaf on the 
spectral cover $C$.\footnote{Note that ${\cal V} \equiv i_{C*}{\cal N}$
is a sheaf on $\mathbb{K}_{U_{\alpha}}$, where 
$i_C: C \hookrightarrow \mathbb{K}_{U_\alpha}$ is 
the inclusion. $V = \pi_{\mathbb{K_{U_\alpha}} *} {\cal V}$, because 
$\pi_{\mathbb{K}_{U_\alpha}} \circ i_C = \pi_C$.}
This principle is sufficient in determining which one 
in (\ref{eq:Jordan-or-split}) describes the matrix valued $\phi$  
at the ramification locus of the spectral cover, as we will see below. 

Since the ramification behavior of the spectral cover surface $C$
can easily be seen in the normal direction to the branch locus 
$s_1^2 - s_2 = 0$, this issue can be discussed by just taking  
the $u=0$ slice. We will abuse notation by writing  
$U_{\alpha} \subset S$ and $C$ 
for the $u=0$ slice of $U_{alpha}$ and 
in the sense of $\pi_C^{-1}(u=0) \subset C$, respectively. 
Furthermore, we treat the sheaf ${\cal N}$ on $C$ as ${\cal O}_C$, 
by taking a trivialization frame locally in ${\cal N}$. 
Thus, $V = \pi_{C*} {\cal O}_C$. 

Let us introduce a local coordinate on the spectral cover $C$ to  
provide a more detailed description of ${\cal O}_C$ and $V$.
Let $\xi^{(1)}$ be a fiber coordinate of the canonical bundle 
on the curve. $\xi^{(1)}$ is related to $\xi$ by 
\begin{equation}
\xi du \wedge dv = [\sqrt{h_{u\bar{u}}} du] \wedge \xi^{(1)} dv.
\end{equation}
%. 
Now the defining equation of the spectral curve is
\begin{equation}
(\xi^{(1)})^2 - \underline{c}^2 M_*^3 v = 0, 
\end{equation}
where $\underline{c} = c/\sqrt{h_{u\bar{u}}}$.
Thus, the local coordinate $v$ of the $u=0$ slice of $U_\alpha$ 
is expressed on $C$ in terms of $\xi^{(1)}$, so that $\xi^{(1)}$ 
can be used as the local coordinate of the spectral cover. 
Using this coordinate $\xi^{(1)}$, 
a section of ${\cal O}_C$ in a local patch around 
the point $\xi^{(1)}= 0$ is given by a polynomial 
$\sum_{i=0}^N f_i (\xi^{(1)})^i$ in $\xi^{(1)}$. 
With the local coordinate $v$ of the base space $U_\alpha$, 
the polynomial is expressed as 
\begin{equation}
 \sum_{i=0}^N f_i (\xi^{(1)})^i = \left(\sum_{j=0} 
     f_{2j} (\underline{c}^2 M_*^3 v)^j \right) 
   + \xi^{(1)} \left( \sum_{j=0} f_{2j + 1} (\underline{c}^2 M_*^3 v)^j \right)
 \equiv f_+(v) + \xi^{(1)} f_-(v),
\end{equation}
and under the pushforward, is mapped to a local picture of 
the rank-2 locally free sheaf $V$ on $U_{\alpha}$. 

The action of $\xi$ and $\xi^{(1)}$ on sections of $i_{C*} {\cal O}_C$
can be realized, respectively, as a matrix representation of 
$2\alpha \varphi_{uv}$ and of $\varphi^{(1)}_v$ acting on sections of $V$.
The coordinate $\xi^{(1)}$ of $\mathbb{K}_{U_\alpha}$ acts, 
as multiplication, on a section $\sum_{i=0}^N f_i (\xi^{(1)})^i$ of 
the sheave $i_{C*} {\cal O}_C$ on $\mathbb{K}_{U_\alpha}$ to yield 
$\sum_{i=0}^N f_i (\xi^{(1)})^{i+1}$.
In terms of $f_+$ and $f_-$, the action of $\xi^{(1)}$ on the section 
is represented by the $2 \times 2$ matrix
\begin{equation}
 (\xi^{(1)} dv \times) :  
  \left( \begin{array}{r} f_+ \\ \underline{c} M_* f_- \end{array}
  \right)
 \mapsto 
  \underline{c} M_* dv \; 
  \left( \begin{array}{cc} & M_* v \\ 1 & \end{array} \right) 
  \left( \begin{array}{r} f_+ \\ \underline{c} M_* f_- \end{array} \right). 
\label{eq:varphi-0-slice}
\end{equation}
Since all the entries of the matrix is holomorphic in the local 
coordinate $v$, the frame that is already chosen implicitly 
turns out to be a holomorphic frame. 
% {it can be regarded as a
% holomorphic frame description $\varphi^{(1)}$ of $\tilde\varphi^{(1)}$.} 
In the holomorphic frame, we have thus found that 
\begin{equation}
\tilde{\varphi}^{(1)} = \underline{c} M_* \; 
  \left( \begin{array}{cc} & M_* v \\ 1 & \end{array} \right) dv.
\end{equation}

This matrix has the eigenvalues $\pm cM_* \sqrt{M_* v}$, and these 
two eigenvalues are non-degenerate for $v \neq 0$. 
At $v=0$, which is on the branch locus when $u=0$,  
the two eigenvalues both become zero, but the matrix itself is not zero, 
since the non-zero entry still remains in the off-diagonal part, 
as can be seen in (\ref{eq:varphi-0-slice}).  
Therefore, of the two possible Jordan normal forms in (\ref{eq:Jordan-or-split}), the first one 
correctly describes $\phi$ at the branch locus.\footnote{The authors 
thank Alexey Bondal for explaining this to us.} 
The spectral cover in this case satisfies the condition of being regular 
in the sense of \cite{Dongai-Gaitsgory}; 
the number of Jordan blocks is the same as the number of distinct eigenvalues. 

Now we have obtained the local behavior of $\tilde{\varphi}$ in 
the holomorphic frame. The Hitchin equations (\ref{eq:Hitchin}), 
however, are written in terms of $\varphi$ in a unitary frame. 
Let us denote a complexified gauge transformation connecting
these two frames, as a $2 \times 2$ matrix, by ${\cal E}$; 
\begin{equation}
\varphi^{(1)} = {\cal E} \, \tilde{\varphi}^{(1)} \, {\cal E}^{-1}.
\label{eq:A}
\end{equation}
Then the gauge field configuration in the unitary frame 
are given by 
\begin{equation}
 iA^{0,1} = {\cal E} \bar{\partial} {\cal E}^{-1}, \qquad 
 iA^{1,0} = {\cal E}^{-1 \dagger} \partial {\cal E}^{\dagger}.
\label{eq:B}
\end{equation}
The complexified gauge transformation ${\cal E}$ can be determined 
by substituting (\ref{eq:A}) and (\ref{eq:B}) to the second one 
of the Hitchin equations (\ref{eq:Hitchin}).

\begin{figure}[tb]
 \begin{center}
     \includegraphics[width=.4\linewidth]{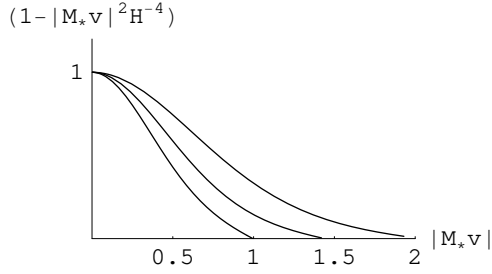}     
  \caption{\label{fig:branch-center} The non-commutativity 
$[\varphi, \overline{\varphi} ]$ decreases for large $|M_* v|$ away from
  the branch locus. $(1 - |M_* v|^2 / {\cal H}^4)$ in
  (\ref{eq:non-comm}) is plotted for $|\underline{c}|^2 = 1.0, 0.5$ and $0.2$ 
(from left to right).}
 \end{center}
\end{figure}

The complexified gauge transformation ${\cal E}$ takes its value 
in $\SU(2)^c = {\rm SL}(2, \C)$. But, Hitchin \cite{Hitchin-1} 
further introduced an ansatz\footnote{This ansatz may seem 
very odd at first sight, but not so much, actually. 
The Hitchin equation was originally 
considered as a dimensional reduction of anti-self-dual equation of 
Yang--Mills field. A BPST anti-instanton solution is given by 
\begin{equation}
  A^a_m = \frac{2 \eta^a_{mn}x^n}{x^2 + \rho^2},
\end{equation}
where $x_m$ ($m=1,2,3,4$) are four coordinates of $\R^4$, 
$\rho$ the instanton size, and $\eta^a_{mn}$ is the $\eta$ symbol 
by 't Hooft, whose definition is $\epsilon_{amn}$ (if $m, n = 1,2,3$) 
and $\delta_{4m}\delta_{an} - \delta_{4n} \delta_{am}$.
Let us introduce complex coordinates $v = (x_3 + i x_4)$ and 
$\zeta = (x_1 + i x_2)$. In this coordinate, the anti-instanton 
configuration (at $\zeta = 0$) becomes 
\begin{eqnarray}
 A_{\bar{\zeta}} = \left(\begin{array}{cc} 
                    0 & 2iv \\ 0 & 0 \end{array}\right)
              \frac{1}{|v|^2 + \rho^2}, & \qquad & 
 A_{\zeta} = \left( \begin{array}{cc} 
                     0 & 0 \\ - 2i \bar{v} & 0 \end{array} \right)
               \frac{1}{|v|^2 + \rho^2}, \\
 A_{\bar{v}} = \frac{iv}{|v|^2 + \rho^2} \tau^3, & \qquad & 
 A_v = \frac{- i \bar{v}}{|v|^2 + \rho^2} \tau^3.
\end{eqnarray}
Of course the dimensional reduction is not the same as setting simply 
$\zeta = 0$; the expressions above at $\zeta = 0$ do not mean anything 
immediately. But, it is surprising to see a close resemblance when 
we compare $A_{\bar{\zeta}}$ with $\varphi^{(1)}_{v}$, 
$A_\zeta$ with $\overline{\varphi}^{(1)}_{\bar{v}}$, and 
$A_{\bar{v}} d\bar{v} + A_v dv$ with $A_{\bar{v}} d\bar{v} + A_v d v$.}
that ${\cal E}$ is of the form 
\begin{equation}
 {\cal E} = \left( \begin{array}{cc}
   \underline{\cal E}^{-1} & \\ & \underline{\cal E}  
  \end{array} \right).
\end{equation}
This can be a solution to the Hitchin equations, because both 
\begin{equation}
i F = \partial \bar{\partial} 
 \ln (\underline{\cal E} \underline{\cal E}^\dagger) \tau^3 
\end{equation}
and 
\begin{equation}
 \left[ \varphi^{(1)}, \overline{\varphi}^{(1)} \right] 
  = - |\underline{c}|^2 M_*^2 
  \left(\underline{\cal E}^\dagger \underline{\cal E} \right)^2 
  \left( 1 - \frac{|M_* v|^2}
                  {\left( \underline{\cal E}^\dagger \underline{\cal E}\right)^4}
  \right) \tau^3 dv \wedge d \bar{v}
\label{eq:non-comm}
\end{equation}
are proportional to $\tau^3$. Introducing 
${\cal H} = \underline{\cal E}^\dagger \underline{\cal E}$, the 
Hitchin equations (\ref{eq:Hitchin}) yield 
\begin{equation}
 \partial_v \bar{\partial}_{\bar{v}} \ln \left( {\cal H}^2 \right)
  = 2 |\underline{c}|^2 M_*^2 {\cal H}^2 \left(1 - 
   \frac{|M_* v|^2}{{\cal H}^4} \right).
\label{eq:Hitchin-2}
\end{equation}
Once we assume a series expansion
${\cal H}^2 = \sum_{m=0}^\infty H_m M_*^{2m} |v\bar{v}|^m$, 
all the coefficients $H_m$ ($m \geq 1$) are determined 
recursively in terms of $H_0$ and $|\underline{c}|$. 
Requiring further that the right-hand side 
as well as the left-hand side of (\ref{eq:Hitchin-2}) go to zero 
at infinity, one can also obtain the value of $H_0$ in terms of 
$\underline{c}$. In numerical calculations, it turns out that 
$H_0 \sim 0.53$ for $|\underline{c}|^2 = 1.0$, 
$H_0 \sim 0.67$ for $|\underline{c}|^2 = 0.5$, and 
$H_0 \sim 0.91$ for $|\underline{c}|^2 = 0.2$. \label{pg:H0-numer}
Figure \ref{fig:branch-center} shows the behavior of 
$(1 - |M_*v|^2/{\cal H}^4)$ for these values of $|\underline{c}|^2$. 
%%%%%%%%%%%%%%%%%%%%%%%%%%%%%%%%%%%%%%%%%%%%%%%%%%%%%%%%%%%
%%%%%%%%%%%%%%%%%%%%%%%%%%%%%%%%%%%%%%%%%%%%%%%%%%%%%%%%%%%%

Although the solution of ${\cal H}^2$ does not determine 
$\underline{\cal E}^2$ and $\underline{\cal E}^{2 \dagger}$ uniquely, 
that is a part of the gauge degree of freedom of compact group $\SU(2)$; 
it is a sensible result that we cannot determine them. 

We are now ready to find out how the $\varphi^{(1)}$-diagonalization 
frame is related to the holomorphic and unitary frames we have been using. 
$\tilde\varphi^{(1)}$ can be diagonalized by this similarity transformation:
\begin{equation}
 \frac{1}{2 \sqrt{M_* v}} 
   \left( \begin{array}{cc} 
     1 & \sqrt{M_* v} \\ -1 & \sqrt{M_* v} 
   \end{array} \right) \widetilde{\varphi}^{(1)} 
 \left( \begin{array}{cc}
    \sqrt{M_* v} & - \sqrt{M_* v} \\ 1 & 1 
 \end{array} \right) =   
 \underline{c} M_* dv \left( \begin{array}{cc} 
   + \sqrt{M_* v} & \\ & - \sqrt{M_* v} 
   \end{array} \right). \label{eq:diagonal}
% \frac{1}{\sqrt{2M_* v}} 
%   \left( \begin{array}{cc} 
%     \underline{\cal E} & \sqrt{M_* v} \underline{\cal E}^{-1} \\ 
%    -\underline{\cal E} & \sqrt{M_* v} \underline{\cal E}^{-1}
%   \end{array} \right) \varphi^{(1)} 
% \left( \begin{array}{cc}
%    \underline{{\cal E}}^{-1} \sqrt{M_* v} & 
%  - \underline{\cal E}^{-1} \sqrt{M_* v} \\ 
%    \underline{\cal E} & \underline{\cal E} 
% \end{array} \right). 
\end{equation}
Note, however, that the similarity transformation matrix 
has a vanishing determinant at $v = 0$, and the diagonalization 
frame is not well-defined at $v = 0$; 
the diagonalization frame can be introduced only in the $v \neq 0$ region. 
The singular behavior of the right-hand side of (\ref{eq:diagonal}) 
at $v = 0$ is regarded as a consequence of a choice of frame that 
is singular at $v=0$. 

It will also become useful later on that the holomorphic frame 
description of the complex conjugate of $\varphi^{(1)}$ is given by 
\begin{equation}
 \widetilde{\overline{\varphi}}^{(1)}
  ={\cal E}^{-1}\left(\varphi^{(1)}\right)^{\dag}{\cal E}
  = \underline{c}^* M_* d\bar{v} \; 
   \left( \begin{array}{cc} 
      & {\cal H}^2 \\ {\cal H}^{-2} (M_* \bar{v}) & 
   \end{array} \right).
\label{eq:phidagg-hol-frame}
\end{equation}

Just like we did in the review on the BPS vortex solution in 
section \ref{ssec:vortex}, we can introduce a function $h(w)$ 
to study the asymptotic behavior of the field theory BPS solution for 
the branch locus. We take 
\begin{eqnarray}
 {\cal H}^2 & = & |M_* v| \exp \left[ + \int^\infty_{|v|^2} 
   \frac{dw'}{2w'} h(w') \right], \\
\underline{\cal E}^2 & = & |M_* v|^{\frac{1}{2}} \exp \left[ + \int^\infty_{|v|^2} 
   \frac{dw'}{4w'} h(w') \right],
\end{eqnarray}
where $h(w)$ is a function satisfying $h \rightarrow 0$ as
$w \rightarrow \infty$, and $h \rightarrow 1$ as $w \rightarrow 0$.
With the behavior $h \rightarrow 0$ at large $w = |v|^2$, 
${\cal H}^2$ approach $|M_*v|$ asymptotically, so that 
$[ \varphi^{(1)}, \overline{\varphi}^{(1)}]$ vanishes in the 
asymptotic region. With the behavior $h \rightarrow 1$, the complexified 
gauge transformation $\underline{\cal E}$ is smooth at $v = 0$. 
Here, we have chosen a gauge by specifying $\underline{\cal E}^2$. 
In terms of the function $h(w)$, the equation (\ref{eq:B}) determines 
the gauge field configuration
\begin{equation} 
 i A^{0,1} = \tau^3 \frac{1-h(|v|^2)}{8} \frac{d \bar{v}}{\bar{v}}.
\end{equation}
All of $\varphi^{(1)}$, $A^{0,1}$ and their Hermitian conjugates in 
the unitary frame\footnote{With a series expansion solution of 
${\cal H}^2$ in \kaku{the} holomorphic frame, we already know that the field
configuration is completely smooth at least in 
\kaku{the} holomorphic frame.} 
are completely smooth at $v=0$. There is no need 
to introduce anything singular in the field-theory description 
of the branch locus.\footnote{If the ramification behavior of the spectral 
surface is regarded as a consequence of taking a quotient by 
\kaku{the} Weyl group 
of the structure group $G'$, as in \cite{HV-Nov08-rev}, and the branch 
locus is regarded as the orbifold fixed point (locus), then it is a natural 
concern whether some kind of ``twisted sector fields'' and some interactions 
have to be introduced at the fixed point (locus). Our observation here, 
however, completely wipes this concern away.}\raisebox{5pt}{,}\footnote{
When $h$ is expressed as a series expansion in $w = |v|^2$, 
$h(w) = 1 + \sum_{m=1}^\infty h_m M_*^{2m} w^m$, $h_m$'s and 
$\tilde{H}_m = H_m/H_0$'s are related by $\tilde{H}_1 = - h_1/2$, 
$4(\tilde{H}_2 - \tilde{H}_1^2/2) = - h_2$ etc. These relations follow 
from $2 i F_{v\bar{v}} = \partial \bar{\partial} [\ln {\cal H}^2 ] \tau^3
 = (- h'/2) \tau^3$.} 
An apparent singularity at $v=0$ in the 
$\varphi^{(1)}$-diagonalization frame in (\ref{eq:diagonal}) is 
due to the fact that the frame behaves singular and is not well-defined 
at $v = 0$; the solution is completely smooth in the unitary / 
holomorphic frame.

In order to determine the asymptotic behavior of this BPS solution 
at large $|v|$, we need to derive a differential equation of $h$, 
just like we have done in the case of \kaku{the} BPS vortex solution. 
The Hitchin equation (\ref{eq:Hitchin-2}) becomes
\begin{equation}
 - \frac{h'}{2} = 2 |\underline{c}|^2 M_*^2 |M_* v| 
   \left( \exp \left[ + \int^\infty_{|v|^2} \frac{dw'}{2w'}h(w') \right] 
     - \exp \left[ - \int^\infty_{|v|^2} \frac{dw'}{2w'}h(w') \right] \right). 
\end{equation}
We expect that $h$ is small at large $|v|$, and we linearize this equation 
with respect to $h$, to find 
\begin{equation}
 \left(\frac{h'}{ w^{\frac{1}{2}}} \right)' \simeq  
  4 |\underline{c}|^2 M_*^3 \frac{h}{w}.
\end{equation}
Thus, the asymptotic behavior turns out to be\footnote{We do not pay 
attention to the overall factor; this factor cannot 
be determined in the linearized analysis.}
\begin{equation}
 h(|v|^2) \simeq 
    \exp \left[ - \frac{8}{3} |\underline{c}| |M_* v|^{\frac{3}{2}} \right]
  = \exp \left[ - \frac{8}{3} \frac{|c|}{\sqrt{h_{u\bar{u}}}} 
                              |M_* v|^{\frac{3}{2}}\right].
\label{eq:asymptotic-ramify}
\end{equation}
This asymptotic behavior tells us how fast 
the gauge field configuration approaches 
a pure gauge form $i(A^{1,0}+A^{0,1}) \sim - \tau^3 i d\theta/4$ 
\kaku{at infinity}, 
% the field strength to zero, 
% ${\cal H}^2$ to $|M_* v|$, 
and $\left[ \varphi^{(1)}, \varphi^{(1)\dagger} \right]$ goes to zero. 
Thus, \kaku{far outside the radius $|M_* v| \sim |\underline{c}|^{-\frac{2}{3}}$,} 
$\varphi^{(1)}$ and $\overline{\varphi}^{(1)}$ almost commute with 
each other; the non-commutativity is exponentially small, 
with the behavior given by (\ref{eq:asymptotic-ramify}). 

%%%%%%%%%%%%%%%%%%%%%%%%%%%%%%%%%%%%%%%%%%%%%%%%%%%%%%%%%%%
\subsubsection{Ramified Spectral Curve with $s_1 \neq 0$}
%%%%%%%%%%%%%%%%%%%%%%%%%%%%%%%%%%%%%%%%%%%%%%%%%%%%%%%%%%%

We have constructed \kaku{the} BPS solution along the $u=0$ slice 
of $U_\alpha$. Let us now generalize the solution slightly 
to obtain a solution along a \kaku{$u=u_0$} slice\kaku{.} 
This, once again, corresponds to assuming that $s_1$ depends very slowly 
on $u$ around $u = u_0$ (that is, $c'$ is small). 

On a slice at 
$u = u_0 \neq 0$, the defining equation of 
the spectral curve is
% \footnote{The comment in footnote
% \ref{fn:2dimTo1dim} also applies here. } 
%
\begin{equation}
 (\xi^{(1)})^2 + 2 \underline{c}' M_*^2 u_0 (\xi^{(1)}) 
   - \underline{c}^2 M_*^3 v = 0, 
\end{equation}
where $\underline{c}' = c'/\sqrt{h_{u\bar{u}}}$.
\kaku{The discussion in the previous subsection \ref{sssec:vortex-s-zero} 
is the same for this case, and in the holomorphic frame, it yields }

\begin{equation}
 \widetilde{\varphi}^{(1)} = \left[ 
  - \underline{c}' (M_* u_0) \; {\bf 1} + \underline{c} 
  \left( \begin{array}{cc}
    & M_* v + (c'/c)^2 (M_* u_0)^2 \\ 1 & \end{array} \right) \right]
      M_* dv.
\label{eq:varphi-u0-slice}
\end{equation}
The spectral curve is regular even for $u_0 \neq 0$. 

In order to construct a solution to the Hitchin equation 
using $\widetilde{\varphi}^{(1)}$ obtained above, we have 
to find a $2 \times 2$ matrix valued complexified gauge 
transformation ${\cal E}$. 
\kaku{As we will see below,} ${\cal E}(M_* v; u=u_0)$ on the 
$u = u_0$ slice 
can be written down easily by using ${\cal E}(M_* v; u=0)$ on the 
$u = 0$ slice. 

The $\widetilde{\varphi}^{(1)}$ in (\ref{eq:varphi-u0-slice}) is 
different from that in (\ref{eq:varphi-0-slice}) in only two points. 
\kaku{One of them is that} the first term of (\ref{eq:varphi-u0-slice}) 
is absent in (\ref{eq:varphi-0-slice}), and 
\kaku{the other is that} the upper-right entry $M_* v$ 
in (\ref{eq:varphi-0-slice}) is shifted by the constant 
$(c'/c)^2 (M_* u_0)^2$ in (\ref{eq:varphi-u0-slice}).
\kaku{In the holomorphic frame, the first of the Hitchin equations 
(\ref{eq:Hitchin}) requires that $\tilde\varphi^{(1)}$ be holomorphic 
in the variable $v$ with no connections, 
and it obviously follows, as can be seen in (\ref{eq:varphi-u0-slice}).}
Since the first term of (\ref{eq:varphi-u0-slice}) does not make 
any difference in the commutator 
$[ \widetilde{\varphi}^{(1)}, \widetilde{\overline{\varphi}}^{(1)}]$ 
or $[ \varphi^{(1)}, \overline{\varphi}^{(1)}]$, we can simply use 
\begin{equation}
 {\cal E}(M_* v; u=u_0) = {\cal E}(M_* v + (c'/c)^2 (M_* u_0)^2; u=0) 
\end{equation}
as the complexified gauge transformation
\kaku{to find a solution to the second one of the Hitchin equations 
(\ref{eq:Hitchin}).} 
The soliton configuration obtained in this way is perfectly smooth 
in \kaku{the} 
holomorphic frame, even at the branch locus 
$M_* v = - (c'/c)^2 (M_* u_0)^2$. It is also smooth in 
\kaku{the} unitary frame, at least when the 
\kaku{gauge $\underline{\cal E}^2 = \sqrt{{\cal H}^2}$} is chosen. 

The non-commutativity between $\varphi^{(1)}$ and
$\overline{\varphi}^{(1)}$ diminishes exponentially 
as in (\ref{eq:asymptotic-ramify}), with $M_* v$ simply replaced by 
$M_* v + (c'/c)^2 (M_* u_0)^2$. 
The two eigenvalues of $\varphi^{(1)}$ are 
\begin{equation}
[- \underline{c}' (M_*u_0) \pm 
  \sqrt{\underline{c}^2 (M_* v) + \underline{c}^{'2} (M_* u_0)^2}] \;
  d(M_* v).
\end{equation}
%

%%%%%%%%%%%%%%%%%%%%%%%%%%%%%%%%%%%%%%%%%%%%%%%%%
\subsubsection{An Approximate Solution on a Complex Surface}
%%%%%%%%%%%%%%%%%%%%%%%%%%%%%%%%%%%%%%%%%%%%%%%%%

We have constructed a local solution to the Hitchin equation on 
each \kaku{constant $u$} slice of a complex surface $U_a$.
This approach was pushed forward under an intuition that 
a collection of \kaku{the} ``solutions'' on 
\kaku{the} individual slices must be 
\kaku{adiabatically sewn to give} 
an approximate solution to the BPS equation on the surface $U_a$, 
if $c'$---the rate of the variation in the $u$ direction---is 
small enough. Let us see that this intuition is correct. 

Let us assume a configuration 
\begin{equation}
 \alpha \widetilde{\varphi} = 
  M_* \left[ - c' (M_* u) \; {\bf 1} + c 
    \left( \begin{array}{cc} 
       & (M_* v) + (c'/c)^2 (M_* u)^2 \\ 1 & 
    \end{array} \right) \right] \; 
  du \wedge dv
\label{eq:vortex-phi-s-nonzero}
\end{equation}
in a holomorphic frame.
% \footnote{Here, $c'$ and $c$ are the original
% ones, not the rescaled ones in footnote \ref{fn:2dimTo1dim}.} 
As a $2 \times 2$ matrix valued complexified gauge transformation, we adopt 
\begin{equation}
 {\cal E}(u, v) = {\cal E}(M_* v + (c'/c)^2 (M_* u)^2; 0); 
\end{equation}
the field configuration $\varphi$ in the unitary frame  
is determined from (\ref{eq:C}), and the gauge field from (\ref{eq:D}). 
All the entries of the matrix representation 
(\ref{eq:vortex-phi-s-nonzero}) are indeed holomorphic in local coordinates 
$(u,v)$. 
The BPS conditions (\ref{eq:BPS-H}, \ref{eq:BPS-G}) 
are now automatically satisfied 
\kaku{in the holomorphic frame.} 
\kesu{, because a holomorphic frame exists almost by definition, 
and all the entries of the matrix 
representation of $\varphi$ in this frame, $\widetilde{\varphi}$ 
above, are holomorphic in local coordinates $(u,v)$.}

In the remaining BPS equation (\ref{eq:BPS-D}), we already know from 
the construction in the preceding subsubsections that the 
field configuration obtained above satisfies 
\begin{equation}
 \left[ \frac{i}{2} h_{u\bar{u}} du \wedge d\bar{u} \right] \wedge
 F - \frac{|\alpha|^2}{2} \left[\varphi, \overline{\varphi} \right] = 0.
\end{equation}
\kaku{Comparing it with the equation (\ref{eq:BPS-D}), one finds 
the term $[(i/2) h_{v\bar{v}} dv \wedge d\bar{v}] \wedge F$ missing here.} 
If $h_{v\bar{v}}$ is very small, then the remaining BPS condition is 
also satisfied approximately. If the coordinate $v$ is rescaled to 
set $h_{v\bar{v}}$ 
\kaku{to be} of order unity, then $c$ in the new coordinate 
becomes very large. Therefore, the rapid change in 
\kaku{the} $v$ direction relatively to the $u$ direction 
justifies our preceding analysis of just taking a slice in the $v$ direction 
to study the BPS field configuration. 

%%%%%%%%%%%%%%%%%%%%%%%%%%%%%%%%%%%%%%%%%%%%%%%%%%%
\subsection{Zero Mode Wavefunction around the Branch Locus}
\label{ssec:ramified-0}
%%%%%%%%%%%%%%%%%%%%%%%%%%%%%%%%%%%%%%%%%%%%%%%%%%%

In section \ref{ssec:ramified-bg}, \kaku{the} 
\kaku{smooth rank-2} solution 
to the BPS equation was constructed. It can be used as a 
background field configuration in models of \kaku{the} local geometry 
where \kaku{an} $E_6$ or $A_6$ 
\kaku{singularity} 
is deformed to $A_4$ as in (\ref{eq:E6toA4}) or (\ref{eq:A6toA4}). 
\kaku{The particles of the low energy effective four-dimensional 
theory may be from the zero mode wavefunctions of the field fluctuations about 
the background.} 
Zero modes are field fluctuations on the background. 
Since a clear distinction was made between  holomorphic, unitary and 
diagonalization frames in the solution, there is no ambiguity left in 
what we mean by ``$\varphi$ background.'' With a branch cut replaced 
by this new smooth background field configuration, we no longer need 
to assume any singular/smooth behavior or boundary conditions 
of field fluctuations at the branch locus without a firm justification.
Therefore, in this section \ref{ssec:ramified-0}, we study 
the zero mode wavefunctions under the rank-2 (doublet) background, 
using the smooth field background obtained above. 
Note that the type $E_6$ codimension-3 singularities generate the 
up-type Yukawa coupling of $SU(5)$ grand unified models, and it is 
one of the main advantages in F-theory compactifications to construct 
realistic grand unified models. 
Therefore, it would be very significant to find the zero modes in 
the background and to calculate the Yukawa couplings among them, for 
F-theory grand unified model building.

\kaku{In practice,} however, the background solution was obtained 
on a local patch of \kaku{the} complex surface only approximately. 
The solution is exact \kaku{only} 
in \kaku{a} certain limit on a surface, or when a slice of the surface is 
taken so that \kaku{the} problem is reduced to that on a curve. In the following, 
we will work on the zero mode \kaku{solutions} 
in the doublet background on the $u = 0$ slice. 
\kaku{It is beyond the scope of this article} 
to find a zero mode solution 
\kaku{on the complex surface, 
and even to find the one on the slice $u = u_0 \neq 0$.} 
\kesu{in $u = u_0 \neq 0$ slice, or to find one on 
the complex surface. }

%%%%%%%%%%%%%%%%%%%%%%%%%%%%%%%%%%%%%%%%%%%%%%%%%%%%
\subsubsection{Profile of the Zero Mode Wavefunctions}
\label{sssec:profile}
%%%%%%%%%%%%%%%%%%%%%%%%%%%%%%%%%%%%%%%%%%%%%%%%%%%%

The zero mode equations on a complex curve \kaku{become}
\begin{eqnarray}
\begin{split}
 i 
% (h_{u\bar{u}}) 
 D' \psi + \rho(\overline{\varphi}^{(1)}) \chi^{(1)} & = & 0, 
\\
 D'' \chi^{(1)} + i \rho(\varphi^{(1)}) \psi & = & 0.
\end{split}
\label{eq:vortex-eom-zeromode}
\end{eqnarray}
These equations are obtained in two independent ways. 
One is to consider the Hitchin 
\kaku{equations} (\ref{eq:Hitchin}) on 
the curve for an \kaku{infinitesimal fluctuation} 
\kaku{$\chi^{(1)}\equiv\delta\varphi^{(1)}$ and $\psi\equiv\delta A^{0,1}$} 
\kaku{about a background} satisfying (\ref{eq:Hitchin}). 
The other is to replace \kaku{$\alpha \chi^{(2,0)}$ }
in the zero mode equations (\ref{eq:0-eq-D}--\ref{eq:0-eq-G}) by 
$[\sqrt{h_{u\bar{u}}} du] \wedge \chi^{(1)}$. Either way, 
the same zero mode equations are obtained. 

Note that 
\kaku{the form of the fluctuations 
\begin{equation}
  \delta \varphi^{(1)} \equiv \chi^{(1)} = i \rho(\varphi^{(1)}) \Lambda, 
\qquad 
  \delta (i A^{0,1}) \equiv i \psi = i D'' \Lambda, 
\label{eq:pure-gauge-A}
\end{equation}
with 
\begin{equation}
 \left[ D' D'' + \rho(\overline{\varphi}^{(1)}) \rho(\varphi^{(1)}) \right] 
    \Lambda  = 0.
 \label{eq:pure-gauge-B}
\end{equation}
trivially satisfy the zero mode equations (\ref{eq:vortex-eom-zeromode}).
}
\kesu{the following form of fluctuations always satisfy the 
zero mode equations:
\begin{equation}
  \delta \varphi^{(1)} \equiv \chi^{(1)} = i \rho(\varphi^{(1)}) \Lambda, 
\qquad 
  \delta (i A^{0,1}) \equiv i \psi = i D'' \Lambda, 
\label{eq:pure-gauge-A}
\end{equation}
with 
\begin{equation}
 \left[ D' D'' + \rho(\overline{\varphi}^{(1)}) \rho(\varphi^{(1)}) \right] 
    \Lambda  = 0.
 \label{eq:pure-gauge-B}
\end{equation}
}
The parameter $\Lambda$ can be regarded as a gauge transformation 
parameter in the off-diagonal components $(U_I, R_I)$, 
like $({\bf 2}, {\bf 10})$ in (\ref{eq:E6toA4}) 
or $({\bf 2}, \bar{\bf 5})$ in (\ref{eq:A6toA4}). 
The expression (\ref{eq:pure-gauge-A}) \kaku{should be} regarded as 
the unitary-frame version of (\ref{eq:f2psi}, \ref{eq:f2chi}), 
with $\Lambda$ the unitary frame version of $\tilde{\Lambda}$, 
and $\tilde{f} = 0$. Since 
\kaku{a charged matter field corresponds to} 
\kaku{a non-trivial holomorphic section} 
$\tilde{f}$ on the matter curve, 
the ``zero mode'' in the form of (\ref{eq:pure-gauge-A}) 
are not really physical zero mode. 
We thus need to subtract 
this type of ``solution'' from 
\kaku{the set of all the solutions satisfying the differential 
equations (\ref{eq:vortex-eom-zeromode}) to obtain the physical spectrum.}
\kesu{what looks like a solution to the zero mode equations.}

The $U(2)$ doublet wavefunctions $\psi$ and $\chi^{(1)}$ 
\kaku{in the doublet background} 
are denoted by $(\psi_\uparrow, \psi_\downarrow)^T$ 
and $(\chi^{(1)}_\uparrow, \chi^{(1)}_\downarrow)^T$, respectively. 
In (\ref{eq:F-Yukawa}), for example, $A,B,C$ 
\kaku{labeled} \kaku{the} weights of 
\kaku{a $U_I$ representation} of a structure group $G'$. 
\kaku{Here, we have} $G' = \U(2)$, $U_I = {\bf 2}$, and 
$A, B = \uparrow, \downarrow$.
Since $\rho(\varphi)$ and $\rho(\overline{\varphi})$ are off-diagonal, 
and $D'$ and $D''$ diagonal in the $2 \times 2$ matrix, 
each of the zero mode equations splits into two. 
One is for \kaku{the pair ($\psi_\uparrow$, $\chi^{(1)}_\downarrow$), } 
and the other for \kaku{the pair ($\psi_\downarrow$, $\chi^{(1)}_\uparrow$).} 
\kesu{a pair of $\psi_\downarrow$ and $\chi^{(1)}_\uparrow$.} 
 
We will use \kaku{the} holomorphic frame, 
because it looks a little \kaku{more} convenient \kaku{than the unitary frame}. 
$\rho(\widetilde{\overline{\varphi}})$ is found 
in (\ref{eq:phidagg-hol-frame}), $i \tilde{A}^{(0,1)} = 0$ and 
\begin{equation}
 i \tilde{A}^{(1,0)} = - \tau^3 \frac{1}{2} \partial \ln {\cal H}^2,
\end{equation}
which follows from (\ref{eq:rel-A10-H}).
For convenience, we denote $M_*v$ as $v$ from now on, making 
everything dimensionless.

\kesu{The first} \kaku{One} set of the zero-mode \kaku{equations} is 
\begin{eqnarray}
  \bar{\partial}(\tilde{\chi}^{(1)}_{\downarrow})_v - 
  \underline{c} (i \tilde{\psi}_{\uparrow})_{\bar{v}}  & = & 0,  
      \label{eq:0eq-A} \\
  \left[ \partial - \frac{1}{2} \partial (\ln {\cal H}^2 ) \right] 
(i \tilde{\psi}_{\uparrow})_{\bar{v}} - 
    \underline{c}^* {\cal H}^2 (\tilde{\chi}^{(1)}_{\downarrow})_v 
  & = & 0. \label{eq:0eq-B}
\end{eqnarray}
%
% This equation is solved completely, by assuming 
% %
% \begin{equation}
%  (\tilde{\chi}^{(1)}_{\downarrow})_v = \sum_{m=0}^\infty A_m (v\bar{v})^m, \qquad 
%  (i \tilde{\psi}^{(1)}_{\uparrow})_{\bar{v}} = v \sum_{m=0} B_m (v\bar{v})^m;
% \end{equation}
% %
% Once $A_0$ is given, all other coefficients are determined successively
% by the zero mode equations. There is no freedom to come in. 
% Since the zero mode equation is linear in the fluctuation fields 
% $(\tilde{\chi}^{(1)}, \tilde{\psi}^{(1)})$, $A_0$ must be free. 
% So, everything is fixed. 
%
For any solutions 
$((\tilde{\psi}_{\uparrow})_{\bar{v}}, (\tilde{\chi}^{(1)}_\downarrow)_v)$ of 
(\ref{eq:0eq-A}, \ref{eq:0eq-B}), one can define 
$\tilde{\Lambda}_\uparrow$ by 
$i \underline{c} \tilde{\Lambda}_\uparrow = (\tilde{\chi}^{(1)}_\downarrow)_v$.  
Now, from (\ref{eq:0eq-A}), $(\tilde{\psi}_{\uparrow})_{\bar{v}} = 
 \bar{\partial}_{\bar{v}} \tilde{\Lambda}_\uparrow$. 
The equation (\ref{eq:0eq-B}) can then be rewritten in terms of 
$\tilde{\Lambda}_\uparrow$, and it becomes a constraint equation on 
$\tilde{\Lambda}_\uparrow$; this constraint equation turns out 
to be exactly the same as (\ref{eq:pure-gauge-B}) 
\kaku{for $\Lambda = (\Lambda_\uparrow, 0)^T$ in the holomorphic frame.} 
\kaku{Since} all the ``zero-mode solutions'' of (\ref{eq:0eq-A}, \ref{eq:0eq-B}) 
are of the form \kesu{of} (\ref{eq:pure-gauge-A}), 
\kaku{the pair 
$((\tilde{\psi}_{\uparrow})_{\bar{v}}, (\tilde{\chi}^{(1)}_\downarrow)_v)$ 
does not contribute to the physical spectrum.} 

\kesu{The second (and the remaining)} 
\kaku{The other} set of the zero-mode \kaku{equations} is 
\begin{eqnarray}
 \bar{\partial} (\tilde{\chi}^{(1)}_\uparrow ) - 
    \underline{c} v (i \tilde{\psi}_\downarrow) & = & 0, \label{eq:0eq-C} \\
 \left[ \partial + \frac{1}{2} \partial (\ln {\cal H}^2) \right]
   (i \tilde{\psi}_\downarrow ) - 
    \underline{c}^* {\cal H}^{-2} \bar{v} (\tilde{\chi}^{(1)}_\uparrow)
   & = & 0. 
    \label{eq:0eq-D}
\end{eqnarray}
Assuming that the zero mode solution is smooth at the origin, $v=0$, 
\kaku{we} consider the most generic form of series expansion 
\begin{eqnarray}
 (\tilde{\chi}^{(1)}_\uparrow)_v  & = & 
  \sum_{m=0}^\infty A_m |v|^{2m} + \sum_{p > 0, \; m=0}^\infty A_{p+m,m}v^p |v|^{2m}
  + \sum_{q > 0, \; m=0}^\infty A_{m,m+q} \bar{v}^q |v|^{2m}, \\
 (\tilde{\psi}_\downarrow)_{\bar{v}}  & = &   
  \sum_{m=0}^\infty B_m |v|^{2m} + \sum_{p > 0, \; m=0}^\infty B_{p+m,m}v^p |v|^{2m}
  + \sum_{q > 0, \; m=0}^\infty B_{m,m+q} \bar{v}^q |v|^{2m}. 
\end{eqnarray}
\kaku{Substituting the series expansions into the zero mode equations 
(\ref{eq:0eq-C}, \ref{eq:0eq-D}) yields one independent set of recursion 
equations for each set of $\{A_m, B_m\}_{m\in\N}$, 
$\{A_{p+m,m}, B_{p+m,m}\}_{m\in\N}$ ($p=1,2,3,\cdots$), 
and $\{A_{m,m+q}, B_{m,m+q}\}_{m\in\N}$ ($q=1,2,3,\cdots$).} 
\kaku{This may be understood by seeing that the derivative 
$\bar\partial$ and the multiplication of $v$ in (\ref{eq:0eq-C}) 
do not relate one of the sets to another, 
and neither do $\partial$ and $\bar{v}$ 
in (\ref{eq:0eq-D}) with the function ${\cal H}$ of $|v|$.}
\kesu{The zero-mode equations (\ref{eq:0eq-C}, \ref{eq:0eq-D}) now 
split into three groups of terms. 
One is for those of $v^p |v|^{2m}$ series for individual $p$'s, 
where coefficients $(A_{p+m,m}, B_{p+m,m})$'s ($m \in \N$, but $p$
fixed) are involved. The second group is the $\bar{v}^q |v|^{2m}$ terms 
for individual $q$'s, and the equation for a given $q$ relates 
$(A_{m,m+q}, B_{m,m+q})$'s ($m \in \N$, but $q$ fixed). Finally, 
the $|v|^{2m}$ terms in the series expansion of 
$\tilde{\psi}_\downarrow$ and $\tilde{\chi}^{(1)}_\uparrow$ 
for a coupled equation, where $(A_m, B_m)$'s ($m \in \N$) are constrained. }

One can see that the $\bar{v}^q |v|^{2m}$ series $(q > 0)$ cannot satisfy 
both (\ref{eq:0eq-C}) and (\ref{eq:0eq-D}).
If $\tilde{\psi}$ begins with $v^n \bar{v}^{n+q}$ for some $n \geq 0$, 
then $v^{n+1}\bar{v}^{n+1+q}$ should be the first term of $\tilde{\chi}$, 
as we see from (\ref{eq:0eq-C}). \kesu{In (\ref{eq:0eq-D}), however,} 
\kaku{However,} 
the first term \kaku{of (\ref{eq:0eq-D})} \kaku{then,} 
starts from either $v^{n-1}\bar{v}^{q+n}$ (if $n \geq 1$) 
or $v^n\bar{v}^{q+n+1}$ (if $n=0$), but the second term of
(\ref{eq:0eq-D}) starts from $v^{n+1}\bar{v}^{q+n+2}$. We can never 
get the lowest-order term \kesu{right in} 
\kaku{satisfying} both (\ref{eq:0eq-C}) and (\ref{eq:0eq-D}).  

For any series solutions with $p > 0$, one can define 
$\tilde{\Lambda}^{(p>0)}_\downarrow$ by 
\begin{equation}
 (\tilde{\chi}^{(1)}_\uparrow)_v|_{p > 0} 
   = v \sum_{p > 0, \; m=0}^\infty A_{p+m,m} v^{p-1}|v|^{2m} 
   \equiv i \underline{c} v \tilde{\Lambda}^{(p>0)}_\downarrow.
\end{equation}
Then, from (\ref{eq:0eq-C}), 
\begin{equation}
 (\tilde{\psi}_\downarrow)_{\bar{v}}|_{p> 0} 
   = \sum_{p > 0, \; m=0}^\infty B_{p+m,m} v^p |v|^{2m} 
   = \bar{\partial}_{\bar{v}} \tilde{\Lambda}^{(p>0)}_\downarrow.
\end{equation}
That is exactly the relation (\ref{eq:pure-gauge-A}) 
in \kaku{the} holomorphic frame.
Finally, (\ref{eq:0eq-D}) becomes a constraint on 
$\tilde{\Lambda}^{(p>0)}_\downarrow$, and turns out to be the same 
as the holomorphic frame version of (\ref{eq:pure-gauge-B}) for 
$\tilde{\Lambda} = (0, \tilde{\Lambda}^{(p>0)}_\downarrow)^T$.
Thus, the $v^p |v|^{2m}$ series with $p > 0$ are of the form 
of (\ref{eq:pure-gauge-A}, \ref{eq:pure-gauge-B}), and are not 
\kaku{physical degrees of freedom.} 
\kesu{the zero mode solutions we look for.} 

Therefore, the zero mode solution must lie within the $|v|^{2m}$ series.
Let us first define $\tilde{\Lambda}^{(p=0)}_\downarrow$ as 
\begin{equation}
 \tilde{\Lambda}^{(p=0)}_\downarrow \equiv 
  \frac{1}{i\kaku{\underbar{c}}} \sum_{m=1} A_m v^{m-1} \bar{v}^{m}, 
 \quad {\rm so~that} \quad 
 (\tilde{\chi}^{(1)}_\uparrow)^{(p=0)}_v = 
     (i \kaku{\underbar{c}} v) \tilde{\Lambda}^{(p=0)}_\downarrow + A_0. 
\end{equation}
Using (\ref{eq:0eq-C}), we find that 
\begin{equation}
(i \tilde{\psi}_\downarrow)^{(p=0)}_{\bar{v}} = 
  i \bar{\partial}_{\bar{v}} \tilde{\Lambda}^{(p=0)}_\downarrow. 
\end{equation}
All the $B_m$'s in the series expansion of $\tilde{\psi}_\downarrow$ 
are given by $A_m$'s in $\tilde{\chi}^{(1)}$. The remaining zero-mode 
equation (\ref{eq:0eq-D}) determines all the $A_m$'s ($m \geq 2$) 
for arbitrary choice of $A_0$ and $A_1$. Thus, the zero mode solution 
\kesu{under} \kaku{in} the doublet background 
on the $u=0$ slice \kesu{have} \kaku{has} two free parameters $A_0$ and $A_1$.

Among the two parameters $A_0$ and $A_1$ of the zero mode solution, 
only one is free for zero mode (hypermultiplet) wavefunctions localized 
around the matter curves. To see this, note that the 
equations (\ref{eq:0eq-C}, \ref{eq:0eq-D}) on the $|v|^{2m}$ series
can be rewritten as 
\begin{equation}
 (\tilde{\chi}^{(1)}_\uparrow)'_v = \underline{c} 
      (i \tilde{\psi}_\downarrow)_{\bar{v}}, \qquad 
 (i \tilde{\psi}_\downarrow)'_{\bar{v}} + 
 \frac{({\cal H}^2)'}{2{\cal H}^2}(i \tilde{\psi}_\downarrow)_{\bar{v}}  = 
   \frac{\underline{c}^* }{{\cal H}^2}(\tilde{\chi}^{(1)}_\uparrow)_v,  
\label{eq:0eq-E}
\end{equation}
where $'$ is a derivative with respect to $w = |v|^2$. 
Combining these two equations, we obtain a rank-2 linear 
differential equation on $ w \in [0, \infty] \subset \R^1$, 
\begin{equation}
  (\tilde{\chi}^{(1)}_\uparrow)^{''}_v + 
  \frac{({\cal H}^2)'}{2{\cal H}^2}(\tilde{\chi}^{(1)}_\uparrow)^{'}_v 
  = \frac{|\underline{c}|^2}{{\cal H}^2} (\tilde{\chi}^{(1)}_\uparrow)_v.
\label{eq:0mode-asymp-eq}
\end{equation}
The two parameters $A_0$ and $A_1$ correspond to two boundary 
conditions 
$(\tilde{\chi}^{(1)}_\uparrow)_v|_{w=0} = A_0$, and 
$(\tilde{\chi}^{(1)}_\uparrow)'_v|_{w=0} = A_1$ at $w = 0$.  
Thus, the freedom in the two parameters $A_0$ and $A_1$ correspond 
to two independent solutions of the rank-2 differential equation linear 
in $(\tilde{\chi}^{(1)})_v$. Since the zero mode (hypermultiplet)
wavefunctions are supposed to be localized along the matter curve 
at $w = 0$, the wavefunctions should become small quickly at large $w$.
Under this boundary condition at large $w$, one can no longer choose 
$A_1$ independently from $A_0$.
% In the large $|v|$ region, we already know that 
% ${\cal H}^2$ approaches $|v| = w^{1/2}$. Thus, in this region, 
% %
% \begin{equation}
%  (\tilde{\chi}^{(1)}_\uparrow) \sim 
%      \exp \left[ \pm \frac{4}{3} |\underline{c}| w^{\frac{3}{4}} \right] 
%  = e^{ \pm \frac{4}{3} \frac{|c|}{\sqrt{h_{u\bar{u}}}} |v|^{\frac{3}{2}} }.
%  \label{eq:asymptotic-chi-1}
% \end{equation}
% %
% are approximately the two independent solutions. 
The ratio between $A_0$ and $A_1$ should be determined so that the 
zero mode solution is normalizable in the $v$ direction (on the $u=0$ slice).
Thus, only one parameter $A_0$ remains free. 

All the series expansion solution combined becomes the form of 
\begin{eqnarray}
(\tilde{\chi}^{(1)})_v & = & \left(\begin{array}{r}
   (\tilde{\chi}^{(1)}_\uparrow )_v \\ (\tilde{\chi}^{(1)}_\downarrow)_v 
   \end{array} \right) 
  = i \rho_{\bf 2} \left( \widetilde{\varphi}^{(1)}_v \right) 
   \left( \begin{array}{c} 
   \tilde{\Lambda}_\uparrow \\ 
   \tilde{\Lambda}^{(p>0)}_\downarrow + \tilde{\Lambda}^{(p=0)}_\downarrow
   \end{array} \right) 
 + \left( \begin{array}{c} A_0 \\ \; \end{array} \right), \\
 (\tilde{\psi})_{\bar{v}} & = & \left(\begin{array}{r}
   (\tilde{\psi}_\uparrow )_{\bar{v}} \\ (\tilde{\psi}_\downarrow)_{\bar{v}} 
   \end{array} \right) 
  = \bar{\partial}_{\bar{v}}    \left( \begin{array}{c} 
   \tilde{\Lambda}_\uparrow \\ 
   \tilde{\Lambda}^{(p>0)}_\downarrow + \tilde{\Lambda}^{(p=0)}_\downarrow
   \end{array} \right) .
\end{eqnarray}
This is an exact solution of the zero mode equations on \kesu{a} 
\kaku{the} complex curve 
given by $u=0$, but it is also regarded as an approximate solution 
\kesu{to the equations on a complex surface on the $u=0$ locus:}
\begin{eqnarray}
(2\alpha \tilde{\chi})_{uv} & \simeq & 2\alpha \left(\begin{array}{r}
   (\tilde{\chi}_\uparrow )_{uv} \\ (\tilde{\chi}_\downarrow)_{uv} 
   \end{array} \right) 
  = i \rho_{\bf 2} \left( 2\alpha \widetilde{\varphi}_{uv} \right) 
   \left( \begin{array}{c} 
    \\  \tilde{\Lambda}^{(p=0)}_\downarrow
   \end{array} \right) 
 + \left( \begin{array}{c} \sqrt{h_{u\bar{u}}} A_0 \\ \; \end{array}
   \right), \label{eq:0mode-branch-chi} \\
 (\tilde{\psi})_{\bar{v}} & \simeq & \left(\begin{array}{r}
   (\tilde{\psi}_\uparrow )_{\bar{v}} \\ (\tilde{\psi}_\downarrow)_{\bar{v}} 
   \end{array} \right) 
  = \bar{\partial}_{\bar{v}}    \left( \begin{array}{c} 
    \\ \tilde{\Lambda}^{(p=0)}_\downarrow
   \end{array} \right) . \label{eq:0mode-branch-psi}
\end{eqnarray}
\kaku{to the equations around the $u=0$ locus on the complex surface.}
Comparing these expressions and (\ref{eq:f2psi}, \ref{eq:f2chi}), 
it is now clear that $(\sqrt{h_{u\bar{u}}} A_0, 0)^T (du \wedge dv)$ 
corresponds to $\tilde{f}$ in (\ref{eq:f2chi}). 
$\tilde{\Lambda}_\uparrow$ and $\tilde{\Lambda}^{(p>0)}_\downarrow$
have been dropped, because they correspond to $\tilde{f} = 0$ and are 
not the solution we are interested in. 

Treating the line bundle ${\cal N}$ on the spectral surface $C$ 
locally as ${\cal O}_C$ means that a trivialization is given 
locally by taking a local generator of ${\cal N}$. Since 
we can always choose $du \wedge dv$ as a local frame of $K_{U_\alpha}$ 
\kaku{with local coordinates $(u,v)$ of $U_\alpha$,} 
\kesu{(when $(u,v)$ are local coordinates of $U_\alpha$), }
\kesu{holomorphic sections } 
\kaku{a holomorphic section} 
$\tilde{f}$ of ${\cal N} \otimes \pi_{C}^* K_{U_\alpha}$ 
on the spectral surface is described locally by 
\kesu{holomorphic functions } 
\kaku{a holomorphic function} 
\kaku{$f_{uv}(\xi, u)$} \kesu{$f_{uv}(\xi, v)$} 
on the spectral surface by using the frame of 
${\cal N} \otimes K_{U_\alpha}$. 
As we saw around (\ref{eq:varphi-0-slice}), $f_{uv}(\xi, u)$ can be 
expanded in a power series \kesu{of} \kaku{in} $(\xi + c' M_*^2 u)$ and $u$; 
the rank-2 vector bundle 
$V \otimes K_{U_\alpha} = \pi_{C*} {\cal N} \otimes K_{U_\alpha}$ 
on the base surface $U_\alpha$ is also given a local trivialization 
by \kesu{having} \kaku{assigning the} terms with even \kesu{power in} 
\kaku{powers of} $(\xi + c' M_*^2 u)$ \kesu{in} \kaku{to} the 
first component $A = \uparrow$ and those 
with odd \kaku{powers} \kesu{power} \kesu{in} \kaku{to} the second 
component $A = \downarrow$. 
\begin{equation}
  \Gamma (U_\alpha; V \otimes K_{U_\alpha}) \ni \tilde{f}
 \longleftrightarrow
  \left(\begin{array}{c} (f_\uparrow)_{uv} \\
            (f_\downarrow)_{uv} \end{array}
  \right) = 
  \left(\begin{array}{r} \sum_j f_{2j}(u) 
   \{ c^2 M_*^2 ((M_* v) + (c'/c)(M_* u)^2 ) \}^j \\
  % (\xi + c' M_*^2 u) 
   \sum_j f_{2j+1}(u) \{ c^2 M_*^2 ((M_* v) + (c'/c)(M_* u)^2 ) \}^j 
   \end{array}
 \right)
\end{equation}
Therefore, the approximate solution (\ref{eq:0mode-branch-chi}) along 
$u = 0$ corresponds to a holomorphic section $\tilde{f}$ of 
$V \otimes K_{U_\alpha}$ with $f_{0}(u = 0) = \sqrt{h_{u\bar{u}}} A_0$.
As we discussed in \cite{Hayashi-2} and in section \ref{ssec:uni.vs.hol} 
of this article, only the holomorphic sections $\tilde{f}$ mod $\xi$ 
correspond to charged matter zero modes. Thus, all other $f_i(u)$'s with 
$i > 0$ are irrelevant. For a zero mode given as a global holomorphic 
section $\tilde{f}$ of 
\kaku{${\cal N} \otimes \pi_{C}^* K_{S}$}
\kesu{$({\cal N} \otimes \pi_{C}^* K_{S})$ }
on a matter curve, a local trivialization patch needs to be taken 
around a $E_6$ or $A_6$ type \kaku{singularity} point; when $\tilde{f}_a$ is the
coefficient holomorphic function in the trivialization patch, then 
its value at \kaku{the codimension-3 singularity point} $u = 0$ 
\kesu{(the codimension-3 singularity point)} is \kesu{used as} 
\kaku{given by} $\sqrt{h_{u\bar{u}}} A_0$; 
$A_0$ then determines all the coefficients 
$A_m$ in $\tilde{\Lambda}^{(p=0)}_\downarrow$ by following the procedure 
we already explained. 

Before we proceed to the next subsubsection, we should make a brief
remark. Reference \cite{Hayashi-2} already studied the zero mode 
solution in the doublet background in great detail, 
\kesu{with a particular focus}
\kaku{exclusively focusing} on \kaku{the} $u = 0$ slice, as we did above. 
The zero mode ``solution'' 
\kesu{there} is not exactly the same as those here. Two possible reasons are 
i) \kaku{a} distinction \kesu{between} \kaku{among} various frames 
were not made clearly in \cite{Hayashi-2}, 
and ii) since the branch locus was a singular 
locus in the description adopted in Ref. \cite{Hayashi-2}, there was 
no rigorous justification for \kaku{a} choice of boundary conditions at the 
singular locus. We suspect that i) is the major reason, but ii) may 
also be relevant. 

Since both of the difference i) and ii) are associated 
with the region near the branch locus (or the soliton background that
replaces it), there should be no difference between the solution in 
\cite{Hayashi-2} and the one in this article in a region away from 
the branch locus. 
Remembering that ${\cal H}^2$ approaches $|v| = w^{1/2}$ at large $w$, 
we can find the asymptotic behavior of the two independent solutions 
to (\ref{eq:0mode-asymp-eq}):
\begin{equation}
 (\tilde{\chi}^{(1)}_\uparrow) \sim 
     \exp \left[ \pm \frac{4}{3} |\underline{c}| w^{\frac{3}{4}} \right] 
 = e^{ \pm \frac{4}{3} \frac{|c|}{\sqrt{h_{u\bar{u}}}} |v|^{\frac{3}{2}} }.
 \label{eq:asymptotic-chi-1}
\end{equation}
The ratio $A_1/A_0$ should be chosen properly so that the coefficient of 
the $(\tilde{\chi}^{(1)}_\uparrow) \propto  
\exp [+ (4/3) |\underline{c}| |v|^{3/2} ]$ 
component vanishes. The asymptotic behavior of the hypermultiplet 
wavefunction is $(\tilde{\chi}^{(1)}_\uparrow) \propto  
A_0 \exp [ - (4/3) |\underline{c}| |v|^{3/2} ] $. 
This is exactly the same as the asymptotic behavior obtained 
in \cite{Hayashi-2}, as expected. 

The wavefunctions obtained rigorously in this article should be
different only in the small $|v|$ region. When the solution 
(\ref{eq:0mode-branch-chi}, \ref{eq:0mode-branch-psi}) is expressed 
in the $\varphi$-diagonalization frame, 
\begin{equation}
 \chi_{uv}|_{u=0} = \frac{1}{2\sqrt{v}} \left( \begin{array}{r} 
   (\tilde{\chi}_\uparrow)_{uv} \\ - (\tilde{\chi}_\uparrow)_{uv} 
    \end{array} \right), \qquad 
 \psi_{\bar{v}}|_{u=0} = \frac{1}{2} \left( \begin{array}{r}
   (\tilde{\psi}_\downarrow)_{\bar{v}} \\  (\tilde{\psi}_\downarrow)_{\bar{v}} 
    \end{array} \right).    
\end{equation}
Now a $1/\sqrt{v}$ singularity appears in $\chi$, whereas the singularity 
appeared in $\psi$ in \cite{Hayashi-2}. \kesu{Analysis} \kaku{The analysis} 
in this article is correct, and hence some of the statements 
in \cite{Hayashi-2} 
on the profile of the zero mode wavefunctions \kaku{near $v=0$} 
need to be superseded by those presented in this appendix. 

%%%%%%%%%%%%%%%%%%%%%%%%%%%%%%%%%%%%%%%%%%%%%%%%%%%%%%
\subsubsection{Metric Dependence of the Zero Mode Wavefunctions}
%%%%%%%%%%%%%%%%%%%%%%%%%%%%%%%%%%%%%%%%%%%%%%%%%%%%%%

The Hitchin equation \kaku{(\ref{eq:Hitchin-2})} 
for the background field configuration \kesu{(\ref{eq:Hitchin-2})} 
depends on the metric along the direction 
of \kaku{the} matter curve $h_{u\bar{u}}$ through 
$\underline{c} = c/\sqrt{h_{u\bar{u}}}$. The zero mode equation also 
depends on the same component of the metric through $\underline{c}$, 
as we see explicitly in (\ref{eq:0eq-C}, \ref{eq:0eq-D}). 
Let us clarify how \kesu{the configuration of} 
the background and 
\kaku{the zero modes} \kesu{fluctuation} 
\kesu{depends} \kaku{depend} 
on this $h_{u\bar{u}}$.

Almost all the information on the background field configuration 
follows from \kaku{the} functions ${\cal H}^2$ and $h(w)$. It is thus 
sufficient to study how those functions depend on $h_{u\bar{u}}$.
Examining (\ref{eq:Hitchin-2}), we find that 
\begin{equation}
 {\cal H}^2 \left(v, |c|^2/h_{u\bar{u}} \right) = 
  (h_{u\bar{u}}/|c|^2)^{1/3} \; {\cal H}^2_* (v (|c|^2/h_{u\bar{u}})^{1/3}).
\end{equation}
Because $H_0$ is the value of ${\cal H}^2$ at $v = 0$, 
we expect $H_0 = H_{0*} (h_{u\bar{u}}/|c|^2)^{1/3}$. 
The values of $H_0$ obtained numerically in p.\pageref{pg:H0-numer} 
satisfy a relation $H_0 \simeq 0.53 \times (1/|\underline{c}|^2)^{1/3}$ 
indeed. 
The asymptotic behavior of $h(w)$ in (\ref{eq:asymptotic-ramify}) 
depends on \kesu{a} \kaku{the} combination 
$(|c|/\sqrt{h_{u\bar{u}}}) \times |v|^{\kaku{3/2}}$;  
this argument is the same as that of ${\cal H}^2_*$ above, and 
is consistent with the discussion above.  

To study the metric dependence of the zero mode wavefunctions, 
(\ref{eq:0mode-asymp-eq}) is the easiest place to start. 
Since 
$|\underline{c}|^2/{\cal H}^2 = |\underline{c}|^{8/3}/{\cal H}^2_*$ 
\kesu{is} \kaku{has} 
the only explicit dependence on the metric $h_{u\bar{u}}$ 
in (\ref{eq:0mode-asymp-eq}), we find that the zero mode wavefunction 
$(\tilde{\chi}_{\uparrow}^{(1)})_v$ depends on the coordinate $w$
only in the form of $w |\underline{c}|^{4/3}$. The asymptotic form of 
$\tilde{\chi}_\uparrow^{(1)}$ indeed depends on this combination 
in (\ref{eq:asymptotic-chi-1}). 
Because $\alpha \tilde{\chi}_{uv} du \wedge dv$ on \kesu{a} 
\kaku{the} complex surface 
needs to be matched onto a holomorphic section $\tilde{f}$ on 
the matter curve $v=0$, $\alpha \tilde{\chi}_{uv}$ should not have 
other overall dependence on $h_{u\bar{u}}$. That is, 
\begin{equation}
 (\alpha \tilde{\chi}_{\uparrow})_{uv}(v,\bar{v},c,h_{u\bar{u}}) 
  = (\alpha \tilde{\chi}_\uparrow)_{uv*}  ( w |\underline{c}|^{4/3})\kaku{.} 
  \kesu{,}  
\end{equation}
Using the first one of the two relations in (\ref{eq:0eq-E}), 
we also find that the zero mode wavefunctions depend on $h_{u\bar{u}}$ 
only through 
\begin{equation}
  (\tilde{\chi}^{(1)}_\uparrow)_v =  \frac{1}{\sqrt{h_{u\bar{u}}}} 
      (\tilde{\chi}^{(1)}_\uparrow)_{v*} (w |\underline{c}|^{4/3}),
      \qquad 
  \kaku{i}(\tilde{\psi}_\downarrow)_{\bar{v}} =
     \frac{|c|^{\kaku{4/3}}}{\kaku{c}(h_{u\bar{u}})^{2/3}}
  (\tilde{\chi}^{(1)}_\uparrow)'_{v*} (w |\underline{c}|^{4/3}).
\label{eq:metric-dep}
\end{equation}
%

%%%%%%%%%%%%%%%%%%%%%%%%%%%%%%%%%%%%%%%%%%%%%%%%%%%%%%%
\subsubsection{Inner Product of Zero Modes at the Branch Points}
\label{sssec:reduced-rmfy-pt}
%%%%%%%%%%%%%%%%%%%%%%%%%%%%%%%%%%%%%%%%%%%%%%%%%%%%%%%

The \kesu{Kinetic} \kaku{kinetic} mixing matrices in the low-energy effective 
theory (\ref{eq:K.F.-truncation}) are expressed as an 
integral over the GUT divisor $S$. The zero mode wavefunctions 
of GUT-charged matter fields, however, are localized along 
\kaku{their} matter curves, and hence we can think of reducing the 
expression (\ref{eq:K.F.-truncation}) into an integral 
over \kesu{a} \kaku{the} matter curve, by carrying out \kaku{the} integration 
in the directions normal to the matter \kaku{curve.} 
\kesu{curves.} 

This was done in section \ref{ssec:KMM-curve}, and the result 
was (\ref{eq:K.F.-curve}). The integration can be carried out 
in the normal directions 
because the zero mode \kaku{wavefunction} \kesu{wavefunctions} 
are known to have \kaku{a} Gaussian profile in \kaku{the} normal directions 
at arbitrary points on \kaku{the} matter \kaku{curve} \kesu{curves} 
that are far away from $A_6$ or $E_6$ type \kaku{singularity} points. 

In this appendix, we have determined the profile of zero mode 
wavefunctions in the normal directions of matter curves, even 
at $A_6$ and $E_6$ type \kaku{singularity} points, 
where the spectral surfaces are 
ramified. Thus, by carrying out the integration in the normal 
directions, we will obtain an expression contributing to the 
effective-theory kinetic mixing matrices arising from such 
ramification points. 

\kesu{Contribution} \kaku{A contribution} to the kinetic mixing matrix 
(\ref{eq:K.F.-truncation}) \kesu{at} \kaku{on} the $u = 0$ slice is 
\begin{equation}
 \Delta K_{i\bar{j}} = 
 \frac{M_*^4}{4\pi} d^2 u h_{u\bar{u}} \; \int_{u=0} d^2v h_{v\bar{v}} 
  \; 2 \left[ 
     (\psi_j)^\dagger_{v} (\psi_i)_{\bar{v}} h^{\bar{v}v} + 
     (\chi^{(1)}_j)^\dagger_{\bar{v}} (\chi^{(1)}_i)_v h^{\bar{v}v}
      \right], 
\label{eq:vortex-delta-kinetic}
\end{equation}
where $d^2 u h_{u\bar{u}}$ and $d^2v h_{v\bar{v}}$ are the volume
\kesu{measure} \kaku{measures} 
in the tangential and normal directions\kaku{, respectively,} 
of the matter curve at $v = 0$. 
Appropriate mass dimensions are restored in the following 
calculations by replacing the argument $w$ in the component-field 
wavefunctions $\psi_{\bar{v}}(w)$ and $\chi^{(1)}_v(w)$.
%  that we obtained in section \ref{sssec:profile}.  
We will omit the overall factor 
$M_*^4/(4\pi)d^2 u h_{u\bar{u}}$ from (\ref{eq:vortex-delta-kinetic})
for a moment, just to save a space.
\kesu{drop $M_*^4/(4\pi)$ and the volume measure along the matter curve. }
The integral in the normal directions becomes 
\begin{eqnarray}
& &  \int_{u=0} d^2 v \; 2 \left[
 (\tilde{\psi}_{\downarrow; j})^\dagger_{v} {\cal H}
 (\tilde{\psi}_{\downarrow; i})_{\bar{v}} + 
 (\tilde{\chi}^{(1)}_{\uparrow; j})^{\kaku{\dagger} }_{\bar{v}} {\cal H}^{-1} 
 (\tilde{\chi}^{(1)}_{\uparrow; i})_v
\right], 
\nonumber\\
& = & (2\pi) \int_0^\infty d w \;  \left[
  \frac{1}{|\underline{c}|^2 M_*^4} 
  (\tilde{\chi}^{(1)}_{\uparrow; j})^{\dagger '}_{\bar{v}} {\cal H}  
  (\tilde{\chi}^{(1)}_{\uparrow; i})^{'}_v + 
 (\tilde{\chi}^{(1)}_{\uparrow; j})^{\kaku{\dagger} }_{\bar{v}} {\cal H}^{-1} 
 (\tilde{\chi}^{(1)}_{\uparrow; i})_v
\right],
\end{eqnarray}
where (\ref{eq:0eq-E}) was used; $'$ is a derivative with respect 
to \kesu{a} \kaku{the} coordinate $w = |v|^2$ with mass-dimension $-2$.
Integrating by parts and using (\ref{eq:0mode-asymp-eq}), we find 
that this $u=0$ contribution becomes 
\begin{eqnarray}
%\begin{split}
& - (2\pi) \frac{1}{|\underline{c}|^2 M_*^4} \left[
      (\tilde{\chi}^{(1)}_{\uparrow; j})^{\dagger }_{\bar{v}} {\cal H}  
  (\tilde{\chi}^{(1)}_{\uparrow; i})^{'}_v  \right]_{w = 0} 
%\\& 
   + (2\pi) \int dw \; \left[ - \frac{1}{|\underline{c}|^2 M_*^4} 
    (\tilde{\chi}^{\kaku{(1)}}_{\uparrow; j})^{\dagger }_{\bar{v}} 
   \left( {\cal H} (\tilde{\chi}^{\kaku{(1)}}_{\uparrow; i})^{'}_v \right)' 
   +(\tilde{\chi}^{(1)}_{\uparrow; j})^{\kaku{\dagger}}_{\bar{v}} {\cal H}^{-1} 
    (\tilde{\chi}^{(1)}_{\uparrow; i})_v \right] 
\nonumber\\
&\hskip-5cm  =   - (2\pi) \frac{1}{|\underline{c}|^2 M_*^4} \left[
      (\tilde{\chi}^{(1)}_{\uparrow; j})^{\dagger }_{\bar{v}} {\cal H}  
  (\tilde{\chi}^{(1)}_{\uparrow; i})^{'}_v  \right]_{w = 0}  
%\\
%   + (2\pi) \int dw \; \left[ 
%  -  (\tilde{\chi}^{(1)}_{\uparrow; j})_{\bar{v}} {\cal H}^{-1} 
%  (\tilde{\chi}^{(1)}_{\uparrow; i})_v 
%  +  (\tilde{\chi}^{(1)}_{\uparrow; j})_{\bar{v}} {\cal H}^{-1} 
%  (\tilde{\chi}^{(1)}_{\uparrow; i})_v    \right]
%& = &
   = (2\pi) \frac{H_{0}^{\kaku{1/2}}}{|c|^2 M_*^2} \tilde{f}^*_j \tilde{f}_i 
   \left( - \frac{A_1}{A_0}\right). 
%\end{split}
\end{eqnarray}
Let us factor out $|\underline{c}|^{4/3}$ from $(A_1/A_0)$ to define 
$(A_1/A_0)_*$ that does not depend on $c$ or $h_{u\bar{u}}$. 
Finally \kesu{a} \kaku{the} result is obtained: 
\begin{equation}
 \Delta K^{(R_I)}_{i\bar{j}} = d^2 u h_{u\bar{u}} \; 
   \frac{M_*^2}{2 |c|} H^{1\over2}_{0*} \left(- \frac{A_1}{A_0} \right)_* \; 
   \left( \tilde{f}_j^* \sqrt{h^{u\bar{u}}} \tilde{f}_i \right).
\label{eq:K.F.-rmfy-pt}
\end{equation}
All of $H_{0*}$, $c$, $(A_1/A_0)_*$ are dimensionless.

%%%%%%%%%%%%%%%%%%%%%%%%%%%%%%%%%%%%%%%%%%%%%%%%%%%%%%%%%%%%%%%
\section{\kaku{$\Z_2$}\kesu{$Z_2$} Transformation on Matter Line Bundles}
\label{sec:Z2-assign}
%%%%%%%%%%%%%%%%%%%%%%%%%%%%%%%%%%%%%%%%%%%%%%%%%%%%%%%%%%%%%%

In this appendix, \kesu{we will present explicitly in Example VII of 
section \ref{ssec:Ex-7}}\kaku{for Example VII of section
\ref{ssec:Ex-7}, we will explicitly illustrate} how \kesu{
the case A}\kaku{the} $\Z_2$ transformation 
on $S = \P^2$ \kaku{in the case A} is lifted to those 
on the matter line bundles ${\cal F}_{({\bf 10})}$ and 
$\widetilde{\cal F}_{(\bar{\bf 5})}$.
In principle, there can be two different ways to lift the $\Z_2$ 
transformation to ${\cal F}_{({\bf 10})}$, and also to 
$\widetilde{\cal F}_{(\bar{\bf 5})}$, and hence there can be 
$2 \times 2$ different ways overall. However, both the $\Z_2$
transformation on ${\cal F}_{({\bf 10})}$ and that on 
$\widetilde{\cal F}_{(\bar{\bf 5})}$ should descend from that on 
a common four-form flux $G^{(4)}$ on a Calabi--Yau 4-fold $X$. 
Thus, it is not obvious whether the $\Z_2$ transformation on the two 
matter line bundles can be chosen independently. In fact, 
we will see in this appendix that there is only one way to lift 
a $\Z_2$ transformation on $S$ and $N_{S|B_3}$ to that of matter 
line bundles ${\cal F}_{({\bf 10})}$ and 
$\widetilde{\cal F}_{(\bar{\bf 5})}$. One will also be able to see 
that the consistent lift is unique not just for the case A choice of 
the $\Z_2$ symmetry, for Example VII, or for the limit
(\ref{eq:E8-limit}) of complex structure moduli.
 
\begin{figure}[tb]
 \begin{center}
     \includegraphics[width=.35\linewidth]{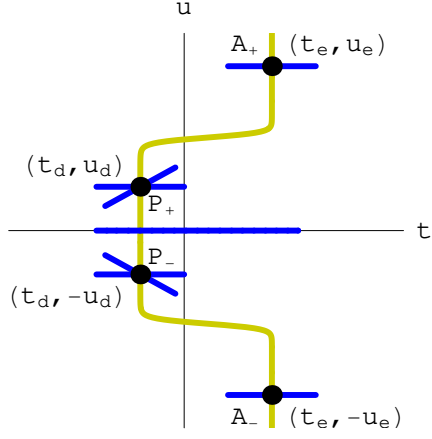} 
\caption{\label{fig:uspatch}(color online) A schematic picture 
of matter curves in the $U_{\check{S}}$ patch in Example VII with 
\kesu{a case A} \kaku{the} $\Z_2\vev{\sigma}$ symmetry \kaku{of the case A}. 
$(T/S, U/S) \equiv (t,u)$ \kesu{can be used as} 
\kaku{denotes} a set of local coordinates in the $U_{\check{S}}$ patch 
of $S = \P^2$. The $\Z_2$ symmetry generator $\sigma$ acts on the 
$U_{\check{S}}$ patch as $\sigma: (t,u) \mapsto (t,-u)$.
The yellow (light gray in black-and-white) curve is meant to be 
$\bar{c}_{({\bf 10})}$, and the blue (dark) one to be 
$\bar{c}_{(\bar{\bf 5})}$. 
A pair of $E_6$ type points $A_\pm$ 
% at \wake{$(t,u) = (t_e, \pm u_e)$} 
and a pair of $D_6$ type points $P_\pm$ 
% at \wake{$(t,u) = (t_d, \pm u_d)$} 
are indicated by blobs. 
}
 \end{center}
\end{figure}

In the Example VII, the GUT divisor is $S = \P^2$, which is 
covered by three patches $U_{\check{S}}$, $U_{\check{T}}$ and 
$U_{\check{U}}$ introduced in section \ref{ssec:Ex-7}. 
A set of local coordinates $(t,u) \equiv (T/S, U/S)$ can be used 
in the $U_{\check{S}}$ patch. A schematic picture of two different types 
of matter \kaku{curve} \kesu{curves,} $\bar{c}_{({\bf 10})}$ and 
$\bar{c}_{(\bar{\bf 5})}$\kesu{,} 
and codimension-3 singularity points of two different kinds in \kesu{the} 
$U_{\check{S}}$ \kesu{are} \kaku{is} shown in Figure~\ref{fig:uspatch}.
Generically, there are ten $E_6$-type \kaku{singularity} points 
in the $U_{\check{S}}$ patch in Example VII, 
two of which are on the $\Z_2$-fixed point locus
at $u=0$, and eight others form four pairs under the $\Z_2$
transformation in \kaku{the} case A.
There are sixteen $D_6$-type \kaku{singularity} points in Example VII in the
$U_{\check{S}}$ patch generically, forming eight pairs 
under the $\Z_2$ symmetry. Only one pair of \kaku{the
$E_6$-type}\kesu{$E_6$} points and 
one pair of \kaku{the} $D_6$ type points are shown in Figure~\ref{fig:uspatch}.
%%%%%%%%%%%%%%%%%%%%%%%%%%%%%%%%%%%%%%%%%%%%%%%%%%%%%%%%%%%%%%%%
%%%%%%%%%%%%%%%%%%%%%%%%%%%%%%%%%%%%%%%%%%%%%%%%%%%%%%%%%%%%%%%%

Let us first take a pair \kaku{$(P_+,P_-)$} of \kaku{the} $D_6$ type points 
\kesu{$P_+$ and $P_-$} in the $U_{\check{S}}$ patch\kaku{.} \kesu{;} 
\kesu{we assume} \kaku{Let us suppose} that they are located at 
$(t,u) = (t_d, u_d)$ and $(t_d, -u_d)$, respectively. 
\kesu{Physics associated
with local geometry of $X$ around these points are described by 
field theory models with an $\SO(12)$ gauge group. 
The $\Z_2\vev{\sigma}$ symmetry transformation in (\ref{eq:map-on-P2}) 
maps the two field theory models to each other.}
For simplicity, let us \kesu{assume} \kaku{suppose} that the spectral surfaces 
\kesu{in the field-theory model around} \kaku{near the singularity point} 
$P_+$ are given by 
\begin{equation}
 {\bf 10}: \quad \xi_s - (t-t_d) \simeq 0, \qquad
 \bar{\bf 5}_1: \quad \xi_s + (u-u_d) \simeq 0, \qquad 
 \bar{\bf 5}_2: \quad \xi_s + [(t-t_d) - (u-u_d)] \simeq 0
\label{eq:spec-surf-D-+}
\end{equation}
\kaku{for a field in the representation ${\bf 10}$ of $\SU(5)_{\rm GUT}$ and 
two in $\bar{\bf 5}$ of $\SU(5)_{\rm GUT}$, 
where $\xi_s \equiv \xi_{tu}$ denotes the fiber coordinate of 
the canonical bundle $K_S$ in the $U_{\check{S}}$ patch, 
when $dt \wedge du$ is taken as the frame.
}
\kesu{the ${\bf 10}$ and two $\bar{\bf 5}$ representation fields. 
Here, $\xi_s \equiv \xi_{tu}$ is the fiber coordinate of 
the canonical bundle $K_S$ in the $U_{\check{S}}$ patch, 
when $dt \wedge du$ is taken as the frame.} 
\kaku{Since the $\Z_2$ transformation maps $(t,u)$ to $(t,-u)$, 
the fiber coordinate $\xi_s$ near the point $P_+$ transforms into 
$-\xi_s$ near the point $P_-$, 
as we already discussed in section \ref{sssec:matter-parity}.}
The spectral surfaces 
\kesu{in the field theory model around} \kaku{near the point} $P_-$, 
then, should be given by 
\begin{equation}
 {\bf 10}: \quad \xi_s + (t-t_d) \simeq 0, \qquad
 \bar{\bf 5}_1: \quad \xi_s + (u+u_d) \simeq 0, \qquad 
 \bar{\bf 5}_2: \quad \xi_s - [(t-t_d) + (u+u_d)] \simeq 0\kaku{.} \kesu{;} 
\label{eq:spec-surf-D--}
\end{equation}
\kesu{this is because $\xi_s$ around $P_+$ is mapped to $-\xi_s$ 
around $P_-$, and $(t,u)$ to $(t,-u)$ under the $\Z_2$ transformation, 
as we already discussed in section \ref{sssec:matter-parity}.}

It follows from (\ref{eq:approx-Gauss-psi}, \ref{eq:approx-Gauss-chi}) 
that the hypermultiplet wavefunctions in the $\SO(12)$ field-theory 
local model around the point $P_+$ are given by 
\begin{eqnarray}
 \alpha \chi_{({\bf 10});P_+} & \simeq & \qquad G(t-t_d) \; 
     (\tilde{f}_{({\bf 10});\check{S}})_{tu}(P_+) \; (dt \wedge du), 
    \label{eq:0mode-D6-1+} \\
  i \psi_{({\bf 10});P_+} & \simeq & - d\bar{t} \; G(t-t_d) \; 
     (\tilde{f}_{({\bf 10});\check{S}})_{tu}(P_+),   \\ 
 \alpha \chi_{(\bar{\bf 5}_1);P_+} & \simeq & \qquad G(u - u_d) \; 
     (\tilde{f}_{(\bar{\bf 5}_1);\check{S}})_{tu}(P_+) \; (dt \wedge du), \\ 
  i \psi_{(\bar{\bf 5}_1);P_+} & \simeq & + d\bar{u} \; G(u-u_d) \; 
     (\tilde{f}_{(\bar{\bf 5}_1);\check{S}})_{tu}(P_+),  \\
 \alpha \chi_{(\bar{\bf 5}_2);P_+} & \simeq & \qquad \qquad \; \; 
     G\left(\frac{(t-t_d) - (u-u_d)}{2^{1/4}}\right) \; 
     (\tilde{f}_{(\bar{\bf 5}_2);\check{S}})_{tu}(P_+) \; (dt \wedge du), \\ 
  i \psi_{(\bar{\bf 5}_2);P_+} & \simeq &
      + \frac{d(\bar{t}-\bar{u})}{\sqrt{2}}
     \, G\left(\frac{(t-t_d) - (u-u_d)}{2^{1/4}}\right) \; 
     (\tilde{f}_{(\bar{\bf 5}_2);\check{S}})_{tu}(P_+), 
     \label{eq:0mode-D6-6+}
\end{eqnarray}
where $G(z) \equiv e^{- |z|^2}$.
Zero mode wavefunctions correspond to holomorphic sections 
of ${\cal F}_{({\bf 10})}$ and $\widetilde{\cal F}_{(\bar{\bf 5})}$, 
and they are expressed locally around $P_+$, by taking trivialization 
frames of ${\cal N}_{({\bf 10})}$ and $\widetilde{\cal N}_{(\bar{\bf 5})}$, 
as
\begin{equation}
 \tilde{e}_{({\bf 10})} \; (\tilde{f}_{({\bf 10}); \check{S}})_{tu} 
(dt \wedge du) \quad {\rm and} \quad 
 \tilde{e}_{(\bar{\bf 5})} \; (\tilde{f}_{(\bar{\bf 5}); \check{S}})_{tu} 
(dt \wedge du), 
\label{eq:trivialize-D6}
\end{equation} 
respectively.
% $(\tilde{f}_{(R);\check{S}})_{tu} (dt \wedge du)$ are the 
% zero mode wavefunctions expressed as the coefficient functions 
% in the trivialization patches near the point $P_+$\kaku{.} \kesu{;} 
% \wake{\kesu{since} \kaku{Since} the 
% matter line bundles ${\cal F}_{({\bf 10})}$ and 
% $\widetilde{\cal F}_{(\bar{\bf 5})}$ are \kaku{the restrictions} 
% \kesu{restriction} of ${\cal N}_{({\bf 10})} \otimes K_S$ and 
% $\widetilde{\cal N}_{(\bar{\bf 5})} \otimes K_S$, respectively, 
% to the \kaku{respective} (covering) matter curves, 
% we have maintained $dt \wedge du$, 
% which is the frame of \kaku{the} trivialization of $K_S$ in $U_{\check{S}}$.}
\kesu{The overlap integration} \kaku{Near the point $P_+$, the overlap form} 
for the down-type Yukawa couplings 
\kesu{from this field theory model around $P_+$} is \kaku{locally given by} 
\begin{eqnarray}
\Delta \lambda^{(d)}(P_+)
& \equiv &  \alpha \left(
\chi_{({\bf 10})} \psi_{(\bar{\bf 5}_1)} \psi_{(\bar{\bf 5}_2)} +
\chi_{(\bar{\bf 5}_1)} \psi_{(\bar{\bf 5}_2)} \psi_{({\bf 10})} +
\chi_{(\bar{\bf 5}_2)} \psi_{({\bf 10})} \psi_{(\bar{\bf 5}_1)} 
  \right)(P_+)  \label{eq:d-Yukawa-+0} \\
&  & \!\!\!\!\!\!\!\!\!\!\!\!\!\!\!\!\!
 = \frac{3}{\sqrt{2}} (-i dt \wedge d\bar{t})(-i du \wedge d\bar{u})
 \;  G G G  \; 
\left[(\tilde{f}_{({\bf 10});\check{S}})_{tu}
(\tilde{f}_{(\bar{\bf 5}_1);\check{S}})_{tu}
(\tilde{f}_{(\bar{\bf 5}_2);\check{S}})_{tu} \right](P_+).
\label{eq:d-Yukawa-+}
\end{eqnarray}
The \kaku{form} \kesu{integral} is localized \kesu{around} \kaku{near} $P_+$, 
because of the three Gaussian factors ``$G G G$''. 
Thus, the Yukawa coupling of \kaku{the} three zero modes 
\kesu{from the region around} \kaku{in the local neighborhood of} 
$P_+$ is determined by the last \kesu{factor, the} product of 
\kaku{the} three $\tilde{f}_{tu}$'s at $P_+$ \kaku{on the right-hand side 
of (\ref{eq:d-Yukawa-+}).}

Similarly, \kaku{the} hypermultiplet wavefunctions in the $\SO(12)$ 
field-theory local model near $P_-$ are \kaku{given by} 
\kesu{in the field theory model around $P_-$. They are }
\begin{eqnarray}
 \alpha \chi_{({\bf 10});P_-} & \simeq & \qquad G(t-t_d) \; 
     (\tilde{f}_{({\bf 10});\check{S}})_{tu}(P_-) \; (dt \wedge du),  
      \label{eq:0mode-D6-1-} \\
  i \psi_{({\bf 10});P_-} & \simeq & + d\bar{t} \; G(t-t_d) \; 
     (\tilde{f}_{({\bf 10});\check{S}})_{tu}(P_-),   \\ 
 \alpha \chi_{(\bar{\bf 5}_1);P_-} & \simeq & \qquad G(u + u_d) \; 
     (\tilde{f}_{(\bar{\bf 5}_1);\check{S}})_{tu}(P_-) \; (dt \wedge du), \\ 
  i \psi_{(\bar{\bf 5}_1);P_-} & \simeq & + d\bar{u} \; G(u + u_d) \; 
     (\tilde{f}_{(\bar{\bf 5}_1);\check{S}})_{tu}(P_-),  \\
 \alpha \chi_{(\bar{\bf 5}_2);P_-} & \simeq & \qquad \qquad \; \; 
     G\left(\frac{(t-t_d) + (u+u_d)}{2^{1/4}}\right) \; 
     (\tilde{f}_{(\bar{\bf 5}_2);\check{S}})_{tu}(P_-) \; (dt \wedge du), \\ 
  i \psi_{(\bar{\bf 5}_2);P_-} & \simeq &
      - \frac{d(\bar{t}+\bar{u})}{\sqrt{2}}
     \, G\left(\frac{(t-t_d) + (u+u_d)}{2^{1/4}}\right) \; 
     (\tilde{f}_{(\bar{\bf 5}_2);\check{S}})_{tu}(P_-).   
     \label{eq:0mode-D6-6-}
\end{eqnarray}
The overlap \kaku{form} \kesu{integral} \kesu{from the region around} 
\kaku{near} $P_-$ becomes 
\begin{eqnarray}
\Delta \lambda^{(d)}(P_-)
& \equiv  &  \alpha \left(
\chi_{({\bf 10})} \psi_{(\bar{\bf 5}_1)} \psi_{(\bar{\bf 5}_2)} +
\chi_{(\bar{\bf 5}_1)} \psi_{(\bar{\bf 5}_2)} \psi_{({\bf 10})} +
\chi_{(\bar{\bf 5}_2)} \psi_{({\bf 10})} \psi_{(\bar{\bf 5}_1)} 
  \right)(P_-)  \label{eq:d-Yukawa--0} \\
&  & \!\!\!\!\!\!\!\!\!\!\!\!\!\!\!\!\! =
 - \frac{3}{\sqrt{2}} (-i dt \wedge d\bar{t})(-i du \wedge d\bar{u})
 \;  G G G  \; 
\left[(\tilde{f}_{({\bf 10});\check{S}})_{tu}
(\tilde{f}_{(\bar{\bf 5}_1);\check{S}})_{tu}
(\tilde{f}_{(\bar{\bf 5}_2);\check{S}})_{tu} \right](P_-).
\label{eq:d-Yukawa--}
\end{eqnarray}

Suppose that the $\Z_2$ transformation is defined on the 
sections of line bundles ${\cal F}_{({\bf 10})}$ and those \kaku{of} \kesu{on} 
$\widetilde{\cal F}_{(\bar{\bf 5})}$ by 
\begin{equation}
 \sigma: (\tilde{f}_{(R);\check{S}})_{tu} \mapsto 
  (\tilde{f}^\sigma_{(R);\check{S}})_{tu} =
  - \epsilon_{(R)} \sigma^* (\tilde{f}_{(R);\check{S}})_{tu}
\label{eq:Z2-def-A}
\end{equation}
with $\epsilon_{(R)} = \pm 1$ that may be different for $R = {\bf 10}$ 
and $R = \bar{\bf 5}$, where 
$\sigma^* (\tilde{f}_{(R);\check{S}})_{tu}$ is the pull-back of 
the coefficient function $(\tilde{f}_{(R);\check{S}})_{tu}$ using 
the local trivialization frame in (\ref{eq:trivialize-D6}). Equivalently, 
for $\tilde{f} \equiv \tilde{f}_{tu} (dt \wedge du)$, 
\begin{equation}
 \sigma: \tilde{f}_{(R); \check{S}} \mapsto 
  \epsilon_{(R)} \sigma^* \tilde{f}_{(R); \check{S}}.
\label{eq:Z2-def-B}
\end{equation}
One can then see from 
(\ref{eq:0mode-D6-1+}--\ref{eq:0mode-D6-6+}) and 
(\ref{eq:0mode-D6-1-}--\ref{eq:0mode-D6-6-}) that the zero mode
wavefunctions $(\psi, \chi)$ for $\tilde{f}$ and 
$(\psi^\sigma, \chi^\sigma)$ for $\tilde{f}^\sigma$ satisfy the relation 
\begin{equation}
 (\psi^\sigma_{(R); P_\pm}, \chi^\sigma_{(R);P_\pm}) =  \epsilon_{(R)}
 \times (\sigma^* \psi_{(R); P_\pm}, \sigma^* \chi_{(R); P_\pm})
\label{eq:Z2-relation}
\end{equation}
for all the representations, $R = {\bf 10}$, $\bar{\bf 5}_1$ and 
$\bar{\bf 5}_2$. The $\Z_2$ transformation on the holomorphic 
sections of ${\cal F}_{({\bf 10})}$ on the matter curve 
$\bar{c}_{({\bf 10})}$ and those of $\widetilde{\cal F}_{(\bar{\bf 5})}$ 
on the curve $\tilde{\bar{c}}_{(\bar{\bf 5})}$ is now translated 
into a $\Z_2$ transformation between the fields of the pair of 
$\SO(12)$ field-theory local models around $P_+$ and $P_-$.
We can also\footnote{Since the field configuration 
$(\psi^\sigma, \chi^\sigma)$ is---by definition---the zero mode 
wavefunction corresponding to a holomorphic section $\tilde{f}^\sigma$, 
it is not necessary to make an extra effort to show that the right-hand 
side is a zero mode, though.} 
see in the language of the $\SO(12)$ field theories 
that zero modes are mapped to zero modes by the $\Z_2$ transformation 
(\ref{eq:Z2-relation}). This is because the 
transformation (\ref{eq:Z2-relation}) is essentially a pull-back 
times $(\pm 1)$, and because the background configuration
$\vev{\varphi}$ is invariant under the pull-back by $\sigma$. 

The $\Z_2$ symmetry transformation on $G^{(4)}$ and $X$ for matter 
parity must become a $\Z_2$ transformation between the pair of 
$\SO(12)$ field-theory local models around $P_+$ and $P_-$. 
The ${\bf adj.}$ representation of $G = \SO(12)$ has an irreducible 
decomposition (\ref{eq:decomp}) with $G' = \U(1) \times \U(1)$ and 
$G'' = \SU(5)_{\rm GUT}$, and fields $A$, $\varphi$, $\eta$, $\psi$ 
and $\chi$ in an irreducible component $(U_I, R_I)$ around $P_+$ are 
mapped to those in the same irreducible component around $P_-$ up 
to $G'$ transformation. The $\Z_2$ transformation (\ref{eq:Z2-relation})
is not just a map from a space of zero mode wavefunctions in $(U_I,
R_I)$ to itself, but it is regarded also as a map from a space of 
all kinds of field configuration in $(U_I, R_I)$ to itself. 
It is easy to see for any choice of $\epsilon_{(R)} = \pm 1$ that 
the transformation (\ref{eq:Z2-relation}) preserves the bilinear 
part of the action (\ref{eq:bosonic}, \ref{eq:fermi-bilin}) of the 
pair of $\SO(12)$ field theories, and we have already seen that the 
zero mode configurations around $P_+$ and those around $P_-$ are mapped 
to each other. However, in order for the $\Z_2$ transformation to be 
a symmetry between the pair of $\SO(12)$ field theory local models, 
all the interactions in (\ref{eq:bosonic}, \ref{eq:fermi-bilin}) 
need to be preserved under the $\Z_2$ transformation, not just the 
bilinear part of (\ref{eq:bosonic}, \ref{eq:fermi-bilin}).

The down-type Yukawa couplings (\ref{eq:d-Yukawa-+0},
\ref{eq:d-Yukawa--0}) originate from trilinear interactions of 
the $\SO(12)$ field theory on 7+1 dimensions.
The $\Z_2$ transformation for matter parity should be a symmetry 
between the pair of the $\SO(12)$ field theories, and hence 
it must be a symmetry between this interaction\footnote{The 
structure constant of $\mathfrak{so}(12)$ should be common to both 
and is omitted from (\ref{eq:d-Yukawa-+0}, \ref{eq:d-Yukawa--0}).} 
at $P_+$ and at $P_-$.
Noting that the trilinear interaction 
$\Delta \lambda^{(d)}(P_+) + \Delta \lambda^{(d)}(P_-)$ 
is transformed under (\ref{eq:Z2-relation}) to  
% Thus, the Yukawa coupling 
% $\Delta \lambda^{(d)}(P_+) + \Delta \lambda^{(d)}(P_-)$ after 
% the $\Z_2$ transformation becomes  
%
\begin{equation}
 \left[ \Delta \lambda^{(d)}(P_+) + \Delta \lambda^{(d)}(P_-) \right]^\sigma
 = \left( \epsilon_{({\bf 10})} \epsilon_{(\bar{\bf 5})}^2 \right)
 \times 
\left[\Delta \lambda^{(d)}(P_-) + \Delta \lambda^{(d)}(P_+) \right], 
\end{equation}
%
% This can also be seen directly by applying (\ref{eq:Z2-def-A})
% to (\ref{eq:d-Yukawa-+}, \ref{eq:d-Yukawa--}).
% \wake{We therefore find that $\epsilon_{({\bf 10})} = +1$ should be used 
% in (\ref{eq:Z2-def-A}, \ref{eq:Z2-def-B}) in defining the $\Z_2$ 
% transformation on the sections of line bundles ${\cal F}_{({\bf 10})}$, 
% in order for this transformation to be a symmetry of F-theory 
% compactifications. [To obtain a non-zero down-type Yukawa coupling.(?)]}
we find that we have to take $\epsilon_{({\bf 10})} = +1$ in order 
for the $\Z_2$ transformation to be a symmetry between the pair of 
$\SO(12)$ field theories at $P_+$ and $P_-$. 

Let us now take a pair \kaku{($A_+$, $A_-$)} of \kaku{the $E_6$-type} 
\kesu{$E_6$} points\kesu{, $A_+$ and $A_-$} in the 
$U_{\check{S}}$ patch\kaku{.}\kesu{;} \kesu{we assume} \kaku{We suppose} 
that they are located at $(t,u) = (t_e, u_e)$ and $(t_e, - u_e)$, respectively. 
\kesu{$E_6$ gauge theories are employed in describing physics associated 
with local geometry around these points. }
For simplicity, let us \kesu{assume} \kaku{suppose} that the spectral surfaces 
\kesu{in the $E_6$ field-theory model around} \kaku{near the point} $A_+$ 
are \kaku{locally} given by 
\begin{equation}
 {\bf 10}: \quad (\xi_s)^2 + 2(u-u_e) (\xi_s) - (t-t_e) \simeq 0, \qquad
 {\bf 5}: \quad \xi_s - 2(u-u_e) \simeq 0, 
\label{eq:spec-surf-E-+}
\end{equation}
for \kaku{the fields in the representation ${\bf 10}$ and ${\bf 5}$, 
respectively, of $SU(5)$.} 
\kesu{the ${\bf 10}$ and ${\bf 5}$ representation fields.} 
Because of the $\Z_2$ symmetry\kaku{,} \kesu{that we assume,} 
the spectral surfaces \kesu{in the field theory model around} 
\kaku{near the point} $A_-$ should be given by 
\begin{equation}
 {\bf 10}: \quad (\xi_s)^2 + 2(u+u_e) (\xi_s) - (t-t_e) \simeq 0, \qquad
 {\bf 5}:  \quad \xi_s - 2 (u+u_e) \simeq 0, 
\label{eq:spec-surf-E--}
\end{equation}

\kesu{Approximate form of} \kaku{The approximate} wavefunctions 
for the zero modes in the ${\bf 10}$ representation is 
\begin{eqnarray}
 \alpha \chi_{({\bf 10}); A_\pm} & \simeq & 
\left( \begin{array}{r}
 F_0(t) \\ 0
       \end{array}
\right) (\tilde{f}_{({\bf 10}); \check{S}})_{tu} (A_\pm) \; (dt \wedge du),
\\
 i \psi_{({\bf 10}); A_\pm} & \simeq &
\left( \begin{array}{r}
 0 \\ F_1(t)
       \end{array}\right) 
(\tilde{f}_{({\bf 10}); \check{S}})_{tu}(A_\pm) \; d \bar{t}, 
\end{eqnarray}
\kesu{for some} \kaku{with the} functions $F_0$ and $F_1$ determined 
by the discussion in the appendix \ref{sec:Hitchin}. 
The zero mode wavefunctions \kesu{for} \kaku{of} 
the fields in ${\bf 5}$ representation, on the other hand, are 
of the form 
\begin{eqnarray}
 \alpha \chi_{({\bf 5}); A_\pm} & \simeq & \quad G(u \mp u_e) \; 
  (\tilde{f}_{({\bf 5}); \check{S}})_{tu} (A_\pm) \; (dt \wedge du), \\
 i \psi_{({\bf 5}); A_\pm} & \simeq & - G(u \mp u_e) \; 
  (\tilde{f}_{({\bf 5}); \check{S}})_{tu} (A_\pm) \; d\bar{u} 
\end{eqnarray}
\kesu{in field theory local models around} \kaku{near the points} $A_\pm$, 
respectively.
Under the $\Z_2$ transformation on the sections of 
${\cal F}_{({\bf 10})}$ \kesu{we have chosen (} 
with $\epsilon_{({\bf 10})} = +1$\kesu{)}, 
\kesu{sections} \kaku{the section} $\tilde{f}_{({\bf 10})}$ 
\kesu{are} \kaku{is} mapped to 
$\tilde{f}^\sigma_{({\bf 10})}$, and \kaku{its} corresponding wavefunctions 
$(\psi^\sigma, \chi^\sigma)$ are related to the original one
by\footnote{The form of the matrix 
$\diag(+1,-1) \times \epsilon_{({\bf 10})}$ can be understood as
follows. Let us take a trivialization of ${\cal N}_{({\bf 10})}$, 
and denote the local frame as $\tilde{e}$. 
$\tilde{e} \otimes (dt \wedge du)$ then becomes a local frame 
of ${\cal N}_{({\bf 10})} \otimes K_S$, and hence of 
${\cal F}_{({\bf 10})}$. Under the $\Z_2$ transformation
(\ref{eq:Z2-def-A}), \kesu{this frame is mapped as in} 
\begin{equation}
 \sigma: \left[ \tilde{e} \otimes (dt \wedge du) \right]|_{(t,u)}
  \mapsto - \epsilon_{({\bf 10})}
  \left[ \tilde{e} \otimes (dt \wedge du) \right]|_{(t,-u)}
\label{eq:tempP}
\end{equation}
for $(t,u) \in \bar{c}_{({\bf 10})} \subset C_{({\bf 10})}$.
At the ramification loci of $C_{({\bf 10})}$, which are 
the \kaku{$E_6$-type} \kesu{$E_6$ type} points in this context, 
the rank-2 bundle $\pi_{C_{({\bf 10})}*} {\cal N}_{({\bf 10})}$ can be described 
locally by  using a frame $(\tilde{e}, \tilde{e} (\xi_s + (u \mp u_e))\kaku{)}$, 
as we already did in \kesu{section} \kaku{appendix} \ref{sec:Hitchin}. 
Since both $\xi_s$ and $u$ are multiplied by $-1$ 
under the $\Z_2$ transformation, we find 
\begin{equation}
 \sigma: \left[ \tilde{e} (\xi_s + (u - u_e)\kaku{)} \otimes (dt \wedge du) 
         \right]|_{(t_e,u_e)}
  \mapsto + \epsilon_{({\bf 10})}
  \left[ \tilde{e} (\xi_s + (u + u_e)) \otimes (dt \wedge du) 
  \right]|_{(t_e,-u_e)}.
\label{eq:tempQ}
\end{equation}
The diagonal entries of $\diag(+1, -1) \times \epsilon_{({\bf 10})}$ 
with the opposite sign come from the opposite sign 
in (\ref{eq:tempP}, \ref{eq:tempQ}).
} 
\begin{equation}
 \left( \begin{array}{c}
  \psi^\sigma_{\uparrow; ({\bf 10}); A_\pm} \\
  \psi^\sigma_{\downarrow; ({\bf 10}); A_\pm} 
	\end{array}\right) = 
 \epsilon_{({\bf 10})} \; 
 \left(\begin{array}{cc}
  1 & \\ & -1
       \end{array} \right)
 \left( \begin{array}{c}
  (\sigma^* \psi_{\uparrow; ({\bf 10})})_{A_\pm} \\
  (\sigma^* \psi_{\downarrow; ({\bf 10})})_{A_\pm} 
	\end{array}\right), 
\label{eq:10-Z2-transf-psi}
\end{equation}
and 
\begin{equation}
 \left( \begin{array}{c}
  \chi^\sigma_{\uparrow; ({\bf 10}); A_\pm} \\
  \chi^\sigma_{\downarrow; ({\bf 10}); A_\pm} 
	\end{array}\right) = 
 \epsilon_{({\bf 10})} \; 
 \left(\begin{array}{cc}
  1 & \\ & -1
       \end{array} \right)
 \left( \begin{array}{c}
  (\sigma^* \chi_{\uparrow; ({\bf 10})})_{A_\pm} \\
  (\sigma^* \chi_{\downarrow; ({\bf 10})})_{A_\pm} 
	\end{array}\right).
\label{eq:10-Z2-transf-chi}  
\end{equation}
Therefore, the $\Z_2$ transformation on the sections of 
${\cal F}_{({\bf 10})}$ is translated to a $\Z_2$ 
transformation (\ref{eq:10-Z2-transf-psi}, \ref{eq:10-Z2-transf-chi})
between the space of field configuration in the $({\bf 2}, {\bf 10})$ 
component of $G' \times G'' = \U(2) \times \SU(5)_{\rm GUT}$ in the 
$E_6$ field theory local model at $A_+$ and the one in the local model 
at $A_-$.
The zero mode wavefunctions \kesu{for} \kaku{in the representation ${\bf 5}$} 
\kesu{the ${\bf 5}$-representation fields} 
satisfy (\ref{eq:Z2-relation}) under the $\Z_2$ transformation 
on $\widetilde{\cal F}_{({\bf 5})}$, with \kesu{an} 
$\epsilon_{({\bf 5})} = \epsilon_{(\bar{\bf 5})} \kesu{= \pm 1}$ 
yet to be determined. 
The $\Z_2$ transformation on the sections of 
$\widetilde{\cal F}_{({\bf })}$ is translated to the $\Z_2$
transformation (\ref{eq:Z2-relation}) between the spaces of field
configuration in the $(\wedge^2 \bar{\bf 2}, {\bf 5})$ component 
near $A_+$ and $A_-$.

Just like we required that the $\Z_2$ transformations on the fields 
in the irreducible components define a $\Z_2$ symmetry transformation 
between a pair of $\SO(12)$ field theory local models at $P_+$ and
$P_-$, we need to require that the $\Z_2$ transformations on the 
fields in the $({\bf 2}, {\bf 10})$ and $(\wedge^2 \bar{\bf 2}, {\bf
5})$ components gives a $\Z_2$ symmetry of the pair of $E_6$ field
theory local models. The $E_6$ field theories contain a trilinear 
interaction that gives rise to the up-type Yukawa couplings. 
It is proportional to 
% \kesu{Contributions to the up-type Yukawa couplings from field-theory local
% models involve an overlap integral} 
\kaku{an overlap from} 
\begin{equation}
 \Delta \lambda^{(u)} \equiv 
  \left(\begin{array}{cc}
   \chi_{\uparrow;({\bf 10})}, & \chi_{\downarrow;({\bf 10})} \\
	\end{array} \right)
  \left(\begin{array}{cc}
   & \psi_{({\bf 5})} \\ - \psi_{({\bf 5})} & 
	\end{array}\right)
  \left( \begin{array}{r}
   \psi_{\uparrow; ({\bf 10})} \\ \psi_{\downarrow; ({\bf 10})}
	 \end{array}\right), 
\end{equation}
which contains the $G' = \U(2)$ part $\epsilon^{AB}$ of the 
structure constant of $E_6$, but the $\SU(5)_{\rm GUT}$ part 
$\epsilon^{abcde}$ is omitted. 
% \kaku{contributes to the up-type Yukawa couplings.}
% \kaku{The overlap forms $\Delta \lambda^{(u)}$ at the points $A_+$ and 
% $A_-$ are summed, and under $\Z_2$ transformation, the sum transforms as} 
% \kesu{When the zero mode wavefunctions $(\psi_{(R)},\chi_{(R)})$ 
% are replaced by their $\Z_2\vev{\sigma}$ transforms, the overlap 
% integral becomes} 
One can see from (\ref{eq:10-Z2-transf-psi}, \ref{eq:10-Z2-transf-chi},
\ref{eq:Z2-relation}) that this interaction at $A_+$ is mapped to 
the same interaction at $A_-$ by the $\Z_2$ transformation and 
vice versa, up to a multiplication constant:
\begin{equation}
 \left[ \Delta \lambda^{(u)}(A_+) + \Delta \lambda^{(u)}(A_-)
 \right]^\sigma
 = \left( - \epsilon_{({\bf 10})}^2 \epsilon_{({\bf 5})} \right) \times 
  \left[ \Delta \lambda^{(u)}(A_-) + \Delta \lambda^{(u)}(A_+)\right].
\end{equation}
We therefore conclude that 
$\epsilon_{(\bar{\bf 5})} = \epsilon_{({\bf 5})} = -1$ should be 
chosen, in order for the $\Z_2$ transformation to be a symmetry 
between the pair of $E_6$ field theory local models at $A_+$ and $A_-$.

%%%%%%%%%%%%%%%%%%%%%%%%%%%%%%%%%%%%%%%%%%%%%%%%%%%%
%

\end{document}